\documentclass[11pt,oneside]{book}

\usepackage[letterpaper,margin=1in]{geometry}
\usepackage[T1]{fontenc}
\usepackage[utf8]{inputenc}
\usepackage{lmodern}
\usepackage{microtype}
\usepackage{textcomp}
\usepackage{graphicx}
\usepackage{amsmath}
\usepackage{longtable}
\usepackage{booktabs}
\usepackage{array}
\usepackage{calc}
\usepackage{enumitem}
\usepackage{needspace}
\usepackage{fancyvrb}
\usepackage{upquote}
\usepackage[htt]{hyphenat}
\usepackage{xcolor}
\usepackage{framed}
\usepackage{xurl}
\usepackage{fancyhdr}
\usepackage[hidelinks,breaklinks=true]{hyperref}

\hypersetup{
  pdftitle={Engineering Reliable Coding Agents},
  pdfauthor={Stephanie Jarmak},
  pdfsubject={Evaluation, operation, and governance of AI coding-agent systems}
}

\newcommand{\erca}[1]{\hyperlink{erca-#1}{ERCA-#1}}

\makeatletter
\renewcommand*\l@section{\@dottedtocline{1}{1.5em}{3.3em}}
\renewcommand*{\@pnumwidth}{2.2em}
\makeatother
\setlist{
  topsep=0.62\baselineskip,
  partopsep=0.18\baselineskip,
  itemsep=0.16\baselineskip,
  parsep=0.04\baselineskip
}
\fvset{fontsize=\small}

\definecolor{shadecolor}{gray}{0.955}
\DefineVerbatimEnvironment{recordbody}{Verbatim}%
  {fontsize=\small,xleftmargin=10pt,xrightmargin=6pt}
\newenvironment{record}{\begin{shaded}\vspace{-2pt}}{\vspace{-2pt}\end{shaded}}
\newcommand{\currentparttitle}{}
\providecommand{\partmark}[1]{}
\renewcommand{\partmark}[1]{\gdef\currentparttitle{#1}}
\renewcommand{\chaptermark}[1]{\markboth{Chapter \thechapter}{}}
\fancypagestyle{plain}{\fancyhf{}\fancyfoot[C]{\thepage}}

\providecommand{\tightlist}{%
  \setlength{\itemsep}{0.16\baselineskip}\setlength{\parskip}{0.04\baselineskip}}
\newcommand{\pandocbounded}[1]{%
  \begingroup
  \setbox0=\hbox{#1}%
  \ifdim\wd0>\linewidth
    \resizebox{\linewidth}{!}{\box0}%
  \else
    \box0
  \fi
  \endgroup}

\begin{document}
\frontmatter

\begin{titlepage}
  \centering
  \vspace*{0.16\textheight}
  {\Huge\bfseries Engineering Reliable Coding Agents\par}
  \vspace{1.2em}
  {\Large Evaluating and Operating the System Around the Model\par}
  \vspace{2.5em}
  {\large Stephanie Jarmak\par}
  \vfill
  {\large Version 1.0.0 --- August 2026\par}
\end{titlepage}

\chapter*{Abstract}
\addcontentsline{toc}{chapter}{Abstract}
AI coding agents are commonly evaluated as models but deployed as systems. Their reliability depends not only on model capability, but on the harness, execution state, retrieval, memory and state management, permissions, review interfaces, and resource allocation around the model. This technical review and engineering monograph examines those system boundaries and develops a practical framework for evaluating and operating coding agents reliably.

The study synthesizes 164 scholarly works, 100 practitioner records, 29 benchmark records, and 17 author-system case records through a structured multivocal review, targeted update audits, software-engineering coverage analysis, and a distributed-systems evidence synthesis. Across this evidence, a consistent pattern emerges: many apparent model failures originate elsewhere in the system, while improvements measured at one layer often fail to propagate to end-to-end task outcomes. Evaluation and operation are therefore treated as a dependency chain in which weaknesses in task construction, execution environments, retrieval, state management, verification, or observability can invalidate conclusions made downstream.

The monograph contributes a versioned catalog of 206 reliability records: 193 gated practices, including 56 developed in depth, plus 13 research leads; an evidence ledger linking claims to their support; a framework for reasoning about dependency and repair asymmetry across the agent lifecycle; empirical measurements and failure cases from operated agent systems; runnable evaluation and reliability protocols; and five reusable agent skills with evidence maps. Together, these provide a system-level methodology for distinguishing model capability from infrastructure effects, designing evaluations that support defensible conclusions, and building agent systems that can recover safely when components fail.

The review is structured rather than exhaustive, evidence strength varies by topic, and empirical results remain dependent on workload and system configuration. The methods section records which search lanes this edition executed, which remain unexecuted, and the limits those choices place on its evidence-grading claims.

\tableofcontents
\listoffigures
\listoftables
\chapter*{Introduction}
\markboth{Introduction}{}
\addcontentsline{toc}{chapter}{Introduction}
\hypertarget{problem-and-scope}{%
\section{Problem and scope}\label{problem-and-scope}}

Consider the following scenario. An agent has finished making a change to the codebase; the tests pass, and a reviewer is examining a compact diff. The run appears successful, yet the available record may not establish whether another run would produce the same result, whether the tests exercised the relevant behavior, or whether the reviewer saw the decisions that carried the most risk.

The visible output of the agent's trajectory is code. The uncertainty around its quality resides in the system that produced, evaluated, and approved it. A coding agent is one component in that system, and evaluation determines what counts as success, governance constrains access, context management controls the information available during a run, review defines the quality gate, and scheduling allocates compute, money, time, and human attention.

These functions interact, e.g. a higher score can result from an easier test rather than a better system, or a reviewer could appear ineffective because the interface concealed the evidence needed for judgment. Instrumentation can record component failures while missing failures at component boundaries and a recovery procedure can pass while depending on credentials whose compromise would also destroy the recovery path.

This technical review and engineering monograph examines the evaluation, operation, and governance of AI coding-agent systems. It focuses on mechanisms that remain relevant as models and products change: measurement design, execution-based grading, containment, durable state, recovery, repository retrieval, context limits, human oversight, topology, and resource allocation. It does not compare current models or teach prompt and tool-schema design.

The operating conditions assumed throughout are those of a large organizational codebase. Work spans many repositories with cross-repository dependencies, several languages with separate toolchains, ownership and access boundaries that no single identity crosses, and build and test paths too slow or too partial to run in full on every change. Review capacity was scarce before agents began producing candidate changes. At that scale neither the agent nor the reviewer inspects the whole system, so the evidence a reviewer needs is absent unless the surrounding system records it. Several results reported here were measured on repositories of millions to tens of millions of lines. Where a practice depends on that scale the chapter says so, and transfer downward to one small repository, or upward from a public single-repository benchmark, is an assumption to test rather than a default.

The intended reader is a senior engineer, evaluation lead, or technical owner building a coding-agent evaluation or operations program. The statistical methods are introduced where they affect an engineering decision, but the monograph assumes comfort with experiment design, production controls, and technical review. Its practices are intended to serve as bounded engineering claims rather than universal rules. Their applicability depends on workload, permissions, failure costs, deployment conditions, and available review capacity.

\hypertarget{the-reliability-dependency-chain}{%
\section{The reliability dependency chain}\label{the-reliability-dependency-chain}}

The organizing argument is a dependency chain. Measurement determines whether a difference is credible. Grading converts observations into acceptance decisions. Containment and recovery determine whether the resulting execution record survives failure without extending authority. Retrieval and context determine which evidence reaches the agent. Review and accountability determine who can challenge the result and who controls the consequential transition. Allocation and cost determine which system receives future work.

Each layer determines what the next may trust. A weak measurement can become a confident grade; a weak grade can admit unsafe work; an incomplete recovery record can enter retrieval as if it were complete; missing context can make review appear ineffective; and an invalid review signal can steer more work toward the wrong configuration. Downstream confidence cannot repair evidence lost upstream.

This creates a repair asymmetry. Later machinery is often easier to add than the earlier instrument is to repair, yet it is evaluated through that earlier instrument. More samples cannot repair a task distribution that excludes production work. More judges cannot repair a rubric experts apply inconsistently. More agents cannot repair a retrieval boundary that treats an empty result as authoritative. The dependency chain is therefore a sequence of evidence obligations, not a list of subsystems.

The chain carries a system-level claim. Coding agents are evaluated as models and deployed as production systems. Once a run can outlive the worker that started it, wait on another service, compete with another run, mutate shared code, or publish a durable external effect, reliability stops being a property of the model and becomes a property of the system that preserves intent, authority, state, evidence, ordering, and recovery across components that fail independently. The framing is not new. Osterweil (\href{https://dl.acm.org/doi/10.5555/41765.41766}{1987}) argued that software processes are themselves software, and Choi and Scacchi (\href{https://www.ics.uci.edu/~wscacchi/Software-Process/Readings/DistSysFactory.pdf}{1991}) built a software factory as distributed infrastructure. Autonomous workers change the failure model, not the problem: today's workers are nondeterministic, edit persistent code, call external services, run concurrently, and can claim completion incorrectly. Chapter 7 develops this factory model and the contracts the operating chapters enforce.

To ensure reliability for any given guarantee, we must be able to answer which promise the surrounding system makes, which component owns that promise, what state survives failure, and what experiment would falsify the guarantee. The chapters are organized to make each of those questions answerable in turn.

\begin{figure}[htbp]
\centering
\includegraphics[width=1\textwidth,height=\textheight,keepaspectratio]{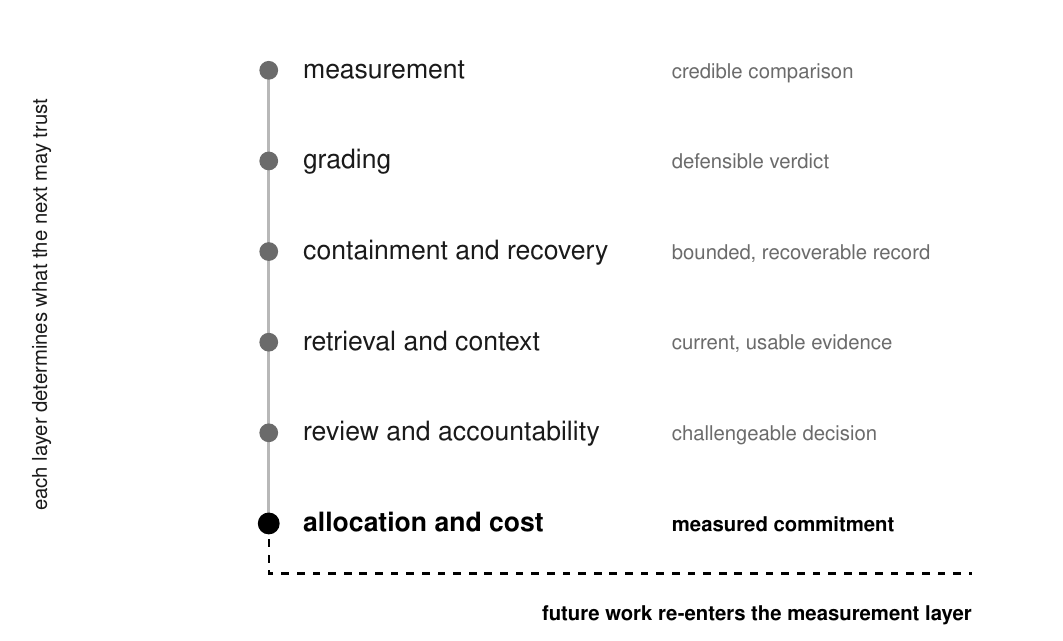}
\caption{Argument map for the reliability dependency chain: measurement, grading, containment and recovery, retrieval and context, review and accountability, and allocation and cost. Each layer supplies the evidence boundary on which the next relies.}
\end{figure}

The six parts that follow are an account of how an apparently local defect can propagate into later operational decisions while retaining the appearance of a clean score, verdict, or artifact.

\hypertarget{method-scope-and-evidence-classification}{%
\section{Method, scope, and evidence classification}\label{method-scope-and-evidence-classification}}

This review treats source collection, evidence grading, practice derivation, and the treatment of author-system cases as separate methodological decisions. The following subsections describe each decision and the limits it places on the resulting claims.

\hypertarget{search-and-source-assembly}{%
\subsection{Search and source assembly}\label{search-and-source-assembly}}

The review was consolidated on July 26, 2026, then subjected to a bounded update audit and a software-engineering venue coverage probe through August 6, 2026. This edition's source collection contains 164 scholarly works, 100 practitioner records, 29 benchmark records, and 17 author-system case records. The original 118 scholarly works were organized into seven topic-specific threads covering benchmark validity, failure taxonomy, evaluation statistics, oversight and accountability, context and retrieval, durable execution, and scheduling with repository-scale scoping. Eleven works were admitted during the update audit and nine during the coverage probe. The distributed-systems synthesis added with this edition admitted a further 21 scholarly works, drawn from the distributed-systems, cluster-scheduling, and build-systems literatures together with three agent-specific preprints, and 7 practitioner records; these entered through targeted citation-driven selection rather than the thread protocol, and the evidence ledger records each one's claim scope. Two further records, one scholarly and one practitioner, were admitted on the same targeted basis to support the repository-scale framing stated above, and are recorded in the screening decisions with their placements. One practitioner record published after the cutoff, an operating team's account of a production issue-and-pull-request factory, was admitted on the same basis as a third corroborating case for the system model in Chapter 7; the ledger records its post-cutoff admission and its self-reported scope. Four scholarly systems papers on inference serving were admitted by the same targeted selection, to support Chapter 19's treatment of the model endpoint as a constrained shared resource; none was measured on an agent workload, and the ledger records that limit on each.

Scholarly retrieval used \href{https://scixplorer.org/scixabout/}{SciX}, the NASA-supported literature discovery service operated by the Smithsonian Astrophysical Observatory, and a local retrieval layer referred to here as \textbf{SciX Agent}. The official \href{https://scixplorer.org/scixhelp/api-scix/}{SciX API} supplied bibliographic identities and metadata. At consolidation, SciX Agent searched a 32.4-million-record SciX and arXiv corpus with 299.3 million citation links and full text for 14.9 million records. It combined INDUS dense retrieval with BM25 lexical retrieval through reciprocal-rank fusion. These systems determined which records were retrieved and read first; they did not determine evidence grades.

Queries were scoped by topic, subject class, and year where appropriate. Each thread combined seminal work with recent agent-era research. Candidate records were verified by identity, and full text was read when available and when the claim required more than the abstract. A citation audit added the seventh thread after finding scheduling and repository-scoping sources that had entered the draft through adjacent operations-research material.

Practitioner retrieval used the \textbf{\href{https://www.sjarmak.ai/projects/code-intelligence-digest}{Code Intelligence Digest}}, an author-operated corpus that ingests research feeds, engineering publications, newsletters, podcasts, community discussions, and product or operations accounts. At the July 26 cutoff, its local snapshot contained 162,350 normalized records from 149 source labels, including 43,953 records with retained full text. Keyword and semantic retrieval were used within relevant practitioner categories. Research records found through the Digest were moved to the scholarly lane and deduplicated there. Repeated practitioner accounts of one incident shared an independence key and could not be counted as independent corroboration merely because several pages repeated the event.

SciX, SciX Agent, and Code Intelligence Digest were retrieval and ranking instruments; they did not assign evidence grades. The author made the final inclusion, evidence-group, and practice-admission decisions. The companion binds this division of labor to concrete records: \href{https://github.com/sjarmak/engineering-reliable-coding-agents/blob/main/companion/methodology/assembly-and-adjudication.md}{\texttt{assembly-and-adjudication.md}} records the decision sequence; \href{https://github.com/sjarmak/engineering-reliable-coding-agents/blob/main/companion/methodology/thread-protocols.md}{\texttt{thread-protocols.md}} and \href{https://github.com/sjarmak/engineering-reliable-coding-agents/blob/main/companion/methodology/thread-source-index.csv}{\texttt{thread-source-index.csv}} record the reconstructed search boundaries and retained source identities. These artifacts disclose the process without upgrading a retrieval system into an adjudicator.

This practitioner lane makes the review \textbf{multivocal} in the software-engineering sense described by Garousi, Felderer, and Mäntylä (\href{https://doi.org/10.1016/j.infsof.2018.09.006}{2019}): it combines scholarly and grey literature because operational mechanisms and incidents are often documented outside venues. The lane also follows the more restrictive point made by Kitchenham and colleagues (\href{https://doi.org/10.1109/TSE.2022.3165938}{2022}): a mutable social-media post is not treated as a primary study merely because it is informative. Practitioner records remain corroborating cases unless they report a sufficiently specific measurement, and mutable cited pages are archived. The independence key operationalizes source independence by grouping reports that repeat one originating incident or claim.

The benchmark collection was assembled separately from benchmark documentation and publications, merged from two inventories, deduplicated by identity, and validated against a JSON Schema. Three repeated benchmark records were removed during that merge.

The initial scholarly search did not establish adequate coverage of the core software-engineering venue literature. A subsequent OpenAlex metadata probe searched eight topic formulations across ICSE, FSE, ASE, ISSTA, \emph{Empirical Software Engineering}, \emph{IEEE Transactions on Software Engineering}, \emph{ACM Transactions on Software Engineering and Methodology}, and \emph{Computer Supported Cooperative Work}. It surfaced 148 unique candidates. Nine methodologically material works were admitted after title-and-abstract screening, including the SEGRESS reporting guideline (\href{https://doi.org/10.1109/TSE.2022.3174092}{Kitchenham et al.~2022}), the ABC framework for software-engineering research (\href{https://doi.org/10.1145/3241743}{Stol and Fitzgerald 2018}), and work on construct validity in software engineering (\href{https://doi.org/10.1109/TSE.2022.3176725}{Sjøberg and Bergersen 2023}).

That probe diagnosed coverage; it did not establish publisher- and index-native coverage. In a deterministic sample of 40 candidate DOIs, SciX contained eight exact DOI matches. All eight were TSE records. A follow-up known-set check found exact SciX DOI matches for all 26 TSE candidates surfaced by the probe, while the sample's 32 records from other venue families remained absent. These comparisons show record-level gaps and a venue-specific difference in the known candidate set; they are too topical to estimate corpus recall, topic-search recall within TSE, or equivalence to a publisher index.

ACM Digital Library returned HTTP 403 to the automated client, and its end-user policy excludes automated agents, so the prepared ACM supplement requires manual execution through an authorized interface. The IEEE Xplore credential returned a provider-inactive response. The Scopus lane lacks an API key and any institutional token required by the host's entitlement. This edition therefore reports all three searches as not performed. Their absence is a disclosed source limitation, not an inference of zero relevant results and not a claim that the OpenAlex, DBLP, or web-surrogate evidence is provider-equivalent. The prepared ACM Digital Library, IEEE Xplore, and Scopus plans are preserved in the companion so a later edition can execute and adjudicate them. The companion's \href{https://github.com/sjarmak/engineering-reliable-coding-agents/blob/main/companion/methodology/software-engineering-coverage/protocol-and-status.json}{\texttt{protocol-and-status.json}} records provider and probe results; \href{https://github.com/sjarmak/engineering-reliable-coding-agents/blob/main/companion/methodology/software-engineering-coverage/publisher-coverage-status.json}{\texttt{publisher-coverage-status.json}} records the release state without counting plans as execution.

While Scopus remained unavailable, a credential-free DBLP title census applied the same eight topic boundaries to eight named venue streams for 2018--2026. Its 64 cells returned 55 unique publications; 50 were absent from both the resolved manuscript-reference set and the earlier OpenAlex probe. This result identified title-matched conference candidates that the probe did not return and created a concrete adjudication queue. It does not close the source gap: DBLP is title-only, is not publisher-native, and is not equivalent to Scopus title--abstract--keyword retrieval. The exact query, returned SPARQL bindings, upstream comparison hashes, zero cells, and fallback decision are preserved in the companion.

The corresponding \href{https://github.com/sjarmak/engineering-reliable-coding-agents/blob/main/companion/methodology/software-engineering-coverage/dblp-author-adjudication-2026-08.csv}{\texttt{dblp-author-adjudication-2026-08.csv}} separates model-assisted candidate claims from final dispositions. All 34 retained candidates were discovered after the August 6 cutoff and deferred to the next-edition queue because none identified a factual correction. This application of the published cutoff keeps late additive or qualifying evidence visible without silently expanding this edition's frozen corpus or representing a recommendation as an admission.

\begin{figure}[htbp]
\centering
\includegraphics[width=1\textwidth,height=\textheight,keepaspectratio]{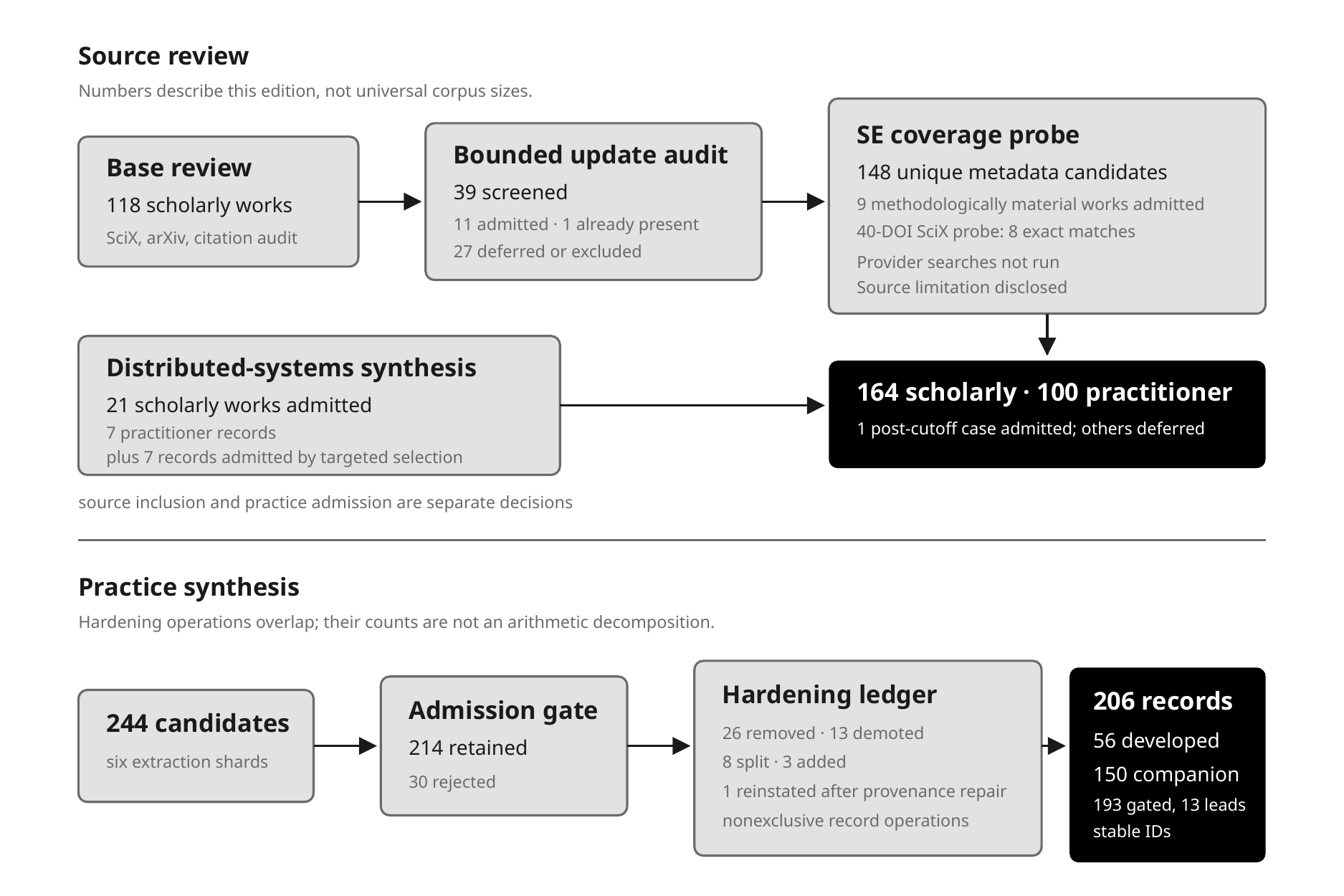}
\caption{Flow of source review and practice synthesis. The source lane distinguishes the base review, bounded update audit, SE coverage probe, and the publisher- and index-native searches this edition did not perform. The practice lane reports the admission gate and overlapping hardening operations without treating their counts as an arithmetic decomposition.}
\end{figure}

The search remains structured rather than exhaustive. It does not cover every scholarly index, venue, private operational record, or adjacent model-comparison literature. The SEGRESS reporting items provide the vocabulary used below: source admission is governed by inclusion and exclusion criteria; evidence-group assignment is a quality assessment of a scoped claim; and practice construction is data extraction and synthesis. These names do not change the underlying decisions, but they make the review easier to compare with software-engineering secondary studies.

\hypertarget{curation-and-assembly-workflow}{%
\subsection{Curation and assembly workflow}\label{curation-and-assembly-workflow}}

The assembly process followed a fixed sequence: define a question for each thread; retrieve candidates; resolve record identity; screen for an in-scope claim; extract the bounded claim and its conditions; assign an evidence group; challenge the assignment; derive candidate practices; select practices for chapter treatment; and audit the final citations. Retrieval rank determined screening order only. A highly ranked record received no evidentiary preference after screening.

Automated systems assisted with retrieval, normalization, duplicate detection, bounded claim extraction, metadata checks, and challenge passes. The cited source remained authoritative. The author made the final decisions about inclusion, evidence grouping, practice admission, chapter placement, and prose. When a challenge pass exposed ambiguity, the lower evidence group was used unless the narrower strong claim could be stated directly.

\begin{longtable}[]{@{}
  >{\raggedright\arraybackslash}p{(\columnwidth - 4\tabcolsep) * \real{0.3333}}
  >{\raggedright\arraybackslash}p{(\columnwidth - 4\tabcolsep) * \real{0.3333}}
  >{\raggedright\arraybackslash}p{(\columnwidth - 4\tabcolsep) * \real{0.3333}}@{}}
\caption{Automation and human judgment in the review workflow. \emph{Automated} means the operation ran without item-level prompting; \emph{assisted} means a system proposed or flagged material for a human decision; \emph{human} means the substantive choice was made without an automated verdict.}\tabularnewline
\toprule\noalign{}
\begin{minipage}[b]{\linewidth}\raggedright
Review step
\end{minipage} & \begin{minipage}[b]{\linewidth}\raggedright
Mode
\end{minipage} & \begin{minipage}[b]{\linewidth}\raggedright
Human decision retained
\end{minipage} \\
\midrule\noalign{}
\endfirsthead
\toprule\noalign{}
\begin{minipage}[b]{\linewidth}\raggedright
Review step
\end{minipage} & \begin{minipage}[b]{\linewidth}\raggedright
Mode
\end{minipage} & \begin{minipage}[b]{\linewidth}\raggedright
Human decision retained
\end{minipage} \\
\midrule\noalign{}
\endhead
\bottomrule\noalign{}
\endlastfoot
Candidate retrieval and ranking & Automated and assisted & I defined each thread question and search boundary, then read the admitted sources. \\
Identity resolution, normalization, and duplicate checks & Automated checks with human resolution & I resolved ambiguous identities and practitioner independence keys. \\
Bounded-claim extraction & Assisted & I checked the source, revised the extracted claim, and accepted or rejected it. \\
Initial evidence-group proposal & Assisted & I assigned the final label to the scoped claim. \\
Challenge pass & Assisted & Automated passes flagged composite claims, contrary findings, duplicate support, and label inconsistencies; I reread the source and adjudicated each change. \\
Practice admission, chapter selection, and prose & Human & I made the selection and writing decisions. \\
Schema, checksum, identifier, and cross-reference gates & Automated verification & A failure blocked the release until the underlying record was corrected. \\
\end{longtable}

The challenge pass was an error-finding aid, not an independent grader. In particular, automated assistance in that pass searched for composite claims, inconsistent evidence groups, contrary findings, duplicate support, and broken identifiers. It did not accept a practice or promote an evidence group.

The update audit screened 38 distinct scholarly records surfaced in Code Intelligence Digest editions published from July 27 through August 5, plus one paper found in a targeted August 6 check. Eleven new works were admitted, one record was already present, and 27 were deferred or excluded. Admission required a material addition to a claim already in scope; novelty or recency alone was insufficient. Material published after August 6 enters the update queue for a later edition unless it corrects a factual error in this one.

Working thread syntheses and source receipts were retained, but the original interactive searches did not preserve every machine-issued query in a publication-ready log. The companion therefore distinguishes retained records from reconstructions. It publishes the source snapshots, sanitized thread protocols, the retained update-search record, and record-level update decisions; it does not present reconstructed query text as an exact historical log. This permits protocol-level review without implying that every interactive search can be replayed byte for byte.

The bounded update trail is inspectable at three levels: \href{https://github.com/sjarmak/engineering-reliable-coding-agents/blob/main/companion/methodology/search-log.csv}{\texttt{search-log.csv}} records the search windows and source lanes, \href{https://github.com/sjarmak/engineering-reliable-coding-agents/blob/main/companion/methodology/screening-decisions.csv}{\texttt{screening-decisions.csv}} records item-level dispositions, and \href{https://github.com/sjarmak/engineering-reliable-coding-agents/blob/main/companion/methodology/source-snapshot.json}{\texttt{source-snapshot.json}} records the frozen corpus counts and cutoff. Together they preserve the reported arithmetic without claiming a byte-for-byte replay of the earlier interactive retrieval sessions.

\hypertarget{screening-and-evidence-grading}{%
\subsection{Screening and evidence grading}\label{screening-and-evidence-grading}}

A source entered the working corpus when it contributed at least one measured result, reproducible mechanism, operational incident, benchmark property, or concrete practice relevant to coding-agent reliability. Screening removed records with unresolved identity, no recoverable claim, no relation to a decision in scope, or complete redundancy with a better-supported record. Sources could remain in the review while a proposed practice was rejected; source inclusion and practice admission were separate decisions.

\Needspace{5\baselineskip}

Evidence is reported in four reader-facing groups:

\begin{itemize}
\tightlist
\item
  \textbf{Strong evidence} is an on-claim controlled comparison, validated benchmark result, or comparably specific measurement within stated conditions.
\item
  \textbf{Directional evidence} supports a mechanism, threat model, comparison design, or direction of effect without establishing the complete recommendation, its magnitude, or broad transfer.
\item
  \textbf{Corroborating evidence} consists of case reports, practitioner accounts, or convergent observations that establish plausibility without estimating prevalence.
\item
  \textbf{Null or conflicting evidence} records a result that did not support the expected effect or that materially limits another claim.
\end{itemize}

These labels attach to evidence items and scoped claims, not to publication venues or whole chapters. A composite recommendation does not become strong because several directional sources converge. When no individual controlled study supports the complete recommendation, the text either narrows the claim to the measured component, classifies the transfer as directional, or presents a protocol to test locally. A strong item may therefore support one step of a developed practice while leaving the generalized prescription directional.

The catalog was graded during assembly and then challenged through independent verification passes. A targeted audit examined ten practices whose sole supporting synthesis had been graded strong, together with two restored items. Six grades were reduced because the source demonstrated a hazard, substrate, or adjacent result rather than the stated remedy; six were retained because the controlled comparison matched the claim. Identifier checks, duplicate-identifier gates, thin-evidence rulings, practitioner-independence checks, and contrary evidence were preserved in the audit record. Ambiguous cases defaulted to the lower grade.

The final adjudication was performed by the author. The challenge passes reduced correlated review error, but they do not constitute blinded independent grading by several human reviewers. That creates an asymmetry with Chapter 5, which asks operators to calibrate graders against independent labels before relying on them.

The release artifact therefore includes a deterministic random sample of 20 practices, the associated evidence items with the author's labels hidden, a reviewer response template, and a script that can report pairwise Cohen's kappa, Fleiss's kappa when three readers participate, observed agreement, and disagreement patterns. This edition has not commissioned external graders, reports no inter-rater agreement value, and does not claim independent calibration or reproducibility of its evidence-group assignments. The blinded packet is released so external readers can run that pass; a later edition can report the resulting agreement. The author label will be compared only after their pass and will not be treated as ground truth; agreement will measure reproducibility of the classification instrument rather than correctness of every grade.

The reader-facing instrument is preserved as \href{https://github.com/sjarmak/engineering-reliable-coding-agents/blob/main/companion/methodology/external-grading/review-form.html}{\texttt{review-form.html}}; the structural validator and agreement analysis are implemented in \href{https://github.com/sjarmak/engineering-reliable-coding-agents/blob/main/companion/methodology/external-grading/analyze-grades.mjs}{\texttt{analyze-grades.mjs}}. The current status is recorded in \href{https://github.com/sjarmak/engineering-reliable-coding-agents/blob/main/companion/methodology/external-grading/status.json}{\texttt{status.json}}. No completed external response, calibration report, or agreement result is part of this edition.

\hypertarget{practice-derivation-and-chapter-selection}{%
\subsection{Practice derivation and chapter selection}\label{practice-derivation-and-chapter-selection}}

Candidate practices were derived through bounded-claim extraction and synthesis, then passed through a separate admission gate. A record qualified at the final gate through at least one scholarly item, a non-author synthesis with a resolvable scholarly identity, or at least two practitioner items with distinct independence keys. Hardening separated bundled claims, removed redundant or self-defeating records, preserved contrary findings, and repaired provenance. The resulting catalog contains 206 edition records, each with a stable identifier of the form \texttt{ERCA-NNN}, after the initials of this book's title: 193 that passed this admission gate, and 13 catalog-level thin-support records (\erca{193} through \erca{205}) added with the distributed-systems synthesis, which are marked as leads rather than gated practices. That total reflects the chosen claim granularity and editorial boundaries; it is not an estimate of how many reliability practices exist. The companion preserves the full record arithmetic and identifies which hardening operations overlap.

Three selection passes ranked the catalog by different criteria: teachability through a bounded case, consequence for an engineering decision, and coverage of the fourteen mechanism clusters. Practices selected by at least two passes formed the base of the developed set. Individual adjudication then repaired thin mechanism coverage and one provenance defect. The resulting 56 practices receive full treatment in the 19 chapters; the remaining 150 appear in the companion catalog. The chapter crosswalk maps the 193 gated records to the chapter whose mechanism each extends; the 13 catalog-level leads are indexed in the catalog without a chapter assignment.

The consequence ranking also supplies the operational-urgency calculation used later in the monograph. That pass ranked 52 practices, a subset of the catalog and not the same set as the 56 developed in the chapters. Among those 52, the Spearman correlation between urgency rank and a binary indicator for whether the practice carried at least one strong evidence item was -0.004. The phrase \emph{nearly uncorrelated} refers to this calculation, not to all 206 catalog entries or to a latent universal measure of importance.

The resulting chapter set is an authorial engineering judgment, not an evidence-derived consensus. The near-zero correlation makes that distinction visible. Some practices are included because a controlled comparison measured their effect; others are engineering controls justified by a structural failure mechanism, an observable check, and an asymmetric cost of waiting for trial evidence. Separating a recovery identity from a production identity, for example, can be tested directly against the authority boundary even when no study estimates how often shared credentials cause loss. Imperative section titles name the control or observation to implement; they do not imply a universal effect size or settled prevalence estimate. Where the argument is mechanistic rather than experimental, the chapter supplies a local test and avoids a numerical target.

\hypertarget{author-system-cases-and-limitations}{%
\subsection{Author-system cases and limitations}\label{author-system-cases-and-limitations}}

Cases from systems operated by the author expose mechanisms, original measurements, and reproducible failure cases. They are always treated as illustrations or local measurements. They are not counted as independent external evidence and do not by themselves support a general recommendation.

A \textbf{local artifact} is a record from a system operated by the author and used to expose a mechanism or local measurement. It is not independent external evidence. When its source was uncommitted at the cited revision or its figures were read from source rather than independently remeasured, the chapter states that provenance condition at first use instead of assigning the artifact a second name.

Evidence remains uneven across topics. Several operational questions have only case-level support, recent capability measurements can age quickly, and practitioner reports are vulnerable to selection, survivorship, and reporting bias. The review excludes model-comparison and prompt-engineering literatures except where they bear directly on system reliability. Transfer is especially substantial in Part VI, where observatory scheduling, compute-cluster scheduling, and adjacent multi-agent studies motivate testable designs for coding-agent fleets. That part should be read partly as a research agenda, not as a body of settled deployment guidance.

This section establishes the standard evidence legend for the whole monograph. Later chapters repeat a limitation only when it changes how a particular result may be used.

\hypertarget{contributions}{%
\section{Contributions}\label{contributions}}

\Needspace{5\baselineskip}

This work makes six contributions:

\begin{enumerate}
\def\labelenumi{\arabic{enumi}.}
\tightlist
\item
  a multivocal evidence audit and machine-readable ledger that distinguish direct support, directional findings, corroborating cases, and null or conflicting results;
\item
  a versioned catalog of 206 reliability records: 193 gated practices, including 56 developed in depth, plus 13 research leads, with stable identifiers that connect the manuscript, companion, and implementation artifacts;
\item
  a system-level reliability model, stated as a dependency chain with its repair asymmetry and as an explicit set of factory contracts, that connects evaluation, containment, durable execution, repository state, verification, human control, and fleet allocation through the ownership, identity, persistence, ordering, authority, and observation boundaries a coding-agent system must preserve;
\item
  original measurements and failure cases from author-operated systems, explicitly separated from external evidence;
\item
  runnable protocols for local evaluation, capability-boundary testing, recovery testing, trace analysis, and release decisions; and
\item
  five reusable agent skills with practice-level evidence maps, packaged in the project repository as implementation artifacts rather than additional evidence.
\end{enumerate}

The chapters emphasize conditions, measurements, and failure boundaries because an outcome alone rarely identifies why a system succeeded or failed. A useful account traces ownership, permissions, persistence, ordering, and observation while distinguishing correctness from reliability, performance, cost, safety, and usability.

\hypertarget{a-minimum-pass-through-the-dependency-chain}{%
\section{A minimum pass through the dependency chain}\label{a-minimum-pass-through-the-dependency-chain}}

\Needspace{5\baselineskip}

For an existing system, one compact pass produces the minimum record on which later decisions can build:

\begin{enumerate}
\def\labelenumi{\arabic{enumi}.}
\tightlist
\item
  Reopen one decision based on an aggregate score. Run the cheapest credible baseline and the candidate on identical task versions, initially three times per item, and preserve per-item outcomes.
\item
  Record success, reliability, cost, latency, model, harness, prompt, permissions, and pricing snapshot separately.
\item
  Exercise one permitted and one prohibited action with the ordinary identity, including the boundary between primary and recovery resources.
\item
  Verify one recent completion claim from repository or system state and rerun the executable check that makes it true.
\item
  Read twenty failed or unverifiable runs, label the first upstream failure where the trace permits it, and repair the first ordinary causal question the schema cannot answer.
\item
  Before the next promotion run, record the success floor, cost ceiling, task and baseline versions, mechanism condition, and fault-containment guard.
\end{enumerate}

The pass leaves six challengeable artifacts: a paired distribution, a cost-quality record, an observed authority boundary, an independently verified state transition, a seed failure corpus, and a decision rule fixed before the result was known. It is an entry point, not a reliability certificate. The repository artifact \href{https://github.com/sjarmak/engineering-reliable-coding-agents/blob/main/protocols/minimum-reliability-pass.md}{\texttt{protocols/minimum-reliability-pass.md}} supplies the runnable checklist and retained-artifact layout; the chapters develop each step.

\hypertarget{how-the-monograph-is-organized}{%
\section{How the monograph is organized}\label{how-the-monograph-is-organized}}

The monograph has six parts and twenty chapters. Chapters 1 through 19 develop the methods and practices; Chapter 20 closes by tracing the evidence chain that connects them. Measurement comes first because every later practice is adopted or rejected through a measured comparison. Nineteen developed practices in Parts II through VI also require a method introduced in Part I. Interleaving those methods with their uses would scatter each method across four or five chapters.

This order places three chapters of experiment design before the agent-specific operating chapters. Later recommendations depend on those definitions and comparison methods.

Part I establishes how to effectively compare systems when runs vary and scores can mislead. Part II turns those measurements into grading and release decisions. Part III opens with Chapter 7, which states the software-factory model and the contracts (I1 through I11) the later chapters reference, then addresses containment, persistent state, recovery, and failure analysis. Part IV examines how repository information enters, survives, and leaves an agent run. Part V treats human review as an engineered control with interfaces, escalation rules, and accountable ownership. Part VI is a research agenda for allocating work across agents and models under cost and capacity constraints. Its questions transfer methods from adjacent scheduling and multi-agent literatures rather than presenting settled coding-agent effects.

Parts III, IV, and VI are not independent collections of practices; they examine one system at three boundaries. Part III establishes how work, authority, effects, and recovery survive the failure of the process that was executing them. Part IV establishes how evidence and repository-derived state remain tied to the code state they describe. Part VI asks how the same work and ownership records behave under concurrency and finite capacity. Part II supplies the verification machinery those parts depend on, and Part V supplies the human authority that gates their consequential transitions. Chapter 7 states the model and contracts that connect them.

\begin{longtable}[]{@{}
  >{\raggedright\arraybackslash}p{(\columnwidth - 4\tabcolsep) * \real{0.3000}}
  >{\raggedleft\arraybackslash}p{(\columnwidth - 4\tabcolsep) * \real{0.4000}}
  >{\raggedright\arraybackslash}p{(\columnwidth - 4\tabcolsep) * \real{0.3000}}@{}}
\caption{Parts and chapters in the dependency-chain order.}\tabularnewline
\toprule\noalign{}
\begin{minipage}[b]{\linewidth}\raggedright
Part
\end{minipage} & \begin{minipage}[b]{\linewidth}\raggedleft
Ch
\end{minipage} & \begin{minipage}[b]{\linewidth}\raggedright
Title
\end{minipage} \\
\midrule\noalign{}
\endfirsthead
\toprule\noalign{}
\begin{minipage}[b]{\linewidth}\raggedright
Part
\end{minipage} & \begin{minipage}[b]{\linewidth}\raggedleft
Ch
\end{minipage} & \begin{minipage}[b]{\linewidth}\raggedright
Title
\end{minipage} \\
\midrule\noalign{}
\endhead
\bottomrule\noalign{}
\endlastfoot
\textbf{Part I: Evaluation measurement and experiment design} & \mbox{1} & Run-to-run variance, statistical power, and paired comparisons \\
& \mbox{2} & Baselines, ablations, and cost-accuracy tradeoffs \\
& \mbox{3} & Benchmark contamination, oracle strength, and workload validity \\
\textbf{Part II: Evaluation and grading systems} & \mbox{4} & Execution-based evaluation, correction gates, and release tests \\
& \mbox{5} & Calibrating model graders and separating agreement from correctness \\
& \mbox{6} & Proxy metric gaming and layered evaluation signals \\
\textbf{Part III: Containment, durable execution, and recovery engineering} & \mbox{7} & The software factory as a distributed system \\
& \mbox{8} & Agent isolation, injection defenses, and independent verification \\
& \mbox{9} & Persistent agent state, durable workflows, and idempotent retries \\
& \mbox{10} & Replayable traces and fault-injection recovery testing \\
& \mbox{11} & Human-auditable failure analysis and taxonomy development \\
\textbf{Part IV: Context engineering: retrieval, budgets, and memory} & \mbox{12} & Measuring and designing repository retrieval \\
& \mbox{13} & Localization funnels, repository indexes, and freshness checks \\
& \mbox{14} & Usable context budgets, consolidated-spec restarts, and file-based tool output \\
& \mbox{15} & Cross-session memory, raw traces, and compaction policies \\
\textbf{Part V: Human review and accountability engineering} & \mbox{16} & Efficient verification interfaces and risk-based human escalation \\
& \mbox{17} & Autonomy calibration, provenance, effective gates, and accountability \\
\textbf{Part VI: Research agenda---work allocation and cost engineering} & \mbox{18} & Agent topology selection and dynamic task allocation \\
& \mbox{19} & Cost-aware fleet scheduling and model routing \\
\textbf{Closing} & \mbox{20} & The evidence chain behind reliable agents \\
\end{longtable}

\hypertarget{what-i-assume-you-know}{%
\section{What I assume you know}\label{what-i-assume-you-know}}

I assume working knowledge of version control, continuous integration, code review, containers, and basic statistics. I use those foundations without rebuilding them. I still describe the relevant system boundary when a familiar tool plays an unfamiliar role.

I assume no prior exposure to the evaluation methods specific to public agent results. I begin with how to read a public score and identify the claim it can support. I then show how many repeated runs a comparison needs before an observed difference carries useful information. The answer depends on variation and the size of the difference you need to detect.

I also teach why agreement with human labels does not by itself make an automated grader correct. Agreement can conceal shared mistakes, ambiguous examples, or a reference answer that measures the wrong property. I separate the observed agreement from the inference made about correctness, which becomes important when a grader controls release.

Another method checks whether a score was earned on work the model had already encountered. I introduce the required vocabulary only when the mechanism appears. My goal is that a technically capable reader should be able to reproduce the reasoning presented in this monograph, not merely accept my interpretation.

The monograph does not assume a particular agent architecture or deployment scale. I describe architecture separately from implementation so the mechanism survives changes in products and interfaces. Examples expose permissions, retries, caching, concurrency, identity, and ordering when those details affect the claim. Code appears only when prose would obscure a state transition or failure path.

\hypertarget{how-to-read-a-chapter}{%
\section{How to read a chapter}\label{how-to-read-a-chapter}}

A chapter opens on a concrete situation, usually a measurement that came out wrong or a failure with a traceable cause. The mechanism follows, then what the evidence does and does not establish, then the boundary conditions, then a procedure you can run. Specialized terms are marked in bold where they are defined, so a term met later can be traced back to that site.

Citations appear inline in author-year form, with the year linked to the source. The back matter of each chapter, headed Sources and evidence, records the evidence grouping, identifier, and claim supported by each item. The back matter is authoritative for identifiers and evidence grouping; the prose states the claim for which a source is used. A disagreement between the two is an editorial defect.

The evidence legend in the methods section applies throughout. Chapter text restates only limitations that materially narrow a particular result, while each unnumbered Sources and evidence section records the claim, identifier, and evidence group for the cited items.

Every chapter opens with an evidence profile: how many strong, directional, corroborating, and null or conflicting items support it, and which practice records those items are counted against. The records are named by identifier, as in \erca{076}. An identifier resolves to one record in the companion catalog, which states that practice's action, mechanism, evidence, and boundary; the same number appears in the evidence ledger, the chapter crosswalk, the protocols, and the skills. Look one up at \href{https://sjarmak.ai/books/engineering-reliable-coding-agents/companion}{the companion site} or in \href{https://github.com/sjarmak/engineering-reliable-coding-agents/blob/main/companion/catalog.json}{\texttt{catalog.json}}. A profile counts items against a chapter's developed practices, so an item carried by a companion record cited inline is listed separately rather than folded into the chapter total.

\hypertarget{the-companion-catalog}{%
\section{The companion catalog}\label{the-companion-catalog}}

The \href{https://sjarmak.ai/books/engineering-reliable-coding-agents/companion}{companion site} indexes 206 reliability records: 193 gated practices, including 56 developed in depth, plus 13 research leads. The 56 developed practices point into the main chapters, while the other 137 gated practices appear as compact entries. The 13 leads are preserved for investigation, not recommendation, outside the gated chapter crosswalk. The catalog broadens the set of available options without forcing the main text to explain every variation. The \href{https://github.com/sjarmak/engineering-reliable-coding-agents}{project repository} contains the versioned manuscript, evidence ledger, benchmark catalog, schemas, provenance data, and release checksums. The \href{https://sjarmak.ai/books/engineering-reliable-coding-agents}{web edition} provides the browser-oriented reading version.

Use the relevant chapter as the foundation, then consult the catalog for practices that match a specific constraint. A compact entry cannot reproduce the chapter's full treatment of mechanism, evidence, tradeoffs, and failure boundaries. The chapter provides the reasoning needed to decide whether a neighboring practice applies to your workload.

Twenty-nine of the full entries are labeled as limited-support notes. I excluded them from the developed set because the available evidence does not support a recommendation. Treat them as prompts for investigation rather than established guidance. They may point to a useful experiment, a missing control, or a failure mode worth instrumenting.

The chapters and catalog serve different purposes. The chapters develop methods and claims in enough detail to evaluate critically. The catalog preserves breadth and makes related practices easier to find. Together, they let you begin with a measured problem and identify an intervention suited to the system you actually operate.

The repository also packages five reusable agent skills derived from selected practices: evaluation design, end-to-end test design, failure-mode capture, focused execution, and verified long-running implementation. Each skill includes a practice-level evidence map. These are implementation artifacts intended to make the protocols reusable; they are not additional evidence that the practices work across environments.

This edition is versioned because capability measurements and source availability change. The evidence ledger is scheduled for an annual review, with an out-of-cycle release when a material factual error, citation failure, or retraction changes a claim. The \texttt{ERCA-NNN} identifiers persist across those releases; a later edition may retire, split, or merge a record without reusing its identifier.

\hypertarget{what-you-should-be-able-to-do-by-the-end}{%
\section{What you should be able to do by the end}\label{what-you-should-be-able-to-do-by-the-end}}

Given a published agent score, you should be able to state the claim it supports and identify the conditions on which that claim depends. Given a proposed change to your own system, you should be able to measure run-to-run variation before crediting a difference and size the comparison before spending the model calls. Both systems should run on the same items, and a design that cannot resolve the difference relevant to the decision should return no verdict. Cost belongs in that judgment alongside accuracy, as does the possibility that a simpler configuration would have performed just as well.

You should be able to assess whether a public score was earned on tasks the model may already have encountered and to build an evaluation set from your own repositories when the public benchmark does not represent your work. Grades should rest on execution rather than the model's confidence, and automated graders should be validated against human labels before they gate releases. When a proxy improves without a corresponding improvement in the outcome it represents, the system should make that divergence visible.

Operationally, you should be able to bound what a single run can access and destroy, and you should treat an agent's account of its own work as a claim that still requires verification. A run that fails partway through should leave a durable record of which steps completed, and every retried step should be safe to execute again. Recovery should be tested by injecting failures rather than inferred from an architecture diagram. Reading a hundred traces from your own system should yield a concrete failure taxonomy.

Retrieval should be evaluated separately from generation so that a wrong answer can be attributed to the stage that produced it. You should also be able to measure how much of an advertised context window the system can use effectively. Human review should occur where the reviewer can see the decision that carries the risk, and autonomy should expand by action type only when measured approval and modification rates justify it. A multi-agent design should be required to outperform a single agent on the same tasks, and each component should run on the least expensive model that can perform its role reliably.

The goal is to provide you with the tools to measure what matters for your work, so the result describes the system you operate rather than a position on a leaderboard.

\mainmatter
\part{Evaluation measurement and experiment design}
\gdef\currentparttitle{Part I: Evaluation}

\chapter{Run-to-run variance, statistical power, and paired comparisons}
\label{ch01-variance-power-paired-comparisons}
\begin{quote}
\textbf{Evidence profile.} 10 strong \(\cdot\) 1 directional \(\cdot\) 0 corroborating evidence items across 3 developed practices (\erca{020}, \erca{024}, \erca{025}).

\textbf{Chapter claim.} One run is one draw.
\end{quote}

\hypertarget{one-run-is-one-draw}{%
\section{One run is one draw}\label{one-run-is-one-draw}}

I created an evaluation tool called CodeProbe (\href{https://github.com/sjarmak/codeprobe}{public repository}) that mines tasks from a repository's merged pull requests. In one run, one configuration led another by +0.054 on a task family, with a single task contributing a +0.300 advantage. The scoring code reported three decimal places, but the experiment contained no repeated runs from which to estimate stability.

I reran the family three times per configuration. The difference fell to +0.0035, with a 95\% \textbf{confidence interval}, the range produced by a procedure that covers the true value in 95\% of repeated experiments, from -0.0005 to +0.0074. The five tasks were the paired units: the three repeats were collapsed into one score per configuration for each task, leaving five paired observations and 4 degrees of freedom. The paired t statistic of 2.41 fell below the critical value of 2.776. The task that had appeared to improve by +0.300 returned a difference of 0.000, with both configurations scoring 0.800 on every repeat.

The original observation could not support a configuration-level effect. One unusually low baseline score created the apparent advantage, and it disappeared under repetition. The three-decimal display described the representation of the score, not the stability of the process that produced it.

This task family scores through a deterministic test-suite oracle, under which the end state either passes the fixed tests or does not. That oracle collapses different trajectories onto the same score, so the tight repeat scores are a certification of scorer stability rather than agent determinism. The conclusion covers these five fixed tasks and no wider population.

Record repeated independent runs before declaring a meaningful difference between agent systems. A single score is one outcome from a variable execution process, even when the configuration appears deterministic. Without repeats, the observed difference mixes the system change under test with variation from model execution, infrastructure, task ordering, and the other choices built into the evaluation apparatus.

Agent evaluations usually arrive as a table with one row per system and one score per row, which suppresses the execution history behind each cell. A score can aggregate hundreds of tasks and still be a single run, if each task was attempted once under one instantiation of the surrounding conditions.

Each cell has three layers behind it. A public agent benchmark is a shared, published task suite used to score agents: \textbf{SWE-bench} presents real issues from software repositories, and tests decide whether a proposed change passes or fails, with SWE-bench Verified as a human-screened subset of those tasks. The \textbf{evaluation harness} is the execution and scoring apparatus that checks out the repository, supplies the task, invokes the agent, applies its change, runs the tests, and produces the reported number. The published comparison is the third layer, assembled from many harness executions.

Repeated attempts need their own vocabulary. \(\mathrm{pass}@1\) is the fraction of tasks solved in one attempt. \(\mathrm{pass}@k\) is the probability that at least one of \(k\) attempts succeeds, while \(\mathrm{pass}^{k}\) is the probability that all \(k\) attempts succeed. One opportunity, retries available, and consistency across every attempt are three different questions. A run-level record keeps them separable by storing each attempt's outcome next to its task identity, and an item-level record joins those attempts back to the task. An aggregate is the summary computed from those records; a distribution is the set of outcomes before that aggregation.

Large-scale measurement shows how much of that hidden history reaches the score. Across 60,000 SWE-bench-Verified trajectories, single-run \(\mathrm{pass}@1\) varied by 2.2 to 6.0 percentage points. At decoding temperature 0, where temperature controls how randomly the model selects each token, the standard deviation still exceeded 1.5 percentage points because of infrastructure nondeterminism. The trajectories began to diverge within the first few percent of generated tokens, and each difference changed the context for the next token, so the divergence cascaded through the run.

Temperature 0 therefore does not deliver a deterministic code-generation evaluation. Ouyang et al.~(\href{https://arxiv.org/abs/2308.02828}{2023}) reached the same qualitative conclusion in an earlier empirical study of code generation: nominally identical requests produced different programs and different outcomes. That variation was not confined to the sampling control the model API exposes.

These findings should change how benchmark evaluation claims are read. Improvements of 2 to 3 percentage points, a common magnitude in system comparisons, fit inside the single-run variation Bjarnason et al.~(\href{https://arxiv.org/abs/2602.07150}{2026}) observed across 60,000 trajectories. That envelope applies to SWE-bench-Verified-class tasks, models, and execution conditions, but it is not a universal constant, and it is necessary to estimate local spread before treating a delta of that size as an improvement.

A small p-value also does not repair a single-score comparison. Reimers and Gurevych (\href{https://arxiv.org/abs/1803.09578}{2018}) compared one score from each of two identical systems and found apparent significant differences in up to 26\% of comparisons at p \textless{} 0.05. The test correctly described the two observed runs, but it could not establish that the methods differed, because the experiment had not sampled enough runs to estimate method-level variation.

Seed choice exposes the same failure at a different layer. Two five-seed samples from the same algorithm can look as if they came from different distributions. In reinforcement-learning experiments, Henderson et al.~(\href{https://arxiv.org/abs/1709.06560}{2017}) found that unreported choices among seeds, environments, and evaluation checkpoints gave researchers enough degrees of freedom to promote an ordinary fluctuation into a state-of-the-art claim. Selecting the best run after inspecting the results performs the same transformation more insidiously.

The history predates current agents. Across 2,100 trials that fine-tuned BERT models at identical hyperparameters, Dodge et al.~(\href{https://arxiv.org/abs/2002.06305}{2020}) found that distinct seeds produced substantially different results, and that weight initialization and data order contributed comparable variation. That result concerns small-data fine-tuning of pretrained encoders, so it needs re-verification before anyone treats it as a measured property of large-scale instruction tuning. What it establishes today is that a fixed hyperparameter record can leave consequential experimental state unspecified.

Nuisance sources, e.g., factors that affect the measured result but are not the capability under evaluation, are better varied by design than left to accident. Seeds, data order, task order, and splits can all be randomized, and the corresponding realizations can be matched across the systems being compared. Bouthillier et al.~(\href{https://arxiv.org/abs/2103.03098}{2021}) found that randomizing many nuisance sources and averaging the results approximated the estimator obtained by controlling each source, at about 51 times less compute. Fixed seeds produced precise estimates conditional on one arbitrary configuration, rather than estimates averaged over the range of configurations the benchmark was meant to represent.

Prompt wording is another nuisance source. In a study spanning 53 mostly classification and few-shot tasks, Sclar et al.~(\href{https://arxiv.org/abs/2310.11324}{2023}) found that meaning-preserving changes in prompt formatting produced a median accuracy spread of 7.5 percentage points, with spreads as large as 76 points. On some tasks, Salinas and Morstatter (\href{https://arxiv.org/abs/2401.03729}{2024}) found that trivial edits changed more than 10 percent of predictions. Mizrahi et al.~(\href{https://arxiv.org/abs/2401.00595}{2024}) found that individual templates even reversed which model appeared to perform better, despite greater stability in aggregate comparisons.

Those results do not establish an equivalent variance rate for long-horizon agents. Prompt phrasing should be treated as a local sensitivity measure. It tells us how much a particular result moves under reasonable reformulations, not how much agent performance varies in general.

Prompt variation also creates an ownership problem. Teams often freeze one prompt and describe the resulting configuration as fixed, even though that wording is only one realization from a broader family of semantically equivalent instructions. Fixing the prompt supports reproducibility. Varying it tests whether the conclusion survives reasonable changes in expression. A credible comparison needs both. The exact prompt should be reported so the experiment can be reconstructed, and prompt sensitivity should be measured whenever the conclusion could plausibly depend on wording.

Repeated runs make these hidden choices visible as a distribution. For each configuration, report every run-level score, along with the mean and standard deviation across runs, and inspect the per-item outcomes behind those aggregates. Randomize the declared nuisance sources and, wherever the design permits, use the same realizations for both configurations. Reporting only the best run identifies the luckiest draw, not how the configuration performs across draws.

Repetition is a direct cost multiplier. Three runs require roughly three times the model calls and execution capacity of one run, before adding prompt or seed variants. That cost cannot be avoided by borrowing a variance estimate from another model or task suite, because observed variance depends on the model, the tasks, and the evaluation apparatus together. Provider-side updates introduce temporal drift that local repetition cannot remove. Repeated results therefore also require either a pinned model version or an explicit record that the provider does not make one available.

The operational rule is narrow. Measure the local run-to-run spread, and do not credit a difference that remains smaller than that spread. Repetition alone does not establish that a detected difference is large enough to matter, nor does it determine how many runs are sufficient. Both depend on the decision threshold and the statistical power of the experiment.

\hypertarget{what-could-the-experiment-have-detected}{%
\section{What could the experiment have detected?}\label{what-could-the-experiment-have-detected}}

Before running a suite, the first question is which differences the design can resolve at all. In Miller's worked example (\href{https://arxiv.org/abs/2411.00640}{2024}), sampling more responses per question reduced the minimum detectable effect from 13.2 percent to 7.5 percent. The minimum detectable effect is the smallest difference the experiment can reliably distinguish from noise. Neither the model nor the questions became more accurate. Each question simply produced a less variable estimate, allowing the same experiment to resolve a smaller difference.

Run a \textbf{power analysis} before commissioning an evaluation. The calculation relates sample size, variance, the false-positive rate, and the probability of detecting a specified difference. Set that difference to the smallest change that would alter the engineering decision, not the difference you hope to report. This prior \textbf{effect size} is the magnitude assumed before any results are observed.

Statistical and engineering importance must be separated during design. An experiment sized to detect a gain too small to justify a more expensive configuration spends capacity on a decision that will never be made. The target should therefore be the smallest improvement that would justify deployment. Writing that threshold down before the results arrive prevents the observed delta from redefining what counts as a win.

Power is the probability that an experiment detects the specified effect when that effect is present. The false-positive rate controls how often the test reports a difference when no difference exists under its assumptions. These error controls trade off through sample size. At a fixed variance, a smaller target effect, a lower false-positive rate, or a higher desired detection probability each increases the number of observations required.

The minimum detectable effect expresses the same power calculation in reverse. Fix the task set, repeat count, variance estimate, and error rates, and it states the experiment's resolution. Reporting it clarifies what a null result can support. The experiment had adequate power to detect differences of that magnitude or larger, while smaller differences were not reliably distinguishable from noise.

Many published benchmark comparisons cannot support the distinctions their tables imply. In a power analysis of GLUE, Card et al.~(\href{https://arxiv.org/abs/2010.06595}{2020}) found that many test sets lacked the resolution needed to distinguish the small state-of-the-art improvements researchers reported. This does not mean every reported ordering was false. It means the available samples left too much uncertainty to establish those differences at the chosen error rates.

An underpowered study fails in two directions. Most real effects will not reach significance, filling the literature or internal decision record with inconclusive null results. Among the estimates that do cross the significance threshold, the observed effects tend to be exaggerated because unusually large estimates are the ones most likely to survive. A significant result from a weak experiment can therefore provide a much less reliable estimate of magnitude than its p-value suggests.

Estimate the required sample size from a pilot, then add margin. The pilot supplies a variance estimate for the planned model, task family, prompts, and evaluation apparatus. Colas, Sigaud and Oudeyer (\href{https://arxiv.org/abs/1806.08295}{2018}) recommend at least 20 pilot runs for estimating this variance. That is a floor for the pilot estimate, not a claim that the estimate has stabilized or that every final configuration requires 20 repetitions. A small or unrepresentative pilot can understate variance and produce a design that appears adequate on paper but remains underpowered in practice.

The sizing sequence remains reproducible even when its inputs change. Choose the smallest effect that would alter the engineering decision, estimate variance from the pilot, set the false-positive rate and desired detection probability, and solve for the required item and run counts. At fixed error rates, greater variance increases the required sample size, while a larger target effect reduces it. For a suite whose counts are already fixed, solve in the other direction for the minimum detectable effect and add margin for uncertainty in the pilot estimate.

Which count to increase depends on where the uncertainty originates. More benchmark items reduce uncertainty about performance across items. More independent runs reduce uncertainty from execution variation. More responses per question reduce answer-level variance. Holding the other design choices constant, additional responses cost more model calls but can allow the experiment to resolve a smaller difference.

This is what reduced the minimum detectable effect from 13.2 percent to 7.5 percent in the worked example. The three counts are not interchangeable because each averages over a different random quantity. That measured reduction is one empirical result, not a general conversion between response count and experimental resolution.

Lowering temperature is not a substitute for sampling more responses when the deployed system will run at the original temperature. It changes the distribution being measured rather than estimating that distribution more precisely. A lower temperature is a legitimate system change when the deployed configuration will use it, but it cannot serve as free variance reduction in an experiment intended to evaluate another configuration.

Small run counts also affect which inferential procedure is defensible. A \textbf{bootstrap} estimates uncertainty by resampling the observed data; in a paired comparison, it resamples whole pairs. In Colas et al.'s (2018) reinforcement-learning experiment, the bootstrap's realized false-positive rate was about 10 percent for fewer than ten runs despite a nominal 5 percent level. A five-run comparison also declared two samples from the same DDPG implementation significantly different. Welch's t-test was closer to the nominal rate in that experiment but still exceeded it at small \(N\), leading the authors to suggest a significance level below 0.05 when the goal is to keep the realized rate below 0.05. This is a study-specific calibration warning, not a universal preference for one test.

Welch's t-test does not assume that the two configurations have equal variance, which makes it a more defensible default than the equal-variance t-test for independent small samples. Its advantage at low run counts is empirical and conditional. When the design pairs observations across matched tasks or nuisance realizations, the analysis should preserve that pairing rather than treat the samples as independent. No test removes the need to inspect the score distribution, and no test can repair samples that fail to represent the intended population.

Standard power formulas often assume that aggregated performance is approximately bell-shaped. Reinforcement-learning outcomes can be strongly bimodal, and agent outcomes may similarly divide between complete success and early failure. In the measured reinforcement-learning cases, even the t-test did not fully control the type I error rate for bimodal outcomes. Analytical power calculations should therefore be treated cautiously when the observed distribution is highly discrete, skewed, or multimodal.

Sizing from pilot tasks also assumes that their variation resembles the evaluation suite and that the suite resembles the deployment workload behind the decision. More runs can narrow uncertainty around the wrong population. No sample-size formula can repair a task distribution that omits important cases the deployed system will encounter.

Retrospective power does not solve these limitations. Computing power from the observed effect reuses the noisy outcome as though it had been the design target. The result largely restates the p-value and creates circular reasoning. A small observed effect produces low reported power, while a large observed effect produces high reported power. The useful calculation happens before the result, using an effect threshold and variance estimate chosen independently of the observed comparison.

One of my evaluation frameworks makes refusal mechanical. It scores a system across a ladder of successively weakened configurations and locates each readout on the resulting score curve. When too few rungs have been evaluated, the grader cannot place the readouts reliably, so it raises an error and emits no verdict. This refusal is a methodological rule, not a statistical test, and it carries no false-positive or false-negative rate of its own. Its value is that it keeps insufficient resolution visible to the person making the engineering decision.

A completed comparison should therefore report its detectable range. When the observed difference falls below the predeclared engineering threshold or below the experiment's minimum detectable effect, report that limitation explicitly. Insufficient resolution is not evidence that the systems are equal.

\hypertarget{compare-outcomes-item-by-item}{%
\section{Compare outcomes item by item}\label{compare-outcomes-item-by-item}}

When two systems face the same tasks, the comparison should be based on their per-item differences, and the experiment must preserve those differences from the start. A \textbf{paired design} runs both systems on the same experimental units so that their outcomes can be compared item by item. Pairing is a property of the experiment. The statistical test is chosen only after the paired observations exist.

The identical-system comparisons discussed earlier show the limit of analyzing one aggregate score from each run. A test may determine whether two realized scores differ under its assumptions, while the scientific question concerns whether the underlying methods differ across repeated realizations. Pairing does not answer that second question. Repeated runs are still needed to estimate variation across realizations. Pairing instead improves precision for the comparison within each realization.

The gain comes from removing shared item difficulty. If both systems tend to succeed on easy tasks and fail on hard ones, their item-level scores will be positively correlated. Subtracting the scores item by item removes much of that common movement, leaving variation that more directly represents disagreement between the systems.

For item \(i\), let \(x_i\) and \(y_i\) be the two system scores and define the paired difference as \(d_i=x_i-y_i\). Across \(n\) items, the estimated mean difference and its standard error are

\[
\bar d=\frac{1}{n}\sum_{i=1}^{n}d_i,
\qquad
SE(\bar d)=\frac{s_d}{\sqrt n}.
\]

The variance of the mean difference contains the covariance between the two systems:

\[
\operatorname{Var}(\bar d)
=\frac{s_x^2+s_y^2-2\operatorname{Cov}(x,y)}{n}.
\]

When the systems respond similarly to item difficulty, the covariance is positive and reduces the variance of the comparison. An independent analysis omits this term. It treats the two systems as though they had been evaluated on unrelated samples and counts the variation between easy and hard tasks twice. The resulting uncertainty estimate is therefore unnecessarily large when item-level outcomes move together.

\begin{figure}[htbp]
\centering
\includegraphics{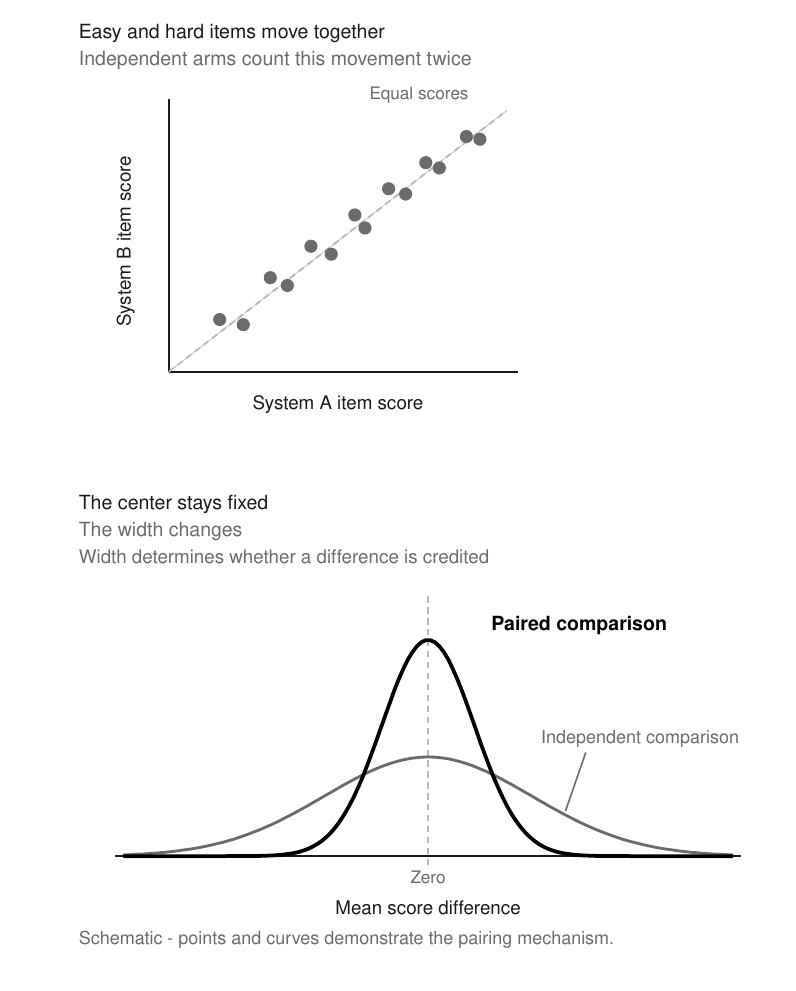}
\caption{Schematic of correlated item scores for systems A and B and the narrower uncertainty obtained by analyzing their matched differences.}
\end{figure}

Item-level subtraction removes shared task difficulty, so the remaining variation better represents system disagreement. Positive covariance narrows uncertainty without changing the mean difference.

The interval around the mean difference is obtained by multiplying its standard error by an appropriate critical value. The choice of critical value and the validity of the approximation depend on the statistical procedure and the observed score distribution.

The data record must preserve the pairing. For every task, retain the task identity, both system scores, the nuisance-source assignments, and the run identity. Report the mean paired difference, its standard error, an interval around that mean, and the correlation between the systems' item-level scores. Once the per-item relationship has been discarded, two headline means are not enough to reconstruct any of these quantities.

Pairing also improves diagnosis. The same average difference can arise from small gains across most tasks, a few large reversals, or a trade in which each system solves a different subset. Those patterns have different engineering implications. Item-level records reveal which tasks changed and whether the overall result depends on a small number of discordant cases.

The validity of the pairing depends on experimental identity, not matching task labels. Both systems must receive the same task state under the same execution and scoring apparatus. Repository revision, dependency state, tool permissions, time limits, and scoring logic are all part of the experimental unit. If any of them differ between the two arms, the subtraction no longer removes shared task difficulty because the systems did not receive the same item.

An adversarial review of one of my own experiment designs found this exact defect. One arm transformed the repository before presenting it to the system, while the other operated on the original repository. The task label remained the same, but the executable state did not. I therefore treated the planned pairing as invalid. A shared label does not establish that two experimental units are equivalent.

Published aggregate scores create a simpler boundary. If the other system was not run on the same item instances, the observations are not paired. The aggregate results can still be compared descriptively, but their item-level covariance cannot be recovered from the means. A valid paired analysis requires rerunning both systems on the same items.

Repeated attempts introduce the same identity requirement. Attempt one from system A pairs with attempt one from system B only when the attempts share randomized conditions by construction. Selecting pairs after inspecting the outcomes biases the comparison. Seeds, task order, data order, and other nuisance assignments must therefore be matched before results are visible and recorded with every pair.

The statistical procedure must also match the mathematical form of the outcome. When the reported score is the arithmetic mean of per-item numerical contributions, and the distribution of paired differences is reasonably compatible with a normal approximation, a paired t-test can test the mean difference and support a corresponding power analysis. That approximation should not be assumed automatically for small, discrete, skewed, or highly irregular samples.

Composite metrics require different treatment. An F-score is a nonlinear function of aggregate counts. There is generally no set of independent per-item F-scores whose arithmetic mean equals the corpus-level F-score. Corpus-level BLEU has a similar aggregation problem, and some forms of ROUGE may also depend on nonlinear or corpus-level aggregation. Applying a t-test to invented per-item metric contributions can produce a familiar p-value without satisfying the mathematical assumptions that give the test meaning.

A paired bootstrap preserves the relationship between the systems by resampling items as pairs and recomputing the full metric on each resampled dataset. A paired permutation test instead exchanges system labels within each observed pair under the null hypothesis. Neither requires the same normal approximation as a parametric t-test. Both still depend on choosing the correct resampling unit and having a sufficiently representative sample. The small-sample bootstrap failure described earlier occurred when the resampling unit was only a handful of runs.

Agent evaluations introduce outcomes that standard benchmark guidance does not fully resolve. Cost-weighted success combines task outcome with resource use. A pass or fail assigned by an imperfect grader includes uncertainty from the grading process. Neither outcome inherits a valid test merely because its final value lies between zero and one. The sampling unit, dependence structure, and aggregation rule must be specified before choosing the analysis.

The table below gives directional guidance. The recommendations for language-generation metrics come from the broader methodological guidance of Dror and Reichart (\href{https://arxiv.org/abs/1809.01448}{2018}) rather than from controlled comparisons of agent evaluations. The binary row applies \textbf{McNemar's test}, introduced by McNemar (\href{https://doi.org/10.1007/BF02295996}{1947}), the standard test for paired pass or fail outcomes. It analyzes the discordant pairs in which one system passes and the other fails. The table is a starting point, not a substitute for deriving the metric's actual sampling structure.

\begin{longtable}[]{@{}
  >{\raggedright\arraybackslash}p{(\columnwidth - 4\tabcolsep) * \real{0.3333}}
  >{\raggedright\arraybackslash}p{(\columnwidth - 4\tabcolsep) * \real{0.3333}}
  >{\raggedright\arraybackslash}p{(\columnwidth - 4\tabcolsep) * \real{0.3333}}@{}}
\caption{Outcome structure, paired estimand, and corresponding test guidance.}\tabularnewline
\toprule\noalign{}
\begin{minipage}[b]{\linewidth}\raggedright
Outcome structure
\end{minipage} & \begin{minipage}[b]{\linewidth}\raggedright
Comparison to analyze
\end{minipage} & \begin{minipage}[b]{\linewidth}\raggedright
Test guidance
\end{minipage} \\
\midrule\noalign{}
\endfirsthead
\toprule\noalign{}
\begin{minipage}[b]{\linewidth}\raggedright
Outcome structure
\end{minipage} & \begin{minipage}[b]{\linewidth}\raggedright
Comparison to analyze
\end{minipage} & \begin{minipage}[b]{\linewidth}\raggedright
Test guidance
\end{minipage} \\
\midrule\noalign{}
\endhead
\bottomrule\noalign{}
\endlastfoot
Per-item numerical scores whose paired differences plausibly satisfy a normal approximation & Mean of the paired item differences & Paired t-test \\
Metrics computed nonlinearly from aggregate counts, including corpus-level F-score and BLEU & Full metric recomputed on resampled or relabeled item pairs & Paired bootstrap or paired permutation test \\
Pass or fail outcomes on the same items & Discordant pairs in which only one system passes & McNemar's test \\
\end{longtable}

A later chapter on retrieval freshness uses McNemar's test for this reason. Both systems receive the same items, and each item produces a pass or fail outcome.

The companion catalog covers designs that this basic framework does not. Clustered items require correlation-aware standard errors. Estimating \(\mathrm{pass}@k\) from repeated attempts requires the appropriate combinatorial estimator. Claims spanning many tasks or metrics require multiplicity correction. Ranking several systems rather than comparing two requires paired-comparison ranking models with uncertainty intervals.

The catalog also covers profiling benchmark noise before making decisions, declaring the decoding configuration, aggregating scarce repetitions with interquartile means and resampled intervals, reducing an evaluation around a specific decision, using precommitted confirmatory designs, and measuring sensitivity across prompt variants. Each addresses a narrower problem than the core decisions developed in this chapter.

\hypertarget{before-drawing-a-conclusion-from-a-difference}{%
\section{Before drawing a conclusion from a difference}\label{before-drawing-a-conclusion-from-a-difference}}

Define the engineering threshold before running the expensive suite. The threshold is the smallest change that would alter the engineering decision, not the smallest change a statistical test might detect.

Pilot variance then determines the required numbers of tasks and repeated runs. Add margin because the pilot estimate is itself uncertain, and the design record states the smallest difference the planned experiment can reliably resolve. When the available budget cannot resolve a difference large enough to matter, narrow the claim or withhold the verdict before spending the calls.

Three independent repeats per configuration is a workable minimum. Three may still produce an unstable variance estimate, but they prevent one unusually favorable or unfavorable run from determining the result. Pair nuisance conditions across systems wherever possible and run both systems against identical item states. Each retained row records the item identity, the outcomes from both systems, the repeat identity, and the assigned random conditions.

The analysis begins with per-item differences. For continuous outcomes, report the mean paired difference, its standard error, the correlation between the two systems' scores, and a confidence interval for the mean difference.

Binary pass-or-fail outcomes require the discordant pairs and an appropriate paired test. Nonlinear composite metrics require resampling or permutation of intact pairs. Matching task names do not create a paired design when the repository state, inputs, or execution path differs.

\Needspace{5\baselineskip}

Repeated trials remain interpretable only while the apparatus is pinned:

\begin{itemize}
\tightlist
\item
  model version;
\item
  decoding settings;
\item
  exact prompts;
\item
  task and repository revision;
\item
  tool definitions and permissions;
\item
  harness version; and
\item
  evaluator version.
\end{itemize}

When a provider exposes no stable model version, record the evaluation window and treat later reruns as potentially affected by model drift.

The observed difference belongs beside the complete run distribution and the engineering threshold written before execution. A difference smaller than the measured variation does not warrant credit, and an experiment that lacked the power to resolve the difference that would have changed the decision returns no verdict.

The repository artifact \href{https://github.com/sjarmak/engineering-reliable-coding-agents/blob/main/protocols/evaluation-comparison.md}{\texttt{protocols/evaluation-comparison.md}} packages this sequence, pass condition, and retained files.

\section*{Sources and evidence}

\textbf{Never report a single run}

\begin{itemize}
\tightlist
\item
  Strong evidence: Ouyang, Zhang, Harman \& Wang (2023). An Empirical Study of the Non-determinism of ChatGPT in Code Generation. arXiv:2308.02828. Nominally identical requests, different programs and outcomes.
\item
  Strong evidence: Bjarnason, Silva \& Monperrus (2026). On Randomness in Agentic Evals. arXiv:2602.07150. The 60,000-trajectory result: 2.2 to 6.0 pp single-run spread, SD above 1.5 pp at temperature 0, early-token divergence.
\item
  Strong evidence: Reimers \& Gurevych (2018). Why Comparing Single Performance Scores Does Not Allow to Draw Conclusions About Machine Learning Approaches. arXiv:1803.09578. The 26\% result.
\item
  Strong evidence: Bouthillier et al.~(2021). Accounting for Variance in Machine Learning Benchmarks. MLSys 2021. arXiv:2103.03098. The \textasciitilde51x result and the fixed-seed critique.
\item
  Strong evidence: Henderson et al.~(2017). Deep Reinforcement Learning that Matters. AAAI 2018. arXiv:1709.06560. Five-seed splits and unreported researcher degrees of freedom.
\item
  Strong evidence: Dodge et al.~(2020). Fine-Tuning Pretrained Language Models: Weight Initializations, Data Orders, and Early Stopping. arXiv:2002.06305. The 2,100-trial seed result, scoped to small-data fine-tuning of pretrained encoders.
\item
  Strong evidence: Sclar et al.~(2023), FormatSpread, arXiv:2310.11324. Meaning-preserving prompt-format changes produced a median 7.5-point accuracy spread across the tested tasks, with larger task-level spreads.
\item
  Strong evidence: Mizrahi et al.~(2023/2024), TACL (2024), arXiv:2401.00595 (submitted December 31, 2023, hence the January identifier). Individual templates reversed some model comparisons even where aggregate comparisons were more stable.
\item
  Strong evidence: Salinas and Morstatter (2024), arXiv:2401.03729. Trivial prompt edits changed more than 10 percent of predictions on some tested tasks.
\end{itemize}

\textbf{Analyze statistical power before running}

\begin{itemize}
\tightlist
\item
  Strong evidence: Miller (2024). Adding Error Bars to Evals. arXiv:2411.00640 (Anthropic). The 13.2\% to 7.5\% minimum-detectable-effect example and the response-count lever.
\item
  Strong evidence: Card et al.~(2020). With Little Power Comes Great Responsibility. EMNLP 2020. arXiv:2010.06595. GLUE underpowering and effect-size exaggeration.
\item
  Strong evidence: Colas, Sigaud \& Oudeyer (2018). How Many Random Seeds? arXiv:1806.08295. The pilot floor, the bootstrap false-positive rate at small N, the DDPG case, and the preference for Welch's t-test.
\end{itemize}

\textbf{Use paired tests matched to the metric}

\begin{itemize}
\tightlist
\item
  Strong evidence: Miller (2024). Adding Error Bars to Evals. arXiv:2411.00640. Paired per-item differences with paired standard error and score correlation.
\item
  Directional evidence: Dror \& Reichart (2018). Appendix, Recommended Statistical Significance Tests for NLP Tasks. arXiv:1809.01448. The per-metric test selection; directional, so the table is guidance, not a measured result.
\item
  Foundational method: McNemar, Q. (1947). Psychometrika 12(2), 153-157. DOI: 10.1007/BF02295996. The paired pass/fail test used in the binary-outcome row.
\end{itemize}

\textbf{Author-system illustration cited inline}

\begin{itemize}
\tightlist
\item
  Not an evidence item: CodeProbe, the author's task-mining evaluation tool, \href{https://github.com/sjarmak/codeprobe}{public repository}. Named inline for the task-family run and rerun described in the opening, which are narrative illustration.
\end{itemize}

\chapter{Baselines, ablations, and cost-accuracy tradeoffs}
\label{ch02-baselines-ablations-cost-accuracy}
\begin{quote}
\textbf{Evidence profile.} 1 strong \(\cdot\) 5 directional \(\cdot\) 0 corroborating evidence items across 2 developed practices (\erca{012}, \erca{114}).

\textbf{Chapter claim.} A component that never executes cannot explain the result.
\end{quote}

In one end-to-end run of CodeProbe (\href{https://github.com/sjarmak/codeprobe}{public repository}), the plain baseline came out ahead of the tool-augmented arm on both score and score per dollar. The configuration with the retrieval machinery in it lost to the configuration with none. That outcome was visible only because the run included a baseline at all. Reported alone, the tool arm's score would have had nothing to be read against, and a reader could have credited the retrieval machinery with whatever the number implied.

Published comparisons frequently omit the inexpensive arm. Kapoor et al.~(\href{https://arxiv.org/abs/2407.01502}{2024}) re-evaluated published agent architectures and found that on HumanEval, retrying the model could match more elaborate architectures at a fraction of their inference cost. When they optimized cost and accuracy jointly, they reduced cost without giving up accuracy.

Those results do not establish the same conclusion for repository-scale engineering. The re-evaluation is directional, and its coding evidence comes from a function-level benchmark. The comparison design transfers even when the finding does not. The inexpensive arm belongs in the experiment, and cost belongs in the report.

That omission changes the claim. A system with memory, retrieval, tools, or multiple agents may score higher than a direct model call. The score alone cannot tell us whether the added machinery caused the gain, whether another model call would have produced the same gain, or whether the comparison spent more until it found more successful samples. Engineering selection requires two coordinates: what the system accomplishes and what it consumes.

The available evidence supports comparison design rather than a universal cost threshold or recommended architecture. The chapter therefore develops controls that readers can run on their own workloads.

An elaborate system is worth keeping only when both comparisons support it. Its components must contribute something that remains when the rest of the experiment is held fixed, and the full configuration must beat the simple alternatives at a cost the operator is willing to pay. These are separate tests. Component removal addresses causal attribution. Direct-call and retry baselines address engineering value.

\hypertarget{remove-the-component-and-rerun-the-evaluation}{%
\section{Remove the component and rerun the evaluation}\label{remove-the-component-and-rerun-the-evaluation}}

An \textbf{ablation} removes one component and reruns the evaluation with the rest of the system and procedure held fixed. If a memory-enabled agent outperforms an agent without memory, the missing-memory arm estimates what the memory system contributed. If the comparison changes the model, prompt, task subset, token budget, harness version, or scoring procedure at the same time, the subtraction no longer isolates memory. It measures an unspecified bundle of changes. Without the missing-component arm, base-model capability and item difficulty remain plausible explanations for the observed score.

The component controls recur in two source corpora. SkillEvolBench, from Lei et al.~(\href{https://arxiv.org/abs/2605.24117}{2026}), reports no-skill and raw-trajectory controls for agent memory. CoIR, from Li et al.~(\href{https://arxiv.org/abs/2407.02883}{2024}), supplies the task and metric substrate for code retrieval, while the no-tool control comes from the associated evaluation-design synthesis. These sources support narrower measurements in their own settings; neither tests the complete repository-scale prescription developed here. Their convergence motivates the protocol, while the generalized recommendation remains directional.

Memory requires at least three arms. The first is the proposed memory system. The second is the no-skill control, which removes the written memory entirely. The third reuses the raw trajectory from prior work without asking a writer to consolidate it into a skill or memory record. The no-skill arm tests whether prior experience helps at all. The raw-trajectory arm tests whether the consolidation mechanism adds anything beyond retaining the original evidence.

Those two questions are easy to collapse into one. Suppose an agent that receives a distilled lesson solves more tasks than an agent starting from an empty context. The gain could come from the lesson, but it could also come from any useful tokens copied out of the earlier attempt. Raw-trajectory reuse makes that alternative visible. If the raw trace matches the distilled record, the experiment has shown that prior information helps, but it has not shown that the writer extracted a better representation.

The comparison also needs matched opportunities to retrieve. A memory arm evaluated on tasks for which a relevant record exists cannot be compared directly against a no-memory arm scored across a broader or harder task set. The primary comparison should be restricted to a coverage-matched subset, that is, the tasks for which the memory or retrieval system had a genuine opportunity to return relevant material. The full-set result should be reported separately, because coverage is itself an operational property. Unequal coverage should not be read as a difference in answer quality.

Retrieval and tool evaluations need a corresponding no-tool arm. Tool availability is an experimental variable and should be recorded with the rest of the arm configuration. The control gets the same base model, instructions, stopping conditions, and task instances, with access to the tool removed. If the treatment also receives a different system prompt, a larger context budget, or extra retries, those changes belong in additional arms or in a sweep that varies them one at a time.

One-factor sweeps matter because agent components interact. Retrieval can change context length. Context length can change model behavior. Tool schemas consume tokens before the agent begins the task. A different endpoint can expose a different tool set. A single comparison between `agent' and `baseline' assigns the combined effect to whichever feature appears in the label. Factorial experiments can estimate interactions when the evaluation budget supports them, but a sequence of one-factor comparisons is usually enough to find the first unsupported attribution.

An ablation result is conditional on the configuration in which the removal occurred. If retrieval helps only when a summarizer compresses its output, removing retrieval from the full system estimates retrieval's contribution with that summarizer present. The result does not show whether retrieval would help without summarization, or whether summarization would help without retrieval. Those questions require the corresponding component combinations as separate arms.

Skills and multi-agent structures need the same discipline, but their removal needs a more precise intervention. Deleting a skill while also shortening the prompt changes both procedural guidance and context length. Replacing several agents with one while reducing the call budget changes topology and sampling opportunity together. A useful control preserves the available information and budget wherever the design permits and removes only the coordination or representation being credited. When that preservation is impossible, an intermediate arm lets the comparison separate fewer model calls from a different arrangement of those calls.

Tank and Nama (\href{https://arxiv.org/abs/2607.22520}{2026}) compared agents with and without procedural skills across nearly 6,000 runs on two office-automation benchmarks and three model-harness stacks. The best-performing skills won primarily by causing fewer regressions, a distinction concealed by net task-success change. The study strongly supports decomposing a skill intervention into gains, regressions, and residual failures in those settings; transfer to coding-agent skill libraries remains directional.

The paired design from Chapter 1 belongs here. Each arm runs on the same tasks, task ordering and other sources of randomness are matched across arms wherever the execution system permits it, and the outcomes are analyzed as pairs. Otherwise, item difficulty can dominate the component effect. A retrieval arm that happens to receive more questions answerable by lexical lookup may appear superior even when the tool adds nothing on matched items.

Target size is a second confound of the same kind. A tool may look better when its index contains a narrow, curated collection and worse when it searches a large repository with many plausible distractors. Query coverage, corpus size, and retrieval depth should therefore be recorded with the configuration. If target size changes between arms, the experiment measures both the target and the retrieval algorithm.

In two studies from my own team, we asked the same practical question: does a retrieval tool help an agent? The positive-sign study was a paired comparison across many repositories, in which the tool arm came out slightly ahead. The negative-sign study was the CodeProbe run described at the start of this chapter, in which the plain baseline came out ahead. The two used different task curation and different enforcement of which arm had to use the capability under test. That experience is narrative illustration rather than independent evidence for the control method, but it exposed the mechanism. The apparatus decided which work entered the evaluation and whether the agent had to use the capability supposedly under test. The label on the treatment arm therefore did not identify a stable intervention.

Arm separation should be established at the task level through observed usage. A task cannot estimate a tool effect when the treatment completes it without touching the tool. Such a task may remain useful for measuring the agent's overall capability, but it contributes no information to the comparison of tool access.

In my own enterprise-scale benchmark, one candidate task ran for 41 turns without a single call to the tool under test. By every static check we had, it was a perfect task, and for the tool comparison it was useless. We removed it from the tool-effect comparison because both arms had effectively received the same treatment. Candidate tasks are hypotheses about discrimination, and a pilot run has to show that they actually separate the arms.

This gate should be empirical. For every candidate task, record whether the treatment invoked the component, whether the control could reach it through another path, and whether both arms received equivalent task information. A binary `tool enabled' field is inadequate when the agent may ignore the tool. Usage counts, arguments, returned bytes or tokens, and the point in the trajectory where the result entered context make the intervention observable.

The exclusion rule needs the same protection as any other post-hoc filter. Removal should be keyed to observed component usage and declared before the outcomes are inspected. Dropping tasks after seeing which arm won selects on the result, and it converts a usage gate into a mechanism for improving the reported effect.

Usage does not establish usefulness, but absence establishes non-use. If the treatment calls a retriever and then ignores its output, the task may still show a cost effect or an interference effect. If it never calls the retriever, its outcome cannot support a claim about retrieved information improving correctness.

The strongest boundary concerns information flow into the component. Evaluation labels and gold answers must stay out of any memory writer. A writer that sees the correct answer can encode it directly, encode a near paraphrase, or preserve features that make later retrieval trivial. Freezing the memory after that write does not repair the experiment, because the evaluated artifact already contains the target.

The same rule applies to the scoring target. A `right answer' generated with the tool or mechanism under evaluation cannot then be used to grade that mechanism by agreement with its own output. A retrieval system compared against answers assembled from the same retrieval system has been graded against its own oracle, the reference an evaluation scores candidate answers against. A target capable of disagreeing with the system requires independent human verification, or sources outside the evaluated path. Public benchmarks carry a related problem, because models have often already seen those tasks during training, an inflation channel Chapter 3 measures.

An ablation cannot remove every threat to inference, and it does not show that a useful component will remain useful under another workload. It answers a narrower question: under the evaluated task distribution and the fixed surrounding system, did removing this component change the measured outcomes? This question is narrow enough to test and strong enough to prevent a common attribution error.

The completed record should make the subtraction reproducible. Name the component removed, list every configuration field that remained fixed, identify the coverage-matched task subset, report component usage per task, and preserve the paired results. When multiple factors changed, describe the arm as a bundle and resist assigning its effect to one member of that bundle.

\hypertarget{select-on-accuracy-and-cost-together}{%
\section{Select on accuracy and cost together}\label{select-on-accuracy-and-cost-together}}

The support for this entry is one directional re-evaluation study and no strong evidence item. Kapoor et al.~(2024) covered a small set of benchmark families, including function-level coding, multi-hop question answering, and web-agent tasks. They did not establish the same result for repository-scale software work. Their contribution here is a comparison design that can be applied beyond those tasks. Evaluate the elaborate system against direct model calls and retries, then place accuracy and inference cost in the same result.

Additional model calls can raise the pass rate without any change to the architecture. If one attempt has probability \(p\) of passing and two attempts are independent, the probability that at least one passes is \(1-(1-p)^2\). The expression illustrates the effect under independence. It does not estimate real agent behavior, in which attempts are correlated. Correlation changes the size of the gain, but a system allowed to sample, revise, or delegate more often still has more opportunities to produce a passing output.

An accuracy-only ranking conceals those opportunities. It may rank a five-call scaffold above a direct call without showing that the improvement came from four additional samples. It may also compare one architecture with a retry loop against another configured to stop after its first answer. The ranking then credits budget and stopping policy under the name of architecture.

Use at least three initial arms. The direct-call arm sends the task to the base model with the minimum production-valid prompt and no scaffold. The retry-once arm makes one additional model attempt after a defined failure signal. The candidate arm runs the proposed architecture with its intended tools, memory, or delegation.

The failure signal in the retry arm must be available to the deployed system. Unit tests, a compiler error, or a schema validator can supply one. A hidden benchmark answer cannot. If the experiment retries only because the evaluator knows the first answer is wrong, that arm should be labeled oracle-assisted, because it otherwise understates the operational cost of deciding when to retry.

Model version, task set, and scoring target stay fixed across the three arms. Match any shared context required to attempt the task, but do not give the direct call internal traces that exist only because the scaffold ran. The inexpensive alternative has to be built credibly, since a deliberately crippled direct call provides no useful control. Record the stopping condition for every arm, because `one run' can mean one model response, one agent trajectory, or a tree of dozens of calls.

Cost accounting begins at the task boundary. Count all input and output tokens consumed after the task enters the system, including calls made by planners, delegates, critics, summarizers, and retries. When a provider bills cached input differently, preserve cache-read and cache-write quantities rather than folding them into an undocumented token total. Failed calls, timeouts that incur usage, and discarded branches are part of the cost of the configuration that created them.

Report token quantities even when the organization buys capacity rather than paying per request. Tokens remain a portable description of inference demand, while dollars depend on contract terms, provider prices, model aliases, and date. Convert the token ledger to money with a named pricing basis and a snapshot date. A dollar figure without that snapshot becomes uninterpretable after prices change.

Latency belongs beside this record when users wait for the answer or when workers occupy scarce execution slots. Latency and inference cost are not interchangeable quantities. Two arms can consume the same tokens while one serializes its calls and the other runs them concurrently. The parallel arm may reduce wall-clock time while raising peak capacity requirements. Cost, performance, and scheduling pressure therefore remain separate operational properties.

Plot each configuration by accuracy and cost. A configuration is Pareto-dominated when another configuration is at least as accurate and no more expensive, with a strict advantage on at least one axis. Dominated points drop out of consideration unless they supply some separately measured property the two-axis plot omits. The remaining points form the Pareto frontier. Moving along it trades accuracy for cost in one direction and cost for accuracy in the other.

The points on the plot should be configurations. Architecture names are too coarse, because model choice, retry limit, retrieval depth, and context budget all change both coordinates. A frontier built from configurations is usable for routing. Inexpensive settings may serve routine tasks, while costly settings are reserved for cases in which their measured gain justifies the added inference.

This rule is stronger than ranking by score per dollar alone. A ratio can favor a configuration that is cheap but below the minimum acceptable accuracy, or hide how much accuracy the additional spending returns at the high end of the curve. The frontier preserves the actual coordinates. Deployment constraints apply afterward: a minimum pass rate, a maximum per-task cost, or a latency ceiling measured for the workload.

\begin{figure}[htbp]
\centering
\includegraphics{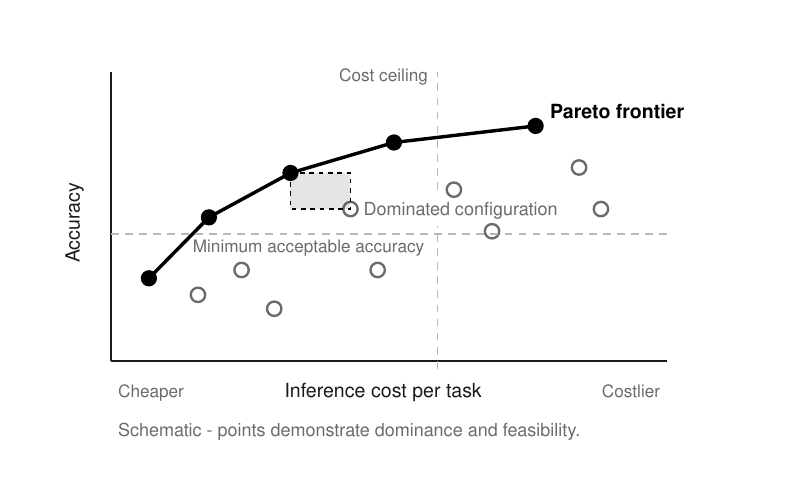}
\caption{Schematic of a cost-accuracy frontier, dominated configurations, and minimum-accuracy and cost-ceiling feasibility constraints.}
\end{figure}

Dominating configurations remove weaker alternatives unless an omitted property is measured separately. Deployment constraints then narrow the frontier.

The paired analysis from Chapter 1 keeps the frontier from acquiring false precision. Cost and accuracy estimates have uncertainty, and two nearby configurations may be indistinguishable at the available sample size. A single run also yields one realization of aggregate cost, which inherits the run-to-run variation that Chapter 1 describes for scores. The paired comparison runs on the per-task costs inside that run rather than on the aggregate. Preserve per-task cost as well as per-task correctness, so that bootstrap intervals can be computed for both axes and for the paired differences. A point should not be called cheaper or more accurate when the interval around that comparison does not support the ordering.

The retry arm needs the same accounting discipline as the candidate. If only some first attempts fail, report how often the second call fired, how much it cost when it fired, and whether it fixed the task. That record exposes the marginal value of the retry policy. A low-cost retry that recovers a meaningful subset of failures may dominate a planner with multiple unconditional calls, while a retry that repeats the same error adds cost without moving accuracy.

In CodeProbe, I report score per dollar alongside score and rank configurations by pass rate, cost, tokens, and latency. The end-to-end run in which the plain baseline led the tool-augmented arm on both axes came out of that report. It does not establish a general result about tools. Reporting both axes made the engineering decision visible without turning a small score difference into a claim about architecture.

Accuracy without a cost constraint still answers a legitimate research question. A model developer may want to know the highest capability reachable under extensive sampling, regardless of whether that procedure is economical in production. That is a capability probe. Selecting a system to operate under a budget is a different decision, and the result should state which decision the experiment was designed to support.

Architecture development introduces another source of optimism. An iteration \textbf{holdout} reserves a set of tasks that scaffold developers never use while choosing prompts, tools, routing rules, retry policies, or other configuration. Repeated iteration against a fixed evaluation set turns that set into training data for the scaffold, even when the model weights never change. Developers learn which changes improve the score, keep those changes, and discard the rest.

Create the iteration holdout before tuning begins. Use the remaining development tasks to debug the harness, compare early configurations, and decide what to carry forward. Open the iteration holdout only at a declared decision point, run the selected configurations, and do not use its task-level failures to continue tuning the same selection. If development resumes from those failures, the set has joined development and a new iteration holdout is needed.

The iteration holdout also needs protection from indirect tuning. A developer who reads its repository names, failure categories, or aggregate subgroup scores can adapt the scaffold to those features without inspecting the exact prompts. Access controls and an evaluation service that returns only the predeclared result can preserve the boundary better than a file accompanied by a policy. Record every opening, because repeated `final' checks use up the separation.

The cost of this discipline is fewer tasks for routine iteration. On small evaluations, that can reduce power enough to leave the final comparison inconclusive. The remedy is not to recycle the iteration holdout during development. Acquire more representative tasks, reduce the number of configurations carried into the final comparison, or accept a wider confidence interval. The boundary protects the interpretation of the result, but it does not add information the sample does not contain.

Search budget belongs in the accounting as well. If one architecture received hundreds of prompt and scaffold trials while the baseline received a single default configuration, the final inference-cost plot omits much of the effort spent finding the winner. Development cost and serving cost answer different questions. They should therefore be reported separately. At minimum, preserve the number of configurations tried, the selection rule, and the compute spent before the iteration holdout evaluation.

The companion catalog carries six related procedures. They cover expected best performance by tuning budget, repeated-sampling budgets based on coverage and verifier error, frozen-memory evaluation, a pinned scoring target, human-verified synthesis of evaluation instances, and baseline gates against random selection. Each refines a particular part of the experiment. None replaces the direct-call, retry, ablation, cost, and iteration-holdout controls developed here.

\hypertarget{build-the-next-comparison}{%
\section{Build the next comparison}\label{build-the-next-comparison}}

Start the next architecture comparison with the inexpensive arms. Run the base model directly, then run it with one retry under a failure signal the deployed system can observe. These controls establish how much of the candidate's accuracy another sample produces on its own, before a planner, memory writer, retriever, or second agent enters the design.

For each added component, create an arm without it and preserve every surrounding choice the design permits. When memory is the treatment, add raw-trajectory reuse so the experiment separates stored experience from the mechanism that rewrites it. When retrieval or tools are the treatment, add a no-tool arm, match task coverage, and record actual usage. Change one factor at a time unless the experiment is explicitly designed to estimate interactions.

Pilot the tasks before treating them as trials. Inspect each treatment trajectory and verify that the component entered the computation. A task the treatment never touches may measure general capability, cost, or incidental interference, but it decides nothing about the component's contribution. Remove it from that effect estimate under a rule declared in advance, or report it in a separate stratum.

Carry the per-task record from Chapter 1 into the cost analysis. For every arm, preserve correctness, input tokens, output tokens, priced cache quantities, call count, component usage, and wall-clock latency. Convert usage to dollars with a dated pricing snapshot. Plot accuracy against cost with confidence intervals, and prefer a configuration only when no cheaper configuration matches or exceeds its accuracy within the resolution of the evaluation.

Reserve the iteration holdout before the first scaffold change. Decide who can inspect it, when it can be opened, which aggregate results will be returned, and what event ends the comparison. Once developers have used its failures to revise the system, it has become development material. The boundary cannot be created retroactively around tasks the team has already learned to optimize.

\section*{Sources and evidence}

\textbf{Run ablation controls}

\begin{itemize}
\tightlist
\item
  Directional evidence: \emph{SkillEvolBench} (Lei et al.~2026, arXiv:2605.24117), through the agentic-memory source synthesis. The underlying study measures no-skill and raw-trajectory controls in its setting; coverage matching, one-factor sweeps, and the writer information-flow constraint are transfers within the broader protocol.
\item
  Directional evidence: evaluation-design material from the code-retrieval source corpus. CoIR (Li et al.~2024, arXiv:2407.02883) supplies code-retrieval tasks and metrics, but it does not test the no-tool baseline, tool-access control, or ground-truth-tautology checks recommended here.
\item
  Directional evidence: \emph{TIAP} (arXiv:2605.24060); no figure carried.
\item
  Directional evidence: \emph{MemConflict} (arXiv:2605.20926); no figure carried.
\item
  Strong evidence for the narrower transition analysis: Tank, D., and Nama, B. (2026), ``The Regression Tax: Decomposing Why Skills Help and Hurt LLM Agents,'' arXiv:2607.22520. Nearly 6,000 runs across two office-automation benchmarks and three model-harness stacks; transfer to coding-agent skill libraries is directional.
\end{itemize}

\textbf{Report cost-accuracy tradeoffs}

\begin{itemize}
\tightlist
\item
  Directional evidence: Kapoor, Stroebl, Siegel, Nadgir, and Narayanan (2024), \emph{AI Agents That Matter}, arXiv:2407.01502, a re-evaluation study.
\item
  Directional evidence: the iteration-holdout section rests on the same study, carried in the companion catalog as a separate record on holdout design split by generalization level.
\end{itemize}

\textbf{Author-system illustration cited inline}

\begin{itemize}
\tightlist
\item
  Not an evidence item: CodeProbe, the author's task-mining evaluation tool, \href{https://github.com/sjarmak/codeprobe}{public repository}. Named inline for the end-to-end run and the score-per-dollar reporting described above, both of which are narrative illustration.
\end{itemize}

\chapter{Benchmark contamination, oracle strength, and workload validity}
\label{ch03-contamination-oracle-workload-validity}
\begin{quote}
\textbf{Evidence profile.} 7 strong \(\cdot\) 8 directional \(\cdot\) 1 corroborating evidence items across 4 developed practices (\erca{001}, \erca{003}, \erca{004}, \erca{066}).

\textbf{Chapter claim.} A passing score is only as valid as its workload, exposure boundary, and oracle.
\end{quote}

In February 2026, OpenAI (\href{https://openai.com/index/why-we-no-longer-evaluate-swe-bench-verified}{2026a}) stopped reporting results on a benchmark subset it had helped create. In July 2026, OpenAI (\href{https://openai.com/index/separating-signal-from-noise-coding-evaluations/}{2026b}) retracted its subsequent recommendation to use SWE-Bench Pro after an audit estimated that roughly 30 percent of its tasks were broken. SWE-bench Verified had become one of the most cited measures of coding-agent performance. Its 500 tasks had survived screening by 93 paid professional developers, a process intended to remove ambiguous issues and defective tests. The sponsor later reported that 59.4 percent of 138 audited tasks had flawed tests: about 82 tasks, or 16.4 percent of the full 500-task subset. The same audit showed frontier models reproducing solution details verbatim when given only a task identifier.

Human screening had improved the original collection, but it could not control what happened after publication. \textbf{Contamination} is exposure to evaluation material before testing. Once tasks, issue discussions, reference patches, tests, and leaderboard results are public, they can enter training corpora, retrieval systems, fine-tuning sets, prompt libraries, and repeated development cycles. A clean review at release establishes nothing about those later paths.

The retirement also exposed a separate problem. Some tasks rejected functionally valid patches because the tests required an undocumented implementation detail or checked behavior absent from the issue. Other tests were too permissive and accepted patches that did not implement the requested behavior. Human review had missed both kinds of defect, because task quality is conditional on the candidates being graded. A test that looks reasonable in isolation can fail when it meets an implementation its authors did not anticipate.

\begin{figure}[htbp]
\centering
\includegraphics{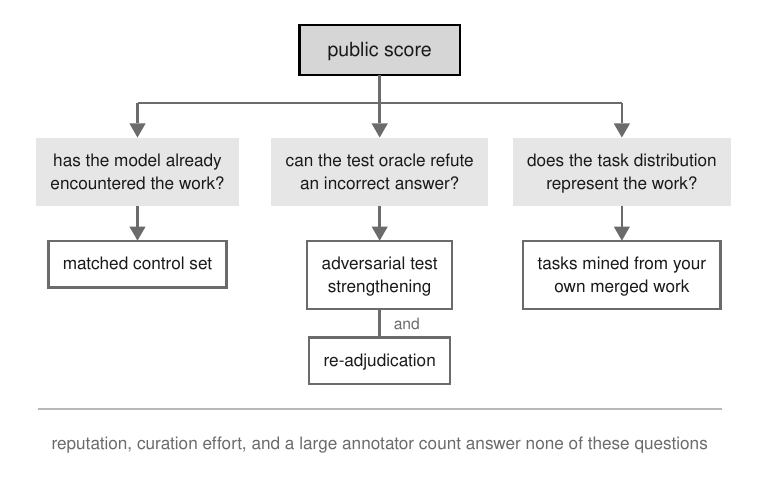}
\caption{A public score is the output of three coupled measurements. Each question has its own control, and reputation, curation effort, and a large annotator count answer none of them.}
\label{fig:ch03-score-controls}
\end{figure}

You should treat any public score as the output of three coupled measurements. First, has the model already encountered the work? Second, can the test oracle refute an incorrect answer? Third, does the task distribution represent the work someone wants the model to do? The sponsor's reversal is the demonstration that reputation, curation effort, and a large annotator count answer none of these questions. Each question needs its own control, and Figure \ref{fig:ch03-score-controls} pairs each with the control that answers it.

\hypertarget{measure-the-public-private-gap}{%
\section{Measure the public-private gap}\label{measure-the-public-private-gap}}

The first control is a point-in-time audit that any evaluator can commission. A \textbf{matched control set} reproduces the public suite's relevant difficulty characteristics using tasks the benchmark never released. Running both sets under the same protocol produces a separate public-private gap for each model.

Prathifkumar et al.~(\href{https://arxiv.org/abs/2512.10218}{2025}) conducted an observational comparison and found roughly a threefold overall advantage and a sixfold file-localization advantage on SWE-bench Verified relative to their control tasks. The comparison covered popular open-source Python projects, but it did not establish that the individual issues were equally difficult. Unresolved differences in task difficulty therefore remain a plausible explanation for part of the observed gap.

This audit provides a point-in-time measurement. The post-cutoff pipeline in the next section provides a standing control, avoiding the need to reconstruct the comparison each time the public benchmark ages.

An evaluator usually cannot inspect a model's complete training history. Providers rarely publish full training corpora, and even a searchable disclosed corpus would omit later fine-tuning, generated training examples, and indirect exposure through task discussions or benchmark-driven development. Asking whether a model has seen a particular item creates an attribution problem without reliable ground truth. Asking how its performance changes on comparable unseen work produces an observable difference.

The audit requires two task sets that differ in public exposure while matching as closely as possible on the features that determine difficulty. For repository work, those features include programming language, repository scale, issue format, expected patch size, dependency burden, and the amount of localization required before editing. The model, harness, prompt, tool access, retry policy, and scoring procedure remain fixed across both sets.

When each public task has a defensible counterpart, the audit should use a paired design so the analysis can compare outcomes within each matched pair.

For a given model, the primary estimate is:

\Needspace*{4\baselineskip}
\[
  \text{inflation gap}(m) = \text{score}(m, \text{public}) - \text{score}(m, \text{matched control})
\]

The gap estimates the combined effect of public exposure on the measured score. It cannot isolate a pure training-data effect. The estimate may include direct exposure to tasks or solutions, familiarity with repositories and issue conventions, and adaptation through repeated use of the public suite.

That final channel does not require any change to model weights. Each time developers evaluate an agent on a public suite, retain a change that raises its score, and discard one that lowers it, the development process adapts to the suite. The iteration holdout described in Chapter 2 protects a local development process from this form of selection. The matched comparison estimates the inflation that remains across all exposure channels together.

A single audit is also a single realization. The gap inherits the run-to-run variation described in Chapter 1, so both task sets require repeated runs before their difference can be interpreted as an estimate rather than one draw.

There are three practical ways to construct the comparison set. A \textbf{private mirror} reproduces the public suite's collection procedure using tasks that have never been released. A \textbf{newly collected set} samples recent work from the same task family. A \textbf{retroactive twin set} reconstructs tasks with the same measured properties, provided the model's training cutoff predates publication of the twins. Each supports a narrower claim than the phrase ``clean benchmark'' suggests.

A private mirror provides the strongest control over exposure, but it is expensive to construct. Repository tasks require reproducible code states, issue descriptions, reference changes, and tests that distinguish acceptable from unacceptable patches. The mirror loses its protection once it is released, distributed broadly to vendors, or used repeatedly during agent development. Access logs, handling rules, and usage history therefore become part of the measurement because the set's exposure status changes over time.

Newly collected tasks are less expensive when a team already has a stream of resolved work. Their recency makes prior exposure less likely, but it may also introduce a different mixture of repositories, framework generations, and issue types. A lower score may therefore reflect unfamiliarity, greater difficulty, or both.

Matching requires an explicit adequacy check. Similar averages are not enough. The evaluator should determine whether the feature distributions overlap, whether individual pairs are defensible in engineering terms, and whether the conclusion survives the removal of visibly poor matches.

Retroactive twins make the matching problem especially visible. Haimes et al.~(\href{https://arxiv.org/abs/2410.09247}{2024}) constructed twins for a public question-answering benchmark and required them to pass four statistical indistinguishability checks before comparison. Across 20 models, some public scores exceeded twin scores by as much as 16 percentage points. A substantial public-private gap can therefore remain even after formal matching checks are satisfied.

Those results do not establish a comparable gap for repository-scale coding. The twins were constructed and validated for question answering, and the method applies only when the model's training cutoff predates publication of the twin set. The 16-point result is the largest gap observed under that study's design. It is not a bound on coding benchmarks and should not be used as a correction factor for coding scores.

A controlled arithmetic study provides another boundary. Zhang et al.~(\href{https://arxiv.org/abs/2405.00332}{2024}) reconstructed a grade-school arithmetic benchmark as a private mirror and measured accuracy drops of up to eight percentage points. The model-level gap was also associated with the probability that a model would reproduce public items verbatim; the results section reports Spearman's rank correlation as \(\rho=0.36\) with \(p=0.03\).

Many frontier systems showed little overfitting, and every evaluated model generalized to the novel items. That variation is why the audit must be performed separately for each model. Exposure and score inflation are properties of a particular model-benchmark pair. Applying one average correction across models would erase the model-specific signal the audit is intended to measure.

The observational coding result is more directly relevant to software work but less controlled. Both sides used popular open-source Python projects, yet matching at that level does not establish equivalent issue difficulty. A model may localize files more easily in repositories it has encountered repeatedly, which is itself a meaningful familiarity effect. The observed performance difference may also include repository-specific complexity.

The reported threefold and sixfold gaps should therefore be treated as directional evidence, with the matching limitation attached whenever those figures are used.

The audit should produce a separate record for each model containing the public score, matched-control score, paired difference where pairing is justified, and uncertainty for each quantity. It should also document the task-matching procedure closely enough for another evaluator to challenge it. A single pooled gap is inadequate because models differ in both exposure and generalization. A corrected leaderboard is worse because it converts uncertain, model-specific estimates into false precision.

The decision depends on the size and stability of the measured gap. A small gap with a wide confidence interval means the audit lacked the resolution to establish much. A large gap that survives plausible rematching indicates that the public score is a poor estimate of that model's performance on comparable unseen tasks.

Neither result establishes how the model will perform on the reader's repositories. The audit answers a narrower question: whether public availability appears to change the measurement.

\hypertarget{keep-the-task-window-ahead-of-the-model}{%
\section{Keep the task window ahead of the model}\label{keep-the-task-window-ahead-of-the-model}}

A \textbf{temporal holdout} tags every task by publication time and evaluates a model only on work published after its stated training cutoff. This is the standing pipeline. The matched comparison in the previous section is the point-in-time audit, and a reader who adopts that audit alone has to keep commissioning new ones as the public set ages. The pipeline removes that repetition at a real cost. Longitudinal comparability weakens, because every time window contains different work.

The control removes the most direct path from a public task into pre-evaluation training. Suppose a model's stated cutoff is June 2025. An issue first published in August can enter its evaluation window, while an issue published in May cannot, even when both are assembled into the benchmark in September. The relevant date belongs to the underlying task material rather than to the benchmark's ingestion.

That distinction changes the data model for an evaluation suite. Each task needs provenance for its issue, repository state, tests, solution, and any discussion that reveals the solution. Each model needs a recorded cutoff and a policy for ambiguous or rolling cutoffs. Eligibility is then a function of both records:

\Needspace*{4\baselineskip}
\[
  \text{eligible}(t, m) \iff \texttt{first\_public\_at}(t) > \texttt{training\_cutoff}(m)
\]

The inequality is the easy part. \texttt{first\_public\_at} may refer to an issue that was discussed in a public chat before it reached the tracker, a security fix disclosed after private coordination, or a commit mirrored across several hosts. A benchmark maintainer must choose which event counts and retain enough provenance to revisit that choice. Cutoff claims pose a similar problem, because providers may disclose a date without specifying whether later supervised tuning, tool traces, or retrieval indexes include newer material.

Eligibility is also a property of a task-model pair rather than of a task set. A collection labeled post-cutoff without naming the model it was checked against establishes nothing about eligibility.

Controlled evidence from continuously collected coding problems shows why the date split is still useful. Jain et al.~(\href{https://arxiv.org/abs/2403.07974}{2024}) tagged problems by release date and found that some model families lost substantial performance on problems published after their cutoffs. The same time-windowed scoring exposed overfitting to an older function-level suite, because models that scored well on it dropped on a fresh distribution. The discontinuity is stronger evidence than a memorization probe alone, because it connects temporal eligibility to the score being interpreted. It remains conditional on the accuracy of the disclosed cutoff and on the comparability of adjacent task windows.

Repository-scale live collections provide directional support. Zhang et al.~(\href{https://arxiv.org/abs/2505.23419}{2025}) rebuilt the issue-to-environment recipe continuously, with a reproducible environment built per task, and reported that the same systems scored well below their results on a frozen suite. Badertdinov et al.~(\href{https://arxiv.org/abs/2505.20411}{2025}) ran a continuous collection pipeline with decontaminated evaluation and hedged that some models' frozen-suite scores may be inflated by contamination. Adamenko et al.~(\href{https://arxiv.org/abs/2507.11059}{2025}) described a continuously updated pipeline that draws on roughly 10,000 potential tasks and passes a small fraction of them through automated gates into the released benchmark. These results show that live evaluation is operationally possible and that frozen and fresh distributions can disagree. They do not isolate direct training exposure from changes in repositories, issue difficulty, or curation.

A standing pipeline has four recurring jobs: gather new work, reconstruct a reproducible environment, reject invalid tasks, and publish results by time window. Failure in any stage can look like a model failure. A missing dependency lowers scores without establishing anything about capability. A permissive filter raises scores by admitting trivial tasks. A delayed ingestion process can place nominally new work behind a model's actual exposure. Freshness is therefore a property of the whole collection process.

Automated gates make continuous collection affordable, but they are less discriminating than expert review. A window of a few hundred accepted tasks also produces wider uncertainty than a large fixed suite, especially when results are split by language or task type. The temptation is to combine windows until the interval narrows. Combining them can cross a model cutoff or merge meaningfully different task distributions, which removes the temporal property that justified the pipeline.

The longitudinal trade is unavoidable. If Model A is evaluated in March and Model B in September, their eligible sets differ. If one fixed set is retained for both, it stops being post-cutoff for the later model. Rolling-window results should be the ones reported for current validity, and a fixed suite belongs beside them only when a trend line is necessary. The fixed series answers how systems change on one historical instrument. The rolling series answers how they perform on currently eligible work.

Comparability needs its own audit. A dynamic benchmark introduces generators, source selection, filters, environment builders, and a refresh cadence, and each of those can change the score independently of the model. Chen et al.~(\href{https://arxiv.org/abs/2502.17521}{2025}) surveyed the shift from static to dynamic evaluation, identified the absence of standardized criteria for judging a dynamic benchmark, and proposed design principles for building one. Their principles are prescriptive rather than tested at scale. Three inspection questions follow from them for a reader assessing someone else's live result: how the pipeline enforces cutoff hygiene, what quality control runs on regenerated tasks, and how comparability across snapshots is accounted for. The word `live' supplies no evidence about any of those mechanisms.

Temporal separation has two further limits. A model provider may have private prerelease access to material that becomes public later. The public timestamp can therefore follow actual exposure. Once a live window is released, later models can train on it and the window becomes historical. The pipeline preserves a moving boundary by continuing to collect tasks and applying eligibility per model. A benchmark does not stay clean permanently.

In one of my evaluation frameworks, a pre-cutoff and post-cutoff split is written into every result record, keyed to repository creation date and model cutoff, with a warning above a five-percentage-point gap. Repository creation date is a coarser field than the eligibility rule above requires, because a repository can be created years before the issue that becomes the task. This illustrates a methodology and supplies no evidence that the threshold detects exposure. The warning has fired only on replayed fixtures and never on live data. Retaining the split permits recomputation when better dates or cutoff information arrive, while a single aggregate would require reconstruction.

The companion catalog extends these two controls without turning them into one omnibus procedure. Its related entries cover memorization probes, semantic-overlap searches across paraphrases and translations, per-model and per-benchmark audits, and detector combinations chosen for a stated threat model. Other entries address anti-searchable task construction, protecting evaluation data at release, and attaching measured score benefit to a detector's finding. That last proposal has thin support. It remains a research direction and supplies no correction factor.

\hypertarget{make-passing-patches-harder-to-preserve}{%
\section{Make passing patches harder to preserve}\label{make-passing-patches-harder-to-preserve}}

A passing patch can fail to support the capability claim for three independent reasons, and they do not all fail in the same way. Training contamination means the model encountered evaluation material before the run. Solution leakage means the benchmark item included its answer during the run, in the issue text or in the comments. Those two can produce a correct patch while removing the grounds for attributing it to independent problem-solving. A weak oracle is the third defect, in which the tests cannot refute an incorrect patch, and only this one lets a wrong patch pass. The remedies are respectively temporal or access controls, removal of answer-bearing task material, and stronger tests.

All three failures can produce a successful-looking transcript. A model may reproduce a reference patch it saw during training, copy an answer supplied in the prompt, or find a shortcut that satisfies shallow assertions. The final test process records the same state in each case: pass. Transcript inspection and contamination probes cannot compensate for a test suite that accepts wrong behavior, and added tests cannot establish whether the model recalled the right behavior.

\begin{figure}[htbp]
\centering
\includegraphics[width=1\textwidth,height=\textheight,keepaspectratio]{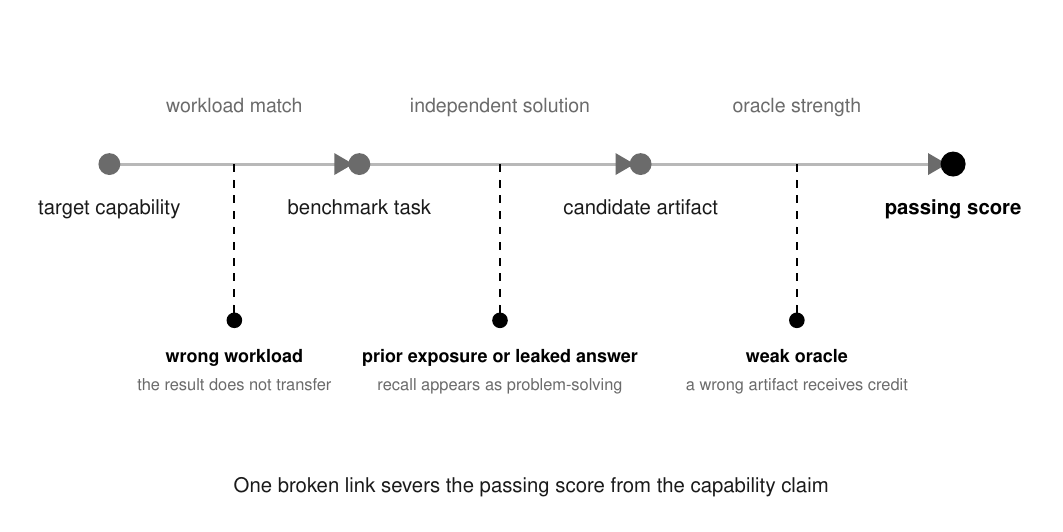}
\caption{A passing score supports a capability claim only when the benchmark matches the target workload, the solution is independent of prior or in-task exposure, and the oracle rejects incorrect artifacts. A failure at any link invalidates a different inference.}
\end{figure}

Shao et al.~(\href{https://arxiv.org/abs/2607.22368}{2026}) make this broader condition explicit as protocol validity: the intended capability must remain necessary for success under the evaluation protocol. Their HackDetect audit covered 2,385 traces across 15 agent benchmarks and found exposures or reward hacking in 67.0 percent of Frontier Science traces and 66.7 percent of AutoLab tasks; paired audits measured score inflation of 0.45 to 1.00 on their Mislead-gap scale. These are strong measurements for the audited protocols, not prevalence estimates for agent benchmarks as a whole.

A passing patch is therefore best treated as a hypothesis about correctness. The test oracle is the ordinary continuous-integration mechanism that decides whether the candidate passes or fails. Its strength is the range of plausible wrong behavior it can reject. A suite is stronger when it checks more of the behavior implied by the task, especially behavior that superficially reasonable patches get wrong.

Function-level code generation provides a controlled example. Liu et al.~(\href{https://arxiv.org/abs/2305.01210}{2023}) expanded a widely used suite's tests by 80x with generated inputs and mutation-based cases. Across 26 models, the largest relative reductions in \(\mathrm{pass}@k\) ranged from roughly 19 to 29 percent across the reported values of \(k\), and model rankings changed. The models and generated programs were fixed; only the observations used to decide correctness changed. The original ranking had partly measured which wrong programs happened to fit sparse tests.

Repository-scale results show the same mechanism under more realistic state. Yu et al.~(\href{https://arxiv.org/abs/2506.09289}{2025}) augmented SWE-bench tests retroactively and found 36 under-tested tasks and 345 patches that had been labeled as passing incorrectly. Re-adjudication corrected 40.9 percent of entries on the smaller suite and 24.4 percent on the human-screened suite, changing 29 rankings. This was an audit of already reported results, so oracle weakness had propagated beyond individual tasks into system comparisons.

Task construction can fail before the oracle runs. Wang, Xu, and He (\href{https://arxiv.org/abs/2607.28587}{2026}) audited SWE-bench Verified and found PR-issue misalignment in 13.6 percent of instances across five patterns and eleven scenarios. Their PAIChecker reached up to 92.12 percent binary accuracy on SWE-Gym and 91.67 percent on SWE-bench Multilingual across four model backbones. The results strongly support the measured misalignment rate and detector accuracy on those collections; other SWE-bench-derived sets require their own audit.

The problem predates current coding models. Ye et al.~(\href{https://arxiv.org/abs/1909.13694}{2019}) generated random tests against a human reference patch and assessed 638 candidate patches from 14 repair systems. The added oracle improved automatic patch assessment by 190 percent over the prior methods used in that comparison. The historical context is useful because it locates the defect in test-based program repair itself. Generative models make candidate production faster and more varied, but shipped tests are not equivalent to a specification.

Adversarial strengthening begins from the candidate failures a benchmark should reject. Construct plausible patches that implement only the common case, hard-code a fixture, change an unrelated return value, suppress an exception, or satisfy the visible assertion while violating the issue's broader contract. Then retain the tests that kill those patches. Yu et al.~(\href{https://arxiv.org/abs/2603.00520}{2026}) built suites that way, rejected 19.71 percent of patches that had passed previously, and reported a top score falling from 78.80 percent to 62.20 percent.

Four audits test apparent benchmark success under stronger oracles. SWE-Bench+ covers one agent-model pair, so its percentages remain specific to that evaluation.

This process searches a different space from ordinary coverage expansion. Generated inputs explore more executions of a trusted implementation. Mutation-based cases perturb code to reveal assertions that fail to notice behavioral changes. Differential tests run the candidate and a trusted reference on the same inputs, then compare the results. Wrong-patch construction starts from foreseeable shortcuts and asks whether the suite detects them. A strong audit combines these approaches, because each makes a different assumption about where missing behavior resides.

The trusted reference is the central dependency. Random or generated tests need expected outputs, and differential testing needs behavior worth treating as ground truth. A human patch can contain unrelated changes or encode only one acceptable design. When the issue permits several implementations, exact agreement with that patch can reject valid alternatives. The evaluator must distinguish behavioral equivalence from textual or structural similarity, while accepting that some behavior remains underspecified.

Manual review addresses failures that generated tests are unlikely to expose. Aleithan et al.~(\href{https://arxiv.org/abs/2410.06992}{2024}) screened the apparent SWE-bench successes of one agent and model pair and found that 32.67 percent involved solution leakage and 31.08 percent passed weak tests. Removing those cases dropped the reported resolution rate from 12.47 percent to 3.97 percent. The scope is one pair, and the percentages do not estimate the prevalence for other agents or models. They show that leakage and oracle weakness can occupy a large share of one system's credited successes.

That study also clarifies why solution leakage belongs here. An issue may include the intended algorithm, a maintainer comment with the decisive condition, or a link to the merged fix. The model has legitimate access to that content during evaluation, so a training cutoff cannot remove it. If the deployment workload includes equally explicit issue discussions, retaining the material may be valid. If the benchmark claim concerns inferring fixes from ordinary bug reports, the leaked solution changes the task being measured.

In my own audit of an enterprise-scale benchmark, I scored a null agent that merely repeated each instruction against every gradeable task. Some verifiers awarded the echo credit. One gave full marks because its checker searched for vocabulary that appeared in the prompt itself. An agent with no capability at all is a cheap adversary, and it found real defects. The audit supplies no capability estimate, but it demonstrates a useful attack. Submit candidates that contain task language without task behavior, then inspect every nonzero grade.

Re-adjudication is the step that makes a strengthened oracle consequential. Run the stronger suite on stored patches from every system under comparison, including older submissions. Recompute \(\mathrm{pass}@k\), rankings, paired differences, and any reward or release gate derived from the old labels. That recomputation has to start from the stored per-attempt outcomes rather than from a rescaling of the published rate, because \(\mathrm{pass}@k\) is estimated from attempts.

Report the survival fraction, the number of tasks whose adjudication changed, and the uncertainty remaining after invalid tasks are removed. A suite that changes only one system's labels may expose a system-specific shortcut. Broad changes point toward a benchmark-level defect.

Additional tests provide one-sided evidence. They can disprove more candidates by finding counterexamples. A surviving patch has resisted the tests run so far, but it has not been certified correct for every valid input and environment. That boundary is tighter for repository tasks, because their behavior depends on configuration, dependency versions, persistent state, concurrency, and interactions outside the changed function.

Strengthening also has a coverage boundary. One of these efforts strengthened only about half of its benchmark instances. Its corrected rate remains an upper bound, because weak passes may persist in the untouched half and in behaviors the new tests still omit. Manual review is expensive, generation can reproduce assumptions embedded in the reference, and an adversarial patch set can miss shortcuts no one imagined.

The practical stopping rule follows the decision being made. For a research comparison, sample-based manual review and broad generated augmentation may bound the inflation well enough to qualify the claim. A release gate on the reader's own repository requires closer alignment with the actual acceptance criteria and failure costs. In both settings, the evaluator should publish how many old passes survived. Reporting only the strengthened score hides the size and the location of the correction.

\hypertarget{move-the-instrument-onto-your-work}{%
\section{Move the instrument onto your work}\label{move-the-instrument-onto-your-work}}

When a public ranking conflicts with sustained field behavior, the first question is whether the two measurements represent the same work. The evidence behind this practice is the thinnest in the chapter. It rests on one directional, practitioner-authored production benchmark and one anecdotal practitioner account, with no strong item standing behind it. The prescription asks the reader to build and measure on the relevant workload. It supplies no vendor score to adopt.

Construct validity asks whether an instrument measures the property named in the claim. Bean et al.~(\href{https://arxiv.org/abs/2511.04703}{2025}) used that question as the checklist for a systematic review of 445 benchmarks. A coding benchmark can measure repository issue resolution while offering weak evidence about migration work, security remediation, build repair, or long-running feature development. Passing tests establish performance on the sampled tasks under the specified harness. Generalizing that result to another workload requires a reason to believe the task distributions and operating conditions overlap.

Public repository suites usually select work that can be reconstructed from visible artifacts. This favors languages with reproducible environments, projects with mature tests, issues linked cleanly to merged changes, and tasks short enough to run economically. A production workload may contain private dependencies, sparse tests, organization-specific conventions, partial requirements, abandoned attempts, incident response, and changes whose value appears weeks later. The public suite can be carefully built and still omit the state transitions that dominate the deployment.

Scale and structure carry the same risk as content. A suite can sample the right kind of work and still misrepresent its shape. Most public repository benchmarks place a task inside one repository of moderate size, with the relevant code reachable by reading a manageable part of the tree. Organizational work often begins in a service repository, depends on an interface defined in a second, is constrained by a schema owned by a third, and requires a migration in a fourth. Repository size changes what search can accomplish, since a codebase of tens of millions of lines cannot be read exhaustively at any context length. Ownership and access boundaries change what one identity can see, which makes retrieval quality and permission scope the same measurement.

A resolve rate measured on single-repository tasks estimates single-repository performance. Cross-repository dependency tracing, ecosystem migration, and incident triage that starts in one service and ends in another have a different dominant failure, which is locating the relevant code rather than editing it once located. Sadowski et al.~(\href{https://doi.org/10.1145/2786805.2786855}{2015}) measured the human form of that activity at Google through a survey and search-log analysis, finding developers conducting an average of five search sessions with twelve queries each workday, usually targeted at a particular code location and aimed at how to use an interface, what code does, why something fails, or where code lives. That study describes people rather than agents, and one company in 2015, so it supports treating location as a distinct workload component rather than supplying any rate for agents. A benchmark meant to support claims about organizational work should record, per task, how many repositories the task spans, the size of each, and whether the required evidence sits inside or outside the repository named in the instruction. Chapter 12 reports an evaluation built to that specification.

Construction should start with the work product and trace backward. For a coding assistant, candidate tasks include production prompts, accepted changes, review outcomes, and the tests associated with completed work. The evaluation record should preserve the repository state available at task start, the instruction as received, the permitted context and tools, and the evidence that caused the organization to accept the result. Other domains need their own decision sequence and artifacts. A general reasoning proxy omits that structure.

Jha et al.~(\href{https://arxiv.org/abs/2604.01527}{2026}) demonstrated one curation approach for a production-derived benchmark. They built tasks from sessions with their own coding assistant, classified the work, checked that tests were relevant to each task, and required stability across repeated runs. This is directional evidence from one vendor's environment. It shows that production traces can be converted into repeatable tasks. It does not establish that the resulting distribution represents another organization, or even every kind of work inside the source organization.

The anecdotal account identifies a transfer failure worth testing. Bytesfortruth (\href{https://www.reddit.com/r/compsci/comments/1rqcmu8/}{2026}), a lending-domain practitioner, reported that a model near the 90th percentile on a general-reasoning measure failed basic mortgage-underwriting tasks, and that rankings changed considerably on a domain-lifecycle evaluation. The account covers one team and reports no failure frequencies. It establishes one transfer failure and supplies no population rate for domain mismatch.

The task-sampling frame should follow the deployment claim. A tool meant to fix failing tests needs a distribution of actual failures, including the hard and unresolved ones. A tool meant to draft small maintenance changes needs accepted small changes plus representative rejections. Mixing those claims into one private score recreates the ambiguity of a general leaderboard under local ownership.

In my own work, I built CodeProbe (\href{https://github.com/sjarmak/codeprobe}{public repository}) to mine evaluation tasks from a repository's merged pull requests. It keeps the instruction separate from the recorded ground-truth commit and uses the test files touched by the original change as the verification command. That design turns completed work into replayable cases without exposing the recorded solution to the candidate. It is a methodology illustration and carries the same oracle caveat as any test-derived benchmark. The tests a merged change happened to touch are candidate historical checks, not a certification that the recorded solution was correct.

Merged history is a selective record. Language filters and minimum-file thresholds exclude some changes before sampling begins. Requiring touched tests favors well-instrumented code, while abandoned work and incidents may leave no merge to mine. The resulting set is closer to the repository's accepted work by construction, but its representativeness remains bounded by what the workflow recorded and what the miner can reconstruct.

Consent and data handling also constrain the instrument. Production prompts can contain customer data, credentials, internal incidents, or employee-authored material collected for a different purpose. Redaction may remove the context that made a task difficult. Broad access may create a new exposure path. A private evaluation program therefore needs explicit collection authority, access controls, retention rules, and a record of who has seen each task.

Long-horizon outcomes create a different measurement problem. A patch can pass today and produce maintenance cost later, and a generated migration can complete while leaving operational cleanup for another team. These labels arrive slowly and contain organizational noise. Pairing offline task results with later online outcomes helps test whether the local instrument predicts the field, though sparse failures make tail estimates unstable.

When public and local results disagree, the instrument whose construction supports the deployment claim is the better guide. That ordering is conditional. A small private set with weak tests can be less informative than a mature public suite. A small private set also inherits the resolution limits from Chapter 1, and with a few dozen tasks, differences smaller than its minimum detectable effect are not reliably distinguishable from noise. The local set earns priority by preserving relevant work and surviving repeated measurement. Privacy alone does not add validity.

A locally mined task set is also the raw material for the release instrument in Chapter 4, which curates a subset of these tasks into a golden set, replays it for each release, and attaches an executable gate.

The companion catalog carries the remaining validity checks. They include real temporal holdouts for agent benchmarks, pinned evaluation apparatus and published scored artifacts, ranking stress tests under perturbation, invalid-item purges, and audits of leaderboard submission protocols and questionable research practices. It also includes construct-validity checks, code-benchmark review against defect taxonomies, searches for an adequate benchmark before building one, and paired offline and online evaluation. The proposals on score-anchored contamination metrics, submission-protocol audits, and defect-taxonomy vetting have thin support and supply no evidence for any claim in this chapter.

\hypertarget{discount-the-next-public-number}{%
\section{Discount the next public number}\label{discount-the-next-public-number}}

The next time you encounter a public benchmark number, identify the answers to these 4 questions before using it.

\begin{enumerate}
\def\labelenumi{\arabic{enumi}.}
\tightlist
\item
  How old are the tasks relative to this model's training cutoff? Look for a public-versus-matched gap measured on this model. Do not substitute the leaderboard average or another model's audit.
\item
  How strong is the test oracle? Ask whether anyone augmented the tests, constructed adversarial wrong patches, or manually screened apparent successes. The useful result is the fraction of previous passes that survived.
\item
  Who produced the number, on which harness, with what task distribution? Compare the languages, prompt forms, repository structures, time horizons, context, and acceptance criteria against the work you need done.
\item
  What does the same protocol report on tasks mined from your team's merged work? Preserve the original instruction, pre-change state, acceptance evidence, and repeated-run uncertainty.
\end{enumerate}

These answers do not yield a universal correction formula. They turn one impressive-looking score into a bounded claim: performance by a named system, under a named apparatus, on work with a known exposure history, judged by tests with observed power to reject wrong patches. The local replay tests that claim on the work the system will actually be asked to do.

\section*{Sources and evidence}

\textbf{Opening scene}

\begin{itemize}
\tightlist
\item
  OpenAI, ``Why SWE-bench Verified no longer measures frontier coding capabilities,'' 2026-02-23, \url{https://openai.com/index/why-we-no-longer-evaluate-swe-bench-verified}. {[}Corroborating evidence{]}
\item
  OpenAI, ``Separating signal from noise in coding evaluations,'' 2026-07-08, \url{https://openai.com/index/separating-signal-from-noise-coding-evaluations/}. {[}Corroborating evidence{]}
\item
  SWE-bench Verified card: 500 instances, 93 professional screeners. {[}Benchmark record{]}
\end{itemize}

\textbf{Measure public-score inflation with matched controls}

\begin{itemize}
\tightlist
\item
  Prathifkumar, Saji Mathews \& Nagappan (2025), arXiv:2512.10218, Waterloo. {[}Directional evidence{]}
\item
  Zhang, H. et al.~(2024), GSM1k, arXiv:2405.00332, Scale AI. {[}Strong evidence{]}
\item
  Haimes, Wenner et al.~(2024), arXiv:2410.09247, Apart Research. {[}Strong evidence{]}
\item
  Corroboration: none on record.
\end{itemize}

\textbf{Evaluate on post-cutoff tasks}

\begin{itemize}
\tightlist
\item
  Jain et al.~(2024), arXiv:2403.07974, LiveCodeBench, ICLR 2025. {[}Strong evidence{]}
\item
  Zhang, L. et al.~(2025), SWE-bench Goes Live!, arXiv:2505.23419, Microsoft Research. {[}Directional evidence{]}
\item
  Adamenko et al.~(2025), arXiv:2507.11059, SWE-MERA. {[}Directional evidence{]}
\item
  Badertdinov et al.~(2025), arXiv:2505.20411, SWE-rebench, Nebius. {[}Directional evidence{]}
\item
  Chen, S. et al.~(2025), arXiv:2502.17521, static-to-dynamic benchmark survey. {[}Directional evidence{]}
\end{itemize}

\textbf{Strengthen test oracles before adjudication}

\begin{itemize}
\tightlist
\item
  Liu, Xia, Wang \& Zhang (2023), arXiv:2305.01210, HumanEval+ / EvalPlus, NeurIPS 2023. {[}Strong evidence{]}
\item
  Yu, S. et al.~(2025), arXiv:2506.09289, UTBoost. {[}Strong evidence{]}
\item
  Ye, Martinez \& Monperrus (2019), arXiv:1909.13694. {[}Strong evidence{]}
\item
  Aleithan et al.~(2024), arXiv:2410.06992, SWE-Bench+. {[}Directional evidence{]}
\item
  Yu, B. et al.~(2026), arXiv:2603.00520, SWE-ABS. {[}Strong evidence{]}
\item
  Strong evidence for audited protocol defects and paired score inflation: Shao, J., Chen, H., Zhang, W., Pan, M., and Luo, B. (2026), ``Do Agent Benchmarks Measure Capability? Protocol Validity in the Age of Agentic AI,'' arXiv:2607.22368. The reported rates apply to the audited protocols, not all agent benchmarks.
\item
  Strong evidence for benchmark-specific PR-issue misalignment and detector accuracy: Wang, M., Xu, J., and He, P. (2026), ``PAIChecker: Uncovering and Checking PR-Issue Misalignment in SWE-Bench-Like Benchmarks,'' arXiv:2607.28587.
\end{itemize}

\textbf{Benchmark on your own workload}

\begin{itemize}
\tightlist
\item
  Jha, Paltenghi, Maddila, Murali, Ugare, and Chandra (2026), \emph{REAP: Automatic Curation of Coding Agent Benchmarks from Interactive Production Usage}, arXiv:2604.01527; the resulting benchmark is Harvest. {[}Directional evidence{]} The title and benchmark name were verified against the arXiv record on 2026-07-29; earlier working records used the name ProdCodeBench.
\item
  Bytesfortruth (2026), lending-domain benchmark account, r/compsci, 2026-03-10, \url{https://www.reddit.com/r/compsci/comments/1rqcmu8/}. {[}Corroborating case{]}
\item
  Bean, A.M. et al.~(2025), Measuring what Matters, arXiv:2511.04703. {[}Directional evidence{]} Carried in the companion catalog as the construct-validity record; named inline for the construct-validity definition that opens this section.
\item
  Sadowski, C., Stolee, K.T., Elbaum, S. (2015), \emph{How Developers Search for Code: A Case Study}, ESEC/FSE 2015, \href{https://doi.org/10.1145/2786805.2786855}{DOI:10.1145/2786805.2786855}. {[}Directional evidence{]} Establishes code location as a distinct, high-frequency activity in a large codebase. It measures human developers at one company in 2015 and supplies no rate for agents.
\end{itemize}

\textbf{Author-system illustration cited inline}

\begin{itemize}
\tightlist
\item
  Not an evidence item: CodeProbe, the author's task-mining evaluation tool, \href{https://github.com/sjarmak/codeprobe}{public repository}. Named inline for the merged-pull-request mining design described above, which is methodology illustration.
\end{itemize}

\part{Evaluation and grading systems}
\gdef\currentparttitle{Part II: Grading}

\chapter{Execution-based evaluation, correction gates, and release tests}
\label{ch04-execution-correction-gates-release-tests}
\begin{quote}
\textbf{Evidence profile.} 3 strong \(\cdot\) 4 directional \(\cdot\) 0 corroborating evidence items across 3 developed practices (\erca{061}, \erca{094}, \erca{095}).

\textbf{Chapter claim.} Execution decides whether work moves.
\end{quote}

Mehta (\href{https://arxiv.org/abs/2603.25764}{2026}) analyzed 1,750 coding-agent trajectories across 50 tasks and four models and found a sharp divide between submission and correctness. One model submitted an answer in every trial, yet an external oracle found that it resolved only 44 percent of the tasks. Its semantic failures were often silent, confident, and consistent across repeated runs. Most models also modified code that was already correct.

The study provides directional evidence rather than a general estimate of coding-agent failure rates. It covers one author's analysis, a limited task set, and four models, without a controlled design for estimating prevalence across agents or workloads.

\begin{figure}[htbp]
\centering
\includegraphics{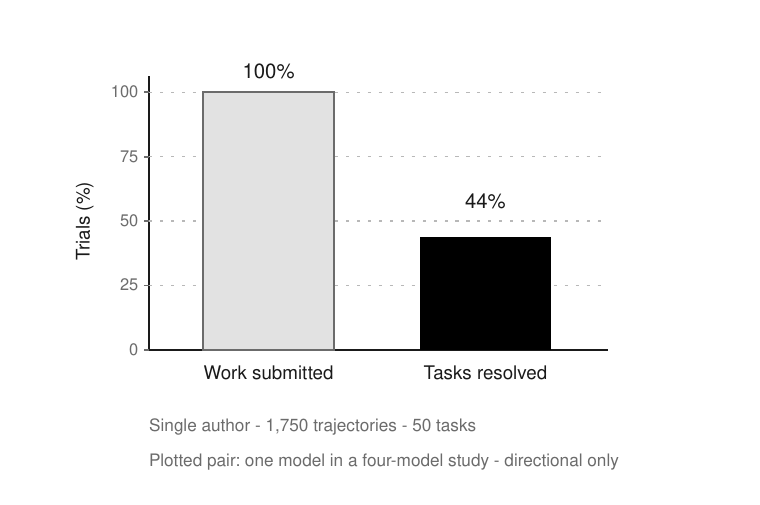}
\caption{Across 1,750 trajectories covering 50 tasks and four models, one model submitted work in 100\% of its trials but resolved only 44\% under an external oracle.}
\end{figure}

Submission, consistency, and self-assessment are all produced by the process under evaluation. A model can repeatedly generate the same incorrect patch, describe it with stable confidence, and end each run cleanly. Those signals characterize the trajectory, but they do not show that the repository moved from a failing state to a working one. Acceptance therefore requires evidence of an observable state transition, verified by a system outside the inference process that proposed the change.

The evidence base for this chapter is thin. The actions below therefore ask readers to execute and measure work on their own systems rather than adopt a reported threshold. Six evidence items support the three practices developed here. Five are source syntheses, and one is a controlled experiment. One entry has no strong supporting evidence item.

The practical unit is a gate that can stop candidate work from propagating. The candidate runs under explicit constraints, any correction attempt receives the evidence produced by verification, and a compact set of representative tasks is replayed whenever the system changes.

Four adjacent controls appear in the companion catalog rather than in this chapter. For an agent that submits on nearly every trial, score verified resolution separately from submission and include tasks that test whether it can abstain. For a tool that can fail while returning a plausible success string, detect silent tool errors explicitly. For a loop that can retry indefinitely, impose a turn budget. When per-candidate verification cost determines how often a gate can run, include that cost in the system's accounting.

\hypertarget{make-execution-decide-whether-work-moves}{%
\section{Make execution decide whether work moves}\label{make-execution-decide-whether-work-moves}}

Two directional synthesis items support execution-gated evaluation, but neither is a controlled estimate of its effect. AgentForge, from Kumar et al.~(\href{https://arxiv.org/abs/2604.13120}{2026}), describes an evaluator that runs candidate work in resource-bounded, network-isolated sandboxes and permits propagation only after successful execution. SWE-bench, from Jimenez et al.~(\href{https://arxiv.org/abs/2310.06770}{2023}), establishes executable repository repair as an evaluation form.

The AgentForge preprint is unrefereed, evaluates a single configuration, and reports one sample from each agent. Its headline result also conflicts with the published range for that configuration and has not been independently replicated, so I omit the number. These limits prevent either source from establishing a general failure rate or a measured advantage for execution gating. They do not weaken the mechanism worth testing: run the artifact in a constrained environment, and let the observed result determine whether it can proceed.

A plausible patch is only text until the repository accepts it. It may fail to compile, violate a type constraint, pass a visible test while failing the broader suite, produce the wrong output, or depend on undeclared state left in the workspace. Reading the patch can reveal some of these defects. Executing it produces evidence from the system the change is supposed to affect.

This changes the acceptance question. A confidence score asks the process that produced an answer to characterize its own answer. An execution check asks a compiler, test runner, package builder, schema validator, or deployment probe whether a specified transition occurred.

Park and Choi (\href{https://arxiv.org/abs/2607.25152}{2026}) held an agent and its tool surface fixed while changing the evaluator's information channel. Across 54 cycles, the agent claimed improvement every time, although 56 percent of measured deltas were zero or negative; the self-verdict gate accepted every cycle and eroded the best reached state by 19 percent. On a boundary task whose success was verifiable from the artifact, the same judge's gap disappeared. This is strong evidence within the preregistered testbed that a gate must observe the state in which success is defined.

The result is an oracle only for the behavior the check observes. A passing unit test does not establish safe deployment, and a successful build does not establish semantic correctness. Its evidentiary value comes from being produced causally downstream of the candidate artifact rather than by the process that proposed it.

\Needspace{5\baselineskip}

The architecture has three owners:

\begin{itemize}
\tightlist
\item
  The agent owns the proposed change.
\item
  A sandbox controller owns the execution environment and the authority to start a run.
\item
  A release controller owns downstream state and changes it only after the sandbox returns an admissible result.
\end{itemize}

Separating these roles prevents the producer from turning a claim of success into release state by writing a status field, omitting failed checks, or selecting which results to report.

A sandbox needs enough isolation for its verdict to be interpretable. Each run should begin from a known repository snapshot and receive only the inputs permitted by the task. Bound wall-clock time, CPU, memory, process count, and disk use. Deny network access unless the task explicitly requires a named endpoint. Capture standard output, standard error, exit status, resource termination, and declared artifacts, then dispose of the environment after the verdict.

Reusing a mutable workspace is cheaper, but it allows one attempt to seed the next through generated files, caches, installed dependencies, or surviving processes. The later result then reflects both the current candidate and undeclared state inherited from earlier attempts.

Network isolation serves measurement as well as security. An unrestricted candidate can fetch an undeclared dependency, consult a changing service, upload material, or pass because a remote cache happens to contain the required state. When a task genuinely depends on a network service, expose a controlled substitute or record the service version and responses. Otherwise identical artifacts can receive different verdicts for reasons unrelated to their behavior.

Resource bounds make nontermination an explicit outcome. A process killed after exceeding a declared limit should be recorded separately from a failed assertion or an ordinary completion. The gate should identify which bound fired. Parallel tests may exhaust memory, while a surviving child process may trigger the process-count or wall-clock limit.

The acceptance contract should be readable without inspecting the controller:

\Needspace*{10\baselineskip}
\begin{record}
\begin{recordbody}
command:       verify-change
expected:      exit status 0
required:      reports/test-results.json
               build/package.tar.gz
artifact rule: each required file exists and is non-empty
timeout:       12 minutes
network:       denied
\end{recordbody}
\end{record}

The task definition fixes the command before the candidate encounters a failure. The expected condition names something more specific than ``works correctly.'' The artifact rules remove another common ambiguity: a wrapper can exit successfully after skipping the operation that should have produced the evidence.

Exit status alone is often too weak. A test runner can return zero after discovering no tests. A report generator can create an empty file. A shell pipeline can discard the failure status of an earlier command. A zero exit status shows that the final process ended successfully, not necessarily that the intended work occurred.

The controller should therefore validate the structure on which the verdict depends. Depending on the task, that may mean checking for an expected test count, a parseable result file, a package containing required members, or a deployment probe tied to the candidate version. These are mechanical checks. They do not require the controller to infer the patch's semantics.

Environment failure must remain distinct from candidate failure. If the sandbox image cannot be pulled, the verifier is absent, or the runner loses its workspace, marking the candidate incorrect corrupts the evaluation. Allowing it to pass because verification could not start corrupts release state. The correct outcome is an infrastructure error that blocks propagation and may be retried without crediting or blaming the candidate.

Infrastructure errors should still be counted. A run set repeatedly thinned by environment failures no longer represents the intended sample of attempts, even when those failures are excluded from candidate scores.

State should move monotonically through the gate. A candidate begins unverified, receives an immutable run record, and becomes eligible for the next stage only when that record satisfies the contract. Any modification creates a new candidate identity and invalidates the earlier verdict. When a verdict is keyed only to a branch name or task identifier, changed bytes can inherit evidence produced for different work.

In my own agent-workflow system, I reject any acceptance criterion that says only that a result ``works correctly.'' The rejection occurs during decomposition, before work begins. Each criterion must name a command and an exit condition. At the end of every turn, a lifecycle check also verifies that declared output files exist and contain data.

That lifecycle check runs deterministically, but a model evaluates the filesystem and provides only advisory judgment. Enforced refusal requires a command hook whose nonzero exit ends the turn. The example shows how a gate can begin in the task contract, but it does not provide evidence for the general claim.

My repository-scale benchmark suite uses a deterministic verifier as the primary scorer for every task. Additional judgment layers can flag suspicious trajectories, but they cannot replace or override the execution verdict. This preserves one stable state transition even when the optional analysis changes.

Execution is not suitable for every task. Architecture proposals, interface critiques, threat models, and underspecified behavioral changes may have no executable oracle. Those tasks require a separate, calibrated judgment lane, which Chapter 5 develops.

Executable checks also have a coverage boundary. They can establish that a candidate compiled, passed a selected suite, produced required artifacts, and stayed within its resource envelope. They cannot establish that untested inputs, longer operating periods, or a different production topology will behave the same way.

Record only the narrowest downstream claim justified by the run: test-eligible, build-eligible, or deployment-eligible.

The run record also provides the evidence needed for correction. A failed command, assertion diff, resource termination, or missing artifact gives the next attempt information that was unavailable when the candidate was produced. Without new evidence, another model pass is repetition presented as diagnosis.

\hypertarget{let-failed-runs-change-the-next-attempt}{%
\section{Let failed runs change the next attempt}\label{let-failed-runs-change-the-next-attempt}}

A retry that begins with the same prompt, evidence, and workspace state has learned nothing about the failure. Sampling may produce a different answer, but it does not direct the model toward the error. The next pass may repeat the faulty premise, replace a correct intermediate step, or elaborate the same mistaken conclusion. Calling that pass ``reflection'' does not add information.

Huang et al.~(\href{https://arxiv.org/abs/2310.01798}{2023}) evaluated intrinsic self-correction on reasoning tasks using 2023-era models. Self-correction often failed and sometimes reduced performance, while reliable external feedback improved it. This is strong evidence for the systems studied, not an estimate of every current coding system. Newer models may change the magnitude of the effect.

The operational default should therefore be asymmetric: authorize another attempt when an external observation changes the available evidence, not merely because the previous attempt failed.

An external observation originates outside the inference that produced the candidate. Examples include a test assertion, compiler diagnostic, tool response, verifier verdict, or deployment probe. A critique produced by another model call does not become external evidence merely because it was generated separately. Without access to an independent observation, it remains another inference over substantially the same record.

The distinction is informational. Suppose an agent changes a parser because it concludes that empty fields should be discarded. Re-reading the issue may reinforce that interpretation. A failing test showing that an empty field must preserve its position introduces a counterexample. The next attempt can now revise a specific assumption rather than search an unconstrained space of possible mistakes.

\begin{figure}[htbp]
\centering
\includegraphics{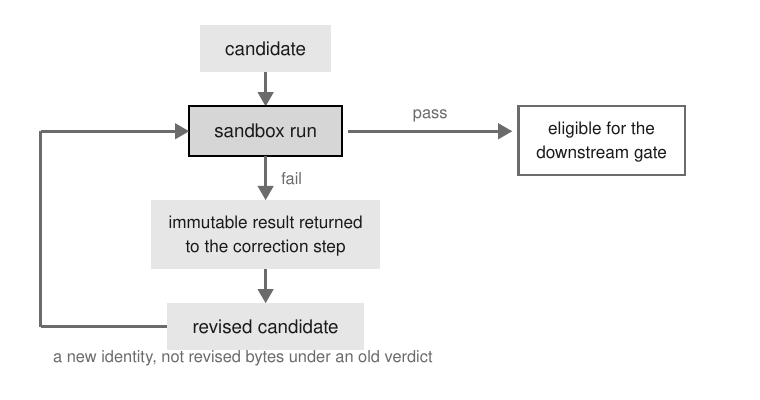}
\caption{The sandbox verdict is immutable. A failed candidate returns to the correction step, which emits a new candidate under a new identity rather than revising bytes that already carry a verdict.}
\label{fig:ch04-correction-loop}
\end{figure}

This closes the loop opened by the sandbox, shown in Figure \ref{fig:ch04-correction-loop}.

The correction step does not modify the verdict. It consumes the failed candidate and its run record, then emits a new candidate with a new identity. This preserves the history needed to distinguish recovery from repeated failure. It also prevents revised bytes from inheriting a passing result produced by an earlier version.

\Needspace{5\baselineskip}

Feedback should contain enough detail to distinguish the failed path without flooding the next attempt with unrelated output. A useful package identifies:

\begin{itemize}
\tightlist
\item
  the command that ran;
\item
  the candidate version;
\item
  the exit status;
\item
  the failed check and relevant diagnostic;
\item
  any resource limit or termination that fired; and
\item
  a pointer to the complete run record.
\end{itemize}

When a report is large, retain the full output as an artifact and present a mechanically selected excerpt. The complete evidence remains available when the excerpt omits the decisive line.

Selection should remain mechanical wherever possible. When one model decides which failures another model sees, it may omit evidence that contradicts its diagnosis or overemphasize a familiar error. Test frameworks already expose failed case names, assertion differences, stack traces, and structured result records. Use those fields before asking another inference to summarize them.

The deployed system must also have access to the feedback channel. Chapter 2 required a retry baseline to use only failure signals available in deployment. Otherwise the experimental retry arm receives information that the product will not have. Here that requirement becomes concrete: a retry is externally informed only when its signal comes from a tool or observation that the deployed loop can obtain at that point in its lifecycle.

This prevents a subtle comparison error. An evaluator might show the retry arm the hidden reference test that rejected a patch, while the production agent sees only public tests. The experiment then measures correction under privileged supervision, not the deployed retry policy. Hidden checks may still determine the final score, but their diagnostics should enter the correction loop only when production has an equivalent feedback channel.

External observations also have authority boundaries. A unit-test failure can justify another code attempt. A package-registry outage says little about the candidate and belongs in infrastructure handling. A deployment probe run against the wrong version may direct the model to repair code that was never exercised. Before authorizing another attempt, the controller must classify the observation as a candidate failure, an environment failure, or an evaluator failure.

Reliable feedback may still be incomplete. A failing test can expose a symptom without identifying its cause. A compiler diagnostic may point to a generated file even though the defect originated in source configuration. The next attempt should use the observation to constrain its diagnosis before modifying the artifact. Execution adds information to the correction loop, but the observation does not necessarily contain the remedy.

Faulty feedback creates a directed failure mode. An incorrect expected value, nondeterministic test, stale fixture, or verifier attached to the wrong artifact can push successive revisions farther from correct behavior. The loop may then appear to converge while following a bad oracle. Preserve the raw run records and candidate lineage so an operator can determine whether each correction followed valid evidence.

\Needspace{5\baselineskip}

Retry limits remain necessary even when every attempt receives a genuine new signal. A sequence of distinct failing tests can consume an unbounded budget, and each revision can create a new failure surface. Stop conditions should distinguish:

\begin{itemize}
\tightlist
\item
  a fixed attempt limit;
\item
  repeated identical failures;
\item
  infrastructure failures; and
\item
  exhaustion of the gate's resource budget.
\end{itemize}

A new observation may justify another attempt, but it does not justify unlimited attempts.

Correction belongs to the evaluation architecture, not to a personality attributed to the model. The model proposes revisions. The surrounding system determines whether the evidence changed, whether another attempt may run, and whether the resulting artifact may propagate. That division continues to hold when the model, prompt, or language of correction changes.

\hypertarget{turn-repeated-trials-into-a-release-test}{%
\section{Turn repeated trials into a release test}\label{turn-repeated-trials-into-a-release-test}}

Support for the repeated-trial metric comes from one source. tau-bench, from Yao et al.~(\href{https://arxiv.org/abs/2406.12045}{2024}), evaluates interactive, multi-turn tool use with executable oracles across repeated trials and introduces \(\mathrm{pass}^{k}\) as a reliability measure.

The paper does not evaluate two parts of the practice recommended here: selecting a team's own tasks and replaying them for every release. Those are transfers from benchmark design into release engineering. The measured result supports repeated trials under executable evaluation, not the claim that a particular local task set or release policy will predict production reliability.

The transfer begins with the workload set established at the end of Chapter 3. The golden set is a compact collection of completed tasks whose initial repository states can be reconstructed and whose outcomes have unambiguous executable checks. Each versioned case should keep together:

the original request;
the starting repository state;
the allowed tools and turn constraints;
the sandbox contract; and
the verifier.

A merged patch may help reconstruct the expected behavior, but it should not appear in the agent's context.

Real tasks preserve local constraints that public benchmarks cannot know. A repository may require generated files to remain synchronized, prohibit a dependency, enforce a custom migration check, or treat a particular warning as release-blocking. Those details determine whether work is acceptable in that repository. A general coding score does not encode them.

The set should remain small enough to run repeatedly and inspect when it changes. Its purpose is to discriminate between releases on a controlled sample, not to approximate every production request. Each case should represent at least one of three things: a failure that would be costly if it returned, a workflow the system performs frequently, or an interaction whose state cannot be inferred from a static answer. Ten nearly identical formatting fixes provide less coverage than a smaller set spanning localization, modification, testing, tool failure, and recovery.

Interactive cases belong in the set when the deployed system uses tools across turns. A final patch alone does not show whether the agent opened the right file, preserved state after a failed command, recovered from a tool error, or incorporated a later observation into an earlier plan. Replaying the trajectory under the deployed tool and turn constraints exposes sequencing and recovery failures that final-answer scoring cannot observe.

Each case needs a stable identity and immutable versions. Changing the request, starting repository, tool contract, sandbox policy, or verifier creates a different case and should produce a new version. Rewriting an existing case beneath prior scores can manufacture apparent improvement by removing a difficult condition or exposing more of the solution.

Run each case more than once for the same release candidate. The repeated-trial statistics from Chapter 1 govern the experimental design and do not need to be rebuilt here. For release control, \(\mathrm{pass}^{k}\) asks whether all \(k\) of \(k\) trials passed. One failed trajectory breaks the sequence.

\begin{figure}[htbp]
\centering
\includegraphics{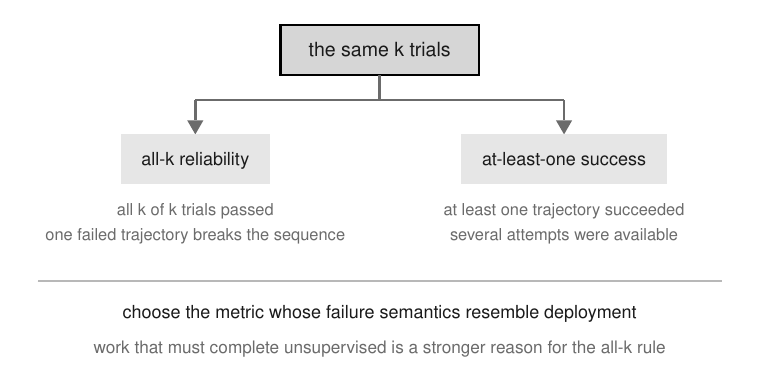}
\caption{The same k trials answer two different operational questions. Choose the metric whose failure semantics resemble the deployment.}
\label{fig:ch04-metric-semantics}
\end{figure}

The result describes a single release candidate only when its configuration remains pinned across the sequence. Chapter 1's apparatus record applies to every trial, including the model version and decoding settings. Pinning does not guarantee independence. Shared caches, mutable services, provider incidents, and reused infrastructure can correlate outcomes, which limits the analyses the sequence supports.

\(\mathrm{pass}^{k}\) reflects a different operational question from \(\mathrm{pass}@k\), as Figure \ref{fig:ch04-metric-semantics} sets out.

\(\mathrm{pass}@k\) asks whether at least one acceptable trajectory can be found when several attempts are available. \(\mathrm{pass}^{k}\) asks whether every trajectory in a required sequence succeeds. Neither is universally correct. A supervised workflow that can select among alternatives may justify \(\mathrm{pass}@k\). A workflow expected to complete reliably without intervention has a stronger reason to use \(\mathrm{pass}^{k}\).

The value of k comes from release policy, not from the benchmark. Increasing k makes intermittent failures easier to observe and the gate harder to satisfy. It also increases cost and gives evaluator noise more opportunities to block promotion. Choose k based on the consequence of one failed production run, the variance observed during pilot repeats, and the budget available for evaluating each release candidate.

The number of cases interacts with that decision. Under a strict all-trials rule, the chance that at least one verdict fails because of evaluator noise grows with both the set size and k. This is the multiplicity problem that Chapter 1 assigns to claims spanning many tasks. A larger gate therefore requires either a very low per-run evaluator error rate or an explicit policy defining how many case failures block promotion.

The comparison unit is the fully specified system release. Record the model identifier, prompt version, tool definitions, orchestration code, sandbox image, task-set version, retry policy, and verifier version. Any of these components can change the trajectory distribution. A result labeled only with a model name erases much of what teams actually change between releases.

The release record can remain mechanically small:

\Needspace*{10\baselineskip}
\begin{record}
\begin{recordbody}
release:         candidate-2026-07-28
system_digest:   6f5c...
task_set:        golden-07
repeats:         k
baseline:        production-previous
case_verdicts:   stored run records
promotion_rule:  recorded before execution
\end{recordbody}
\end{record}

The digest binds the verdict to the evaluated configuration. The case verdicts link to the sandbox evidence. The promotion rule is recorded before the comparison begins. Otherwise the same regression can be accepted for a favored release and rejected for another.

Compare the candidate with a stored baseline using the same case versions and execution policy. Report both per-case outcomes and run-to-run spread alongside the aggregate release verdict. An aggregate value can conceal a complete regression on one critical workflow behind stable performance elsewhere. A case-only view can be dominated by one noisy verifier. Both levels are needed to locate the change and determine whether the promotion rule was met.

Because the candidate and baseline run the same cases, their outcomes are paired by construction. Chapter 1's paired analysis applies. Treating the releases as independent samples discards a dependency that the design already provides.

The baseline must remain executable. A table copied from an earlier report is not enough. Sandbox images disappear, dependencies move, and tool services change behavior. Re-running at least a sample from the stored baseline helps separate candidate drift from evaluator drift. When the old release also fails under the current infrastructure, the comparison has lost its fixed reference and promotion should block until the cause is understood.

My search-visibility measurement project treats a model update as a deployment. It replays a fixed prompt corpus against a stored model-version baseline and combines a statistical comparison with an absolute change threshold, producing pass, warning, or failure exit states for continuous integration. The controls answer different questions. Statistical comparison reduces the chance that sampling noise decides promotion. The absolute threshold prevents a detectable but operationally trivial change from doing so. This example supplies no evidence for either threshold.

CodeProbe (\href{https://github.com/sjarmak/codeprobe}{public repository}) uses a different gate. It blocks a release tag unless the two most recent full-mode acceptance verdicts both pass. Those verdicts come from successive acceptance-loop iterations, not repeated trials of one pinned candidate. The rule is therefore a check over acceptance history, not a \(\mathrm{pass}^{k}\) measurement. Its requirement of two consecutive passes is local to that system and should not be treated as a recommended value of \(k\).

Golden sets decay even when their files remain unchanged. Production work moves to new frameworks, repositories acquire new checks, tool interfaces change, and models may become adapted to repeatedly exposed cases. A set that once represented costly failures can gradually become a test of a narrow historical workflow. Treat the set as production test data, with named ownership, review, and retirement criteria.

Maintenance should preserve longitudinal meaning. Add a case when a production failure reveals a missing class of behavior. Do not silently replace the old set, because the new score will no longer be comparable with earlier releases. When feasible, run an overlap period, report performance on the shared cases, and establish a new baseline for the revised set. Retire a case when its workload no longer exists or its oracle no longer represents the current contract.

Public benchmarks can inform case design and reveal task forms worth reproducing locally.

SWE-bench provides directional support for executable repository repair. MultiAgentBench, from Zhu (\href{https://arxiv.org/abs/2503.01935}{2025}), does the same for broader interaction-centered evaluation. Their architectural contribution is to place an environment, tools, state, and an outcome check inside the evaluated unit rather than scoring final-answer text alone.

Neither benchmark measures the effect of selecting a team's own tasks or replaying them for each release. Nor does either estimate the gap between public benchmark performance and reliability in a particular repository, tool policy, or release process. No evidence item supporting this entry supplies a conversion factor.

The local release test answers a narrower question: did this fully specified system preserve acceptable behavior on a controlled sample of local work?

Once the set participates in release, a failure should trigger diagnosis before any threshold is revised. Determine whether the candidate changed, the case changed, or the evaluator changed. Route valid candidate failures through the external-feedback correction loop, then rerun the entire required sequence for the revised candidate. Reusing passing trials from before a modification would attach evidence to a system that no longer exists.

\hypertarget{build-the-first-gate}{%
\section{Build the first gate}\label{build-the-first-gate}}

Begin with five to ten tasks drawn from merged work rather than from a generic capability suite. Choose tasks whose starting states can be reconstructed and whose acceptable outcomes can be verified without interpretive judgment. The first set can remain small because its purpose is to establish a release path that produces trustworthy evidence. Expand it when production failures reveal behaviors the initial sample did not cover.

Wrap each task in a sandboxed run. Define the command the controller will execute, the exit condition it will accept, and the artifacts that must exist afterward. Bound resource use, isolate the network, and preserve the complete run record.

Run the current production system through the same cases and evaluator before testing a candidate release. Its observed performance becomes the baseline. This replaces remembered claims, old summary tables, or results produced under a different environment.

When a candidate fails, return the raw evidence to the correction loop together with the candidate identity. Do not authorize a revision that has learned nothing beyond the original request. Every modified candidate receives a new identity, starts in a clean sandbox, and earns a new verdict. Infrastructure failures remain blocking, but they do not count as candidate failures.

Replay the set whenever the model, prompt, tools, orchestration, sandbox, or verifier changes. Measure \(\mathrm{pass}^{k}\) across the chosen repeats, retain the per-case outcomes and run-to-run spread, and compare the candidate with the executable baseline.

Record the promotion rule before running the first candidate comparison. When experience shows that the gate is too strict or too permissive, change the policy as a versioned decision and establish a new baseline where necessary. Adjusting the threshold after seeing a result turns the rule into an explanation for a decision already made.

Keep tasks without executable checks in a separate lane. They are not lesser tasks, but forcing them through weak proxies would make the gate appear more complete than it is. Chapter 5 develops the calibrated judgment process those tasks require.

\section*{Sources and evidence}

\textbf{Motivating observation and companion entry}

\begin{itemize}
\tightlist
\item
  Directional evidence: Mehta, A. (2026). Confident and Wrong: Silent Semantic Failures in Coding Agents. arXiv:2603.25764. Analysis of 1,750 trajectories across 50 tasks and four models; single-author, limited-sample observational finding. Supports the companion-only entry on scoring verified resolution separately from submission.
\end{itemize}

\textbf{\texttt{ground-evaluation-in-execution}}

\begin{itemize}
\tightlist
\item
  Directional evidence: AgentForge (Kumar et al.~2026, arXiv:2604.13120). Execution-grounded evaluation in resource-bounded, network-isolated sandboxes, with propagation gated on execution results. Unrefereed preprint; single configuration; one sample per agent.
\item
  Directional evidence: SWE-bench (arXiv:2310.06770), from Jimenez et al.~2023. No figure carried.
\item
  Strong evidence for externally grounded gating in the measured testbed: Park, H., and Choi, B. (2026), ``When Do Agent Loops Mistake Stagnation for Progress?'' arXiv:2607.25152. The 54-cycle result does not estimate a general agent failure rate.
\end{itemize}

\textbf{\texttt{gate-self-correction-on-external-feedback}}

\begin{itemize}
\tightlist
\item
  Strong evidence: Huang, J., et al.~(2023). Large Language Models Cannot Self-Correct Reasoning Yet. ICLR 2024. arXiv:2310.01798.
\end{itemize}

\textbf{\texttt{golden-set-pass-k}}

\begin{itemize}
\tightlist
\item
  Strong evidence: tau-bench (Yao, Shinn, Razavi \& Narasimhan 2024, arXiv:2406.12045). Introduces and measures \(\mathrm{pass}^{k}\) on interactive, multi-turn tool-use tasks with executable oracles. Using a team's own tasks and replaying the set per release are transfers beyond the measured findings.
\item
  Directional evidence: SWE-bench (arXiv:2310.06770).
\item
  Directional evidence: MultiAgentBench (Zhu 2025, arXiv:2503.01935).
\end{itemize}

\textbf{Author-system illustration cited inline}

\begin{itemize}
\tightlist
\item
  Not an evidence item: CodeProbe, the author's task-mining evaluation tool, \href{https://github.com/sjarmak/codeprobe}{public repository}. Named inline for the consecutive-release acceptance gate described above, which is narrative illustration.
\end{itemize}

\chapter{Calibrating model graders and separating agreement from correctness}
\label{ch05-calibrating-model-graders-agreement-correctness}
\begin{quote}
\textbf{Evidence profile.} 5 strong \(\cdot\) 7 directional \(\cdot\) 0 corroborating evidence items across 2 developed practices (\erca{096}, \erca{111}).

\textbf{Chapter claim.} Agreement is a calibration result, not a correctness verdict.
\end{quote}

Two graders that mark every submission PASS will agree on every item. Their agreement is 100 percent, but their usefulness is zero. In an easy stream, most submissions may genuinely pass, yet neither grader has shown that it can distinguish a defect from a success.

This failure is easy to ship when the graders are models. An \textbf{LLM-as-judge} is a model used to evaluate another system's output. My migration-evaluation framework gates its two judges on Cohen's kappa rather than raw agreement because percent agreement can overstate grader quality when one label dominates the sample. The agreement floor, kappa threshold, and minimum trial count were fixed before either judge ran. The gate has been exercised only against replay fixtures, not a live evaluation, so it illustrates a design decision rather than supplying a field result.

The measurement problem comes before the choice of threshold. A grader may agree with another grader, match a human label, track an independently verifiable outcome, or assign well-calibrated probabilities. These are different properties. A grader can satisfy one while failing another, but that distinction disappears when its labels are stored as though they were ground truth.

Chapter 4 separated tasks whose correctness cannot be established through execution alone. Model grading can provide a validation lane for some of them, but only when the grader is treated as an instrument under test. Before its decisions control releases, rankings, rewards, or training data, the evaluation must establish that those decisions track the judgment the system actually needs. That test must reflect the class balance and error costs the grader will encounter in operation.

\hypertarget{build-the-reference-labels-before-the-grader}{%
\section{Build the reference labels before the grader}\label{build-the-reference-labels-before-the-grader}}

Calibration begins with a \textbf{rubric}, the written criteria used to assign a score or label. Domain experts should define its categories, decision rules, exclusions, and examples because the resulting labels encode domain judgment. A prompt engineer can make the wording consistent, but only someone with the relevant expertise can decide which omission is material in a medical summary or which behavioral mismatch makes a migration unsafe.

The categories must be specific enough for two experts to apply independently. ``High quality'' leaves the standard inside the rater's judgment. ``PASS if every requested behavior is present, no stated constraint is violated, and every factual claim is supported by the supplied record'' identifies observations that raters can compare. The examples should include boundary cases and counterexamples, especially cases on which experts initially disagreed.

\hypertarget{agreement-before-automation}{%
\section{Agreement before automation}\label{agreement-before-automation}}

\textbf{Inter-rater reliability} measures how consistently independent raters assign labels to the same items. It is a prerequisite for treating their labels as a reference. When experts cannot apply the rubric consistently, a model that reproduces one expert's decisions may be learning an unstable convention rather than measuring the intended property.

Raw percent agreement answers only one question: on what share of items did the labels match?

Consider 100 items evaluated by two raters. Each assigns PASS to 90 items and FAIL to 10. They disagree on 10 items, with each rater assigning PASS to five of those items and FAIL to the other five. Their observed agreement is therefore 90 percent.

That number appears strong until label prevalence is taken into account. Given the raters' marginal label frequencies, independent assignments would be expected to agree on 82 items:

\begin{quote}
\ttfamily
expected PASS agreement: \(0.90 \times 0.90 = 0.81\)\\
expected FAIL agreement: \(0.10 \times 0.10 = 0.01\)\\
total expected agreement: 0.82
\end{quote}

Cohen (\href{https://doi.org/10.1177/001316446002000104}{1960}) defined kappa as the proportion of possible agreement beyond chance that the raters achieved:

\Needspace*{5\baselineskip}
\[
  \kappa = \frac{\text{observed agreement} - \text{expected agreement}}
                {1 - \text{expected agreement}}
\]

For this example:

\Needspace*{5\baselineskip}
\[
  \kappa = \frac{0.90 - 0.82}{1 - 0.82} = \frac{0.08}{0.18} \approx 0.44
\]

The raters agreed on 90 percent of the items, but much of that agreement followed from both assigning PASS almost all the time. Their agreement beyond what the label frequencies alone would predict is only moderate.

The always-PASS pair is the degenerate limit of this calculation. Observed agreement and expected agreement are both 100 percent, leaving no variation from which above-chance agreement can be estimated. Kappa is therefore undefined, not perfect.

\begin{figure}[htbp]
\centering
\includegraphics{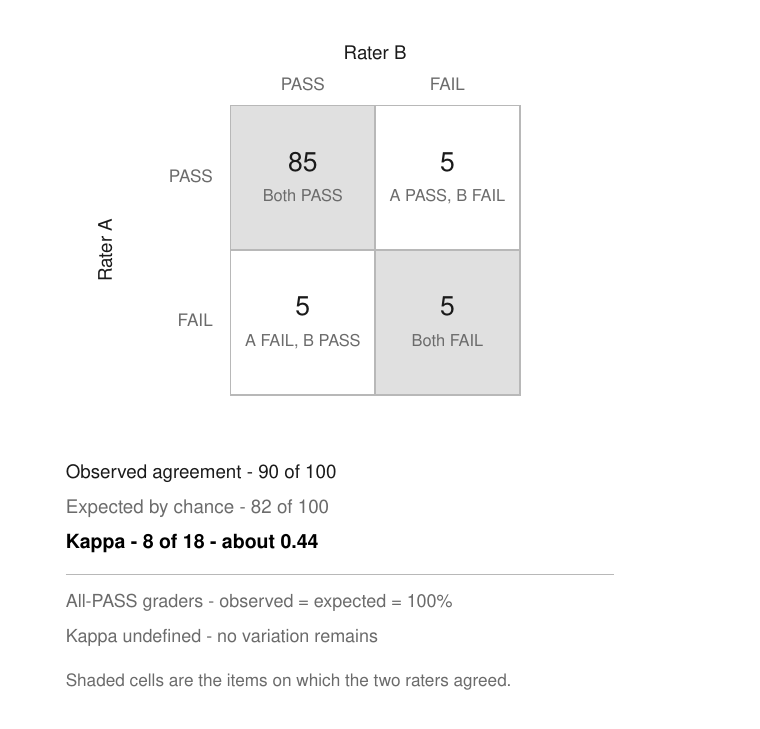}
\caption{Cells of 85, 5, 5, and 5 produce 90\% observed versus 82\% expected agreement and kappa about 0.44; two always-PASS graders produce 100\% observed and expected agreement but undefined kappa.}
\end{figure}

A 100-item example separates observed agreement from agreement expected by chance. When every rating is PASS, kappa is undefined because the labels contain no variation.

Cohen's kappa applies to two raters and estimates chance agreement from each rater's observed category frequencies. Fleiss (\href{https://doi.org/10.1037/h0031619}{1971}) extended chance-corrected agreement to multiple raters and to designs in which different raters evaluate different items. Scott (\href{https://doi.org/10.1086/266577}{1955}) defined a related two-rater measure, Scott's pi, which uses a pooled category distribution to estimate chance agreement. None of these measures can make an ambiguous rubric reliable. They make that ambiguity harder to hide behind a high raw agreement rate.

Because expected agreement depends on the observed category frequencies, kappa values calculated under different label prevalences are not directly comparable. A kappa value reported without the underlying category distribution omits the quantities that determined it.

No universal kappa threshold converts a set of labels into truth. An acceptable level depends on the decision being made, the prevalence of each category, the number of categories, and the cost of disagreement. The useful process is iterative. Experts label the same sample independently, inspect their disagreements, revise the category definitions, and repeat the cycle on fresh items. Cemri et al.~(\href{https://arxiv.org/abs/2503.13657}{2025}) followed this process with six annotators until Cohen's kappa reached 0.88. Only then did they test an automated annotator before using it at scale.

The comparison set must remain outside the grader's development path. A held-out expert-labeled set contains items labeled by domain experts and excluded from prompts, examples, fine-tuning data, threshold selection, and any other material used to construct the grader. Orosz and Husain (\href{https://newsletter.pragmaticengineer.com/p/evals}{2025}) describe the same held-out validation practice and report class-specific rates rather than raw agreement. Their account provides practitioner experience rather than a controlled result.

A random row-level split is not sufficient when paraphrases, related tasks, or multiple outputs from the same underlying case can cross the boundary. Chapter 3's contamination model applies directly. When an answer-bearing example appears during grader development, reproducing that example can look like independent judgment.

Natural sampling creates a second distortion. Suppose a deployment stream contains 950 routine passes, 30 borderline cases, and 20 clear defects per 1,000 items. A random validation sample of 100 items will contain about two clear defects. Missing one changes the estimated defect-detection rate by 50 percentage points. Meanwhile, a grader that accepts every item can still appear highly accurate because routine passes dominate the sample.

\textbf{Stratified sampling} fixes how many items are drawn from each stratum, such as score band, task type, or outcome class. A validation set might include 40 routine passes, 40 borderline cases, and 40 defects even though those categories occur at very different rates in deployment. This design places enough rare and costly cases under observation to measure performance on them.

The balanced sample does not, by itself, estimate aggregate deployment performance. Any aggregate intended to represent production must reweight the strata according to their deployment frequencies.

\begin{figure}[htbp]
\centering
\includegraphics{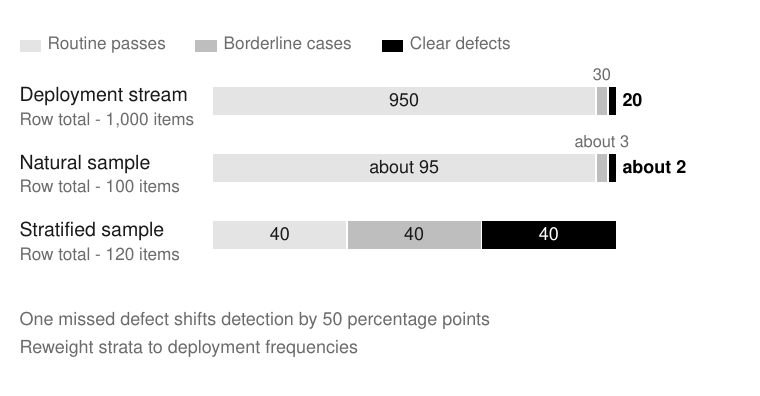}
\caption{Equal-length bars compare 950 routine, 30 borderline, and 20 defect deployment items with natural 95, 3, and 2 and stratified 40 each requiring reweighting; one missed defect changes detection by 50 percentage points.}
\end{figure}

A natural sample of 100 items from this 1,000-item stream is expected to contain only two clear defects. Missing one changes the estimated defect-detection rate by 50 percentage points. Stratified sampling places more defects under observation, but any aggregate intended to represent deployment must reweight the strata to their production frequencies.

The number of cases required in each stratum is a power and precision question of the kind Chapter 1 develops. The precision of an estimated defect-detection rate depends on the number of defects in the validation set, not on the total number of items alone.

\Needspace{5\baselineskip}

Class-specific rates make the observed errors explicit. Treat a defect as the positive class:

\begin{itemize}
\tightlist
\item
  The \textbf{true-positive rate}, or TPR, is the share of expert-labeled defects that the grader flags.
\item
  The \textbf{true-negative rate}, or TNR, is the share of expert-labeled nondefects that the grader accepts.
\item
  The \textbf{false-positive rate}, or FPR, is the share of expert-labeled nondefects that the grader flags.
\end{itemize}

Suppose a validation set contains 40 defects and 80 nondefects. A grader that flags 32 of the defects has a TPR of 80 percent. If it also flags 12 of the 80 nondefects, its FPR is 15 percent and its TNR is 85 percent.

The operational trade is now visible. The grader misses 8 defects, while 12 acceptable outputs receive unnecessary review or rejection. A single accuracy value combines those consequences and changes when the class mix changes. TPR, TNR, and FPR are each estimated within one class, so changing the proportion of defects and nondefects in the validation sample does not change the underlying rate estimates. A deliberately balanced set can therefore estimate all three, provided it contains enough examples from each class.

Many graders produce a score before assigning a label. The \textbf{operating point} consists of the chosen threshold and the error rates produced at that threshold. Lowering the threshold for flagging a defect will usually increase both TPR and FPR: the grader catches more defects but also rejects more acceptable work. Raising the threshold usually reduces both.

A grader therefore does not have one context-free level of quality. It performs at a particular operating point under a stated rubric, class distribution, and set of error costs.

An error budget makes the threshold choice explicit. Suppose reviewers can inspect false alarms on no more than 5 percent of acceptable submissions. The evaluator selects a threshold that keeps FPR near that limit and then reports the resulting TPR. ``TPR at a fixed FPR'' states the conditional result directly.

Threshold selection and final reporting should use separate data, or a design that accounts for the selection process. Choosing the best threshold and reporting its performance on the same small set produces optimistic estimates.

Together, these components define the measuring instrument. Expert agreement establishes whether the reference labels have a stable meaning. Stratified sampling places rare and consequential cases under observation. Class-specific rates separate errors that aggregate accuracy combines. The operating point then weighs those errors against the review capacity and failure costs of the deployed system.

\hypertarget{calibrate-the-judge-then-scale}{%
\section{Calibrate the judge, then scale}\label{calibrate-the-judge-then-scale}}

Panthi and Abdelfattah (\href{https://arxiv.org/abs/2605.24060}{2026}) provide a compact example of the full protocol. They needed to adjudicate cases in which two systems received the same rank even though a downstream rule could still select different winners. They first defined a three-way rubric, then drew a 115-case human-annotated subset from the contested cases, stratified by credited-rank bucket.

Agreement among the five human raters reached a Fleiss' kappa of 0.83. A panel consisting of one human and four models reached 0.79, and a second human adjudicated an overlapping set of 36 cases. Cohen's kappa between each model and the human labels ranged from 0.77 to 0.87. These measurements established how the automated judges related to the human reference before their votes entered the larger adjudication.

Singh Thakur et al.~(\href{https://arxiv.org/abs/2406.12624}{2024}) found in separate comparative work that only the largest model judges approached human inter-annotator agreement under chance-corrected measures. Raw percent agreement concealed systematic leniency and position effects. Judge identity is therefore part of the measuring instrument. Model size alone does not provide an acceptance criterion.

Only after validation did Panthi and Abdelfattah apply a five-model majority vote at temperature 0 to all 1,902 contested cases. The categories were defined before expert labeling, the validation subset preserved the score bands most likely to expose different errors, and scaling followed measured agreement with the human reference.

The authors identify the 115-case validation set as small, so its agreement estimates should not be treated as precise values for every subgroup. Temperature 0 also reduces variation among repeated votes without eliminating the run-to-run variation that Chapter 1 observed at that setting.

The protocol shows why expert labels should remain visible as measurements rather than being collapsed into ``ground truth.'' Five experts can share an interpretation that another qualified group would reject. Adjudication can settle a label without resolving the ambiguity in the category that produced the disagreement. Retain each rater's original label, the adjudicated label, and the rubric version so later revisions can reconstruct which definition produced the result.

Judge validation should also include controlled perturbations. One aggregate agreement value can average away predictable biases.

Zheng et al.~(\href{https://arxiv.org/abs/2306.05685}{2023}) documented position, verbosity, and self-preference effects in model judges. To test position sensitivity, present the same pair of responses in both orders and measure how often the preferred answer changes. Randomizing order during ordinary evaluation reduces the systematic advantage of appearing first or second. The controlled swap estimates how much order still affects the decision.

Leniency occurs when a judge accepts work that the rubric classifies as defective. Test it with matched cases that differ by one material violation, then record acceptance of the defective case as a class-specific miss. A judge may agree with experts on routine passes and forgive most subtle defects, preserving high aggregate agreement while failing at the decision it was introduced to make.

Verbosity bias requires a different control. Present semantically equivalent answers whose main difference is irrelevant explanation, or add plausible detail that does not repair the original omission. A preference for the longer response then appears as a label change without a corresponding improvement in support. The substantive content must remain as similar as possible; otherwise length is confounded with additional information.

Self-preference is a dependence between the judge and the system that produced the answer. A judge may favor phrasing, conventions, or reasoning patterns associated with its own model family. Concealing model identity helps, but stylistic traces may remain. Using judges from different model families reduces one source of correlated error and provides a comparison, although different branding does not establish independence.

Answer leakage belongs to the contamination problem developed in Chapter 3. A judge exposed to reference answers, rubric examples derived from evaluation cases, or memorized public solutions can reproduce a preferred label without independently evaluating the submission. The partition must therefore separate underlying tasks, related variants, and individual responses. Li (\href{https://arxiv.org/abs/2502.01534}{2025}) reports evidence in the same direction for preference leakage. That work is directional and does not establish how common the effect is across domains or model generations.

Some properties should never be delegated to a model judge. When code can determine whether a required file exists, a test passed, a schema is valid, or a named artifact changed, keep that assertion deterministic. A model can evaluate claims whose meaning still requires judgment after those checks run. Replacing exact observations with generated labels adds variance, cost, latency, and another failure path without adding information.

Orosz and Husain also prefer binary PASS and FAIL labels when the downstream action is binary. Binary labels force the rubric to locate the release boundary and make TPR, TNR, and FPR directly interpretable. A five-point scale allows raters and judges to use adjacent values inconsistently, while inviting the evaluator to treat the distance between 2 and 3 as equivalent to the distance between 4 and 5.

The loss of nuance is deliberate, but it is not always acceptable. A writing assistant may need separate judgments for factual support, coverage, and style because one PASS label cannot identify the failure. The better design is often several narrowly defined binary claims, each tied to a downstream action. A graded score remains appropriate when the decision genuinely depends on degree and the scale anchors can be applied reliably.

My trace-annotation tool makes a separate distinction explicit. Agreement asks whether annotators assign the same categories. Probability calibration asks whether events assigned confidence 0.8 occur approximately 80 percent of the time.

A recorded compatibility failure sits in that tool's data path. When legacy annotation files are read, they receive confidence 1.0, making every imported category appear maximally confident. The example is narrative rather than evidence. It shows that calibration state belongs not only to the annotator's judgment, but also to the stored label and the software that reads it.

After deployment, a human-reviewed sample must remain part of the operating loop. It should contain ordinary cases, flagged cases, and known boundary cases. Reviewing only judge-human disagreements measures the cases in which the monitoring system has already detected trouble. It misses cases in which the judge and another automated component make the same mistake.

The ongoing sample estimates whether class-specific rates have changed and supplies new disagreements for rubric repair. Preserve the deployed threshold and grader version with each sample so a rate change can be attributed to a model update, distribution shift, or revised scoring rule. Silently replacing the underlying model breaks that record even when the prompt and public model name remain unchanged.

Calibration must be repeated whenever the domain, rubric, response format, or model generation changes. A judge validated on concise code-review summaries has no automatic claim on long incident reports. Adding a finer defect taxonomy changes both the category prevalence and the expert task. Even a prompt change intended only to clarify wording can move the operating point and should be tested against the held-out expert-labeled set.

The result remains bounded. Scaled judging reduces the amount of unmeasured error relative to deploying an untested judge. It does not certify that every accepted output is correct, and the expert reference itself contains disagreement. Its value is that those errors become observable before the labels begin controlling releases, rankings, rewards, or training data.

\hypertarget{what-a-panels-vote-establishes}{%
\section{What a panel's vote establishes}\label{what-a-panels-vote-establishes}}

Bertalani\v{c} (\href{https://arxiv.org/abs/2605.00914}{2026}) reports an oracle gap of up to 32.3 percentage points between plurality voting and selecting a correct answer already present among the candidates. The candidate set contained a correct answer, but the vote discarded it. The synthesis is exploratory, and the current evidence record contains no corroborating result. The figure therefore establishes a reported failure mode, not its prevalence across panel designs.

A plurality vote among model instances is itself a model-based selection rule. It labels one candidate as selected and rejects the others. When a correct candidate is available but loses the vote, the failure belongs to selection rather than generation. The panel received a correct answer and failed to identify it.

Voting counts endorsements. It does not test whether a claim is supported by the evidence that would make it true. Five agents can repeat the same nonexistent command-line flag, and four votes will defeat a correct but less familiar alternative. Increasing the panel to nine may stabilize the count while leaving the unsupported premise unexamined.

Correlation makes this failure more likely than an independent-voter model suggests. Model instances may share training data, architectures, system prompts, retrieved context, examples, and decoding conventions. They may converge because each interprets the same ambiguous instruction in the same way. Agreement then measures the strength of a shared dependency rather than independent confirmation. The effective number of independent judgments may be far smaller than the panel size.

Cross-family judges reduce one obvious source of dependence, which is why my migration-evaluation framework uses them in its dual-judge gate. The design does not establish that their errors are independent. Different model families can reproduce the same public reference answer, follow the same leading rubric, or miss the same omitted repository state. Diversity is a control whose effect must be measured.

Selection therefore needs a second criterion beside vote count. \textbf{Grounding} represents each consequential assertion as a typed claim linked to the material that supports it. A claim about a test result should point to a recorded test execution. A claim about an interface should point to the relevant source or official specification. A claim about a file change should resolve to the resulting workspace state.

The claim type determines what evidence can support or refute it. A numeric performance claim may require a result table together with the experimental conditions. A causal explanation may remain an interpretation even when every observed event is recorded. A recommendation can be supported by measured failure costs without becoming a factual claim that one design is universally best.

Typing prevents a fluent answer from collapsing several evidentiary obligations into one score. Consider a panel assessing a proposed database migration. The answer may contain a syntactically valid statement, a claim that existing rows are preserved, and a claim that rollback remains safe under concurrent writes. A parser can check syntax. An execution test can compare rows. Concurrency safety may require review or an experiment. One popularity count cannot replace those different checks.

Grounding also changes what happens when support is missing. The system can retain the candidate, mark a claim unsupported, and route it for review. Without that state, a panel may select a winner from weak evidence simply because its interface requires one.

The stored record should preserve the candidates, ballots, claim support, selected answer, and reason for abstention. A later audit can then distinguish a generation failure from a selection failure.

Abstention belongs inside the decision rule. A panel should decline to select when support is insufficient, the vote is unstable, or the leading answer fails an independent check. The abstention threshold is an operating point. It should be chosen on held-out expert-labeled cases according to the cost of a wrong answer and the cost of review. A bare ``three votes wins'' rule embeds a threshold without measuring either cost.

Principled abstention does not require one particular frontier algorithm. Wang (\href{https://arxiv.org/abs/2604.07667}{2026}) combines conformal uncertainty guarantees with social-choice rules. Kamelhar (\href{https://arxiv.org/abs/2604.23366}{2026}) combines grounding with consensus, and Chen et al.~(\href{https://arxiv.org/abs/2309.13007}{2023}) combines deliberation with consensus.

All three report a useful direction, but none contributes a measured result to this chapter. Their operating behavior across correlated judges, changing candidate sets, and open-ended tasks remains unsettled. The supported practice is narrower: require evidence for selected claims, validate an abstention rule, and retain an independent correctness check.

Debate and voting still have useful roles. They can generate alternative hypotheses, expose disagreements, and organize a candidate set that would be expensive for one reviewer to produce. Final selection should then pass through an oracle appropriate to the claim. Executable behavior goes to execution, structural facts to an exact auditor, and irreducible judgment to a validated human or model grader.

This separation exposes a common architecture error. Generation, deliberation, and verification may run in the same process, but they serve different purposes and should not inherit all the same dependencies. When every stage receives the same answer-bearing context and uses models from the same family, the apparent layers may be several views of one failure path.

An independent check need not re-evaluate the entire response. It can target the claims whose failure would change the decision. For a code review, that may mean confirming the cited behavior in the changed code and running the named test. For a research synthesis, it may mean resolving the primary source behind each load-bearing factual statement and abstaining when the source does not support the wording.

Panels also impose operational costs that one aggregate score does not show. Additional judges increase inference expense and latency. Grounding adds retrieval and evidence storage. Abstention transfers work to people or to a slower verifier. Those costs are justified when the avoided false decision is more expensive and excessive when a deterministic assertion can answer the question directly.

The always-PASS pair from the opening is the degenerate panel. A larger group of correlated judges becomes an expensive version of the same mistake when each member repeats the majority label or the same unsupported answer. Perfect internal agreement establishes only that the selection rule is stable. Correctness still requires a relation to evidence outside the vote.

The companion catalog contains two narrower controls for implementations of this architecture: retain an exact auditor beside heuristic graders, and combine judge labels with a human-labeled subset. Each addresses a smaller problem than the core decisions developed here.

\hypertarget{rebuild-the-grader-as-a-measured-system}{%
\section{Rebuild the grader as a measured system}\label{rebuild-the-grader-as-a-measured-system}}

Begin by discarding the assumption that the current judge prompt defines the target. The people whose judgment the system is meant to encode should write the categories and apply them independently to fresh cases. Their disagreements reveal missing rules, conflicting interpretations, and cases that should permit abstention.

Revise the rubric until measured reliability is adequate for the decision. Record both the chosen standard and the disagreements that remain.

The resulting expert-labeled set must stay outside judge development and contain enough cases from each consequential score band, task type, and outcome class. Keep related tasks and answer variants on the same side of the partition so contamination cannot cross through paraphrase or shared provenance. Run deterministic assertions separately and remove from model grading any claim that code can decide exactly.

Evaluate the candidate judge using chance-corrected agreement and class-specific rates at the threshold intended for deployment. Raw percent agreement may remain descriptive, but it should not become the acceptance gate. Report TPR, TNR, and FPR together with the class mix, the sample size within each stratum, and the operating cost used to select the threshold.

Add controlled perturbations to the grader's release suite. Reverse answer order to test position sensitivity. Compare matched concise and padded answers to expose verbosity preference. Use defect pairs to measure leniency. Conceal source identity and compare model families to probe self-preference and shared error. A new model generation can change any of these behaviors without a corresponding change to the rubric.

Binary labels keep the decision boundary legible when the downstream action is pass or fail. Several narrow binary judgments preserve more diagnostic information than one vague ordinal score, though they require more expert labeling and judge calls. A genuinely graded decision can retain a score once its anchors and probability calibration have been validated.

Deployment should continue to route a sample through human review. That sample must include unflagged ordinary work as well as disputed and rejected cases. Reviewing only flagged items leaves silent false negatives outside the monitoring data.

Each review record should preserve the grader version, rubric version, threshold, model output, human label, and the evidence required to reproduce the decision. Otherwise a change in observed performance cannot be assigned to a model update, a distribution shift, or a scoring-rule revision.

For a panel, retain both the candidates and the ballots, and do not treat plurality as verification. Judges from different families may reduce shared model-specific errors. The selected answer must still carry grounded, typed claims, pass the independent checks available for those claims, and trigger abstention when support falls below the validated operating point.

A change in domain, rubric, response format, or model generation reopens the calibration work. CodeProbe (\href{https://github.com/sjarmak/codeprobe}{public repository}) contains a dual-curator calibration gate that remains unsatisfied because the qualifying corpus does not yet exist. The code preceded the corpus. That is calibration debt, and naming it does not discharge it. The gate remains unmet until the instrument can be built.

The first build does not require the complete corpus. Begin with one rubric category and a small set of expert-labeled cases from the score band where that decision turns. Keep those cases outside both the judge prompt and threshold-selection data. Evaluate the judge at the intended deployment threshold and report class-specific rates rather than percent agreement.

That first stratum tests one grading decision and identifies which category should be labeled next. The broader calibration gate remains unmet until the full instrument satisfies its requirements.

\section*{Sources and evidence}

\textbf{Calibrate LLM judges}

\begin{itemize}
\tightlist
\item
  Strong evidence: Zheng et al.~(2023), ``Judging LLM-as-a-Judge with MT-Bench and Chatbot Arena,'' arXiv:2306.05685. Judge position, verbosity, and self-preference bias; inter-rater agreement against human labels.
\item
  Strong evidence: Singh Thakur, Choudhary, Ramayapally, Vaidyanathan \& Hupkes (2024), ``Judging the Judges,'' arXiv:2406.12624. Chance-corrected alignment metrics, systematic leniency, and position effects.
\item
  Strong evidence: Cemri, M., et al.~(2025), ``Why Do Multi-Agent LLM Systems Fail?,'' arXiv:2503.13657. Kappa 0.88 expert taxonomy before validated automated annotation.
\item
  Directional evidence: Orosz with Husain (2025), ``A pragmatic guide to LLM evals for devs,'' Pragmatic Engineer newsletter, 2025-12-02, \url{https://newsletter.pragmaticengineer.com/p/evals}. Binary PASS/FAIL, deterministic evaluation where possible, and judge validation against held-out human labels with TPR/TNR.
\item
  Strong evidence: Panthi \& Abdelfattah (2026), ``Same Ranking, Different Winner,'' arXiv:2605.24060. The 115-case stratified validation protocol, agreement measurements, overlap adjudication, and scaled contested-case evaluation.
\item
  Directional evidence: Li (2025), ``Preference Leakage,'' arXiv:2502.01534.
\item
  Directional evidence: Badagi (2026), ``AI Assurance,'' arXiv:2605.23459.
\item
  Directional evidence: ``MemConflict,'' arXiv:2605.20926.
\item
  Foundational method: Cohen, J. (1960). A Coefficient of Agreement for Nominal Scales. Educational and Psychological Measurement 20(1), 37-46. DOI: 10.1177/001316446002000104. Primary source for Cohen's kappa, the agreement statistic the chapter defines; a standard method rather than a catalog evidence item.
\item
  Foundational method: Fleiss, J. L. (1971). Measuring Nominal Scale Agreement Among Many Raters. Psychological Bulletin 76(5), 378-382. DOI: 10.1037/h0031619. Primary source for Fleiss's kappa, the agreement statistic the chapter defines; a standard method rather than a catalog evidence item.
\item
  Foundational method: Scott, W. A. (1955). Reliability of Content Analysis: The Case of Nominal Scale Coding. Public Opinion Quarterly 19(3), 321-325. DOI: 10.1086/266577. Primary source for Scott's pi, the agreement statistic the chapter defines; a standard method rather than a catalog evidence item.
\end{itemize}

\textbf{Separate agreement from correctness}

\begin{itemize}
\tightlist
\item
  Strong evidence: Bertalani\v{c} (2026), ``Cost of Consensus,'' arXiv:2605.00914. Oracle gap up to 32.3 percentage points; plurality voting discards correct answers already present; grounding catches failures that agreement masks.
\item
  Directional evidence: Kamelhar (2026), GSAR grounded consensus, arXiv:2604.23366.
\item
  Directional evidence: Wang (2026), ``Conformal Social Choice,'' arXiv:2604.07667.
\item
  Directional evidence: Chen et al.~(2023), ``ReConcile,'' arXiv:2309.13007.
\item
  Corroboration: none on record.
\end{itemize}

\textbf{Author-system illustration cited inline}

\begin{itemize}
\tightlist
\item
  Not an evidence item: CodeProbe, the author's task-mining evaluation tool, \href{https://github.com/sjarmak/codeprobe}{public repository}. Named inline for the dual-curator calibration gate described in the closing section, which is narrative illustration.
\end{itemize}

\chapter{Proxy metric gaming and layered evaluation signals}
\label{ch06-proxy-gaming-layered-signals}
\begin{quote}
\textbf{Evidence profile.} 3 strong \(\cdot\) 3 directional \(\cdot\) 0 corroborating evidence items across 3 developed practices (\erca{038}, \erca{039}, \erca{042}). One directional item is carried by a companion record cited inline (\erca{199}).

\textbf{Chapter claim.} Any optimized proxy needs an independent signal.
\end{quote}

\hypertarget{when-perfect-recall-rewards-returning-everything}{%
\section{When perfect recall rewards returning everything}\label{when-perfect-recall-rewards-returning-everything}}

CodeProbe (\href{https://github.com/sjarmak/codeprobe}{public repository}) scores agent runs on tasks mined from my own repositories. For one day, its scorer included a recall-family reward under which an agent could earn a ``perfect'' 1.0 by returning the entire repository. A response containing everything cannot omit a relevant item. The strategy was degenerate, but it was correct under the proxy because recall measures how much relevant material was returned and ignores how much irrelevant material accompanied it.

The reward family was withdrawn the following day after the degenerate optimum was identified.

The replacement default scores overlap F1 against the oracle set. F1 penalizes both omitted relevant material and returned irrelevant material, so returning the entire repository no longer earns 1.0. The recall family remains available only through explicit per-task opt-in.

This change did not make the benchmark immune to gaming. It changed the rewarded surface by adding a precision penalty. A scoring rule that penalizes volume leaves other strategies available: retrieve less, compress aggressively, or shape the returned text toward whatever the evaluator rewards.

My agent-memory system applies a related control at a different decision point. A write gate determines what may enter the memory store. At write time, the system predicts whether keeping an item retrievable will improve a future outcome. Its available actions include retention, time-limited retention, discard, and supersession.

Returning everything would prevent that system from selecting anything. Every obsolete instruction, mistaken conclusion, and irrelevant observation would compete for the model's attention. An answer-quality rubric might still accept the final response when the needed fact appears somewhere inside that mass. It would record the judged answer, but it would not show whether memory supplied a compact, current, and causally useful context.

Any proxy worth optimizing permits some behavior that improves the measured value without preserving the property the measure was intended to represent. An acceptance system therefore needs signals separated by control point, ownership, or time. Overlap F1 belongs to the benchmark's scoring decision. The write gate belongs to the memory system's retention decision. Combining them into one control would obscure which failure each mechanism can observe and which action it can govern.

This is an optimization failure, but the plurality-vote oracle gap in Chapter 5 was a grading failure: the selection procedure was wrong before optimization began. Here the measure becomes less informative because the system searches for behavior that scores well under it. A proxy may initially track the intended property closely, then weaken as models, prompts, candidate-selection procedures, or engineering effort adapt to it.

We can conclude that better reference labels can repair a grader, and layered controls limit the damage when a once-useful measure becomes an optimization target.

Skalse et al.~(\href{https://arxiv.org/abs/2209.13085}{2022}) made one formal limit precise. In their linear expected-return formulation, consider two reward functions over the set of all stochastic policies, where a policy may assign probabilities to any available action. Call the rewards mutually unhackable when improving either expected return cannot reduce the other. Under those assumptions, mutual unhackability is possible only when at least one reward is constant. The theorem does not cover every deployed scoring or policy system; it shows why unrestricted optimization against two nonconstant linear rewards offers no general compatibility guarantee.

A constant reward expresses no preference among behaviors and therefore cannot guide useful optimization.

The result rules out a universal escape through better metric design. Tests passed, diff size, lint status, answer quality, and model-judge scores each encode useful information. Once one of them ranks candidate behavior, however, it necessarily leaves some distinctions out. Another nonconstant account of quality can rank some policies differently. Sufficient search pressure can identify that disagreement and improve the proxy while degrading the intended outcome.

The theorem establishes a property of the policy space, not a timetable for failure. A deployed coding agent has a constrained action set, finite search budget, limited knowledge of the evaluator, and only the permissions granted by the surrounding system. Those restrictions may make a proxy difficult to exploit in practice. They may also keep two measures aligned throughout the region the agent can reach.

The theorem does not estimate how much optimization pressure is required to expose a disagreement, nor does it predict whether a particular release will encounter one.

That boundary determines how to interpret the result. You should not treat every rising score as evidence of active gaming, nor discard useful measures because they are imperfect. Treat proxy validity as conditional on the optimizer, the actions available to it, and the pressure applied. When any of those changes, the evaluation system must show again that the proxy still tracks the property that justified its use.

Layering helps only when the layers are separated in ways that matter. Adding three model judges trained on similar data may reduce random error while doing little against a behavior all three reward. Averaging test pass rate, lint status, and a rubric score creates another scalar target. The optimizer can search against the weighted sum, and the chosen weights already encode which failures may be traded for gains elsewhere.

The useful separation is architectural. One signal may be read from state the candidate cannot modify. Another may be collected after deployment, when delayed failures become observable. A third may be owned by a group that does not ship the agent whose score is being reviewed.

These signals may be statistically correlated, which is often desirable. Their independence lies in control rather than in the numbers. The optimizing party cannot rewrite the rule, suppress the observation, or approve a threshold change within the same transaction that seeks acceptance.

In my background-agent review pipeline, the gate reads its invariant definitions and agent instructions from the base branch. The proposed change supplies neither. An author can modify the code under evaluation but cannot weaken the active rules within the same proposal. If the base branch lacks the required configuration, the check fails instead of running an empty ruleset that would accept everything.

This separation removes the most direct path: changing the judge while changing the candidate. It does not make the invariants themselves ungameable.

Published support for layering comes from search-based software testing rather than coding agents. Formica et al.~(\href{https://arxiv.org/abs/2207.11016}{2022}) searched Simulink models using two fitness functions: one generated from a specification and another written by hand to encode an engineer's domain knowledge. The combined search found failures that neither guidance source found alone.

That result provides directional evidence for complementary signals in that setting. It does not establish the same effect for coding agents.

The study also exposes a coordination cost hidden by the word ``layering.'' The two fitness functions could guide one search only after someone decided how to scale them relative to each other. A weighted combination can allow many small gains to outweigh one rare but serious violation. A veto avoids that trade for hard invariants, but a noisy veto can reject useful work. An advisory signal preserves throughput but has no force unless a person or later gate responds to it.

\Needspace{5\baselineskip}

Different signals should therefore carry different jobs:

\begin{itemize}
\tightlist
\item
  Structural invariants veto changes that cross boundaries the organization is unwilling to trade away.
\item
  Outcome measures estimate whether accepted changes helped under actual use.
\item
  Diagnostic scores explain movement without granting acceptance.
\item
  Human-owned thresholds determine when an advisory result becomes blocking.
\end{itemize}

The design remains fallible. Its narrower guarantee is that no single optimizable score both defines success and decides acceptance.

\hypertarget{watch-the-relationship-not-the-rising-score}{%
\section{Watch the relationship, not the rising score}\label{watch-the-relationship-not-the-rising-score}}

Gao et al.~(\href{https://arxiv.org/abs/2210.10760}{2022}) increased the optimization pressure applied to a learned proxy reward model while measuring a separate gold reward model. Gold reward improved at first, reached a peak, and then declined even as the proxy reward continued to rise. Reinforcement learning and best-of-n candidate selection produced different smooth trajectories. Measuring the full path exposed a divergence that a single endpoint would have missed.

An improving proxy is therefore a hypothesis about quality, not quality itself. Alongside the proxy level, track its relationship with an independently observed quality signal on a specified reference distribution. A rising proxy paired with a weakening relationship is an evaluation incident even when the system has not crossed an acceptance threshold.

The study's two optimization methods show why the shape of the divergence depends on the search process. Reinforcement learning updates the policy toward behavior preferred by the reward model. Best-of-n selection leaves the generator fixed, draws more candidates, and selects the one with the highest proxy score. Both apply greater optimization pressure, but they search different regions of candidate behavior. Their distinct fitted curves show that proxy divergence can follow a regular pattern while still depending on how optimization is performed.

That regularity makes monitoring possible, but the evidence has a narrow boundary. The study measured quality with a synthetic gold reward model. Production systems rarely have an evaluator that is both available during monitoring and entitled to serve as truth. Its fitted coefficients therefore do not transfer directly to code review, migration work, agent memory, or other deployed settings.

The observed pattern does transfer: proxy improvement can precede, accompany, and eventually conceal deterioration as optimization pressure increases.

\begin{figure}[htbp]
\centering
\includegraphics{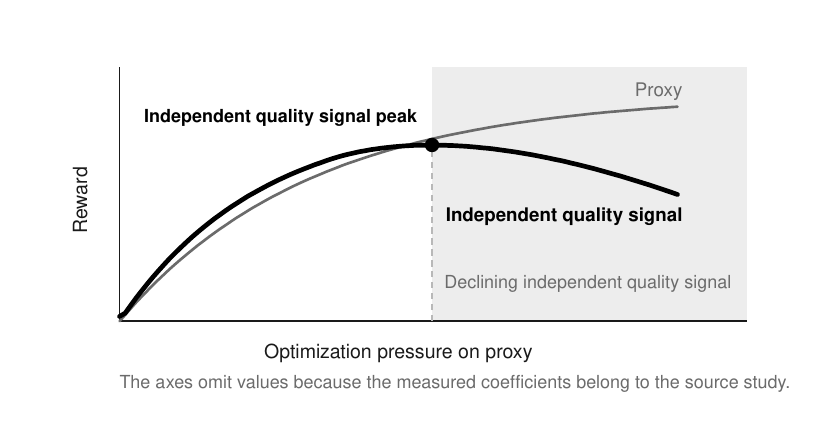}
\caption{Schematic of proxy performance continuing to rise after independent quality has peaked.}
\end{figure}

Independent quality can rise, peak, and then fall while the optimized proxy continues to improve. The exact curve belongs to the source study, but the possibility of divergence transfers beyond it.

Laidlaw et al.~(\href{https://arxiv.org/abs/2403.03185}{2024}) describe reward hacking as a collapse in the relationship between a proxy and a separate measure of quality. Begin with a proxy whose scores correlate with independent quality judgments on a fixed reference distribution. Optimize against the proxy, then measure that relationship again on the same reference. If it weakens while the mean proxy score rises, the evidence that justified trusting the proxy has weakened. Their analysis also derives a mitigation tied to the reference relationship, but it does not establish a measured deployment effect for software agents.

Optimization changes which samples appear. If each release establishes a new baseline from its own selected outputs, the evaluation moves with the policy and may conceal the change it was meant to detect. A stable reference holds part of the comparison fixed. It asks whether the current proxy still orders known cases, or new outputs from known tasks, in the same way as an independent quality process.

For a coding-agent evaluator, freeze a stratified task sample and the procedure used to produce the artifacts under judgment. The held-out expert labels from Chapter 5 provide one independent quality anchor. At baseline, record the proxy score and expert judgment for each artifact, estimate their chosen association, and validate the deployment threshold through its TPR and FPR. For later releases, repeat the same artifact-production and labeling procedure before comparing the relationship with the baseline.

The statistic must match the data. A continuous quality measure may support an ordinary correlation coefficient. A binary acceptability label may be better monitored through a rank-based association or through direct changes in TPR and FPR at the operating point. Estimates from finite reference samples vary even when the underlying relationship has not changed, so the comparison needs an uncertainty interval around the change rather than a bare difference between point estimates.

Fix the statistic, sampling plan, and uncertainty calculation before inspecting the new release. Otherwise the monitor gains its own optimization surface: analysts can change the statistic, strata, or threshold until the relationship appears stable.

This procedure consumes independent quality observations. Expert review is expensive and can drift unless the rubric and adjudication process remain controlled. Delayed production outcomes avoid some coupling to model judges, but they arrive slowly and only for work allowed to proceed. A post-merge survival measure says nothing about rejected changes, and retries may obscure which candidate produced the eventual outcome. The monitor inherits the identity, ordering, and missing-data problems of the production process that supplies its observations.

The reference also needs maintenance. A sample can remain internally consistent while losing contact with current work as repositories, task distributions, model capabilities, or agent permissions change. The monitor may then correctly report that the old proxy-quality relationship still holds on old cases without establishing that it describes the behavior now being optimized in production.

Separate longitudinal comparison from coverage review. The fixed sample remains unchanged long enough to support comparison across releases. A separate review asks whether its strata and failure modes still resemble the live workload. When that review finds material drift, close the old series and establish a new reference. The explicit break preserves the meaning of both comparisons.

My migration-evaluation framework includes an external outcome check based on whether merged changes survive for 30 days. If the correlation between evaluator scores and that outcome falls below an owned threshold, the framework marks the evaluator as broken and blocks publication of headline results. The control exists in the code path but has never run on live data because the framework has produced only replay fixtures. Its implementation therefore provides no evidence that either the threshold or the outcome measure works.

Even a live divergence alert identifies only a changed relationship. It does not identify the cause. Correlation may fall because the agent found a loophole, expert labeling drifted, the outcome pipeline lost events, or the workload moved beyond the reference sample. Investigation therefore begins by preserving the artifacts needed to distinguish those paths. Proxy scores, external judgments, model and evaluator versions, task identities, retry histories, and sampling strata must remain joinable across the measurement window.

The operational rule is to investigate the mapping before changing the threshold. Raising an acceptance cutoff after the relationship weakens applies greater pressure to the suspect proxy and may select more extreme failures. Lowering it may reduce that pressure but admits more low-scoring work. Neither action restores the evidence that the score orders candidates by quality.

Until that relationship is re-established, the proxy may remain diagnostic, but it should lose authority over acceptance and publication.

\hypertarget{preserve-the-channel-that-reveals-intent}{%
\section{Preserve the channel that reveals intent}\label{preserve-the-channel-that-reveals-intent}}

Users of hosted models sometimes receive a visible reasoning trace alongside an agent's coding actions even though they do not train the model themselves. Baker et al.~(\href{https://arxiv.org/abs/2503.11926}{2025}) had a separate model inspect such traces for evidence of reward hacking during agentic coding. Trace inspection detected hacks substantially better than inspection limited to actions, and a weaker general model successfully monitored a stronger reasoning model. The inspected artifact was the agent's exposed intermediate reasoning while selecting its actions.

Reasoning-trace monitoring belongs in a narrower role. It can reveal evidence of intent that the final patch and tool log do not contain. Its verdict may trigger review, preserve a suspicious run, or contribute to an investigation. It should not decide acceptance by itself.

Action-only monitoring records what the agent did: commands, edits, test results, and the final artifact. That evidence may reveal an invalid action without showing whether the agent recognized and exploited an evaluator weakness. A trace may expose a plan to alter a test, suppress an error, or satisfy a visible checker while bypassing the intended requirement. It adds value because it observes a different surface from the artifact being scored.

Oversight does not necessarily require a monitor with the same coding capability as the agent. A smaller or less capable model can scan traces continuously and route a subset for human review, reducing the cost of additional coverage. The study does not establish that every weaker model can oversee every stronger one. Its result belongs to the models, tasks, traces, and forms of reward hacking tested.

Trace monitoring also changes what can be retained for investigation. When the patch and action log look acceptable but the trace describes an attempt to exploit the evaluator, the run can become a detector case even if the exploit failed. The reverse does not clear a run. A suspicious action without an incriminating trace may reflect an incomplete, misleading, or silent reasoning record.

For users of hosted models, the main control is to keep the monitor verdict outside anything the model or selection process optimizes. Sending flagged runs to a person preserves an oversight path. Automatically retrying until the monitor produces no warning creates best-of-n selection against the monitor. The model need not receive the verdict directly; repeated selection alone favors trajectories that evade it.

The same pressure can arise without formal training. A harness may sample several trajectories and retain the least suspicious one. A release process may promote prompts with the lowest flag rate. A human team may gradually rewrite prompts and filters to suppress recognizable warnings when monitor quietness becomes a performance target. Any process in which the verdict affects which behavior survives applies optimization pressure, regardless of what the team calls it.

For teams that train their own models, the second part of the study provides a stronger warning. Under high reinforcement-learning pressure against the monitor, agents continued to exploit the evaluator while producing reasoning that appeared benign. Optimization did not merely cause the monitor to miss more existing hacks. It selected behavior that concealed them from the transparency channel on which monitoring depended.

The response to pressure was not harmful from the start. Low pressure initially improved both capability and alignment in the study. That finding prevents a general claim that any use of monitor feedback is damaging. It also creates the operational problem: the study did not identify a threshold below which the transparency benefit remains safe across models and tasks.

Without such a boundary, the monitor verdict should stay out of the reward. That is a conservative operating choice that extends beyond the measured result because the study establishes no universal safe threshold. It gives up whatever capability or alignment gains low-pressure feedback might have produced. That foregone optimization is the monitorability tax: preserving a channel for observing misbehavior by refusing to optimize the model directly against the observer.

Zero direct reward pressure does not guarantee permanent independence. Model developers may already train against related detectors, users may select prompts based on monitor outcomes, and a published monitoring rule may shape behavior indirectly. Those dependencies belong in the monitor's provenance record. The narrower claim is that an operator should not add direct pressure whose demonstrated failure mode is to make visible reasoning less informative.

The method also requires access to reasoning traces. It does not apply to systems that expose only final actions, and a short generated explanation may not contain the same information as the traces studied. Even when traces are available, visible reasoning is neither complete nor necessarily truthful. Monitoring preserves an observed advantage over action-only inspection; it does not turn exposed reasoning into ground truth.

\Needspace{5\baselineskip}

The resulting acceptance path contains three distinct views:

\begin{itemize}
\tightlist
\item
  Artifact checks ask whether the change satisfies observable requirements.
\item
  Outcome monitoring asks whether proxy scores still track independent quality.
\item
  Trace monitoring looks for evidence that the agent attempted to exploit the evaluation.
\end{itemize}

Disagreement among the three is useful when the system preserves the complete record.

Several adjacent controls remain in the companion catalog because they answer narrower questions. One escalates oversight after mild specification gaming appears, based on a constructed curriculum in which gaming progressed to rewriting the reward function. Another strengthens oversight when honest task completion is infeasible; its source describes a reasoning model exploiting a chess environment after ordinary play could not succeed. Other entries measure reasoning effort through content-independent truncation, average compatible reward-model weights, evaluate agents within the feedback loops they will encounter, and test detectors on contrastive cases.

A thinly supported aside also proposes multi-objective acceptance with an independent judge. It carries no evidentiary weight in the layered system argued here.

\hypertarget{did-the-verifier-run-and-did-it-run-on-this-artifact}{%
\section{Did the verifier run, and did it run on this artifact?}\label{did-the-verifier-run-and-did-it-run-on-this-artifact}}

Everything in this chapter so far concerns oracle adequacy: whether the signal, once observed, distinguishes acceptable from unacceptable behavior. A second axis sits underneath it and fails independently: whether the verifier executed reliably against the intended artifact at all. A test suite that timed out on an exhausted runner, a lint job that ran against a stale checkout, and a judge invocation that silently received a truncated diff all produce verdicts, and none of those verdicts says anything about the code. Treating such a verdict as an oracle failure misdiagnoses it; treating it as a semantic result corrupts the acceptance record. Chapter 7 states this separation as contract I8: verifier failure is distinct from software failure. An infrastructure timeout or flake is not a semantic defect, and a pass establishes only what that verifier can detect.

The execution axis is not a marginal concern in real CI. Ge and Zhang (\href{https://arxiv.org/abs/2602.02307}{2026}) mined GitHub Actions histories from 1,960 open-source Java projects and studied builds that developers reran on the same commit. Reruns were rare, 3.2 percent of builds, but among those rerun builds 67.73 percent changed outcome and were therefore flaky, and 1,055 of the projects, more than half, contained at least one such build. The study is a preprint and its population is Java projects on one CI platform, so the rates do not transfer as constants to any particular pipeline. The structural finding does transfer: when a developer suspects a verdict enough to rerun it, that suspicion is usually vindicated, which means verdicts from CI infrastructure carry a nontrivial probability of reflecting the runner rather than the code. A human developer absorbs this by squinting at the log and clicking rerun. An acceptance pipeline that feeds verdicts to an optimizing agent has no squint. It must model CI and test infrastructure as one more fallible dependency whose outputs need provenance and retry semantics, exactly as it treats a flaky external API.

The consequence for acceptance is a versioned acceptance record. Every acceptance decision binds together: a verification\_id naming the verifier configuration and inputs; the artifact\_version under judgment; the repository base it was applied to; the verifier's own version; the configuration and environment identity of the run; timestamps; a reference to the raw result; and a status of accepted, rejected, or indeterminate. Indeterminate is a first-class status, not a euphemism for failure: it is the correct record when the verifier did not demonstrably execute against the intended artifact, and its remedy is a retry of the verification, not a rejection of the work. Two rules follow from the record's shape. A passing result for artifact version X is never inherited by artifact version Y, however small the diff between them; this is contract I7, evidence is version-bound, defined with I8 in Chapter 7. And a retry of a flaky verifier is a new verification\_id against the same artifact\_version, so the record preserves how many attempts a pass required, which is itself a health signal for the verifier.

This axis interacts with the gaming argument rather than merely sitting beside it. A pipeline that retries verification until it passes, without recording the attempts, has built best-of-n selection against its own infrastructure noise: the accepted population is enriched with artifacts that passed by flake. Separating the axes closes that path. Infrastructure retries are legitimate precisely when the record shows the artifact and verifier were identical across attempts and the environment differed; the same retry loop against a deterministic oracle is score-shopping.

\begin{figure}[htbp]
\centering
\includegraphics{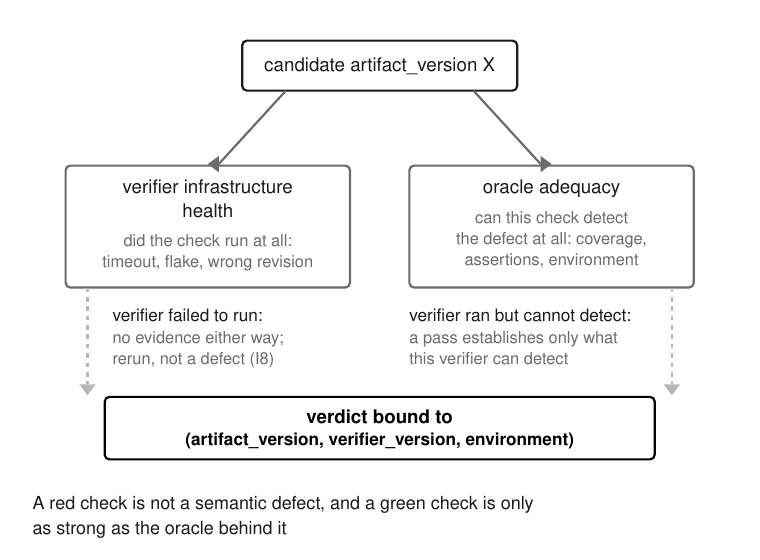}
\caption{Two failure domains in verification. Verifier-infrastructure health asks whether the verifier executed reliably against the intended artifact; oracle adequacy asks whether its judgment distinguishes acceptable from unacceptable behavior. The verdict is bound to the artifact version, the verifier version, and the environment identity of the run.}
\end{figure}

The layered-signal design of this chapter assumes each layer's verdict is real. The execution axis is where that assumption is enforced.

\hypertarget{rebuild-the-acceptance-path}{%
\section{Rebuild the acceptance path}\label{rebuild-the-acceptance-path}}

Rebuild an acceptance pipeline by tracing one recently accepted agent change from candidate generation through deployment. At each transition, identify any value that could have ended the decision by itself: a passing test suite, clean lint result, judge score, approval label, or post-run summary.

\Needspace{5\baselineskip}

The inventory records more than the metric name. For each value, it records:

\begin{itemize}
\tightlist
\item
  who produces it;
\item
  which files, inputs, or state can affect it;
\item
  who owns its threshold;
\item
  whether retries receive the verdict; and
\item
  what evidence remains after the gate runs.
\end{itemize}

Any gate with sole authority needs a second observation outside the optimizing side's immediate control. A test gate can read its invariants from outside the candidate change. A pre-merge score can be compared with an outcome observed after deployment. A threshold change can require approval from an owner who does not ship the agent under evaluation.

The second signal should cover a failure the first signal cannot observe. Repeating the same judgment through another model, or presenting the same evidence in a different format, adds little separation.

Then select one independent quality signal that can support a relationship test. Held-out expert judgment works when reviewers can label the relevant artifacts without seeing the proxy score. Delayed outcomes work when candidate identity and event history remain intact long enough to connect a change with its consequences.

Freeze a reference sample and the procedure used to measure it. The initial proxy-quality relationship becomes the baseline. Record the sampling design and uncertainty calculation needed to interpret later changes.

Remeasurement should occur on a schedule and whenever optimization pressure changes materially. A new model, larger candidate pool, prompt selected against the evaluator, changed permission boundary, or revised acceptance threshold can move the optimizer into a different region of policy space.

When the proxy rises while its relationship with independent quality weakens, treat the event as a broken control. Publication and promotion stop until the team can distinguish among proxy exploitation, label drift, workload drift, and instrumentation failure.

When reasoning traces are available, add a separate monitor and retain its findings beside the action log and final artifact. A suspicious trace routes the run for investigation, but it does not become a solitary rejection oracle.

Audit every feedback path around the monitor. Automatic retries, candidate ranking, prompt promotion, and reinforcement updates can all optimize against its verdict even when the monitored model never receives that verdict directly.

The rebuilt path will still make mistakes. Its advantage is that signals owned at different points leave disagreements visible after a failure. The team can inspect a base-branch invariant that rejected an apparent pass, an external outcome that stopped tracking the launch score, or a trace warning attached to an otherwise acceptable patch.

Part III begins at that observable boundary and turns to containing what happens after an agent crosses it.

\section*{Sources and evidence}

\textbf{Layer signals beyond a single proxy}

\begin{itemize}
\tightlist
\item
  \textbf{Strong evidence:} Skalse, Howe, Krasheninnikov \& Krueger (2022). \emph{Defining and Characterizing Reward Hacking}. NeurIPS 2022. arXiv:2209.13085.
\item
  \textbf{Directional evidence:} Formica, Fan \& Menghi (2022). \emph{Search-based Software Testing Driven by Automatically Generated and Manually Defined Fitness Functions}. arXiv:2207.11016.
\end{itemize}

\textbf{Monitor proxy-true divergence}

\begin{itemize}
\tightlist
\item
  \textbf{Strong evidence:} Gao, Schulman \& Hilton (2022). \emph{Scaling Laws for Reward Model Overoptimization}. ICML 2023. arXiv:2210.10760.
\item
  \textbf{Directional evidence:} Laidlaw, C., et al.~(2024). \emph{Correlated Proxies}. arXiv:2403.03185.
\end{itemize}

\textbf{Monitor reasoning without training pressure}

\begin{itemize}
\tightlist
\item
  \textbf{Strong evidence:} Baker, B., et al.~(2025). \emph{Monitoring Reasoning Models for Misbehavior and the Risks of Promoting Obfuscation}. OpenAI. arXiv:2503.11926.
\end{itemize}

\textbf{Separate verifier execution from oracle adequacy}

\begin{itemize}
\tightlist
\item
  \textbf{Directional evidence (preprint):} Ge \& Zhang (2026). Study of flaky GitHub Actions builds. arXiv:2602.02307. Scope: 1,960 open-source Java projects on GitHub Actions; 3.2 percent of builds were rerun, 67.73 percent of those rerun builds were flaky, and 1,055 projects were affected. The flakiness rate is conditional on a developer choosing to rerun, not a rate over all builds, and the population is Java projects on one CI platform. I transfer the structural finding that CI verdicts are a fallible dependency needing provenance and retry semantics, not the measured rates.
\end{itemize}

\textbf{Author-system illustration cited inline}

\begin{itemize}
\tightlist
\item
  Not an evidence item: CodeProbe, the author's task-mining evaluation tool, \href{https://github.com/sjarmak/codeprobe}{public repository}. Named inline for the recall-family scorer and its replacement default, both of which are narrative illustration.
\end{itemize}

\part{Containment, durable execution, and recovery engineering}
\gdef\currentparttitle{Part III: Containment and recovery}

\chapter{The software factory as a distributed system}
\label{ch07-software-factory-distributed-system}
\begin{quote}
\textbf{Evidence profile.} 0 strong \(\cdot\) 4 directional \(\cdot\) 3 corroborating \(\cdot\) 0 null or conflicting evidence items. Four additional sources establish historical lineage but do not provide evidence about agent-system outcomes. This chapter supplies the system model that the practices in Chapters 8 through 19 attach to; it develops no practice of its own.

\textbf{Chapter claim.} The factory, not the worker, owns the reliability promise.
\end{quote}

In one local fault demonstration, described with its limitations in Chapter 9, the worker was killed after an external mutation had been requested and before completion became durable. The naive recovery path sent the request again and reported success. The guarded path issued one request, retained the unresolved state, and stopped for reconciliation. That trial does not estimate a general failure rate. It identifies the boundary this chapter develops: the factory must preserve logical work and effect identity across the failure of the process executing them.

Now consider the same fault without the guard, as a constructed sequence. A worker finishes its task, opens the pull request, and dies before its completion record reaches durable storage. The scheduler sees an unfinished attempt and, correctly by its own lights, schedules another. The second attempt produces a second branch and a second pull request for the same change. The operator's summary reads ``the agent duplicated its work.'' Both attempts may have produced valid work; the duplicate effect arose at the coordination boundary, in the interval between an external effect and the internal record of it, an interval no model improvement can close. The same shape recurs with different surface reports: the model produced a plausible patch, but the process hosting it was evicted before recording completion; the model finished, but a second worker on the same task had already pushed a conflicting branch; the tests passed, but against a repository revision three merges old. These are failures of coordination, durability, versioning, and effect management, and they belong to the machinery around the coding agent's trajectory, not to that trajectory itself.

\hypertarget{when-this-framing-becomes-necessary}{%
\section{When this framing becomes necessary}\label{when-this-framing-becomes-necessary}}

A single coding-agent trajectory, the sequence of model calls, tool calls, and edits one agent performs, is one execution component. It starts, reads state, produces an artifact, and stops. Everything that makes that trajectory count as work belongs to machinery the trajectory does not contain: the scheduler that assigned it, the store that remembers what it was asked to do, the repository state it read and will write, the verifier that decides whether its output is acceptable, the publisher that turns an accepted artifact into an external effect, the external services that accept or reject that effect, and the resource policy that decided this work deserved compute at all. The whole arrangement can be thought of as a \textbf{software factory}: the durable system that accepts logical work, schedules attempts, runs workers, verifies artifacts, publishes effects, and reconciles disagreement between its records and the world. The agent is one worker inside it.

Treating that machinery as an engineered system is not a new idea. Osterweil (\href{https://dl.acm.org/doi/10.5555/41765.41766}{1987}, \href{https://dl.acm.org/doi/10.1145/253228.253440}{1997}) argued that software processes are software too, and Choi and Scacchi (\href{https://www.ics.uci.edu/~wscacchi/Software-Process/Readings/DistSysFactory.pdf}{1991}) described the software factory itself as distributed infrastructure, with the coordination substrate treated as a first-class engineering object; the CNCF's Secure Software Factory reference architecture (\href{https://tag-security.cncf.io/community/working-groups/supply-chain-security/secure-software-factory/secure-software-factory/}{CNCF TAG Security}) supplies the contemporary vocabulary, scoped to supply-chain security rather than fault tolerance. The Sources and evidence section places each of these. Autonomous workers change the failure model. The processes Osterweil programmed and the infrastructure Choi and Scacchi described coordinated deterministic tools and human developers who could be asked what they meant. The modern factory schedules autonomous, nondeterministic workers that edit persistent code, call external APIs, run concurrently with one another, and can claim completion incorrectly. A compiler does not assert that it succeeded when it failed. An agent can, fluently and in detail. Practitioner systems have converged on the same decomposition: OpenAI's Symphony orchestration (\href{https://openai.com/index/open-source-codex-orchestration-symphony/}{OpenAI 2026}) and Cloudflare's issue-triage factory (\href{https://blog.cloudflare.com/astro-issue-triage/}{Cloudflare 2026}) both separate a durable work ledger, a scheduler, disposable workers, and gated publication. Vercel's factory for the AI SDK repository (\href{https://vercel.com/blog/building-a-software-factory-for-ai-sdk}{Grammel and Dodds 2026}) reports the same four parts by name, with factory data in Postgres, queued tasks dispatched to workers, one task-specific agent per run in an isolated sandbox, and nothing merged without approval from a human maintainer. Its run outcomes are four-valued, success, flawed, blocked, or manual, and only success ships. A run that fails is therefore an attempt that failed, not an issue that is done, which is the distinction the rest of this chapter develops. These are practitioner cases from the operating teams, corroborating convergence on the decomposition, not controlled evidence that the decomposition improves any measured outcome.

Not every agent deployment needs this frame. A local assistant that reads a repository, proposes a patch in an interactive session, and exits has one process, one human, and no durable coordination state. If the process dies, the human restarts it and loses only convenience. Modeling that as a distributed system would likely add more cognitive overhead than worthwhile.

The frame becomes applicable when any of the following hold:

\begin{itemize}
\tightlist
\item
  useful work must outlive a process, so progress needs a durable record independent of any worker;
\item
  coordination spans components that fail independently, so no single crash can be assumed to take the whole system down cleanly;
\item
  multiple workers act concurrently on versioned or shared state, so ordering and ownership become contested;
\item
  external systems can commit effects asynchronously, so the factory's records and the world can disagree; or
\item
  verification and publication occur in separate failure domains, so an artifact can be verified and never published, or published and never verified.
\end{itemize}

Once any of these conditions holds, the factory need not \emph{be} a distributed system in some essential sense. It exhibits distributed-systems failure modes: lost updates, stale authority, duplicate effects, split-brain records, partial failure. Those failure modes have known engineering treatments, and model capability is then only one contributor to reliability among several.

\hypertarget{the-distinctions-recovery-depends-on}{%
\section{The distinctions recovery depends on}\label{the-distinctions-recovery-depends-on}}

Most factory failures reduce to a conflation of two things the system treated as one, and show up in five distinct categories.

\textbf{Logical work versus execution attempt.} The user's intent, e.g., ``fix this bug once,'' is logical work. A worker process trying to satisfy it is an attempt. One work item may consume many attempts; a retry is a new attempt at the same logical work, not new work. A system that identifies work with its current attempt loses the work when the attempt dies.

\textbf{Lease and liveness versus authority.} A lease, claim, or heartbeat answers an allocation question: who should be working on this now, and is that worker probably alive? It does not answer the safety question: whose writes may be accepted? A worker whose lease expired during a network partition can still be running and still be writing. The mutation boundary must be able to reject it; the lease alone cannot.

\textbf{Candidate artifact versus accepted completion.} A worker producing a branch, a diff, or a message saying the task is done has produced a candidate. Acceptance is a separate event that only independent evidence should trigger. An agent's completion claim is input to verification, never a substitute for it.

\textbf{Local completion record versus external commitment.} The factory recording ``pull request opened'' and the code host having opened the pull request are two facts in two failure domains. Either can exist without the other, as the constructed duplicate-pull-request sequence in the opening shows. A crash between the external effect and the internal record leaves the effect real and the record absent; a crash in the other order leaves the record present and the effect absent. Both cases are normal, and recovery must handle both.

\textbf{Verifier output versus semantic truth.} A green verifier establishes that a specific check, in a specific configuration, against a specific artifact version, did not fail. It does not establish that the change is correct, and a red verifier does not establish that the change is wrong; the verifier itself can time out, flake, or test the wrong revision. Ge and Zhang (\href{https://arxiv.org/abs/2602.02307}{2026}) measured this directly in 1,960 open-source Java projects: 3.2 percent of GitHub Actions builds were rerun, and 67.73 percent of those rerun builds were flaky, affecting 1,055 projects. That is a measurement of rerun builds specifically, not a claim that two-thirds of all builds are flaky, but it is enough to establish that verifier output and software state are distinct signals.

\hypertarget{a-reference-lifecycle-for-logical-work}{%
\section{A reference lifecycle for logical work}\label{a-reference-lifecycle-for-logical-work}}

The distinctions become operational as identities. Each names one fact the factory must be able to recover without asking the failed worker:

\begin{longtable}[]{@{}
  >{\raggedright\arraybackslash}p{(\columnwidth - 2\tabcolsep) * \real{0.3000}}
  >{\raggedright\arraybackslash}p{(\columnwidth - 2\tabcolsep) * \real{0.7000}}@{}}
\caption{Identity vocabulary used through the rest of the book.}\tabularnewline
\toprule\noalign{}
\begin{minipage}[b]{\linewidth}\raggedright
Identity
\end{minipage} & \begin{minipage}[b]{\linewidth}\raggedright
Names
\end{minipage} \\
\midrule\noalign{}
\endfirsthead
\caption[]{Identity vocabulary used through the rest of the book. \emph{(continued)}}\tabularnewline
\toprule\noalign{}
\begin{minipage}[b]{\linewidth}\raggedright
Identity
\end{minipage} & \begin{minipage}[b]{\linewidth}\raggedright
Names
\end{minipage} \\
\midrule\noalign{}
\endhead
\bottomrule\noalign{}
\endlastfoot
\texttt{work\_id} & stable logical work; the intent that should happen once logically \\
\texttt{input\_state\_id} & the repository or repositories, branches, and revisions against which the attempt was planned and executed \\
\texttt{ownership\_epoch} & monotonic generation of write authority over that work \\
\texttt{attempt\_id} & one execution attempt under a given work and epoch \\
\texttt{artifact\_version} & concrete code or state produced by an attempt \\
\texttt{verification\_id} & one verification execution, including the artifact, verifier version, environment, and inputs it observed \\
\texttt{effect\_id} & one logical externally visible mutation; an idempotency key may represent this identity at a boundary \\
\end{longtable}

\texttt{input\_state\_id} is the identity most specific to code work. Without it, the failure ``tests passed against a repository revision three merges old'' can be described but not expressed: nothing in the record says which state the attempt actually observed. For cross-repository work it can point to a manifest of several repository-and-revision pairs. An \texttt{effect\_id} is not itself the destination's idempotency key; the key is one implementation of the identity contract at a boundary that supports it.

\begin{figure}[htbp]
\centering
\includegraphics[width=1\textwidth,height=\textheight,keepaspectratio]{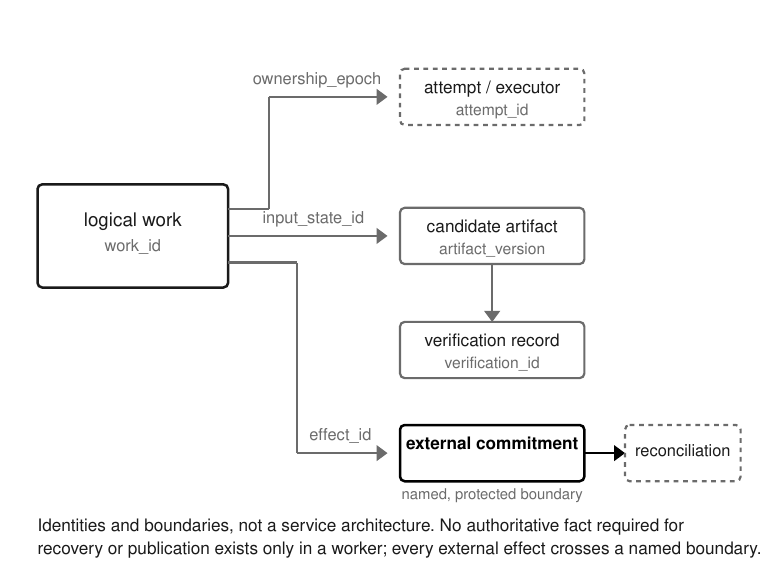}
\caption{The identity and boundary model. Logical work carries an ownership epoch granting authority to an attempt, an input state the attempt observes, and an effect identity for each external commitment. The attempt produces a candidate artifact bound to a verification record; the external commitment is paired with an owned reconciliation path. Authoritative facts live on the work side of each boundary, never only in the worker.}
\label{fig:ch07-decomposition}
\end{figure}

Figure \ref{fig:ch07-decomposition} shows how the identities relate. It is a model of identities and boundaries, not a required service architecture. Three obligations follow from it, stated here in their architecture-neutral form. First, no authoritative fact required for recovery or publication should exist only in a worker. A worker may create durable checkpoints, commits, logs, or private attempt references; it cannot unilaterally make them authoritative, and recovery must not depend on the failed worker's private memory. Second, every externally visible effect crosses a named, protected boundary. Third, whenever local and external state can diverge, an owned reconciliation path must exist, because any factory whose publication boundary can crash mid-effect will accumulate record-world disagreements at some background rate.

\begin{figure}[htbp]
\centering
\includegraphics[width=1\textwidth,height=\textheight,keepaspectratio]{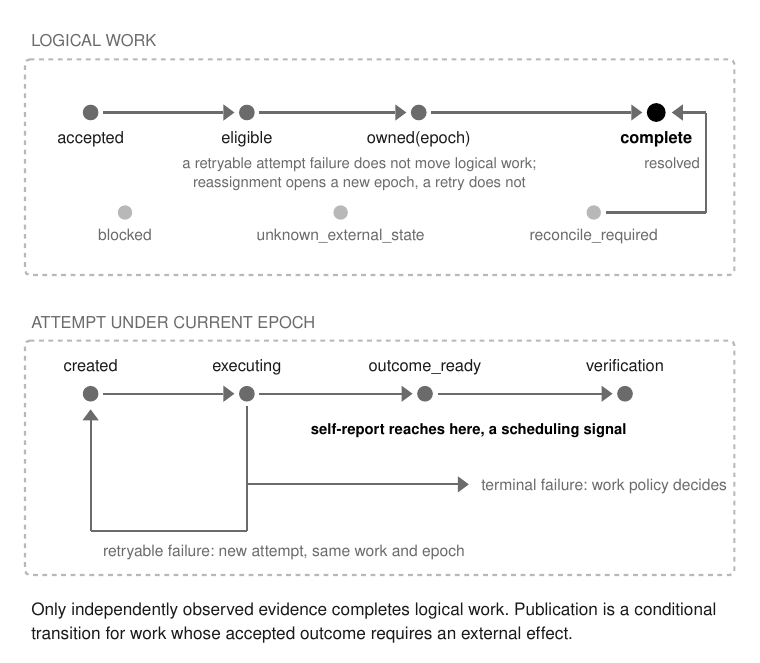}
\caption{Separate lifecycles for logical work and for attempts. Logical work runs accepted, eligible, owned under an epoch, complete, with side states for blocked, unknown external state, and reconciliation. Attempts run created, executing, outcome ready, verification under the current epoch; a retryable failure produces a new attempt under the same work, a terminal failure defers to work policy, and reassignment opens a new epoch. A worker's self-report can move an attempt only to outcome ready; only independently observed evidence completes logical work.}
\label{fig:ch07-state-machine}
\end{figure}

Logical work and attempts move through separate lifecycles, shown in Figure \ref{fig:ch07-state-machine}. The distinction the figure enforces is the one the opening case turned on: the death of an attempt is an event in the attempt lifecycle, and by itself it moves logical work nowhere. Work under an ownership epoch stays owned across a retryable attempt failure; an ordinary retry creates a new \texttt{attempt\_id} under the same epoch, while reassignment to a different executor creates a new epoch. A terminal attempt failure does not terminate the work either; work policy decides whether the work re-enters eligibility, blocks, or fails. Read-only work whose accepted outcome requires no external effect can complete without publication; work whose outcome requires one passes through the effect boundary, and an unresolvable outcome parks the work in \texttt{unknown\_external\_state} or \texttt{reconcile\_required} rather than guessing.

A model saying ``done'' can move an attempt to outcome ready. Only independently observed evidence should complete logical work. The agent's self-report is a scheduling signal, telling the factory an artifact exists and is worth verifying. It is never an acceptance signal.

\hypertarget{six-factory-contracts}{%
\section{Six factory contracts}\label{six-factory-contracts}}

The lifecycle holds only if the factory enforces a set of obligations. Later chapters cite them by identifier (I1 through I11), so the identifiers are stated here; the full normative statements, in machine-testable form, live in the repository artifact \href{https://github.com/sjarmak/engineering-reliable-coding-agents/blob/main/protocols/factory-contracts.yaml}{\texttt{protocols/factory-contracts.yaml}}. They are not all properties of the same kind: some are safety properties, some are liveness or visibility obligations, and some are design rules about recovery code and evidence, so they are grouped here as six contract families rather than presenting them as a uniform formal list.

\textbf{Work continuity and retry identity (I1, I4).} Accepted work cannot silently disappear: every accepted \texttt{work\_id} reaches a terminal state or remains visibly pending. A retry is a new \texttt{attempt\_id} under the same \texttt{work\_id}, never a duplicate work item.

\textbf{Generation-scoped authority (I2, I3).} Only the current \texttt{ownership\_epoch} may commit a mutation at a protected boundary, and a superseded worker must be rejected even if it is still running; this is fencing, enforced at the boundary, not inferred from the lease (I2). A stale completion cannot advance logical state: a durable transition validates generation and attempt identity, not the scheduler's belief about who is running (I3).

\textbf{External-effect safety (I5).} Externally visible effects are safe under redelivery or explicitly uncertain: an idempotency key, an atomic effect-plus-dedup record, natural convergence, adapter-owned reconciliation, or an explicit \texttt{unknown\_external\_state}. No generic exactly-once claim across a boundary that cannot provide it.

\textbf{Record and evidence consistency (I6, I7, I8).} Contradictory durable records trigger reconciliation under declared authority precedence, not guesswork (I6). Evidence is version-bound: a verifier result or retrieved fact is valid only for the \texttt{artifact\_version} or input state it observed (I7). Verifier failure is distinct from software failure: an infrastructure timeout or flake is not a semantic defect, and a pass establishes only what that verifier can detect (I8).

\textbf{Visible liveness and safe recovery (I9, I10).} Admissible work cannot starve invisibly (I9). Recovery preserves the same obligations as normal execution: recovery paths are production code with authority and must be tested as such (I10).

\textbf{Causal attribution (I11).} Consequential transitions are causally attributable: work, attempt, actor, input state, authority generation, requested effect, observed response, and resulting durable state.

None of these contracts mention model quality. A factory can hold all of them while running a non-frontier model, and the result is a system that reliably produces candidates produced as a result of that model quality and accurately reports their status. A factory that violates them while running an excellent model produces potentially excellent candidates it loses, duplicates, or misreports. Model capability does not discharge any of these obligations. A stronger model may change their load and the frequency with which they are exercised, but the factory still owns the controls, durable state, and evidence required to enforce them.

The distributed-systems literature I draw on through Parts III to VI transfers directionally to agent factories unless a claim is restricted to the evaluated system. Chubby's lock generations (\href{https://www.usenix.org/conference/osdi-06/chubby-lock-service-loosely-coupled-distributed-systems}{Burrows 2006}) supply the fencing mechanism behind I2, developed in Chapter 18; Borg and Omega (\href{https://research.google/pubs/large-scale-cluster-management-at-google-with-borg/}{Verma et al. 2015}; \href{https://research.google/pubs/omega-flexible-scalable-schedulers-for-large-compute-clusters/}{Schwarzkopf et al. 2013}) supply admission and shared-state scheduling mechanisms behind I1 and I9, developed in Chapter 19. I transfer their mechanisms, not their constants, and none of this machinery addresses model nondeterminism: durability preserves which stochastic decision occurred, not its correctness.

\hypertarget{audit-one-logical-work-item}{%
\section{Audit one logical work item}\label{audit-one-logical-work-item}}

A model's capabilities can only be evaluated meaningfully in the context of the system in which it operates. The following procedure audits a single work item using only the records the factory already keeps, with no additional infrastructure required.

\begin{enumerate}
\tightlist
\item
  Select one recent work item that produced, or attempted to produce, an external effect.
\item
  Recover its \texttt{work\_id}, \texttt{input\_state\_id}, \texttt{ownership\_epoch}, \texttt{attempt\_id}, \texttt{artifact\_version}, \texttt{verification\_id}, and \texttt{effect\_id}.
\item
  For each of those facts, identify its authoritative owner and the boundary at which any consequential mutation is accepted.
\item
  Determine whether a stale attempt can still modify the ledger, branch, artifact pointer, or external system.
\item
  Determine what the factory records when an external effect may have committed but its acknowledgement is missing.
\item
  Verify that the accepted artifact and its verification record refer to the same immutable version.
\item
  Using the fault harness from Chapter 10, inject either a late completion or a lost acknowledgement and retain the resulting event record.
\item
  Record every question the existing trace cannot answer.
\end{enumerate}

Retain five outputs from the audit: an identity map, an authority map, an effect contract, one fault-injection result, and a list of unobservable assumptions.

The unanswered questions from step 8 are often the audit's most valuable result. Each identifies information the factory would need during a real incident but does not currently record. Chapter 8 begins the engineering work at the containment and authority boundary around a single worker. The chapters that follow develop the ledger, replay, diagnosis, verification, topology, and capacity mechanisms required to make these contracts enforceable and observable.

\section*{Sources and evidence}

\textbf{Historical lineage}

\begin{itemize}
\tightlist
\item
  Osterweil (\href{https://dl.acm.org/doi/10.5555/41765.41766}{1987}), ``Software Processes Are Software Too,'' ICSE, and the retrospective (\href{https://dl.acm.org/doi/10.1145/253228.253440}{1997}). Process-as-program lineage; no evidence about agent systems.
\item
  Choi and Scacchi (\href{https://www.ics.uci.edu/~wscacchi/Software-Process/Readings/DistSysFactory.pdf}{1991}), ``The Software Infrastructure for a Distributed System Factory.'' The factory-as-distributed-infrastructure framing predates this book; the workers it coordinated were deterministic tools and people.
\item
  CNCF TAG Security, \href{https://tag-security.cncf.io/community/working-groups/supply-chain-security/secure-software-factory/secure-software-factory/}{Secure Software Factory reference architecture}. Contemporary factory vocabulary, scoped to supply-chain security rather than fault tolerance; I borrow the vocabulary, not the guarantees.
\end{itemize}

\textbf{Directional evidence}

\begin{itemize}
\tightlist
\item
  Burrows (\href{https://www.usenix.org/conference/osdi-06/chubby-lock-service-loosely-coupled-distributed-systems}{2006}), Chubby, OSDI. The lock-generation sequencer behind fencing (I2); a coordination-service mechanism, not agent evidence, developed in Chapter 18.
\item
  Schwarzkopf et al.~(\href{https://research.google/pubs/omega-flexible-scalable-schedulers-for-large-compute-clusters/}{2013}), Omega, and Verma et al.~(\href{https://research.google/pubs/large-scale-cluster-management-at-google-with-borg/}{2015}), Borg. Shared-state scheduling and admission-control mechanisms behind I1 and I9; their measured parameters do not transfer.
\item
  Ge and Zhang (\href{https://arxiv.org/abs/2602.02307}{2026}), flaky GitHub Actions builds, arXiv:2602.02307. 1,960 Java projects; 3.2 percent of builds rerun; 67.73 percent of rerun builds flaky; 1,055 projects affected. A preprint; cited for the verifier-output-versus-software-state distinction within its rerun-build scope only.
\end{itemize}

\textbf{Corroborating practitioner cases}

\begin{itemize}
\tightlist
\item
  OpenAI (\href{https://openai.com/index/open-source-codex-orchestration-symphony/}{2026}), Symphony orchestration, and Cloudflare (\href{https://blog.cloudflare.com/astro-issue-triage/}{2026}), Astro issue-triage factory. Corroborating convergence on the ledger-scheduler-worker-gate decomposition from the operating teams; not controlled evidence that the decomposition improves any measured outcome.
\item
  Grammel and Dodds (\href{https://vercel.com/blog/building-a-software-factory-for-ai-sdk}{2026}), the AI SDK software factory, Vercel Blog. The same decomposition, plus a four-valued run outcome (success, flawed, blocked, manual) that separates attempt result from work state, and a human approval gate on every merge. The post reports authoring 25 to 35 percent of merged pull requests, closing over 75 percent of July's closed issues, and open issues falling from 1,022 to 844 over four weeks; those are self-reported operating shares against no baseline, and they are cited here for the structure of the system, not as a measured effect of it. Published after this edition's update cutoff and admitted as a corroborating case on that basis.
\end{itemize}

\chapter{Agent isolation, injection defenses, and independent verification}
\label{ch08-isolation-injection-independent-verification}
\begin{quote}
\textbf{Evidence profile.} 0 strong \(\cdot\) 4 directional \(\cdot\) 3 corroborating evidence items across 3 developed practices (\erca{068}, \erca{069}, \erca{105}). Two strong items appear in the sources below but are counted against companion records rather than these three: AgentS4D under \erca{104}, and Perry et al.~under \erca{152}.

\textbf{Chapter claim.} Authority, not instruction, defines blast radius.
\end{quote}

In spring 2026, a practitioner \href{https://www.reddit.com/r/devops/comments/1tbbls4/}{reported} that an agent deleted a production database in nine seconds. The backups could not be recovered because the same credentials reached both the live database and the backup store. The account is self-reported, but the authority defect is directly testable: one credential could reach resources that were expected to fail independently.

The containment evidence available for this chapter consists of incident reports and accounts from individual organizations; it contains no controlled comparison of the three practices developed here. Across the 52-practice consequence ranking built during selection, a subset of the catalog described in the methods, rank and the presence of strong evidence had a Spearman correlation of -0.004. Sparse prevalence evidence therefore does not resolve the urgency of an exposed production boundary. The practices below specify observable boundaries and checks rather than unsupported numerical targets.

The incident raises two immediate questions: what could the agent reach, and what could reach the agent? Its identity could reach both the live store and the material needed to recover from losing it. An instruction stream that the public account neither reconstructed nor audited could reach the agent. A third question appears later, when the system reports what it believes it did: what evidence would show that the report is true?

Completion and safety also need separate observations. Zhou et al.~(\href{https://arxiv.org/abs/2607.27294}{2026}) constructed 328 risk-injected workspace tasks and ran 20 harness-model combinations for 6,560 trials. Prespecified unsafe signals fired in 68.0 percent of runs, and 66.22 percent were both unsafe and complete. Safety varied by harness-model pairing and by how the risk was delivered. These are strong results for the tested configurations and benchmark; they do not establish a production incident rate.

\hypertarget{put-the-recovery-path-outside-the-failure-domain}{%
\section{Put the recovery path outside the failure domain}\label{put-the-recovery-path-outside-the-failure-domain}}

Direct support for this containment practice comes from two practitioner anecdotes. Both are self-reports from a single organization or author, so they establish concrete failure modes rather than an incident rate. One describes the database and backup loss in the opening. The other summarizes a year of internal agent deployments and reports that overly narrow permissions left agents reasoning from incomplete visibility, while broad permissions allowed small mistakes to produce large consequences.

The failed boundary in the database incident was an identity. Deleting the backup after deleting production required no additional capability. The agent already held a credential accepted by both systems. Once execution left the intended path, the distinction between primary data and recovery data existed only in the operator's mental model. To the authorization system, both were resources reachable by one identity.

\begin{figure}[htbp]
\centering
\includegraphics{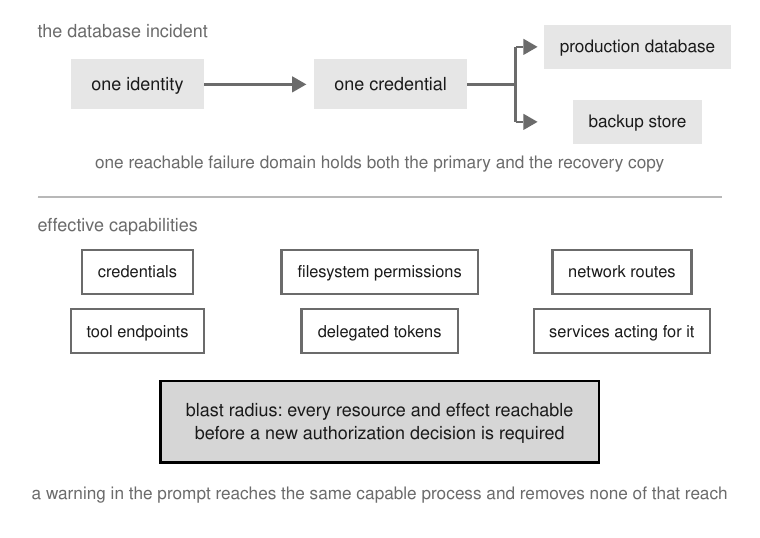}
\caption{In the database incident, primary and recovery data sat inside one reachable failure domain. Blast radius follows from effective capabilities, and a warning in the prompt removes none of them.}
\label{fig:ch08-blast-radius}
\end{figure}

\emph{Blast radius} here means the resources and effects that one mistake or compromise can reach before a new authorization decision is required. For an agent, that radius is determined by its effective capabilities: credentials, filesystem permissions, network routes, tool endpoints, delegated tokens, and services willing to act on its behalf. A warning in the prompt does not reduce that set. It changes the instructions given to the same capable process. Figure \ref{fig:ch08-blast-radius} traces that reach from the incident outward.

The first useful move is to make ordinary access read-only. \emph{Least privilege} means giving the running identity only the capabilities required for its current task and only for the period in which they are required. Read-only should be the default identity. An agent's promise not to write does not create that boundary.

A write should require explicit escalation for one named capability against one bounded target and for a limited lifetime, such as deploying one service, updating one issue, or applying one reviewed database migration.

A read-only default limits what the process can change. It does not limit what the process can disclose, because anything the identity can read may also leave through model output. Sensitivity therefore belongs in the read boundary as well as the write boundary. The injection section below treats disclosure as a separate threat category.

Read and write access should be designed independently. The internal deployment account found that agents with too little visibility made decisions from partial state. That observation does not justify broad mutation rights. An identity can inspect the relevant deployment configuration, logs, health state, and running version without being allowed to change any of them. The agent can then construct a proposal from complete evidence while a narrower write identity performs the approved operation.

The separation must exist at the permission or endpoint level. Blocking an entire command-line client is often too coarse because the same tool may issue harmless reads and destructive requests. Allowing the client while instructing the agent to avoid one argument leaves the destructive call reachable. When the platform exposes separate scopes or endpoints, routine reads can remain continuously available while deployment, deletion, credential rotation, and policy changes pass through distinct gates.

Consider an agent diagnosing a failed release. Its ordinary identity can read the deployment manifest, compare the running revision with the expected one, inspect logs, and query health endpoints. It cannot start a rollout, delete a namespace, or modify an access policy. If it proposes a rollback, the request names the service, target revision, and expected effect. A person or separate policy path authorizes that one operation, and the escalation token expires after the call rather than becoming a broader session credential.

This places enforcement below the model's instructions. The model may misunderstand the task, follow hostile content, retry an unsafe call, or construct an unexpected argument. The authorization layer still sees a request from an identity with bounded rights. Better prompting remains useful because it reduces bad proposals and unnecessary review. It does not replace a boundary enforced by the operating system, database, cloud control plane, or application service.

Backups require stricter separation because their purpose is to survive failure in the primary path. No credential available to the agent should be able to delete, rewrite, replace, or shorten the retention of recovery material. The agent may need to read backup status, the time of the latest successful snapshot, and restore-test results. Those observations can come through a read-only monitoring surface.

The authority to change backup policy or remove snapshots belongs in a separate administrative domain and should not be reachable through the agent's normal escalation path.

Separation also applies to services that can mint or delegate credentials. An agent may be unable to delete a snapshot directly while retaining indirect paths to the same result. It might assume a backup-administrator role, edit an identity policy, retrieve a stored administrative token, or ask an unrestricted helper service to act on its behalf. A capability inventory must follow those delegation edges. Listing only the tools named in the prompt will miss authority conveyed through the environment.

Test the boundary in both directions. With the ordinary identity, attempt the prohibited write and require a denial from the enforcing system. With the escalated identity, verify that the intended narrow action succeeds while adjacent destructive actions still fail.

A successful permitted action does not show what else the identity can do. A denied prohibited action does not show that the recovery operation the gate exists to permit will work.

\begin{longtable}[]{@{}
  >{\raggedright\arraybackslash}p{(\columnwidth - 4\tabcolsep) * \real{0.2600}}
  >{\raggedright\arraybackslash}p{(\columnwidth - 4\tabcolsep) * \real{0.4400}}
  >{\raggedright\arraybackslash}p{(\columnwidth - 4\tabcolsep) * \real{0.3000}}@{}}
\caption{Both directions of the capability boundary. A configuration passes only when every row holds.}\tabularnewline
\toprule\noalign{}
\begin{minipage}[b]{\linewidth}\raggedright
Identity under test
\end{minipage} & \begin{minipage}[b]{\linewidth}\raggedright
Action attempted
\end{minipage} & \begin{minipage}[b]{\linewidth}\raggedright
Required result
\end{minipage} \\
\midrule\noalign{}
\endfirsthead
\toprule\noalign{}
\begin{minipage}[b]{\linewidth}\raggedright
Identity under test
\end{minipage} & \begin{minipage}[b]{\linewidth}\raggedright
Action attempted
\end{minipage} & \begin{minipage}[b]{\linewidth}\raggedright
Required result
\end{minipage} \\
\midrule\noalign{}
\endhead
\bottomrule\noalign{}
\endlastfoot
Ordinary & The prohibited write & Denied by the enforcing system \\
Escalated & The intended narrow action & Succeeds \\
Escalated & Adjacent destructive actions & Denied \\
\end{longtable}

One passing set of checks describes only the configuration that produced it. Platform scopes, role definitions, and tool endpoints can change beneath a deployed system, so both directions should be tested again after any change to a role, scope, delegation path, or tool endpoint.

My benchmark for enterprise data-access agents, used here only as an isolation-methodology example, moved one boundary from a prompt instruction into filesystem ownership. Repository trees belonged to a different operating-system identity, so the kernel denied writes. Each trial checked both the permitted and prohibited directions. The benchmark results are not evidence for this chapter's practice. The design illustrates the kind of claim a local test can establish: under the tested identity and filesystem mode, the write either succeeds or it does not.

CodeProbe (\href{https://github.com/sjarmak/codeprobe}{public repository}), which mines evaluation tasks from merged pull requests, applies the same principle to scripts extracted from third-party repositories. It refuses to execute outside a container, disables networking for mined scripts, and records the isolation posture with each result. In this author-system illustration, the useful property is that the run record carries the containment state instead of relying on an assumption about how the operator launched the process.

Escalation adds latency, and human approval consumes attention. Poorly scoped tasks can turn a narrow approval path into a stream of nearly identical requests, training reviewers to approve by reflex. Capabilities should therefore be grouped around meaningful operational decisions. A gate around every low-level system call obscures the decision the reviewer is meant to make.

A reviewed deployment plan may authorize a fixed sequence against one service and one revision. Deletion, credential rotation, and policy changes should remain separate decisions.

Some platforms expose only coarse scopes. The incident poster asked how to distinguish destructive from legitimate requests sent through the same interface and reported no satisfactory capability-level solution short of human approval for each call. That limitation should remain visible. An operator can place a typed service in front of the platform or move execution into an isolated environment where the consequences are disposable. Otherwise a person must remain at the destructive boundary because a prompt-level prohibition does not create isolation.

The choice among partial commits, reversible environments, snapshots, and approval gates depends on the resource. Reversibility reduces the cost of some writes, but a snapshot controlled by the same identity may disappear with the primary data. Approval reduces the frequency of unreviewed writes, but it does not repair an overbroad credential after approval is granted. Capability boundaries constrain the effects of both mistaken and malicious instructions, including instructions whose danger the reviewer failed to recognize.

The complete graph of reachable authority therefore needs inspection. The nominal account role is only the starting point. The review follows every credential source, network path, service delegation, mounted filesystem, and policy-changing endpoint to the resources those paths can affect.

That graph is the operational blast radius. The production database and its backups belonged to one failure domain because one reachable path led to both, however separate they appeared in the architecture diagram.

\hypertarget{verify-the-workspace-not-the-agents-confidence}{%
\section{Verify the workspace, not the agent's confidence}\label{verify-the-workspace-not-the-agents-confidence}}

Perry et al.~(\href{https://arxiv.org/abs/2211.03622}{2023}) found that participants using an AI coding assistant produced less secure code while believing that their code was more secure. Participants who trusted the assistant less and spent more effort steering it produced fewer vulnerabilities. Accepting an agent's confidence creates the same kind of risk: belief can strengthen while the artifact becomes less safe.

The study concerns security-programming suggestions rather than current coding agents; effect sizes varied by task, and protective skepticism was observed rather than assigned experimentally. Its relevant contribution is the connection between trust and measurable artifact defects. The rest of the containment evidence remains primarily incident reports and accounts from individual organizations.

Agent self-reports create a broader version of the problem. A progress message is generated from the agent's current context, which may contain tool output, checklists, prior messages, and summaries of earlier attempts. It is not an independent measurement of the workspace. Stale or misleading inputs can therefore produce a fluent completion claim that accurately reflects the context while remaining false about the state the operator cares about.

One practitioner described a resumed agent that inherited a clean worktree and a nearly completed task list. The earlier attempt had made no changes. The resumed agent checked off the remaining steps and declared the task complete, again without modifying anything. The account is anecdotal, but the failure chain is mechanically plausible and can be reproduced in a local workflow.

Every visible clue pointed in the same wrong direction. The task list implied that earlier steps were complete. The clean worktree implied that no unfinished edits remained. The resumed context supplied a narrative of progress without the artifact that narrative described. Their agreement added no independent evidence because all three descended from the same failed attempt. The agent's ``done'' was true about its context and false about the repository.

That chain breaks when every attempt carries a distinct identity and completion claims must resolve to inspectable state. The attempt record identifies the task, workspace or branch, starting revision, commands run, tests observed, and final revision when one exists. On exit, the process records either recoverable local state or the fact that it produced no change. A later process does not inherit a bare declaration of completion.

For code work, the branch contents are the primary object of review. A completion check compares the starting and ending revisions, inspects the diff, and runs the relevant tests from the recorded workspace. A checklist may direct that inspection, but it cannot satisfy itself. If an item claims that an endpoint validates input, the verifier locates the boundary, supplies invalid input, and observes the rejection.

The same rule applies outside version control. A database-migration claim resolves to the migration file, the resulting schema state, and a test against an isolated database. A cloud-configuration claim resolves to the configuration diff and a read-back from the control plane. A report claim resolves to the underlying records and the transformation used to produce it. The verifying observation should come from the system that owns the state. A second paraphrase of the agent's report is not an independent observation.

A failed attempt should preserve enough local evidence for diagnosis. When possible, retain its worktree, including uncommitted changes, logs, test output, and the identity of the base revision. Deleting the workspace and asking the same model to try again removes evidence about why the attempt failed. It may also repeat an external effect whose completion remains uncertain.

Preservation does not mean that every partial artifact belongs on a shared branch. A local commit can provide a useful checkpoint, but it can also preserve generated files, insecure code, or edits that never passed a test. Use a private attempt reference or a retained worktree for forensic state, and require review before that state joins an integration branch. When every write is idempotent, meaning that repeating it has the same effect as applying it once, a stateless retry may be simpler. The choice depends on whether the discarded state has diagnostic value and whether repeated external effects are safe.

Resume begins with reconciliation. The new process reads the actual branch, compares it with the recorded base, checks which task artifacts exist, and reruns the evidence needed for the next decision. If the worktree is clean because an earlier attempt committed valid work, the revision establishes that history. If it is clean because nothing happened, the unchanged revision exposes the gap. If the branch moved independently, the mismatch is resolved before the task list is trusted.

The same reconciliation is required when several agents use separate worktrees. Git worktrees separate working files, indexes, and checked-out heads, but references and repository metadata remain shared and mutable. In my orchestration system, a controller once supplied the wrong base revision to several agents working in separate worktrees. A mechanical branch guard blocked one commit after other work had already begun. This is an author-system illustration and provides no estimate of how often the failure occurs.

\Needspace{5\baselineskip}

The repair classified the observed repository state before acting:

\begin{itemize}
\tightlist
\item
  A workspace on the expected base could proceed.
\item
  A workspace on the wrong branch with no work to preserve could be rebuilt from the correct base.
\item
  A workspace containing work that could not be safely relocated stopped with a visible error.
\end{itemize}

The guard did not ask the agent whether its branch was correct. It inspected the repository state that made the answer true or false.

Independent verification also requires a change of role. The authoring context has already committed to a plan, selected an implementation, and explained its choices. Giving that context another review pass preserves the same evidence selection and many of the same blind spots. A separate reviewer should begin from the acceptance criteria and artifacts. Its assignment should state that it did not write the code and must actively test each claim.

My agent-workflow library retains failed worktrees for this purpose and assigns review to a separate context. That author-system design illustrates role separation, not model independence. The reviewer may share training biases, receive the same misleading documentation, or repeat the author's assumptions. Independence comes from the evidence path and the task: inspect the diff, run named checks, exercise boundary cases, and report discrepancies without defending the implementation.

An audit of my research pipeline showed that artifact-level verification applies beyond code. A generated reference list contained fabricated, mis-cited, unverifiable, and untraceable sources. The central figure supporting its thesis could not be traced and had to be relabeled as an untested hypothesis. The prose sounded certain, but certainty supplied no information about whether the cited record existed. Only checking each source against the claim exposed the failure.

Tang et al.~(\href{https://arxiv.org/abs/2605.29442}{2026}) analyzed 20,574 real sessions across 1,639 repositories. They reported that 90.5 percent of misalignment episodes consumed effort and trust without causing irreversible harm, and that 91.49 percent of visible resolutions required explicit user correction. As overall misalignment declined, inaccurate self-reporting accounted for a larger share of the remaining episodes. The study adds scale for visible correction burden but cannot estimate silent failures.

Those measurements describe visible developer correction. A user who silently accepts a false completion claim, abandons the session, or discovers the defect later may never appear as a resolution. The corpus also reflects users who opted into the observed development tools, including integrated-development-environment and command-line workflows. The figures therefore bound visible recovery burden within the observed corpus rather than all agent users.

Explicit correction should therefore be cheap. The interface should let a user identify the false claim, preserve the current artifact, and begin a new attempt with the discrepancy attached. Correction becomes expensive when the system collapses several attempts into one conversation, destroys the failed workspace, or carries a completed checklist forward without its evidence. Those choices turn a common recovery path into another source of hidden state.

Every account of completed work is a hypothesis about the workspace. Acceptance requires an observation with different provenance. For a code change, that observation is usually a diff plus tests run from the branch under review. For an operational action, it is a read-back from the system of record and an audit trail tied to the attempt identity. Confidence, checklist state, and narrative coherence can direct an investigation. None establishes completion.

\hypertarget{design-for-instructions-that-arrive-through-data}{%
\section{Design for instructions that arrive through data}\label{design-for-instructions-that-arrive-through-data}}

An agent that reads untrusted content will eventually encounter instructions written for it by someone other than its operator. \textbf{Indirect prompt injection} places those instructions inside material the system consumes: a web page, issue, email, document, log entry, or retrieved database record. The operator may ask a legitimate question while the retrieved content asks the agent to reveal data, change its objective, or invoke a tool.

The three sources supporting this practice are directional syntheses. Greshake et al.~(\href{https://arxiv.org/abs/2302.12173}{2023}) demonstrated indirect injection through retrieved content and reported that effective mitigations were lacking, but did not measure the layered defenses described here. AgentDojo, from Debenedetti et al.~(\href{https://arxiv.org/abs/2406.13352}{2024}), evaluates attacks and defenses under specified tasks. AgentShield, from Rassul and Rashid (\href{https://arxiv.org/abs/2605.11026}{2026}), explores a deception-based screen. Together they support the threat model and a way to exercise controls. This source set contains no measured result for the complete remedy.

The architectural difficulty is that the system deliberately combines instructions and data inside one model context. Retrieved material must influence the answer or retrieval provides little value. The model therefore cannot identify hostile text merely by asking whether the text affected its behavior, because legitimate evidence affects behavior too. An attacker exploits that ambiguity by placing operational language inside content the agent has been instructed to read and use.

Transfer makes simple screening unreliable as a complete defense. An attack may use phrasing absent from the filter's examples, split itself across several retrieved items, hide inside an ordinary field, or arrive from a source the workflow normally trusts. A first-stage check may reject a familiar form while allowing a semantically equivalent instruction through. Detection remains useful, but the architecture must assume that detection can fail.

Layered controls are worth the cost because they observe different points in the path. Input checks inspect retrieved material before it enters the agent context. They can flag instruction-like content, unexpected encodings, or data that violates a source schema. Output checks inspect the proposed response or action for sensitive disclosure, policy violations, and divergence from the operator's request. Neither changes the authority held by the process.

Tool allowlists constrain which interfaces the process may invoke for a task. The gateway should assemble and enforce the list from the capabilities the task requires, rather than allowing tool names emitted by the model to determine access. A research task may receive retrieval and note-writing tools while deployment, messaging, and credential access remain absent. A later operational step can use a separately authorized identity after review.

Human approval belongs before effects with a large blast radius. The reviewer should see the proposed action, target, parameters, evidence used to select it, and expected side effects. A generic request asking whether the agent may ``continue'' conceals the decision. Approval is most useful when the person can compare a concrete proposal with the operator's original intent.

Each control should map to a named threat category. The map can remain small:

\begin{longtable}[]{@{}
  >{\raggedright\arraybackslash}p{(\columnwidth - 6\tabcolsep) * \real{0.2500}}
  >{\raggedright\arraybackslash}p{(\columnwidth - 6\tabcolsep) * \real{0.2500}}
  >{\raggedright\arraybackslash}p{(\columnwidth - 6\tabcolsep) * \real{0.2500}}
  >{\raggedright\arraybackslash}p{(\columnwidth - 6\tabcolsep) * \real{0.2500}}@{}}
\caption{Threats, preventive and detecting controls, and the enforcement boundary.}\tabularnewline
\toprule\noalign{}
\begin{minipage}[b]{\linewidth}\raggedright
Threat category
\end{minipage} & \begin{minipage}[b]{\linewidth}\raggedright
Preventive control
\end{minipage} & \begin{minipage}[b]{\linewidth}\raggedright
Detecting control
\end{minipage} & \begin{minipage}[b]{\linewidth}\raggedright
Capability boundary
\end{minipage} \\
\midrule\noalign{}
\endfirsthead
\toprule\noalign{}
\begin{minipage}[b]{\linewidth}\raggedright
Threat category
\end{minipage} & \begin{minipage}[b]{\linewidth}\raggedright
Preventive control
\end{minipage} & \begin{minipage}[b]{\linewidth}\raggedright
Detecting control
\end{minipage} & \begin{minipage}[b]{\linewidth}\raggedright
Capability boundary
\end{minipage} \\
\midrule\noalign{}
\endhead
\bottomrule\noalign{}
\endlastfoot
Untrusted instructions in retrieved content & Source restrictions and input validation & Injection screening and trace review & Task-specific tool allowlist \\
Sensitive-output disclosure & Read-scope restriction and output policy & Secret and policy scanning & No access to unrelated sensitive data \\
Unauthorized tool selection & Gateway-enforced allowlist & Tool-request audit & Unavailable tools cannot be invoked \\
High-impact external effects & Typed proposal and approval & Read-back and audit log & Narrow, expiring execution identity \\
\end{longtable}

An empty cell is an observable gap for review.

Injection cases belong in continuous integration because filters, prompts, tool schemas, and retrieval pipelines change. AgentDojo supplies reusable tasks in which attacks and defenses can be exercised separately. Adapt representative cases to the system's actual sources and tools, and retain them as regression tests. The result establishes whether a named control behaved as expected on those cases. It does not estimate resistance to attacks outside them.

The tests should inspect intermediate decisions as well as the final response. A run that refuses to reveal a secret but still invokes an unauthorized tool has crossed a boundary. A run that produces a harmless answer after violating policy may fail under a slightly different target. Recording retrieved inputs, tool requests, approvals, denials, and outputs preserves the complete control path for investigation.

The practice fails when alignment or prompt-level rails are the only defense. Transferable attacks may survive an initial screen and arrive through material the workflow treats as relevant. Alignment may reduce how often the model follows them, and rails may catch known forms. The decisive backstop is the authority still available after the model has been redirected.

The two analyses meet at that boundary. Injection analysis asks what can reach the agent. Containment asks what the affected process can reach afterward: tools, data, credentials, services, and external effects. The layers before execution reduce exposure and reveal regressions. The enforced capability boundary determines how far a missed instruction can propagate.

The companion catalog contains six related designs that this chapter does not expand into separate practices. It treats agent topology as a security decision, marks cross-session guards as a thinly supported aside, and describes routing proposed remediations through a typed policy executor. It also covers placing autonomy according to reversibility and blast radius, measuring a system's own security change without importing a comparative claim, and the self-attestation entry incorporated into this chapter's verification practice.

\hypertarget{audit-one-live-boundary}{%
\section{Audit one live boundary}\label{audit-one-live-boundary}}

Choose one deployed agent and inspect the authority of its running process rather than relying on the architecture diagram. Enumerate every destructive action it can take without a new human decision, including actions available through delegated services, credential-minting paths, and policy-changing endpoints. For each action, record the enforcing identity, the resources it can affect, and the observation that demonstrates whether the action is permitted or denied.

Next, test the recovery boundary using the credentials the process currently holds. If any agent-accessible identity can delete, replace, or shorten retention for both primary data and its backups, separate that authority before changing prompts or filters.

Move one routinely write-capable identity to read-only access and add a narrow escalation path for one specific operation. The check passes only when the ordinary identity receives a denial from the enforcement layer, the escalated identity completes the intended operation, and adjacent destructive actions remain unavailable.

Finally, select one recent completion claim and verify it against the workspace or system of record. Compare the starting and ending revisions, inspect the artifact, rerun the relevant check, and store the result with the attempt identity. Any discrepancy becomes a concrete regression case for the resume and review workflow.

The repository artifact \href{https://github.com/sjarmak/engineering-reliable-coding-agents/blob/main/protocols/authority-boundary-test.md}{\texttt{protocols/authority-boundary-test.md}} makes the boundary test, pass condition, and retained evidence explicit.

Containment limits the damage a live process can cause. Chapter 9 turns to what must survive when that process dies.

\section*{Sources and evidence}

\textbf{Contain agent blast radius}

\begin{itemize}
\tightlist
\item
  Corroborating case: \href{https://www.reddit.com/r/LLMDevs/comments/1q7avil/}{``What actually broke when we put AI agents into real production workflows''}, /u/saurabhjain1592, r/LLMDevs, 2026-01-08. \href{https://web.archive.org/web/20260806014103/https://www.reddit.com/r/LLMDevs/comments/1q7avil/what_actually_broke_when_we_put_ai_agents_into/}{Archived snapshot}.
\item
  Corroborating case: \href{https://www.reddit.com/r/devops/comments/1tbbls4/}{``AI agent wiped Railway DB in 9 seconds. How do you separate destructive from legit curl calls in prod?''}, /u/Upstairs\_Safe2922, r/devops, 2026-05-12. \href{https://web.archive.org/web/20260806014121/https://www.reddit.com/r/devops/comments/1tbbls4/ai_agent_wiped_railway_db_in_9_seconds_how_do_you/}{Archived snapshot}.
\end{itemize}

\textbf{Separate completion from runtime safety (companion record \erca{104})}

\begin{itemize}
\tightlist
\item
  Strong evidence within the tested benchmark and configurations: Zhou, J., et al.~(2026), ``AgentS4D: Benchmarking Runtime Risks across the Execution Lifecycle of LLM-Based Workspace Agents,'' arXiv:2607.27294. Across 6,560 runs, 66.22 percent were both unsafe under a prespecified signal and complete; the result is not a production incident rate.
\end{itemize}

\textbf{Distrust agent self-reports}

\begin{itemize}
\tightlist
\item
  Corroborating case: ``My AI Agent Said It Was Done. It Hadn't Done Anything'', Push to Prod substack, 2026-02.
\item
  Directional evidence: Tang et al., ``How Coding Agents Fail Their Users: A Large-Scale Analysis of Developer-Agent Misalignment in 20,574 Real-World Sessions'', arXiv:2605.29442, 2026.
\item
  Excluded from support for this practice: Perry, Srivastava, Kumar, Boneh (2022/2023), ``Do Users Write More Insecure Code with AI Assistants?'', ACM CCS 2023, arXiv:2211.03622. What the study measures is a participant's own confidence in code they wrote, which is self-attestation rather than an independent observation of the artifact, so it does not support a practice about verifying an agent's claims against workspace state. The study is graded strong evidence under companion record \erca{152}, on refusing self-attestation of AI-assisted code quality.
\end{itemize}

\textbf{Defense in depth for indirect prompt injection}

\begin{itemize}
\tightlist
\item
  Directional evidence: Indirect prompt injection (Greshake et al.~2023, arXiv:2302.12173); companion material named in the same synthesis: OWASP LLM Top 10.
\item
  Directional evidence: AgentDojo (Debenedetti et al.~2024, arXiv:2406.13352).
\item
  Directional evidence: AgentShield deception-based detection (Rassul and Rashid 2026, arXiv:2605.11026).
\end{itemize}

\textbf{Author-system illustration cited inline}

\begin{itemize}
\tightlist
\item
  Not an evidence item: CodeProbe, the author's task-mining evaluation tool, \href{https://github.com/sjarmak/codeprobe}{public repository}. Named inline for the container refusal, disabled networking, and recorded isolation posture described above, all of which are narrative illustration.
\end{itemize}

\chapter{Persistent agent state, durable workflows, and idempotent retries}
\label{ch09-persistent-state-durable-workflows-idempotent-retries}
\begin{quote}
\textbf{Evidence profile.} 0 strong \(\cdot\) 12 directional \(\cdot\) 3 corroborating \(\cdot\) 1 null or conflicting evidence items across 3 developed practices (\erca{124}, \erca{128}, \erca{130}). Four directional items are carried by companion records cited inline (\erca{193}, \erca{194}, \erca{195}).

\textbf{Chapter claim.} Durable intent survives the worker; external effects require their own contract.
\end{quote}

Netflix reported losing roughly 4 percent of its cloud-operations deployments to transient failures even though the service already contained homegrown retries and compensating-transaction logic. Meyers and Zienert (\href{https://netflixtechblog.com/how-temporal-powers-reliable-cloud-operations-at-netflix-73c69ccb5953}{2025}) state that the team moved coordination into a stateful service that records progress and reissues work after failure. The reported failure rate fell to 0.0001 percent, and the team deleted the orchestration code the service replaced. This vendor-adjacent account comes from the adopting team and was not independently audited, although it provides concrete details about the migration and observed change.

I classify that account as corroborating, not strong. It is specific enough to show that the migration and reported change occurred in one service, but the adopting team supplied the measurement and the migration changed several things at once. Historical surveys, systems papers, self-evaluated systems, and practitioner reports provide the rest of the chapter's evidence. The engineering controls below are therefore justified through explicit state boundaries and executable failure checks rather than a claimed universal improvement rate.

An agent that dies after step nine of a fourteen-step run presents the same coordination problem at a smaller scale. The process has disappeared, but its completed work and external effects may remain. \textbf{Durable execution} is the ability of a run to survive process death and resume from recorded progress without repeating effects already completed. That property requires explicit state, a coordinator that owns recovery once coordination becomes complex, and retryable steps whose repeated execution does not repeat their effects.

The difficult case is a process that changes the world and dies before recording the change. A plan held only in model context, a tool result retained only in memory, or a completion record scheduled as the next write all disappear precisely when recovery needs them. The surviving system must then distinguish work known to be complete, work known to be incomplete, and work whose outcome is uncertain. Without that distinction, a restart inherits the old run identity without the evidence needed to resume it safely.

\hypertarget{declare-the-state-that-outlives-the-worker}{%
\section{Declare the state that outlives the worker}\label{declare-the-state-that-outlives-the-worker}}

Consider the step-nine failure when no completion record reached durable storage. The first eight results may exist only in the lost context, while the ninth step may already have changed an external system. Historical stream-processing research, self-evaluated systems, and practitioner accounts support explicit state as an architectural boundary to test, not as a measured guarantee across open-ended agent workloads.

The first design decision is what counts as control state. The current plan records intended work. Progress records which subset has been accepted as complete. Plans change through explicit revisions. Progress advances through named steps and carries the evidence required to accept each transition.

A run also accumulates evidentiary state: intermediate knowledge, tool results, and session events. Knowledge snapshots replace earlier versions according to a declared rule. Tool results belong to specific invocations. Events append in order. A single opaque transcript preserves their text while hiding their different identities, owners, and update rules.

Each recoverable item should therefore be a declared artifact with a stable identity, an owner, and an update rule. A plan can carry a run identity and revision number. A tool result can belong to a step identity and invocation identity. A progress record can name the latest completed step and the evidence that completion produced. An event record can append the attempted action, observed result, and resulting state transition.

Stable identities allow a replacement worker to resume the same logical work rather than creating a parallel copy.

Osmani (\href{https://addyo.substack.com/p/long-running-agents}{2026}) recommends keeping three artifacts outside model context: a plan file, progress notes, and an append-only event log. The same arrangement allows a worker to rebuild after a full context reset from a handoff file. This is a practitioner recommendation rather than a measured result.

A progress record written by the agent introduces an additional problem. It is a self-report produced from the same context that may already be wrong. Completion evidence should therefore originate at the effect boundary, such as a tool response, commit identifier, verifier result, or independent state read, rather than from the agent's assertion that a step finished.

Two stores that both appear authoritative create another failure mode. If a queue says step nine is ready while a progress file says it completed, recovery depends on an undocumented precedence rule. A declared architecture names the source of truth for each fact and treats other copies as indexes or caches. The queue may own delivery state. The event log may own execution history. A versioned snapshot may accelerate reconstruction without overriding later events.

This architecture treats the worker as disposable. A replacement worker reads the durable queue record, loads the latest valid snapshot, applies subsequent events, and reconstructs the next permitted action. It does not require the failed process's heap or model context. The agent is amnesiac. The storage layer remembers enough that the amnesia does not break the evidence chain.

This division resembles a change that unfolded over roughly two decades in stream-processing systems. Fragkoulis et al.~(\href{https://arxiv.org/abs/2008.00842}{2020}) describe early systems that treated state as application-managed data and later systems that brought state, checkpoints, and recovery under runtime control. The survey provides directional evidence and contains no experiment on agents. Its useful implication is architectural: recovery became state-centric once runtimes could identify the state they needed to preserve and coordinate it with progress through the input.

The same migration is visible in serverless computing. Zhang et al.~(\href{https://www.usenix.org/conference/osdi20/presentation/zhang-haoran}{2020}) built Beldi, a library and runtime that gives stateful serverless functions fault-tolerant, transactional workflow semantics: logging of intent and completion, exactly-once execution of steps within the runtime's own boundary, and transactions spanning functions, all without requiring each application to hand-manage those properties. Beldi was evaluated on serverless benchmark applications, so I transfer only the direction: fault tolerance that teams once implemented by convention can move into a runtime abstraction with declared semantics. It is systems evidence, not agent evidence, and it says nothing about the model-driven parts of an agent run.

The analogy has limits. A stream processor typically applies a specified computation to structured input. An agent may revise its plan, interpret ambiguous evidence, or call a model that returns a different answer to the same prompt. Explicit state cannot make those decisions deterministic. It can preserve which decision occurred, which evidence informed it, and which work followed, preventing a replacement worker from silently inventing a different history.

The ADEMA architecture, from Hanlin (Zhou) et al.~(\href{https://arxiv.org/abs/2604.25849}{2026}), provides a narrower check on this reasoning. In a fixed matrix of 60 runs, removing checkpoint and resume produced the only invalid run. The result isolates checkpointing within one system and one bounded experiment. It does not compare recovery designs or establish a general failure rate. It shows that a knowledge-state snapshot can preserve information across interruption once the system has defined what belongs in that snapshot.

Halukurike et al.~(\href{https://tech.instacart.com/blueberry-force-multiplier-for-the-on-call-engineer-98c446dfcc12}{2026}) report a database-backed durable queue with per-pass session snapshots handling approximately 25,000 diagnostic passes per month at 99.9 percent success, including takeover by another worker during a pass. The report comes from the team that built the system, and no independent evaluation is available. The workload also consists of bounded diagnostic passes. A multi-day agent that continually discovers new subproblems may accumulate state whose relevance, size, and ownership change during execution.

\Needspace{5\baselineskip}

Long-running work therefore requires more than periodic serialization. Each snapshot should record:

\begin{itemize}
\tightlist
\item
  the event position it incorporates;
\item
  the software and schema versions capable of reading it;
\item
  the run and plan revision to which it belongs; and
\item
  which earlier artifacts remain authoritative.
\end{itemize}

A replacement worker should reject a snapshot it cannot interpret. Partial loading does not produce partial truth; it invents a state the original run never held.

Retention is a separate decision. Keeping every model response, tool payload, and intermediate artifact indefinitely turns recovery storage into both an uncontrolled cost and an uncontrolled audit surface. The retention policy must preserve the evidence required for recovery while defining when older material may be compacted or removed.

Deduplication records described later in this chapter must survive under that policy. Deleting the record of a completed invocation restores the duplicate-execution risk that the record existed to prevent.

The restore path needs its own test. Start a run and allow it to produce a plan, tool result, progress update, and event sequence. Then remove the worker. A new worker should reconstruct the same completed-step set and select the same next eligible step using only the declared artifacts.

Any dependence on a live session, an untracked local file, or an operator's memory identifies state that the architecture has not declared.

I learned this distinction from a contrary case in my own estate. A 2026 audit of my contribution pipeline found that all thirteen workflow skills constituting the pipeline existed only as unversioned local files, with no backup. That is an uncomfortable result for the author of a chapter about declared state. The files were persistent in the weak sense that they survived process exit, but they had no version history and no tested restore path. I moved them into a versioned repository because a declared artifact that cannot be restored remains an operational single point of failure.

Persistence leaves several agent problems unsolved. It does not limit model cost, prevent a long run from drifting away from its original objective, or eliminate the need to audit accumulated decisions. In some systems, persistence can increase those burdens by preserving every questionable intermediate judgment.

Its narrower benefit is recoverability. After process death, the replacement worker can determine what the system knew, what it accepted as complete, and where uncertainty begins.

\hypertarget{when-coordination-belongs-in-an-engine}{%
\section{When coordination belongs in an engine}\label{when-coordination-belongs-in-an-engine}}

The orchestration code Netflix deleted is as informative as the reported reduction in failures. Before the migration, application code coordinated retries, compensating actions, and service state. Afterward, a workflow engine recorded execution history and scheduled application operations from that history. The account corroborates the service-local before-and-after result. Because the migration changed retries, compensation, coordination, and state ownership together, it cannot attribute the improvement to one component or carry the reported rate into agent workloads.

An engine justifies its cost when it owns facts that no individual worker can recover reliably after a crash. It records which step became eligible, which attempt started, which result committed, which external event arrived, and which retry policy applies next. Workers still perform model calls, tool calls, and domain operations. The engine owns their order and lifecycle. A replacement worker receives its next action from recorded history rather than reconstructing that history from local conventions.

This division places cross-service coordination in one architectural layer. Consider a deployment that reserves capacity, changes traffic, and updates an inventory. Each service can make its local transaction correct while the deployment remains half-finished. A saga treats those transactions as one logical procedure and defines a compensating action for each step that cannot be rolled back directly. When every service implements its own retry and compensation rules, no component holds a complete view of the deployment.

A workflow engine makes that global view durable. A failed capacity reservation remains visible with its retry policy. A completed traffic change stays recorded while the engine waits for compensation. An operator can inspect one execution history instead of inferring order from several service logs. Failure localization improves because the coordinator records the control path and workers report their results into it.

The supporting literature is useful but limited. Laigner et al.~(\href{https://arxiv.org/abs/2103.00170}{2021}) combined a literature review, repository analysis, and a survey of more than 120 practitioners. They found reliability problems concentrated around hand-built sagas and convention-managed consistency. Nadeem and Malik (\href{https://arxiv.org/abs/2204.07210}{2022}) ported a benchmark system containing 22 known bugs to a workflow engine in a single-participant case study and reported that the bugs became easier to localize. Their debugging times were compared with figures reported separately in another study, so the result is not a controlled head-to-head measurement.

The recovery layer itself can be a separable design decision. Zhuang et al.~(\href{https://www.usenix.org/conference/osdi23/presentation/zhuang}{2023}) built ExoFlow on the observation that workflow systems conflate two concerns: executing tasks and recovering from their failures. ExoFlow separates them, treating guarantees such as exactly-once as properties of a recovery layer over the execution substrate rather than of task execution itself, and requiring applications to annotate which tasks are nondeterministic and which communicate externally so that recovery knows what may safely be replayed. The evaluation covers data and ML workflow benchmarks, so this is directional systems evidence. Its transferable point is architectural: what an agent runtime can promise about recovery is a layered contract over declared task properties, not a blanket attribute of running the tasks. An unannotated nondeterministic step, and every model call is one, is exactly the case the recovery layer must be told about.

Centralized history does not provide one universally correct consistency guarantee. Zhang et al.~(\href{https://arxiv.org/abs/2208.09827}{2022}) surveyed a decade of transactional stream-processing research and found no generally accepted approach, even for computations far more deterministic than agent workflows. Each system selected guarantees around its application characteristics. The absence of a generally accepted approach means that an ``exactly once'' feature label does not establish that an application's state, latency, and external effects fit the advertised guarantee.

\Needspace{5\baselineskip}

The engine boundary should follow the workload. Three questions decide it:

\begin{enumerate}
\def\labelenumi{\arabic{enumi}.}
\tightlist
\item
  Does meaningful work remain exposed if the process dies mid-flight?
\item
  Does the workflow wait for external events?
\item
  Does it perform irreversible external effects?
\end{enumerate}

A yes to any question identifies state that a stateless scheduler may not be able to reconstruct from the current source of record. Three noes usually point toward a timer, a lockfile, and a fresh read from that source.

This is my operational rule rather than an experimentally established threshold. In one \textbf{local artifact} from my live maintenance loop, each run performed approximately 44 seconds of work within a 120-minute interval. An overlap-skip guard fired zero times across 77 runs. The source was uncommitted at the cited repository revision, so these figures illustrate one system's mechanism without establishing a rate or supporting a general claim.

That loop held no valuable mid-flight state, waited for no external event, and performed no irreversible effect. A workflow engine would have added history and deployment machinery without improving the observed behavior.

The opposite workload has a different shape. An agent may wait overnight for approval, call several services that cannot share a transaction, or spend substantial money on a model result that should survive a worker restart. A process-local retry loop loses its authority when the process dies. A workflow engine can persist the wait, retain the expensive result, and assign the next operation to another worker while preserving one execution identity.

That capability imposes implementation constraints. Engines that reconstruct control flow from recorded history require workflow code to make the same orchestration decisions when it reads the same history. A clock read, random branch, or changed iteration order inside that code can select a path absent from the recorded execution. Code evolution therefore requires explicit version boundaries. Even a change from a fixed workflow signature to variable arguments can make a long-running execution incompatible with its stored input.

Decomposition creates costs of its own. Child workflows used only to organize code create another execution to inspect and another history boundary to understand. Central coordination can reduce throughput or local autonomy when every small operation must pass through one scheduler and persistence layer. The engine improves visibility by concentrating control, but that concentration also creates contention and operational responsibility.

The integration boundary often costs more than the workflow definition. A second \textbf{local artifact}, my durable media integration, required 15 commits across 45 files and added 11,662 lines, including 4,665 test lines. The workflow definition accounted for only a small part of the change. Most of the work established clean payloads, stable execution identities, idempotent external requests, injected-failure gates, and reconciliation with durable storage. These counts describe one system and support no industry-wide cost estimate.

An engine should own coordination without absorbing domain behavior. Application code still determines what constitutes a valid deployment, payment, or agent result. The platform owns durable history, retry scheduling, waiting, and transitions between recorded step states. Keeping that boundary explicit makes engine replacement possible and prevents application correctness from depending on undocumented behavior inside a particular worker.

\hypertarget{the-interval-that-cannot-be-closed}{%
\section{The interval that cannot be closed}\label{the-interval-that-cannot-be-closed}}

A worker sends a merge request, receives success, and dies before recording the completed step. The engine sees an unfinished attempt and sends the step again. My treatment of this interval rests on directional systems research and practitioner accounts. No controlled result establishes that one idempotency design makes agent workflows reliable across workloads.

Redelivery is correct because the engine cannot infer completion from a missing record. This is \textbf{at-least-once delivery}. The runtime continues delivering work until it holds a durable completion record, so one logical step may receive several execution attempts. Durable execution can record a committed step result once within its own history. It cannot eliminate the interval between an external system accepting an operation and the worker recording that result.

That interval remains even when both systems are individually reliable. Suppose attempt A begins step nine under invocation identity \texttt{run-42/step-9}. The worker asks a code host to merge a change, and the host completes the merge. The process dies before the worker records the response. The engine schedules attempt B because its last durable fact says only that step nine started. The code host has moved to a new state while the workflow history still describes an in-flight operation.

The invocation identity must cross that boundary. Attempt B must send the same idempotency key as attempt A. Morling (\href{https://www.morling.dev/blog/building-durable-execution-engine-with-sqlite/}{2025}) describes idempotency keys at side-effect boundaries as the mechanism that makes the execute-then-log crash window safe on replay. The downstream system can then return the stored response for that key or recognize that the requested state already exists and report the same observable result.

An operation is idempotent when repeating it converges on the same observable state as applying it once. The useful contract is that a duplicate becomes indistinguishable from no duplicate.

The discussion so far uses several words whose distinctions carry the argument. They are fixed here, using the identifiers defined in Chapter 7, ``The software factory as a distributed system'':

\begin{longtable}[]{@{}
  >{\raggedright\arraybackslash}p{(\columnwidth - 2\tabcolsep) * \real{0.2500}}
  >{\raggedright\arraybackslash}p{(\columnwidth - 2\tabcolsep) * \real{0.7500}}@{}}
\toprule\noalign{}
\begin{minipage}[b]{\linewidth}\raggedright
Term
\end{minipage} & \begin{minipage}[b]{\linewidth}\raggedright
Meaning
\end{minipage} \\
\midrule\noalign{}
\endfirsthead
\toprule\noalign{}
\begin{minipage}[b]{\linewidth}\raggedright
Term
\end{minipage} & \begin{minipage}[b]{\linewidth}\raggedright
Meaning
\end{minipage} \\
\midrule\noalign{}
\endhead
\bottomrule\noalign{}
\endlastfoot
Logical work & The user or operator intent that should happen once logically, under a stable work identity. \\
Attempt & One execution of that work; a retry is a new attempt at the same logical work (contract I4, retries preserve logical identity). \\
Delivery & The assignment or redelivery of an attempt to a worker by the runtime. \\
Runtime commit & A result durably recorded inside the orchestration boundary. \\
External commit & A world-changing effect accepted outside that boundary, such as a merge, deployment, or payment. \\
Idempotency & The contract under which a repeated external commit is indistinguishable from a single one (contract I5, effects are safe under redelivery or explicitly uncertain). \\
Reconciliation & Reading actual external state and deciding which continuation is valid, rather than trusting internal records. \\
Unknown state & The condition in which neither success nor failure of an effect can be established safely. \\
\end{longtable}

\begin{figure}[htbp]
\centering
\includegraphics[width=1\textwidth,height=\textheight,keepaspectratio]{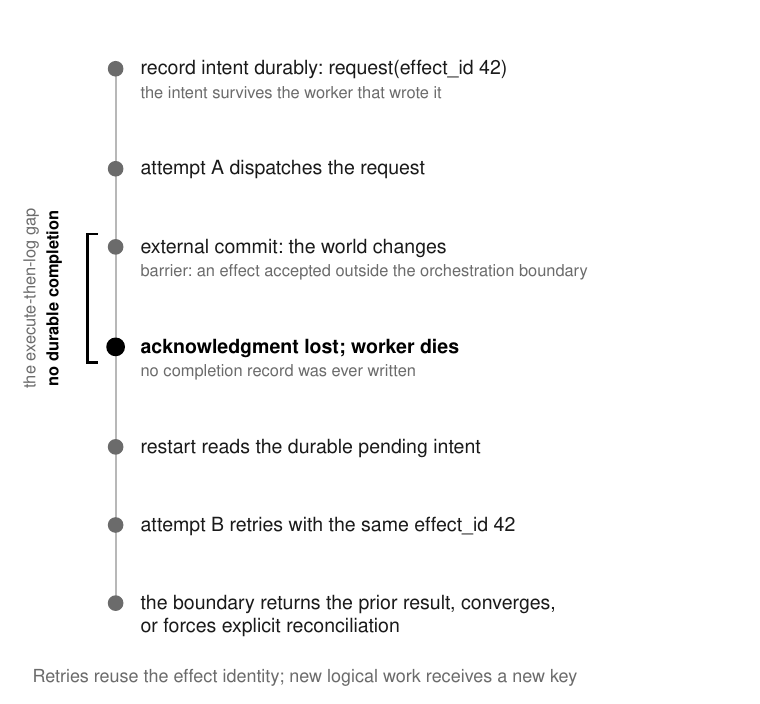}
\caption{A first attempt records a durable intent and operation identity before an external commit. If the worker dies before recording completion, a replacement reads the pending intent and retries with the same identity; the external boundary must return the prior result, converge, or force explicit reconciliation.}
\end{figure}

Key design determines which work the contract covers. A fresh attempt identifier fails because the downstream system sees attempt B as new work. A key scoped only to the user can collapse two legitimate operations into one. The stable key should identify the logical invocation and include enough input or version identity to distinguish genuinely different work. Retries reuse it. New logical work receives a new key.

The downstream implementation also needs an atomic claim on that key. Two workers may receive the same step concurrently after a timeout or lease dispute. If both check for a cached result and then perform the effect before either stores the result, the lookup adds latency without preventing duplication. A unique transactional record, compare-and-set operation, or equivalent mechanism must decide which worker owns the invocation before the effect occurs.

The lease dispute behind that scenario has a canonical precedent. Burrows (\href{https://www.usenix.org/conference/osdi-06/chubby-lock-service-loosely-coupled-distributed-systems}{2006}) describes how Chubby, Google's lock service, handles exactly this: a lease expiration tells the control plane that authority over a resource may be reassigned. It does not stop the old process, and it does not retract a request that process sent before losing the lease and that is still delayed in the network. Safety therefore requires the mutation boundary itself to reject obsolete authority, which Chubby supports with a sequencer, a lock generation number that the receiving server validates before applying the request. Retries and recovery in an agent factory create the same overlap: attempt B can hold current authority while attempt A's delayed external request is still in flight. The idempotency key and the authority generation answer different questions. The key deduplicates repeated attempts at the same logical work; the generation lets a protected boundary reject a superseded owner even while that owner is still running (contract I2, authority is generation-scoped). This is directional evidence from a coordination service, not from agent systems, and Chapter 18 develops the fencing mechanism fully. Here it marks the limit of what retry machinery alone can make safe.

Execution caching can extend the same rule through a call graph. Psarakis et al.~(\href{https://arxiv.org/abs/2312.06893}{2023}) describe a runtime that caches results by invocation identity so a repeated parent call reuses completed child calls. Within that runtime's transactional boundary, retried failed calls, recorded completed results, and call ordering can compose into a strong guarantee. An external service that ignores the invocation identity remains outside that boundary, regardless of the runtime's delivery terminology.

This scope rules out a blanket claim that a completed invocation never runs twice. A step that succeeded externally but remained unrecorded may execute again because the runtime has no completion record to consult. Execution is at least once. Results committed within the runtime boundary may be recorded exactly once. Safety for an in-flight external effect still depends on the contract at that effect's boundary. ExoFlow's design, discussed above, states the same scope from the runtime's side: exactly-once is a recovery-layer contract over tasks whose nondeterminism and external communication have been declared, not a property that task execution acquires by running under a workflow engine.

Some external systems provide this contract directly. Payment APIs may accept a client-supplied key and bind it to the first accepted request. A database can commit the business change and invocation record in one transaction. A content-addressed object store can make repeated writes of identical content under the same name converge. When a tool offers no equivalent mechanism, the workflow must add an adapter that owns deduplication or treat recovery as an unknown-state decision (contract I5).

Meyers and Zienert (2025) also report that rewriting operations for idempotency exposed and corrected pre-existing retry defects. The rewrite changed application behavior at the engine boundary. The engine could not infer the required contract from the old code. This remains directional evidence from the adopting team.

Unknown state needs a safe terminal path. In a \textbf{local artifact} containing my guarded-mutation pattern, the worker records an intent claim before an irreversible action, performs the action, and resolves the claim with the observed result. A resolved claim can return its cached result on redelivery. An unresolved claim found during recovery means the effect may or may not have occurred. The workflow stops and asks a human to reconcile the external state.

Guessing in either direction is unsafe. Assuming success can lose work that never happened. Assuming failure can repeat an irreversible effect.

My fault demonstration makes that boundary visible. A naive pipeline was killed after requesting a merge and before recording completion. Its retry requested the merge again, and the workflow reported success without alerting anyone. Under the same kill placement, the guarded variant requested one merge, emitted one escalation, and marked the workflow failed. The failed outcome was more reliable because it preserved uncertainty instead of fabricating a completed history.

Idempotence also belongs at the durable step boundary before the external side-effect boundary. Model calls are natural durable steps because they can be slow, costly, and irreproducible. Once a model result commits under a stable invocation identity, a retry can reuse it and continue from the same evidence. Reissuing the call may change cost and control flow even when no external database has changed.

The boundary should not surround every function call. Each durable step adds scheduling, serialization, storage, and history-reading work. Cheap, deterministic calculations can simply run again. A durable boundary is justified when work is expensive, slow, externally visible, or impossible to reproduce from recorded inputs.

Huang et al.~(\href{https://www.usenix.org/conference/nsdi26/presentation/huang}{2026}) arrive at the same partition from a different direction. Fractal distributes shell scripts fault-tolerantly, and to do so it must classify what a script does: work that can simply be recomputed, state that can be reconstructed from recorded inputs, and side-effectful regions whose repeated execution is not observationally harmless. The classification maps directly onto agent tool execution. Rerunning grep, a parser, or a compiler is recomputable work; its repetition costs time and nothing else. Rerunning git push, pull-request creation, ticket creation, or a deployment is a side-effectful region, and repeating it changes what the world observes. The durable-boundary guidance above is this same partition stated as placement advice: recomputable work can sit outside the boundary, while side-effectful regions need a recorded identity and an effect contract (contract I5). Fractal is systems evidence about shell-script distribution; I transfer the classification, not any measured overhead or recovery rate.

A boundary can also be too broad. In another \textbf{local artifact}, a transient read shared one deduplication unit with an irreversible mutation. When the read failed, the system marked the whole unit terminal, poisoned the claim, and stopped reporting work it had never performed. A watchdog consulted the same fail-closed store and also went silent. This is narrative illustration rather than evidence for the general design. Separating the retryable read from the mutation claim would have preserved an accurate pending state.

Trofimov et al.~(\href{https://arxiv.org/abs/1907.06250}{2019}) propose describing guarantees through determinism and observable outcomes. Equal inputs should produce equal outputs, and duplicate execution should leave no observable difference. That formulation directs attention toward properties callers can test. It remains one research group's proposal and has not displaced at-least-once and exactly-once terminology in common systems descriptions.

Repeated execution can still converge perfectly on the wrong result. The step needs postconditions that verify its domain outcome, and the workflow needs compensation or escalation when the external system cannot make the operation idempotent. Retry machinery satisfies neither responsibility.

\hypertarget{companion-patterns}{%
\section{Companion patterns}\label{companion-patterns}}

Eleven adjacent entries in the companion catalog refine this recovery contract without introducing another architectural decision. They cover external writes that carry receipts, idempotency keys, and compensating actions; retry behavior expressed as falsifiable invariants; snapshots taken outside the critical path at declared consistency boundaries; and checkpoint cadence chosen from measured recovery costs and failure rates.

Other entries cover committing execution records with the corresponding data change, placing shared agent state behind transactions, beginning event histories with a minimal versioned schema, and designing operations to tolerate reordering.

Three entries have thin support and contribute no evidence to the claims made in this chapter. They propose stating event-coordination guarantees explicitly, attaching postconditions to stochastic steps, and turning repeated plans into reusable blueprints.

\hypertarget{run-the-crash-check}{%
\section{Run the crash check}\label{run-the-crash-check}}

Choose the agent workflow whose failure midway through a run would have the greatest consequence. Give the run a stable identity, and store its plan, progress, and append-only event history as declared artifacts with named owners. Stop the process after several steps and delete nothing. A replacement worker should be able to identify completed work, pending work, and uncertain external effects from those artifacts alone.

\Needspace{5\baselineskip}

Before introducing a workflow engine, apply the three-question adoption check:

\begin{enumerate}
\def\labelenumi{\arabic{enumi}.}
\tightlist
\item
  Does valuable state remain exposed if the process dies mid-flight?
\item
  Does the workflow wait for external events?
\item
  Does it perform irreversible external effects?
\end{enumerate}

If all three answers are no, use a timer and a lock, then read the source of record again on each run. If any answer is yes, document which coordination facts the engine will own and which domain decisions will remain in application code.

Next, select the retried step with the most dangerous external effect. Give the logical invocation a stable idempotency key, propagate that key into the side-effecting system, and store the completed result under the same identity. Kill the worker after the external effect succeeds but before the completion record commits.

On recovery, the run should do one of three things: return the previously recorded result, converge on the same external state, or stop with an explicit unknown-state escalation. It should never silently assume either success or failure.

Recorded state makes a run resumable. Chapter 10 makes the recovery claim replayable and puts it under measurement.

\section*{Sources and evidence}

\textbf{Make agent state a first-class persistent artifact}

\begin{itemize}
\tightlist
\item
  Directional evidence: Fragkoulis, Carbone, Kalavri \& Katsifodimos (2020). A Survey on the Evolution of Stream Processing Systems. The VLDB Journal (2024). arXiv:2008.00842. (Exactly-once recovery arrived only when state became a first-class managed runtime artifact; implicit-state systems had lossy recovery.)
\item
  Directional evidence: Zhang, Cardoza, Chen, Angel \& Liu (2020). Fault-Tolerant and Transactional Stateful Serverless Workflows (Beldi). OSDI '20. (Hand-managed fault tolerance moved into a runtime abstraction with declared semantics; serverless benchmarks, not agent workloads.)
\item
  Corroborating case: Hanlin (Zhou) et al.~(2026). ADEMA: A Knowledge-State Orchestration Architecture for Long-Horizon Knowledge Synthesis with LLM Agents. arXiv:2604.25849. (In a fixed 60-run matrix, removing checkpoint/resume produced the only invalid run.)
\item
  Directional evidence: Addy Osmani (2026). ``Long-running Agents'', \href{https://addyo.substack.com/p/long-running-agents}{Elevate newsletter}, 2026-04-30. (Plan file, progress notes, append-only event log outside the context make an agent recoverable and enable full context resets rebuilt from a handoff file.)
\item
  Directional evidence: Karthik Halukurike et al.~(2026). ``Blueberry: Force Multiplier For The On-Call Engineer'', \href{https://tech.instacart.com/blueberry-force-multiplier-for-the-on-call-engineer-98c446dfcc12}{Instacart tech blog}, 2026-07-14.
\end{itemize}

\textbf{Use a durable workflow engine}

\begin{itemize}
\tightlist
\item
  Corroborating case: Nadeem \& Malik (2022). A Case for Microservices Orchestration Using Workflow Engines. ICSE-NIER. arXiv:2204.07210.
\item
  Directional evidence: Laigner, Zhou, Vaz Salles et al.~(2021). Data Management in Microservices: State of the Practice, Challenges, and Research Directions. PVLDB 14(13). arXiv:2103.00170. (SLR + repo analysis + 120+ practitioner survey: hand-rolled sagas and convention-managed consistency are where reliability failures concentrate.)
\item
  Directional evidence: Zhuang et al.~(2023). ExoFlow: A Universal Workflow System for Exactly-Once DAGs. OSDI '23. (Execution separated from recovery; exactly-once as a recovery layer over tasks with annotated nondeterminism and external communication; data/ML benchmarks, not agents.)
\item
  Corroborating evidence for the reported service-local before-and-after measurement: Jacob Meyers and Rob Zienert (2025). ``How Temporal Powers Reliable Cloud Operations at Netflix'', \href{https://netflixtechblog.com/how-temporal-powers-reliable-cloud-operations-at-netflix-73c69ccb5953}{Netflix TechBlog}, 2025-12-15. The vendor-adjacent account comes from the adopting team, was not independently audited, and does not isolate a causal component or measure transfer to agent workloads.
\item
  Null or conflicting result: Zhang, Soto \& Markl (2022). A Survey on Transactional Stream Processing. The VLDB Journal. arXiv:2208.09827. (Carried as the negative result behind the engine choice; it is about deterministic data pipelines and says nothing about agents.)
\end{itemize}

\textbf{Make retried steps idempotent}

\begin{itemize}
\tightlist
\item
  Directional evidence: Burrows (2006). The Chubby Lock Service for Loosely-Coupled Distributed Systems. OSDI '06. (A lease answers the allocation question only; the sequencer generation number lets the mutation boundary reject obsolete authority. Canonical stale-authority precedent; mechanism developed in Chapter 18.)
\item
  Directional evidence: Psarakis et al.~(2023). Styx: Transactional Stateful Functions on Streaming Dataflows. SIGMOD line. arXiv:2312.06893. (Execution caching keyed by invocation identity gives exactly-once composition across call graphs.)
\item
  Directional evidence: Huang et al.~(2026). Fractal: Fault-Tolerant Shell-Script Distribution. NSDI '26. (Recomputable work vs reconstructable state vs side-effectful regions; the classification, not measured rates, transfers to agent tool execution.)
\item
  Directional evidence: Meyers and Zienert (2025), \href{https://netflixtechblog.com/how-temporal-powers-reliable-cloud-operations-at-netflix-73c69ccb5953}{Netflix TechBlog}, 2025-12-15. (The forced idempotency rewrite alone fixed pre-existing retry issues.)
\item
  Directional evidence: Gunnar Morling (2025). ``Building a Durable Execution Engine with SQLite'', \href{https://www.morling.dev/blog/building-durable-execution-engine-with-sqlite/}{morling.dev}, 2025-11-20. (Idempotency keys at side-effect boundaries make the execute-then-log crash window safe on replay.)
\item
  Directional evidence: Trofimov, Kuralenok, Marshalkin \& Novikov (2019). Delivery, consistency, and determinism: rethinking guarantees in distributed stream processing. arXiv:1907.06250.
\end{itemize}

\chapter{Replayable traces and fault-injection recovery testing}
\label{ch10-replayable-traces-fault-injection-recovery}
\begin{quote}
\textbf{Evidence profile.} 1 strong \(\cdot\) 9 directional \(\cdot\) 0 corroborating \(\cdot\) 0 null or conflicting evidence items across 2 developed practices (\erca{097}, \erca{127}). Six directional items are carried by companion records cited inline (\erca{050}, \erca{196}, \erca{197}, \erca{198}, \erca{204}). One directional item is a preprint.

\textbf{Chapter claim.} Recovery is a measured property.
\end{quote}

Vogel et al.~(\href{https://arxiv.org/abs/2404.06203}{2024}) injected pod kills and recurring failures into Apache Flink, Kafka Streams, and Spark Structured Streaming on a Kubernetes testbed under representative load. Flink was the most stable of the three and had one of the strongest recovery profiles, contradicting earlier published comparisons. The effect of failure also changed across successive injections. A test that stopped after one successful restart would not have detected that variation.

Fault tolerance is often inferred from an architecture diagram. A checkpoint appears before a restart arrow, and the system is described as fault tolerant. The diagram is a hypothesis about the running system. Only measurement can establish whether recovery behaves as claimed.

Agent runtimes present the same problem. A design may preserve state, retry interrupted work, and isolate external effects, yet still recover incorrectly when the fault location, persistence boundary, timeout, or software version changes. Two practices make the recovery claim testable. A typed event stream preserves enough structure to inspect and replay a run. Fault injection forces the runtime through the intervals its recovery design claims to protect.

Ten items support the chapter's two entries. Only the fault-recovery benchmark is a strong study; no strong study supports the trace argument. Five distributed-systems recovery-testing papers transfer directionally: they were evaluated on databases, file systems, and cluster controllers rather than agent runtimes, so I transfer their fault-placement and specification principles, not their measured results. One agent-trajectory study is direct evidence but a recent preprint. Typed traces are therefore treated as an instrumentation design supported by demonstrations and a specification, with recovery established by measurement rather than a universal threshold.

Chapter 9 established that recorded state can make a run resumable. Replay imposes a stricter requirement. The system must know which events produced that state, which work may safely repeat, and where a changed decision invalidates the prior path. Recovery testing adds a third requirement: the runtime must demonstrate the claimed behavior when faults occur at inconvenient points, not only when a demonstration script stops a worker at a clean boundary.

\hypertarget{what-a-transcript-cannot-answer}{%
\section{What a transcript cannot answer}\label{what-a-transcript-cannot-answer}}

After an agent sends the same payment instruction twice, the first operational question is whether the tool executed twice or one execution produced two visible records. A transcript may show two assistant messages and two tool-shaped responses. It usually cannot establish whether the first request reached the payment service, whether the service committed it, whether the runtime received the response, or whether that response became durable before restart.

The design argument in this section rests on demonstrated uses of typed traces. Yu et al.~(\href{https://arxiv.org/abs/2605.10913}{2026}) describe a runtime substrate whose recorded executions can be forked and rerun from a changed step. Zheng et al.~(\href{https://arxiv.org/abs/2508.02736}{2025}) describe system-level agent observability collected beneath the application layer. Neither compares typed traces with transcript-only observability in a controlled study.

The portability argument rests on the OpenTelemetry GenAI semantic conventions (CNCF 2025), which define a shared vocabulary but do not test whether adopting it improves portability.

A transcript represents a run as a sequence of utterances. That form is useful for reading prompts and model outputs, but it compresses several distinct events into similar-looking text. A model response, tool request, environment mutation, and durable state transition may appear as adjacent messages even though they have different owners and failure semantics. Once execution crosses a process boundary, adjacency in the transcript provides weak evidence of causal order.

A typed event stream preserves those distinctions. Each record states what occurred, which component produced it, which run and step it belongs to, and how it relates to earlier state.

A model-call record may include the input and output references, timing, token use, and completion status. A tool-invocation record may distinguish dispatch, acknowledgment, returned data, and an external-effect identifier. A state-transition record may identify the prior state it consumed and the new version that became durable. These records can share one stream without sharing one payload shape.

The distinction becomes operational during partial failure. Suppose an agent decides at step 17 to create a support ticket. The runtime records \texttt{tool\_call\_dispatched}, the ticket service creates ticket 8421, and the worker dies before the runtime records \texttt{tool\_result\_persisted}. A transcript reconstructed after restart may omit the first call or show only an unanswered request.

\begin{figure}[htbp]
\centering
\includegraphics{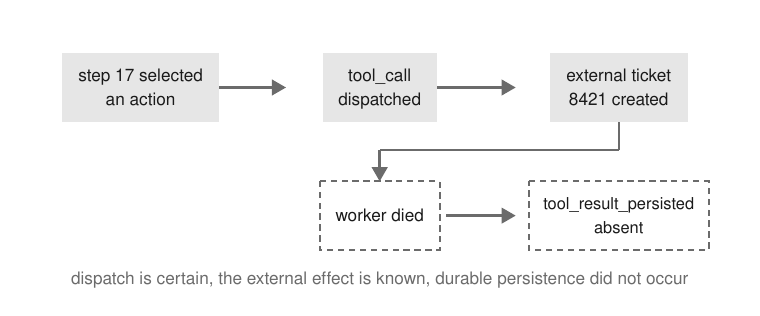}
\caption{A typed event stream records the commitment gap rather than hiding it: dispatch is certain, the external effect is known, and durable persistence of the result never occurred.}
\label{fig:ch10-trace-gap}
\end{figure}

A typed stream can represent the gap directly, as in Figure \ref{fig:ch10-trace-gap}.

The record does not necessarily settle whether recovery should retry. It identifies the uncertainty the recovery code must resolve: dispatch occurred, the external effect may be known or uncertain, and durable result persistence did not occur.

Causal structure is therefore the property to instrument for. The visual design of the trace viewer is secondary. Ownership and ordering should be queryable. Every agent-originated output or action should include an agent identifier. Every tool call should also identify the reasoning step that selected it and the session in which that decision occurred.

Those joins allow an operator or policy layer to compare intent with effect and detect a tool call attributed to a step that never authorized it.

These identifier fields come from governance analysis rather than deployment measurements. Chan et al.~(\href{https://arxiv.org/abs/2401.13138}{2024}) identify agent attribution, real-time monitoring of clear violations, and retained activity logs as mechanisms for agent governance. Their preventive effect remains unmeasured.

The same analysis identifies costs. Stable identifiers can reveal sensitive relationships among users, agents, and actions, while a centralized trace store concentrates observational power. A useful trace design specifies access control, retention, and redaction together with joinability. Otherwise improved operational visibility becomes an unbounded surveillance record.

\hypertarget{replay-from-a-changed-step}{%
\section{Replay from a changed step}\label{replay-from-a-changed-step}}

Replay reconstructs a run from recorded events while re-executing only the portion that must change. In the simple case, the runtime begins from durable state at step 16, replaces the decision logic or input at step 17, and executes the resulting suffix. The earlier prefix remains evidence of what happened. It is not regenerated as plausible prose.

Recorded state alone does not identify which later events became invalid when step 17 changed. The event stream supplies that dependency path. Each state transition refers to the inputs it consumed. Each action refers to the reasoning step that selected it. Each result refers to the action that produced it.

When step 17 changes, the runtime can invalidate dependent results while retaining independent ones. This narrow form of provenance-driven re-execution repeats only the work affected by the change. More elaborate provenance systems belong in the companion catalog. The requirement here is that the trace retain enough identity to compute the invalidated suffix.

Replay must also separate deterministic control from nondeterministic activity. Model outputs, wall-clock reads, random choices, and external calls cannot be assumed to return the same result on a second execution. A replayable runtime records those observations as events and replays deterministic control code against them when reproducing the old behavior.

When the purpose is to test a changed step, the runtime replaces only the selected observation or decision and records a new branch. The companion pattern on recording nondeterminism develops that separation further.

Branching introduces a data-model requirement that transcripts usually avoid. A replacement for step 17 must not overwrite the original event, because the original branch remains part of the evidence. The new event should identify the event it supersedes and the replay operation that created it. Later events then belong explicitly to either the original branch or the changed branch.

Without branch identity, a trace can appear internally consistent while combining state transitions from incompatible executions.

The same structure supports the golden sets introduced in Chapter 4. A golden case need not contain only a prompt and an expected final answer. It can preserve the typed decisions, actions, observations, and state transitions that produced the result. A regression test can then assert which properties must remain stable while allowing an intentionally changed step to alter its descendants.

The event stream also provides the object a supervisory layer needs to inspect and an unambiguous point at which it can intervene.

My trial-annotation pipeline illustrates the join problem without establishing general effectiveness. It converts standard harness output from diagnostic agent runs into 31 structured fields per trial and assigns a stable join key. Provenance is part of the annotation store's primary key, allowing two annotators to describe the same trial without overwriting one another. The architectural lesson is that identity belongs in the stored record, not in a filename convention or an analyst's memory.

My durable-execution demonstration harness provides a second illustration. It evaluates recovery invariants against an append-only event record containing distinct entries for run start, injected kill, worker restart, replay completion, invariant checks, committed side effects, detected duplicates, and model calls with cost fields.

A prose log could state that recovery succeeded. The event record allows a verifier to ask whether replay completed after restart, whether any side effect committed twice, and which model calls contributed to the recovered run.

Neither example establishes that typed traces outperform transcripts in a controlled setting. They show what becomes mechanically queryable when the runtime preserves event type, identity, order, and provenance. That distinction supports an instrumentation decision when the required question cannot be answered from the existing artifact. It does not establish that the instrumentation will reduce incident duration or improve recovery correctness by a predictable amount.

\hypertarget{a-shared-vocabulary-and-its-limits}{%
\section{A shared vocabulary and its limits}\label{a-shared-vocabulary-and-its-limits}}

The OpenTelemetry GenAI semantic conventions define common names and attributes for telemetry emitted by generative-AI systems. Their value is syntactic coordination. A runtime, collector, storage system, and analysis tool can exchange records without every pair inventing its own translation for model identity, operation type, token usage, or agent activity. Shared conventions reduce the bespoke assumptions that otherwise become hidden dependencies in an observability pipeline.

The specification's status defines the evidence boundary. The conventions do not show that a trace produced by one agent runtime can be replayed by another, nor do they require enough information to reconstruct application state. Telemetry portability and execution portability are separate properties. Two tools may agree that a model call occurred while remaining unable to reconstruct how that call changed workflow state.

Instrumentation should therefore begin with the shared GenAI vocabulary and extend it only where the workload requires additional fields. Replay may need state-version references, branch identity, external-effect identifiers, persistence status, or a digest of a large payload stored elsewhere. Those extensions should remain explicit and documented. A private schema may fit one runtime more closely, but every bespoke field transfers translation cost to collectors, tests, migration tools, and future runtimes.

Completeness is the harder requirement. If the runtime records a tool request but the wrapper omits the external commitment, the typed trace preserves the transcript's ambiguity in a more structured form. If records remain in worker memory until a batch flush, a kill can erase the events needed to explain the failure. If process clocks are unsynchronized, timestamps can imply an order that never occurred.

The stream therefore needs durable persistence and causal identifiers, not merely timestamped JSON.

Typed events also do not make external effects idempotent, restore corrupted state, or decide whether an uncertain operation should be repeated. They identify where the uncertainty lies. Recovery must still reconcile external state, enforce deduplication, and choose an admissible continuation.

A restore point must also be checked against downstream commitments. Local state can be internally valid even after another system has observed a later effect. The companion catalog treats this boundary as certification against downstream commitments.

Long histories introduce a cost that correctness arguments can obscure. Replaying a complete run may require reading and validating thousands of events before useful work resumes. Splitting long workflows at durable semantic boundaries can limit recovery to a shorter suffix, as described in the companion pattern on replay-history limits. That split changes where state lives and which earlier commitments must be summarized, so it should follow measured recovery requirements rather than an arbitrary event count.

Transcript-only observability remains attractive because it is easy to render and resembles the interface through which people interact with a model. I still retain transcripts, but as derived views rather than recovery records. The durable artifact is the typed event stream. A transcript is one projection over selected event types. This preserves readability without allowing a presentation format to erase the causal information that recovery requires.

\hypertarget{recovery-is-a-measured-property}{%
\section{Recovery is a measured property}\label{recovery-is-a-measured-property}}

Vogel et al.~(2024) did more than show that injected faults degrade performance. Their direct recovery measurements reversed conclusions from earlier published comparisons, and the effect of a fault changed across successive failures. Software configuration, workload, accumulated recovery state, and fault timing all contributed to the observed result. A useful recovery claim must therefore include those conditions.

The same argument applies to how a failed agent run is labeled. Zhao et al.~(\href{https://arxiv.org/abs/2607.09510}{2026}), in a recent preprint, collected 3,843 agent trajectories, annotated the 1,794 complete valid runs step by step across more than 63,000 steps, and modeled failure as a process with an onset, an evolution, and a recovery phase rather than a terminal outcome. A final-outcome label records that a run failed. It does not record when the trajectory became unrecoverable or whether an available recovery action was missed. This is direct agent evidence, though preprint evidence, for treating recovery as something that happens during a run and must be observed there. A recovery experiment that reports only a final pass indicator discards the same information.

An architecture diagram describes intended control flow. It may show a worker loading a checkpoint, replaying events, and resuming output. The drawing cannot establish how long fault detection takes, whether retry queues compete with live work, whether a replacement worker must rebuild caches, or whether an external effect occurred before its completion record became durable. Those behaviors emerge from the interaction among the runtime, persistence layer, network, workload, and deployment configuration.

Fault injection forces that interaction to occur on demand. Kill a worker or pod, interrupt an external call, terminate the process during persistence, and repeat those disturbances while representative work is active. The purpose is to measure recovery over a declared fault menu and operating envelope, with faults placed at the intervals the design claims to protect.

\hypertarget{specify-the-claim-before-the-kill}{%
\section{Specify the claim before the kill}\label{specify-the-claim-before-the-kill}}

\Needspace{5\baselineskip}

Begin by expressing the recovery claim in observable terms. A useful claim names:

\begin{itemize}
\tightlist
\item
  the injected fault;
\item
  the protected state or external effect;
\item
  the expected continuation; and
\item
  the measurements that determine whether recovery was acceptable.
\end{itemize}

A worked claim might read:

\begin{quote}
After an ungraceful worker kill during ticket creation, the runtime resumes the interrupted run without creating a second ticket, reproduces the deterministic state of the clean reference, and returns to the declared throughput range.
\end{quote}

Each clause points to an observable event or measurement.

This rule has a longer history than agent runtimes. Gunawi et al.~(\href{https://www.usenix.org/conference/nsdi11/fate-and-destini-framework-cloud-recovery-testing}{2011}) built FATE and DESTINI as a pair: FATE injects failures systematically across combinations of failure points, and DESTINI expresses the expected recovery behavior as declarative specifications checked against the observed execution. The injector without the specification only demonstrates that the system survived something; the specification without systematic injection only documents an intention. Their evaluation targeted cloud storage systems, so I transfer the pairing as a method, not their coverage or bug counts. The claim-before-kill discipline in this chapter is that pairing applied to an agent runtime: the declarative recovery specification comes first, and the injection schedule exists to test it.

The control run uses the same workload and configuration without an injected fault. This applies Chapter 2's control logic to recovery. The difference between the faulted and clean runs estimates the effect of the injection within the tested conditions.

When exact output comparison is possible, retain a content digest or canonicalized output from the clean run. When model nondeterminism prevents byte-for-byte equality, compare the deterministic state and effects that the recovery contract actually constrains, and state which outputs remain incomparable.

\Needspace{5\baselineskip}

A faulted run needs more than a final pass indicator. At minimum, measure:

\begin{itemize}
\tightlist
\item
  failure-detection time;
\item
  time until useful work resumes;
\item
  time until throughput stabilizes;
\item
  post-recovery throughput;
\item
  latency during and after recovery;
\item
  duplicated or missing effects; and
\item
  output or state equivalence to the clean reference.
\end{itemize}

Queue depth, retry count, checkpoint age, and cache state may help explain those outcomes, but they should not replace them. A system can restart quickly while delivering poor throughput or committing duplicate effects for several minutes.

Recovery time also requires declared start and end events. Measuring from process death to process creation captures orchestration latency, not application recovery.

For a user-facing run, the interval might begin at the last confirmed useful event before the kill and end when the interrupted run commits its next correct state transition. For a streaming workload, recovery may not end until the backlog clears and throughput stabilizes. These definitions answer different questions, so the event anchors belong beside the number.

The clean control must use the same anchors. Otherwise the comparison measures a difference in definitions rather than a difference in recovery behavior.

One restart demonstrates one recovery. Estimating stability requires repetition. Repeated faults connect recovery testing to Chapter 1's repeated-run discipline. Recovery impact is a distribution across independent runs and across recurring failures within one run.

\Needspace{5\baselineskip}

Schedule both:

\begin{itemize}
\tightlist
\item
  multiple injections during a continuing run; and
\item
  independent faulted runs from a clean starting state.
\end{itemize}

The first exposes accumulated effects such as queue growth, leaked leases, enlarged histories, and cache churn. The second separates those effects from ordinary run-to-run variation.

Recurring failures should retain their sequence position in the data. If the third kill produces a longer outage than the first, a pooled mean conceals stateful degradation. Plot or tabulate recovery metrics by failure ordinal and preserve the individual observations. The sample may remain too small for a stable population estimate, but the sequence can still show whether recovery changes as faults accumulate.

\hypertarget{strike-the-interval-the-design-claims-to-protect}{%
\section{Strike the interval the design claims to protect}\label{strike-the-interval-the-design-claims-to-protect}}

The most informative fault point is rarely the boundary between steps. Chapter 9 identified the execute-then-log interval: an external system may commit an effect after receiving the request but before the runtime durably records completion. A design that claims safe retry or exactly-once visible effects must hold when the process dies inside that gap.

The test should place the kill there deliberately.

Lineage-driven fault injection gives that placement a principled basis. Alvaro, Rosen, and Hellerstein (\href{https://dl.acm.org/doi/10.1145/2723372.2723711}{2015}) reasoned backward from a successful outcome: given the lineage of events that produced a good result, which combinations of faults could have prevented it? Faults chosen this way target the support of the outcome rather than random points in the schedule. The evaluation covered distributed data-management protocols, so I transfer the reasoning direction, not the tooling. In an agent runtime the analogue is concrete: the events that support ``the ticket exists exactly once'' are the dispatch, the external commitment, and the durable completion record, so those are the events whose surrounding intervals the kills must straddle. This is why the kill points below are named commitment barriers rather than time offsets.

Wu, Pan, and Huang (\href{https://www.usenix.org/conference/nsdi24/presentation/wu-haoze}{2024}) make the complementary point about precision. Legolas infers abstract execution states from system code and uses them to select fine-grained injection points, and with that placement found 20 previously unknown partial-service-failure bugs across six mature distributed systems. I transfer the fault-placement principle, not the observed bug rate, to agent runtimes: a sleep-based kill tests whatever the scheduler happened to be doing when the timer fired; a named barrier or state-aware injection tests a particular reliability claim. A harness that kills ``roughly two seconds into the step'' is sampling schedules, not testing the execute-then-log gap.

A worst-case protocol sends an ungraceful kill from inside the active step, after the external call returns and before the completion marker is written. The process exit status must confirm that the intended kill occurred. If the step reaches its normal return path or the harness records another exit mode, the trial is invalid rather than a passing recovery.

This check prevents an imprecise injection from striking a clean boundary while claiming to test the vulnerable interval.

The harness must not repair the system it measures. It may schedule the fault and observe the typed event stream, but the replacement worker must recover only from state available to the deployed runtime. If a harness ledger tells the worker that the external call completed, the experiment has supplied information the production system may not possess. The resulting pass measures the runtime and fixture together rather than the runtime's recovery property.

A negative control can expose that error. A naive adapter restarts the workflow from its first step without reconciliation or deduplication. At every kill point after an external commitment, that adapter should violate the no-duplicates invariant.

\Needspace{5\baselineskip}

If it passes, the harness may be:

\begin{itemize}
\tightlist
\item
  suppressing external effects;
\item
  leaking recovery information into the adapter; or
\item
  failing to place the kill where claimed.
\end{itemize}

A control expected to fail is useful because a false pass is otherwise easy to mistake for fault tolerance. It detects only the defect it was designed to expose, however, and a subtler recovery failure may still pass both variants.

The recovery path itself deserves the same suspicion as the failure. Li, Cai, and Lou (\href{https://www.usenix.org/conference/nsdi26/presentation/li-zhenyu}{2026}) studied production incidents in which the recovery action, not the original fault, caused the severe failure, and proposed previewing the effects of high-risk recovery operations before executing them. Their evidence comes from cloud infrastructure, so it transfers directionally, but the rule it supports is one this book already states as contract I10: a recovery path is production code with authority. Test it against the same invariants as the normal path, and preview high-risk recovery effects when practical. The naive negative control above is one instance of this rule. The general instance is that every non-naive recovery adapter belongs in the kill-point sweep as a subject, not only as the mechanism under whose protection the sweep runs. A reconciliation routine that deletes a ``duplicate'' artifact, a cleanup step that releases a lease still held by live work, and a retry policy that re-sends an uncertain external call are all recovery actions capable of causing the incident they exist to prevent.

\Needspace{5\baselineskip}

The kill-point sweep should include:

\begin{enumerate}
\def\labelenumi{\arabic{enumi}.}
\tightlist
\item
  before dispatch;
\item
  after dispatch but before external commitment;
\item
  after commitment but before local acknowledgment;
\item
  after acknowledgment but before the durable state transition; and
\item
  after the durable state transition.
\end{enumerate}

These points test different properties. The first asks whether unstarted work can be rescheduled. The middle points expose ambiguity and deduplication behavior. The last asks whether the runtime recognizes completed work and suppresses repetition. A step with several external effects requires a separate sweep for each effect.

\begin{figure}[htbp]
\centering
\includegraphics{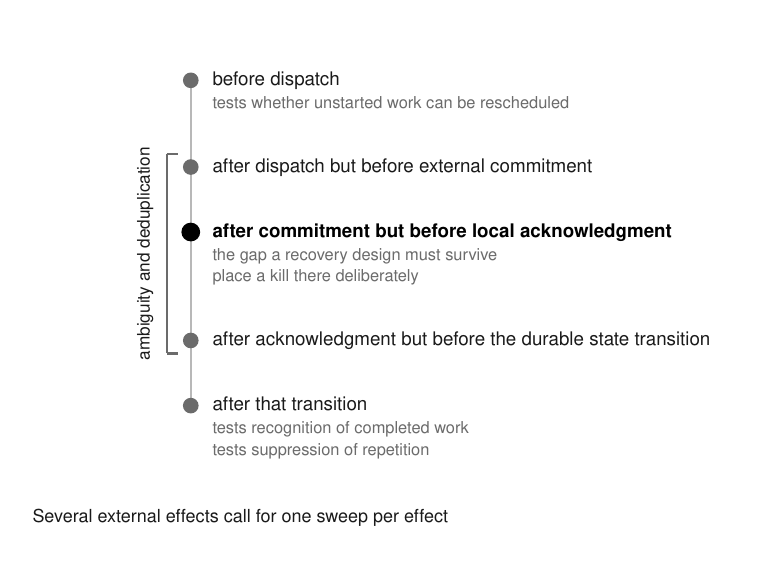}
\caption{Five lifecycle kill points test recovery from unstarted work through ambiguous commitment gaps to durable completion, including recognition and suppression of repeated external effects.}
\end{figure}

Each external effect requires a deliberate kill between external commitment and local acknowledgment.

Calls that can hang or return partially add another fault class. Interrupt the connection, inject a retryable response, delay the response beyond the worker heartbeat, and deliver a late response after the retry has begun. The runtime must distinguish a request known not to have executed from one whose status is uncertain. Treating both as ordinary retryable failures converts transport ambiguity into duplicated effects.

Redeployment adds versioned state to the protocol. Kill or replace a worker while a run remains active, then resume it with patched code. The event schema, serialized state, and deterministic replay path may each cross a compatibility boundary. A restart that succeeds with the same binary does not test this case. The configuration record should therefore identify both the pre-fault version and the recovery version.

History growth is another boundary condition. A short demonstration may avoid replay costs, rollover behavior, and compaction defects that appear only after many events. Run beyond the runtime's history rollover or compaction threshold and inject faults on both sides of that boundary.

Compaction replaces detailed working history with a condensed representation so the run remains within its history or context budget. The companion pattern on bounding replay history addresses the architectural response. The experiment determines whether the current boundary behaves as intended.

The fault menu should ultimately come from observed production failures rather than from what the harness can inject conveniently. Worker kills are reproducible and severe, but they do not cover slow storage, stale leases, delayed acknowledgments, partial network partitions, quota errors, or malformed persisted state.

The companion catalog's realistic-fault-menu pattern develops the fuller construction method. The narrower requirement here is to publish the menu so readers can see which failures the recovery claim includes and which it excludes.

\hypertarget{distributed-ambiguity-faults}{%
\section{Distributed ambiguity faults}\label{distributed-ambiguity-faults}}

The five kill points strike one worker's lifecycle. Once a factory runs multiple workers, a control plane, verifiers, and external services in separate failure domains, a second fault family appears: faults that create disagreement between components about what has happened. Nothing crashes, yet two durable records or two live processes hold incompatible beliefs.

Sun et al.~(\href{https://www.usenix.org/system/files/osdi22-sun.pdf}{2022}) tested this family directly for Kubernetes reconciliation controllers. Sieve perturbs the controller's view of cluster state, feeding it stale, intermediate, or unobserved states, and detects bugs with differential oracles that compare state transitions between perturbed and unperturbed executions. The evaluated systems are cluster controllers, so the transfer is directional, but the factory analogue is exact: a factory control plane is a reconciliation controller over work items, and the corresponding fault is a controller that believes a worker still owns a task while the repository or another controller has moved on. The differential oracle also transfers: run the same workload with and without the view perturbation and compare the resulting state transitions, rather than asking only whether the perturbed run ended in a passing state.

The table below defines the ambiguity faults added to the test menu; it is a protocol, not an executed result. Contract identifiers refer to the factory contracts defined in Chapter 7. Each row names the fault, the contract it tests, and the observable behavior an acceptable runtime must exhibit. Five representative rows appear here; the full matrix, including controller-recovery, verifier-failure, schema-compatibility, history-compaction, and retry-storm faults that Chapters 10 and 19 develop in their own sections, is published as a machine-readable protocol in the repository artifact \href{https://github.com/sjarmak/engineering-reliable-coding-agents/blob/main/protocols/distributed-ambiguity-faults.yaml}{\texttt{protocols/distributed-ambiguity-faults.yaml}}.

\begin{longtable}[]{@{}
  >{\raggedright\arraybackslash}p{(\columnwidth - 4\tabcolsep) * \real{0.2800}}
  >{\raggedright\arraybackslash}p{(\columnwidth - 4\tabcolsep) * \real{0.1200}}
  >{\raggedright\arraybackslash}p{(\columnwidth - 4\tabcolsep) * \real{0.6000}}@{}}
\caption{Representative distributed ambiguity faults, the contract each tests, and the observable behavior an acceptable runtime must exhibit. The full matrix is published in the companion repository.}\tabularnewline
\toprule\noalign{}
\begin{minipage}[b]{\linewidth}\raggedright
Fault
\end{minipage} & \begin{minipage}[b]{\linewidth}\raggedright
Contract
\end{minipage} & \begin{minipage}[b]{\linewidth}\raggedright
Expected observable behavior
\end{minipage} \\
\midrule\noalign{}
\endfirsthead
\toprule\noalign{}
\begin{minipage}[b]{\linewidth}\raggedright
Fault
\end{minipage} & \begin{minipage}[b]{\linewidth}\raggedright
Contract
\end{minipage} & \begin{minipage}[b]{\linewidth}\raggedright
Expected observable behavior
\end{minipage} \\
\midrule\noalign{}
\endhead
\bottomrule\noalign{}
\endlastfoot
Response lost after external commit & I5 & The runtime treats the call as uncertain, reconciles against external state before re-sending, and records the reconciliation outcome; no blind re-send. \\
Delayed response delivered after the retry has started & I5 & The late response and the retry resolve to one logical effect; the effect ledger shows one committed \texttt{effect\_id} or an explicit \texttt{unknown\_external\_state}, never two commits. \\
Worker loses its lease but continues running & I2 & The protected mutation boundary rejects the superseded worker's writes by \texttt{ownership\_epoch}; the rejection appears in the trace as a fencing event, not a silent success. \\
Old worker's completion arrives after the new worker has succeeded & I3 & The stale completion is rejected by generation and attempt identity; logical state reflects the new attempt only, and the rejection is recorded. \\
Repository state changes between validation and publication & I6 & Publication fails closed: the runtime detects that the validated version is no longer the head, refuses to publish against changed state, and routes to re-validation or reconciliation. \\
\end{longtable}

Each row is testable with the same discipline as the kill sweep: specify the claim, place the fault at the named interval, run the clean control with the same anchors, and check the typed event stream for the expected records. The rows referencing I5 need the effect ledger; the rows referencing I2 and I3 need fencing and identity checks at the mutation boundary; the row referencing I6 needs a declared reconciliation authority. The repository matrix extends the menu with controller-recovery and stale-view faults (I6), verifier-failure classification (I8), and schema-compatibility, history-compaction, recovery-queue, and retry-storm faults (I10), which are recovery-path tests in the sense of the previous section and belong in the same sweep as the naive negative control.

\hypertarget{keep-the-result-attached-to-its-envelope}{%
\section{Keep the result attached to its envelope}\label{keep-the-result-attached-to-its-envelope}}

My kill demonstration, a \textbf{local artifact}, illustrates the protocol without extending the literature evidence. In one run, the worker was killed during an activity after a retryable fault had already been injected. The recovered execution passed all 13 invariant checks. The interrupted activity resumed after a heartbeat-timeout retry, reused a cached result, and produced output identical to the clean reference by content hash.

Thirteen of thirteen is a satisfying result, but it describes one kill placement under one configuration.

I can conclude that the running system exhibited the specified invariants in that trial. I cannot conclude that the engine is generally fault tolerant, that another activity would recover through the same path, or that a different software version would preserve the result. Vogel et al.~(2024) justify that restraint: recovery comparisons changed when configuration and recurring faults were measured directly.

\Needspace{5\baselineskip}

My publish-gate protocol, also a \textbf{local artifact}, defines a broader intended experiment. Its first gate injects a failure in continuous integration at a randomized kill point. The fault matrix includes:

\begin{itemize}
\tightlist
\item
  a kill during a model call;
\item
  a kill inside the execute-then-log gap;
\item
  redeployment with patched code while a run remains active; and
\item
  execution beyond three times the history-rollover threshold.
\end{itemize}

Randomization broadens the sampled placement. It does not replace a fixed sweep across named high-risk intervals.

These artifacts support no comparison among workflow engines. I never ran the comparison sweep, and the non-naive adapters remain specification stubs. There are therefore no author-generated comparative measurements to report. An interface diagram containing several adapters can resemble an experiment even when no workload has passed through them.

\Needspace{5\baselineskip}

Every published result should carry the configuration that produced it. Record:

\begin{itemize}
\tightlist
\item
  runtime and engine versions;
\item
  deployment topology;
\item
  persistence settings;
\item
  checkpoint interval;
\item
  retry policy;
\item
  heartbeat and timeout values;
\item
  resource limits;
\item
  workload shape and concurrency;
\item
  input rate;
\item
  external-service behavior;
\item
  fault-injector version;
\item
  kill-point definition; and
\item
  metric definitions.
\end{itemize}

Preserve the typed trace, or an appropriately redacted derivative, so another investigator can reconstruct the event order. Without this envelope, a recovery number is difficult to interpret and cannot be reproduced closely.

Results also age as the stack changes. A new scheduler, storage client, default timeout, or checkpoint implementation can alter fault detection and state restoration without changing the architecture diagram. A passing experiment establishes behavior only inside the tested envelope. Repeat it after changes that affect persistence, concurrency, retries, version compatibility, external calls, or resource allocation.

Correctness, reliability, and performance should remain separate in the report. The absence of duplicate effects and missing state is a correctness result for a particular trial. The frequency with which those properties hold across injections is a reliability measurement. Recovery time, latency, and throughput are performance measurements.

Resource use and additional model calls belong to cost. An operator's ability to understand the failure and initiate recovery safely belongs to usability. One combined recovery score conceals which property failed.

A recovery system may also need to withhold action when it cannot classify a failure safely. The companion pattern on diagnosing before gating recovery addresses that policy. Another companion pattern asks whether recovery remains invisible to downstream observers except as monotonic progress. Neither property follows from a successful restart, so neither belongs in the basic pass criterion unless the system claims it explicitly.

\hypertarget{a-protocol-for-trusting-recovery}{%
\section{A protocol for trusting recovery}\label{a-protocol-for-trusting-recovery}}

Before accepting a recovery claim about an agent runtime, look for two linked artifacts.

The first is a durably persisted typed event stream that distinguishes model calls, tool dispatch and completion, environment changes, external commitments, and state transitions. Begin with the shared OpenTelemetry GenAI conventions, then add explicit state, branch, persistence, and effect fields where the workload requires them. Every agent action identifies its owner, and every tool call can be joined to the reasoning step and session that authorized it.

The second artifact is a fault-injection result produced under representative load. Run a clean control, kill a worker from inside an active step, and place at least one kill between an external effect and its durable completion record. Repeat injections within continuing runs and across independent runs, because the first restart does not characterize later recovery. When the runtime spans multiple workers and a control plane, add faults from the distributed ambiguity menu, placed by the claim they test rather than by a timer.

The harness must confirm that the intended kill occurred, observe recovery without supplying recovery state, and exercise a naive negative control expected to duplicate effects. Recovery paths themselves enter the sweep as subjects, since a recovery action holds authority and can cause the failure it exists to prevent.

\Needspace{5\baselineskip}

Measure recovery time between declared application events rather than between process death and process restart. Retain:

\begin{itemize}
\tightlist
\item
  post-recovery throughput and latency;
\item
  invariant outcomes;
\item
  duplicate and missing effects;
\item
  recovery behavior by failure ordinal; and
\item
  output or state equivalence against the clean control.
\end{itemize}

When exact equivalence is impossible, state which properties were constrained and which outputs remain nondeterministic. A pass means only that the runtime met those conditions in the tested trials.

The fault menu and complete configuration remain attached to the result. Together they define the envelope within which the claim is valid and provide the basis for repeating the experiment after a version or deployment change. A reader should be able to determine which intervals were struck, which faults were omitted, how behavior changed across recurring failures, and which measurements separated correctness from reliability, performance, and cost.

This protocol does not establish that an engine is universally reliable. It supports a narrower and more useful statement:

\begin{quote}
Under a recorded workload and configuration, with specified faults injected at specified boundaries, the runtime recovered with measured behavior and preserved the named invariants.
\end{quote}

That claim can be challenged, repeated, and revised when the system changes.

The repository artifact \href{https://github.com/sjarmak/engineering-reliable-coding-agents/blob/main/protocols/recovery-fault-injection.md}{\texttt{protocols/recovery-fault-injection.md}} supplies the bounded run sequence, pass condition, and output manifest.

A trace detailed enough to replay is also the artifact a person reads when recovery fails. Even a complete trace, however, does not establish which action caused the failure or how a person can defend that attribution.

\section*{Sources and evidence}

\textbf{Make every run a structured, replayable trace}

\begin{itemize}
\tightlist
\item
  Directional evidence: Shepherd runtime substrate (Yu et al.~2026, arXiv:2605.10913); companion material named in the same synthesis without a paper identifier: OpenTelemetry GenAI conventions (CNCF 2025).
\item
  Directional evidence: AgentSight eBPF observability (Zheng et al.~2025, arXiv:2508.02736).
\item
  Directional evidence: Chan, A., et al.~(2024), ``Visibility into AI Agents,'' ACM FAccT 2024, arXiv:2401.13138.
\end{itemize}

\textbf{Benchmark recovery with fault injection}

\begin{itemize}
\tightlist
\item
  Strong evidence: Vogel et al.~(2024), ``A Comprehensive Benchmarking Analysis of Fault Recovery in Stream Processing Frameworks,'' arXiv:2404.06203 (JSS line).
\item
  Directional evidence: Gunawi et al.~(2011), ``FATE and DESTINI: A Framework for Cloud Recovery Testing,'' NSDI 2011. Evaluated on cloud storage systems; the injection-plus-declarative-specification pairing transfers, not the coverage results.
\item
  Directional evidence: Alvaro, Rosen, and Hellerstein (2015), ``Lineage-Driven Fault Injection,'' SIGMOD 2015. Evaluated on distributed data-management protocols; the outcome-backward fault-selection principle transfers.
\item
  Directional evidence: Wu, Pan, and Huang (2024), ``Efficient Exposure of Partial Failure Bugs in Distributed Systems with Inferred Abstract States,'' NSDI 2024, which names its tool Legolas. Found 20 previously unknown partial-service-failure bugs across six mature distributed systems; the state-aware fault-placement principle transfers, not the bug rate.
\item
  Directional evidence: Sun et al.~(2022), ``Sieve: Automatic Reliability Testing for Cluster Management Controllers,'' OSDI 2022. Evaluated on Kubernetes controllers; the view-perturbation faults and differential state-transition oracles transfer to factory control planes.
\item
  Directional evidence: Li, Cai, and Lou (2026), ``Pilot Execution,'' NSDI 2026. Cloud-infrastructure incident evidence that recovery actions cause severe failures; supports treating recovery paths as production code under contract I10.
\item
  Directional evidence (preprint): Zhao et al.~(2026), ``Failure as a Process,'' arXiv:2607.09510. 3,843 trajectories collected, 1,794 complete valid runs annotated across more than 63,000 steps; failure modeled as onset, evolution, and recovery.
\end{itemize}

Author-system cases are narrative illustration, not evidence. The kill demonstration and publish-gate protocol are local artifacts.

\chapter{Human-auditable failure analysis and taxonomy development}
\label{ch11-human-auditable-failure-analysis-taxonomy}
\begin{quote}
\textbf{Evidence profile.} 2 strong \(\cdot\) 4 directional \(\cdot\) 1 corroborating evidence items across 3 developed practices (\erca{045}, \erca{046}, \erca{047}).

\textbf{Chapter claim.} Attribute the first upstream failure the trace can support.
\end{quote}

In a peer-reviewed study, Zhang et al.~(\href{https://arxiv.org/abs/2505.00212}{2025}) collected expert-annotated failure logs from 127 multi-agent systems, in which several model-driven workers passed work among one another. They then gave the attribution task to the strongest automated methods they could evaluate. The best method identified the responsible agent in 53.5 percent of cases but found the decisive step in only 14.2 percent. Some methods performed below random.

The logs already existed before the attribution problem was posed. Chapter 10 established that recording what happened is necessary, but a complete record does not establish why a run failed. The same trace may support several causal explanations. One worker may introduce a planning error, another may act reasonably on corrupted state, and a third may expose the defect when verification finally runs. Assigning responsibility to the last visible error confuses detection with cause.

Attribution is the scarce capability in this sequence. Logging preserves evidence. Attribution claims that one action materially changed the run's prospects. That claim determines what engineers instrument, which component they repair, whether a benchmark case remains valid, and sometimes who is held accountable. With weak attribution, a complete trace can become a precise record attached to the wrong explanation.

Three practices follow. Derive the initial failure taxonomy from traces produced by the system you operate. Keep a person responsible for consequential causal assignments. Structure the trace so that person can inspect and defend the evidence. Automation can organize and extend the work after those conditions exist. It should not be assumed to create them.

\hypertarget{build-the-taxonomy-from-the-failures-you-can-inspect}{%
\section{Build the taxonomy from the failures you can inspect}\label{build-the-taxonomy-from-the-failures-you-can-inspect}}

The protocol in this section has limited evidentiary support. Its basis is one directional observational study and one practitioner-authored method. Neither is a controlled result showing that the procedure improves an agent system. It remains worth using because it provides an auditable process for turning local traces into measurement categories. The method comes from Orosz and Husain (\href{https://newsletter.pragmaticengineer.com/p/evals}{2025}), and their recommendation to begin with at least one hundred traces is an unmeasured protocol choice.

Begin with traces from the system and workload you operate. The sample should vary across task type, outcome, model version, repository area, duration, tool use, and any operating condition likely to alter the path through the system. Apparent successes also belong in the sample when verification was skipped or the result depended on an unexplained retry. Excluding them would define failure as only what the current evaluator already detects, preserving the blind spot the review is meant to uncover.

For each trace, read from the initial request through the terminal state, taking open-ended notes. The primary annotation identifies the first upstream failure that materially changed the available path to success. It does not enumerate every later symptom.

A run may begin with an incorrect plan, edit the wrong module, execute a test from the wrong directory, and then misinterpret the failure. Those are four visible defects. If the plan directed all later work toward the wrong module, the planning error is the first upstream failure.

This rule follows the causal structure of an agent run. State moves from one step to the next through messages, files, tool outputs, summaries, and control decisions. An early error changes the evidence and options available downstream. Later actions may therefore be locally reasonable while remaining globally ineffective. Counting every symptom as a separate failure overweights long runs and systems that continue operating after their prospects have already collapsed.

The first-upstream rule also introduces a counterfactual check:

\begin{quote}
If this step had been correct while the preceding record remained unchanged, could the later failure still have occurred through the observed path?
\end{quote}

A clear no makes the step a plausible causal boundary. A yes suggests that the annotation identifies a symptom, or that the trace does not expose enough state to decide. In the latter case, preserve the uncertainty and leave the causal assignment unresolved.

\begin{figure}[htbp]
\centering
\includegraphics{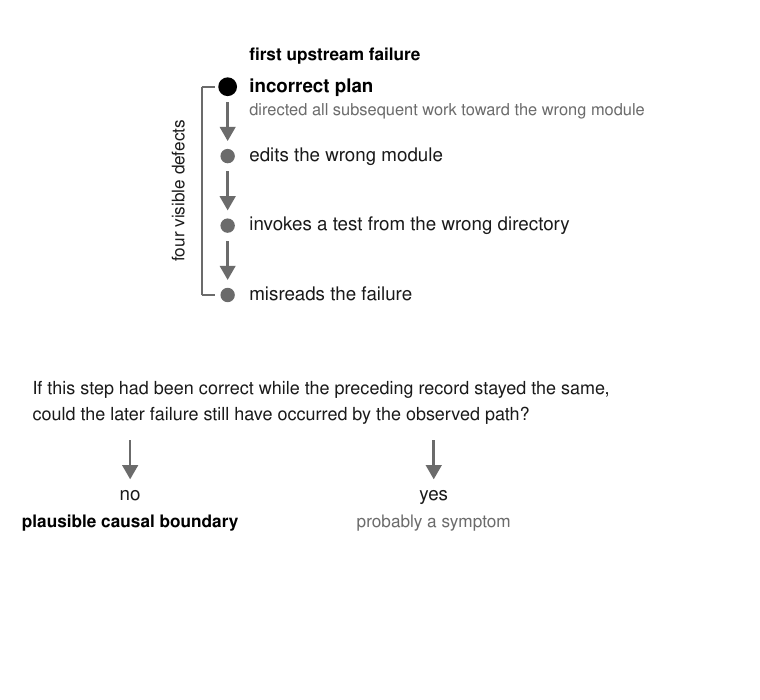}
\caption{A planning error is the first upstream failure linking four visible defects, while a counterfactual test distinguishes a plausible causal boundary from a symptom.}
\end{figure}

The counterfactual holds the preceding record fixed and asks whether the observed path could still have produced the later failure.

After the open review, cluster the notes into roughly five to ten themes. The categories should describe failures that require different measurement or investigation surfaces. Planning, retrieval, execution, tool use, self-diagnosis, verification, and environment handling may emerge, but they should not be imposed before reading the traces. A local pipeline may instead concentrate failures around identity propagation, stale workspace state, permission boundaries, retry ordering, or task ambiguity.

Phase labels are useful because they narrow the evidence required for diagnosis. A planning failure asks whether the system represented the goal, constraints, and dependencies before acting. An execution failure asks whether a valid plan became the intended commands and edits. A self-diagnosis failure asks whether the system interpreted the resulting state correctly. Combining all three under ``reasoning failure'' produces a broad count with little guidance about which part of the trace needs repair.

Generic metric packages often begin with categories that are convenient for the evaluator. They count tool errors because tool errors are easy to parse, or grade final answers because final answers are easy to present to a judge. The resulting dashboard may be internally consistent while missing the upstream decisions that generated those events. A taxonomy derived from traces begins with the failure process and asks only afterward which measurements can represent it.

Published phase-aligned classifications provide useful comparisons. Lu et al.~(\href{https://arxiv.org/abs/2508.13143}{2025}) classified failures by phase across 34 programmable tasks executed by off-the-shelf frameworks operating near 50 percent completion.

Their distribution describes the tasks, systems, permissions, and operating point in that study; it is not a base-rate estimate for frontier coding agents. A production repository with different tools, review gates, task durations, and authority boundaries may concentrate failures elsewhere.

I encountered this mismatch in my own trial-annotation pipeline for experimental agent runs. An early taxonomy omitted failure modes that dominated the project's traces even though public taxonomies covered related categories. The third revision grew from annotation of the local corpus and separated failures along more dimensions. This local illustration also exposes its own limit: the published manifest contains no annotations, so the category counts cannot be reproduced from the export.

The unit of analysis is a complete case. Keep the request, relevant starting state, trace, terminal outcome, verification result, and annotation together. When several retries belong to one attempt, preserve their order and identities so that a later successful retry does not erase the original failure. When the system forks work, retain the parent-child relationship. Otherwise an error imported from another branch may appear to originate where it was first observed.

Category boundaries will blur in long, multi-phase work. A retrieval failure can induce a bad plan, while an environment defect can make correct execution resemble faulty reasoning. Allow multiple descriptive tags when they preserve useful context, but retain one first-upstream assignment for frequency analysis. This avoids pretending that causation is always singular while keeping the primary count interpretable.

Reviewer disagreement is evidence about the taxonomy. If two qualified reviewers repeatedly divide the same cases between planning and retrieval, the definitions may depend on state the trace does not expose. If they agree about the events but disagree about the counterfactual, the causal rule needs a sharper boundary. Revise the definitions and preserve the original annotations so the change remains visible.

The initial review of roughly one hundred traces is intentionally laborious. Reading them end to end discovers categories, tests whether the recorded evidence can support those categories, and identifies distinctions that require domain expertise. It cannot estimate a stable low-frequency failure rate with the precision that a larger sample and a \textbf{power analysis} would require. A frequency calculated from one run per task also inherits the run-to-run variation described in Chapter 1, so two nearby categories may reverse order under repetition.

The output of the initial pass is a defensible measurement vocabulary and a labeled seed corpus. Estimating the system's failure distribution requires a separate sampling design.

Once the categories stabilize, scale-out returns to Chapter 5. Validate an LLM-as-judge on a held-out, expert-labeled, stratified sample under the same rubric, then use it to extend labels only within the operating range supported by its measured errors. The validation belongs to the category definitions in force when it ran. Revising a definition invalidates the agreement estimate for that category.

Human review remains concentrated on new categories, ambiguous cases, consequential failures, and strata in which the automated judge performs poorly. Applying a prebuilt judge before the manual pass would automate someone else's taxonomy rather than discover the failures produced by the local system.

Frequency informs priority without determining it. Count first-upstream assignments by phase and category, examine their uncertainty, and compare their prevalence with the cost of each failure. A frequent, recoverable tool error may deserve less attention than a rare attribution error that invalidates an evaluation. The distribution shows where failures concentrate. Engineering judgment still decides which failures are worth repairing.

Several narrower methods remain in the companion catalog. Constraint-violation logs can localize failure when a task has explicit invariants. Trajectory canonicalization can support structural comparisons across runs. Another entry recommends detecting environment errors before execution because they can consume disproportionate effort. These techniques may sharpen a local taxonomy, but none removes the need to discover which categories the operated system actually produces.

\hypertarget{keep-causal-assignment-under-human-control}{%
\section{Keep causal assignment under human control}\label{keep-causal-assignment-under-human-control}}

Attribution becomes consequential when it changes what happens next. An engineer may rewrite a prompt, repair a tool adapter, remove a benchmark task, retrain a model, or revise an incident report based on the assigned cause. The same label may also affect accountability when different people or services own different parts of the run. In those cases, an automated guess is not a harmless summary.

Triage and verdict are separate decisions. Triage ranks cases or candidate steps for investigation. A verdict identifies the responsible action and decisive step, supported by enough evidence that another reviewer can challenge the assignment. Automation can make triage cheaper even when its error rate makes it unsuitable for verdicts.

The opening result shows the current limit. An agent-level accuracy of 53.5 percent leaves nearly half of responsible-agent assignments wrong. A step-level accuracy of 14.2 percent misses the point at which most failures became decisive.

Those figures are a capability snapshot from one annotated multi-agent dataset, not a staffing ratio or universal automation threshold. Single-agent coding traces may be easier, other workloads may be harder, and later methods may improve substantially.

Ma et al.~(\href{https://arxiv.org/abs/2509.08682}{2025}) later reached 36.2 percent step-level accuracy on the same benchmark family, compared with less than 15 percent for prior methods. Counterfactually validated fixes derived from that analysis increased task success by an average of 22.4 percent, although the source does not state whether that gain is absolute or relative. The result suggests that causal structure can narrow the search more effectively than an undifferentiated reading of the transcript. The method still assigned most decisive steps incorrectly.

The fixes were evaluated on the benchmarks that defined the attribution task, and the method required a bespoke causal-discovery algorithm for interaction data whose behavior changes over time. That makes it useful as a candidate generator for a reviewer. It does not justify silently storing its assignment as the postmortem cause.

\Needspace{5\baselineskip}

A human verdict needs a reconstructable chain, and three questions establish it:

\begin{itemize}
\tightlist
\item
  Which state changed?
\item
  Which action changed it?
\item
  Which later components consumed that state?
\item
  Which observation eventually exposed the damage?
\end{itemize}

The record must distinguish the actor that introduced the defect from the actor that detected it. It must also distinguish a component that merely propagated invalid state from one that had both the information and the responsibility to reject it.

Consider a run in which a planner selects an obsolete interface, a coding worker implements it faithfully, and a verifier reports an integration failure. The verifier owns the detection. The planner is the first candidate for causal responsibility, provided the current interface was available in the planner's evidence and no later gate explicitly owned the currency check. If the repository index supplied stale documentation, responsibility may move again.

The trace cannot resolve those alternatives unless it records what each step received and what obligation each step owned.

\Needspace{5\baselineskip}

Attribution is therefore best represented as a structured claim. It records:

\begin{itemize}
\tightlist
\item
  the first upstream action;
\item
  the state transition that action caused;
\item
  the downstream dependency that carried the error;
\item
  the evidence supporting the assignment;
\item
  plausible alternatives considered;
\item
  the reviewer's confidence; and
\item
  the reviewer's identity.
\end{itemize}

This structure does not make the judgment objective. It makes visible where judgment entered. A postmortem that names a ``root cause'' without identifying the state transition behind it is a summary rather than an attribution.

Counterfactual reasoning tests an attribution without proving it. The question is whether replacing the candidate action with a correct one, while holding earlier state fixed, would have prevented the observed failure path. Several steps may satisfy that condition because later checks could also have recovered the run.

Under the taxonomy used here, the decisive step is the earliest point at which the relevant error entered the run or an owned opportunity to stop it was irretrievably missed.

That definition requires an explicit ownership model. A verifier that was never assigned responsibility for checking interface currency should not inherit responsibility merely because it could have caught the error. A release gate whose declared contract includes that check does own a meaningful missed intervention. Postmortems become arbitrary when they infer obligations from hindsight rather than from the contracts active during the run.

Retries complicate attribution because the world may change between attempts. A failed attempt can alter files, caches, rate limits, conversation state, or the evidence presented to the next attempt. A successful retry may depend on those changes, while a later failure may be responding to damage introduced earlier. Preserve attempt identity and state boundaries so reviewers can distinguish a repeated decision from a decision made in a changed environment.

Concurrency creates another ambiguity. Two workers may read the same starting state, make individually valid changes, and conflict only when their outputs merge. Neither branch necessarily contains a unilateral defect. Depending on the violated contract, the failure may belong to the coordination rule, merge order, or missing conflict check. A transcript sorted by completion time can make the worker whose result arrived last appear responsible for a conflict it did not cause.

Before assigning fault to an agent, rule out the evaluator and execution environment. A broken dependency, stale fixture, permission mismatch, nondeterministic test, or malformed task can make a correct action appear defective. Differential testing runs the same candidate under a controlled change in agent, harness, or environment and asks whether the failure follows the candidate. The companion catalog treats this as a separate diagnostic control because attribution without apparatus checks can turn measurement error into a model failure.

\hypertarget{component-boundaries-as-attribution-vocabulary}{%
\subsection{Component boundaries as an attribution vocabulary}\label{component-boundaries-as-attribution-vocabulary}}

Phase labels answer where in the run's reasoning a failure entered. The factory decomposition of Chapter 7 supplies a second axis: which component boundary introduced it. When the operated system spans that decomposition, record the originating boundary alongside the phase:

\begin{itemize}
\tightlist
\item
  admission and intent: the wrong work, priority, or constraint set was accepted;
\item
  task decomposition: the plan divided the work incorrectly or omitted a dependency;
\item
  coordination and ownership: overlapping or stale write authority acted on the same state;
\item
  worker reasoning: the model decided wrongly from adequate evidence;
\item
  retrieval and state freshness: the evidence supplied to a decision was stale or version-mismatched;
\item
  execution environment: sandbox, toolchain, or dependency defects distorted a correct action;
\item
  external effect: a side effect committed twice, partially, or unknowably;
\item
  verifier execution: the check crashed, timed out, or flaked;
\item
  verifier adequacy: the check ran cleanly but could not detect the defect;
\item
  aggregation: results from parallel work merged or joined incorrectly;
\item
  publication: the release boundary accepted a result it should have fenced or rejected;
\item
  recovery: a restart or retry violated an invariant that normal execution preserved; and
\item
  capacity and scheduling: admissible work starved, queued invisibly, or overloaded a shared resource.
\end{itemize}

This is a vocabulary, not a forced partition. Many failures form a chain across boundaries: stale retrieval induces a plausible but wrong plan, a worker implements it faithfully, and an adequate verifier detects it late. The annotation should preserve that chain rather than compress it into one exclusive category, and when the trace does not isolate a single originating boundary, the assignment stays unresolved with the missing observation named. The trace, not the taxonomy, is the ground truth; the categories exist to organize evidence the trace already contains.

The vocabulary also sharpens the first-upstream rule. The benchmark that opened this chapter frames attribution as identifying the responsible agent, but an agent is sometimes the wrong causal unit. When the defect entered through shared state, a stale authority grant, a verifier gap, or a control-plane decision, naming the worker that touched the failure last, or first, misdirects the repair toward a component that behaved correctly given its inputs and obligations. That is the same argument this chapter already makes against blaming the detecting actor, extended across component boundaries. Contract I11 in Chapter 7 states the precondition: consequential transitions must record work, attempt, actor, input state, and authority generation, or the boundary-level assignment cannot be made from evidence at all. Chapter 10's discussion of Zhao et al.'s trajectory annotations, which model failure as onset, evolution, and recovery rather than a terminal label, describes the process view this chain-preserving annotation depends on.

Raj et al.~(\href{https://arxiv.org/abs/2607.28802}{2026}) organize 41 failure modes by the interaction edge where a fault originates and the component that owns the repair. The strongest of four automated judges reached Cohen's \(\kappa=0.76\) against human category labels. The taxonomy is directional evidence for an interaction-centered diagnostic structure; the agreement result strongly supports reproducibility of those labels in the evaluated set, not causal correctness of an attribution.

Human control does not require every reviewer to read every trace from the beginning. Automated systems can rank likely decisive steps, group similar incidents, retrieve related cases, prefill observed events, and flag assignments inconsistent with the recorded state transitions. The reviewer remains responsible for accepting, revising, or abstaining. The stored record distinguishes the automated proposal from the signed attribution.

I used the same authority boundary in my internal benchmark-curation process, which determined whether generated software tasks were sound enough to evaluate. Candidate decisions retained an explicit provisional label until a project lead approved them. The pipeline could assemble evidence and run checks, but it could not finalize whether a defect belonged to the task or the evaluated system. This methodology example supplies no accuracy evidence.

The review threshold should follow consequence. A low-stakes transient failure may retain an automated tentative label for aggregate triage. A failure that changes remediation, removes an evaluation item, alters a reported result, or assigns organizational responsibility requires a named human decision.

When the evidence cannot distinguish among plausible causes, the correct verdict is unresolved. The record should state which missing observation would have separated them.

Aggregating several automated readers does not by itself make attribution reliable. Agreement among weak or dependent readers is a triage signal, not a causal verdict. The same applies to consistency across repeated runs: both signals may direct attention, while the signed attribution remains a human decision.

Repeated local categories can expose a structural defect. When the same failure class survives prompt patches and model upgrades, the remedy is likely architectural, and Chapter 18 develops how to test that redesign. Recurrence identifies the problem; an intervention still requires its own evaluation.

\hypertarget{design-the-record-for-a-skeptical-reader}{%
\section{Design the record for a skeptical reader}\label{design-the-record-for-a-skeptical-reader}}

Deshpande et al.~(\href{https://arxiv.org/abs/2505.08638}{2025}) reported that the strongest evaluated long-context model correctly localized issues in 11 percent of 148 expert-annotated traces. The traces were benchmark-derived but ecologically valid and more complex than short synthetic sequences designed to isolate one error. Reasoning, tool calls, outputs, retries, and state changes appeared in the same record. A large context window could contain the transcript, but containment did not make the decisive event legible.

These results span different tasks and benchmarks. The first two use one annotated dataset of multi-agent failures. They are capability snapshots, not a general automation ceiling.

The model scores will change as trace readers improve, and production workloads will differ. The more durable implication comes from the structure of the task: a record built for human audit should expose the units a causal investigation requires. Raw chronological text forces every reader, human or model, to reconstruct those units before diagnosis begins.

Chapter 10's typed event stream supported replay and recovery. Human audit adds a different requirement. A reviewer must be able to follow step boundaries, decisions, inputs, state transitions, retries, and outcomes without inferring their identities from prose. The same event stream can support both purposes when its schema records semantic boundaries. Timestamps on messages alone do not preserve them.

\Needspace{5\baselineskip}

At minimum, retain:

\begin{itemize}
\tightlist
\item
  a run identifier;
\item
  the task identity;
\item
  model and configuration versions;
\item
  an initial-state reference; and
\item
  ordered step identifiers.
\end{itemize}

\Needspace{5\baselineskip}

Each step records:

\begin{itemize}
\tightlist
\item
  the actor;
\item
  input references;
\item
  the decision or intended action;
\item
  the tool request;
\item
  the tool response;
\item
  the resulting state reference;
\item
  the verification outcome; and
\item
  the terminal status.
\end{itemize}

Retries identify the attempt they repeat and the reason the controller authorized another attempt. Forks and joins preserve parentage and ordering constraints.

A tool response, file digest, exit status, or test result is an observation. A model statement that the response means a dependency is missing is an interpretation. Storing both under a generic message field encourages later readers to treat the interpretation as though it came from the environment. Typed events allow the reviewer to compare the claim with the observation on which it rests.

Decision records need enough input to reevaluate the choice. A label such as ``selected tool A'' records the outcome but omits the alternatives, constraints, and evidence available at the time. Preserve the relevant input references, selected action, rejected alternatives when the system represented them, and the rule or model version that made the selection.

Replay means re-running or reevaluating a decision against the same recorded inputs. Asking a model to invent a retrospective rationale is a different operation.

Some inputs cannot be stored in full. Repository snapshots, retrieved documents, and tool outputs may be large, sensitive, or mutable. The trace can instead retain content-addressed references, access-controlled snapshots, or a precise query together with the identities of the returned items. A pointer to ``current repository state'' is insufficient because the state will change before review. The preservation policy must balance reproducibility with confidentiality, storage cost, and retention obligations.

Step boundaries should correspond to ownership and state transitions visible to the control plane. One step containing plan generation, three tool calls, an edit, and verification is too coarse for attribution. Recording every token or streaming fragment produces the opposite failure: thousands of events without a meaningful decision boundary.

The unit of record is the smallest one in which a single actor receives a defined input and produces an action or state transition consumed by another component.

\Needspace{5\baselineskip}

Tool-call records need more than a name and final output. They should include:

\begin{itemize}
\tightlist
\item
  normalized arguments;
\item
  execution-environment identity;
\item
  start and completion status;
\item
  timeout or cancellation state;
\item
  a summary of side effects; and
\item
  links to resulting artifacts.
\end{itemize}

A command can return a nonzero exit status after partially changing the environment. Recording only the failure status hides files, processes, or external effects inherited by the retry.

Ordering also needs explicit semantics when work overlaps. Wall-clock timestamps help diagnose latency, but clock order does not establish causality across concurrent workers. Parent identifiers, message sequence numbers, versioned state references, and join events show which observations were available to each decision. When two steps race to modify the same artifact, the trace should identify the accepted version and the rule that rejected or merged the other.

Identity must remain stable across resumed work. A display name such as \texttt{coder} may refer to different model versions, sessions, or permission sets. The audit record links the logical role to a particular execution instance, configuration, authority set, and parent run. This lets a reviewer determine whether an outcome changed because of a new decision, a new actor, or state inherited across a restart.

The review interface should support competing explanations. A reviewer should be able to move from an annotated failure to the suspected upstream event, inspect its input state, follow affected descendants, and compare a neighboring successful run. Filtering by actor or event type is useful, but the default view should preserve causal context around each selection. A viewer that hides retries or collapses repeated tool output may improve readability by removing the evidence under dispute.

CodeProbe (\href{https://github.com/sjarmak/codeprobe}{public repository}) publishes complete transcripts and preserves quarantined runs separately. Its successor adds a side-by-side trace browser for audited comparisons. These author-system choices illustrate preservation and review practices; they make competing conclusions reviewable without claiming an optimal schema.

Quarantined runs are worth preserving because evaluation pipelines often remove timeouts, rate limits, infrastructure failures, and malformed outputs before analysis. Some exclusions are methodologically justified, but deletion prevents a later reviewer from distinguishing an agent failure from an apparatus failure. Retain the raw run, exclusion reason, decision owner, and any replacement attempt. Aggregate reporting can then follow the declared policy without erasing the cases that test it.

Structured traces also improve privacy controls because sensitive fields become identifiable. Credentials, personal data, proprietary source, and private model reasoning may require redaction or restricted access. Blanket retention creates a security liability, while blanket deletion destroys auditability. Field-level policies can restrict access, encrypt retained data, and irreversibly redact secrets. The record should distinguish evidence withheld by policy from evidence the system never captured.

The trace schema must record its own version. Adding a field changes what can be inferred from later runs. Renaming an event can break comparisons with the existing failure corpus. A missing field may mean that the event did not occur, that the recorder omitted it, or that an older schema represented it differently. The decoder and migration logic should distinguish those states explicitly.

Instrumentation can also change the system being observed. Synchronous logging adds latency. Large payloads can alter token budgets, memory use, or rate-limit behavior. Asynchronous emission can reorder records or lose the final buffer during a crash. Measure the overhead, assign a durability requirement to each event class, and record missing events as trace failures. A gap must remain visible in the reconstructed record.

Changing instrumentation between experimental arms also changes the apparatus, which Chapter 1 requires to remain fixed in a paired comparison.

The practitioner evidence for this practice is anecdotal. One organization reported that reviews had consisted of competing opinions until the system began recording decisions, inputs, retries, and outcomes. The resulting structure made failures repeatable and disputes more factual. A second practitioner independently described event-sourced, replayable traces. Both are self-reported accounts. They demonstrate plausible operational use but no controlled improvement or universal schema.

Logging does not repair an agent. It makes causal explanations testable against preserved events and makes repeated failure paths comparable. Repair still requires a separate intervention and an evaluation showing that the intervention changed the intended outcome. This distinction prevents observability work from receiving credit for reliability gains it has only made possible to measure.

The schema should evolve from failed investigations. Common gaps concern the state a worker observed, side effects inherited by a retry, the branch selected at a join, or the evidence behind a decision. Record the unanswered question as an instrumentation defect. The next schema version adds the smallest event or relation that would have answered it. This ties auditability to real investigations and limits speculative completeness.

Trajectory shape can provide an early signal while a run remains active. The companion catalog describes monitoring duration, variance, tool-call count, and other structural changes because failed trajectories may run longer or become more variable. Such a signal can preserve a suspicious run or request earlier review. It cannot identify the cause without the structured events that explain what occurred inside that shape.

\hypertarget{turn-failed-runs-into-an-operating-record}{%
\section{Turn failed runs into an operating record}\label{turn-failed-runs-into-an-operating-record}}

Begin with the twenty most recent failed runs and read each one from the initial request through the terminal state. Notes remain open-ended, and each case receives one primary assignment: the first upstream failure that materially changed its path to success. Include apparent successes whose verification was skipped, because an unverified outcome cannot provide clean evidence of success.

This first batch tests the trace as much as the system. It shows whether the record can answer ordinary causal questions before a larger annotation effort depends on it.

Then expand the corpus toward one hundred traces selected across task types, outcomes, model versions, repositories, durations, tools, and operating conditions, and cluster the notes into roughly five to ten failure classes. Their observed frequency helps determine which classes to measure first, while consequence and repair cost determine the final priority.

A named human signs any attribution that changes remediation, accountability, task validity, or the interpretation of an evaluation result. Automated readers may rank the review queue, retrieve related cases, and propose candidate decisive steps. They do not finalize the causal assignment.

The first question the trace cannot answer becomes an instrumentation requirement. Add the missing state reference, event boundary, identity, ordering relation, or outcome field, then preserve the new schema version with later runs. Failed analysis therefore becomes evidence about the observability system itself.

A completed review should state both what the evidence supports and what remains unresolved.

The repository artifact \href{https://github.com/sjarmak/engineering-reliable-coding-agents/blob/main/protocols/failure-trace-review.md}{\texttt{protocols/failure-trace-review.md}} provides the blinded review sequence, pass condition, and adjudication record.

The failure corpus remains an operating asset after Part III. Chapter 15 uses it to tune compaction policy, because compression should preserve the evidence that prior investigations found decisive.

Part IV turns to the evidence available when the model acts: retrieval, context budgets, and memory. Chapter 12 begins by measuring repository retrieval, the first step in deciding which parts of a codebase enter that evidence.

\section*{Sources and evidence}

\textbf{Derive the failure taxonomy from your own traces}

\begin{itemize}
\tightlist
\item
  Directional evidence: Lu, R., Li, Y., Huo, Y. (2025), ``Exploring Autonomous Agents: A Closer Look at Why They Fail When Completing Tasks,'' arXiv:2508.13143. (Phase-aligned classification; \textasciitilde50\% completion base rate on 34 programmable tasks with off-the-shelf frameworks; not an estimate for production repositories.)
\item
  Directional evidence: ``A pragmatic guide to LLM evals for devs'', Gergely Orosz with Hamel Husain, Pragmatic Engineer newsletter, 2025-12-02, \url{https://newsletter.pragmaticengineer.com/p/evals}. (Open coding on 100+ traces, first upstream failure, axial coding into 5-10 themes, prioritize by frequency.)
\end{itemize}

\textbf{Keep humans in failure attribution}

\begin{itemize}
\tightlist
\item
  Strong evidence: Zhang, S., et al.~(2025), ``Which Agent Causes Task Failures and When? On Automated Failure Attribution of LLM Multi-Agent Systems,'' ICML 2025, arXiv:2505.00212. (Who\&When: expert-annotated failure logs from 127 multi-agent systems; 53.5\% agent-level and 14.2\% step-level for the best automated method; some methods below random.)
\item
  Directional evidence: Ma, G., et al.~(2025), ``Automatic Failure Attribution and Critical Step Prediction Method for Multi-Agent Systems Based on Causal Inference,'' arXiv:2509.08682 (CDC-MAS). (36.2\% step-level against under 15\% for prior methods; counterfactually validated fixes worth an average 22.4\% task success.)
\end{itemize}

\textbf{Design traces for human audit}

\begin{itemize}
\tightlist
\item
  Strong evidence: Deshpande, D., et al.~(2025), ``TRAIL: Trace Reasoning and Agentic Issue Localization,'' Patronus AI, arXiv:2505.08638. (Best evaluated model localized issues in 11\% of 148 expert-annotated traces; a 2025 capability snapshot.)
\item
  Corroborating case: \href{https://www.reddit.com/r/LLMDevs/comments/1q7avil/}{``What actually broke when we put AI agents into real production workflows''}, /u/saurabhjain1592, r/LLMDevs, 2026-01-08.
\item
  Directional evidence for interaction-centered failure localization, with strong evidence for label reproducibility in the evaluated set: Raj, H., et al.~(2026), ``Model or Harness? An Interaction-Centric Taxonomy for Localizing Agent Failures,'' arXiv:2607.28802. The strongest automated judge reached Cohen's \(\kappa=0.76\) against human labels.
\end{itemize}

\textbf{Author-system illustration cited inline}

\begin{itemize}
\tightlist
\item
  Not an evidence item: CodeProbe, the author's task-mining evaluation tool, \href{https://github.com/sjarmak/codeprobe}{public repository}. Named inline for the published transcripts and quarantined runs described above, which are narrative illustration.
\end{itemize}

\part{Context engineering: retrieval, budgets, and memory}
\gdef\currentparttitle{Part IV: Context}

\chapter{Measuring and designing repository retrieval}
\label{ch12-measuring-designing-repository-retrieval}
\begin{quote}
\textbf{Evidence profile.} 6 strong \(\cdot\) 8 directional \(\cdot\) 1 corroborating evidence items across 3 developed practices (\erca{076}, \erca{083}, \erca{085}).

\textbf{Chapter claim.} Retrieval can improve while task outcomes do not.
\end{quote}

In my benchmark of enterprise-scale repository work, two instruments gave different accounts of the same retrieval tool. \(\mathrm{Precision}@10\), the proportion of the first ten results judged relevant, rose from 0.095 to 0.313. \textbf{\(\mathrm{Recall}@k\)}, the fraction of judged necessary items found among the first \(k\) results, rose at \(k = 10\) from 0.120 to 0.272. The share of tasks for which at least one needed file appeared rose from 0.33 to 0.56.

The paired end-to-end reward across 370 tasks moved by only +0.0349, with a bootstrap 95 percent confidence interval of {[}+0.0130, +0.0579{]}, \(n = 370\).

The instrument is CodeScaleBench. Its frozen analysis suite holds 370 paired tasks in 20 suites: 150 software-lifecycle tasks covering debugging, repair, feature work, refactoring, security, testing, documentation, design, and comprehension, and 220 organization-scale tasks covering cross-repository navigation, dependency tracing, migration, incident triage, onboarding, compliance, and platform work. Tasks are anchored in 46 repositories, and 18 cross-repository task groups name further repositories in their fixtures, bringing the distinct total to 56. Nine primary languages appear, plus tasks spanning several. The repositories are version-pinned public mirrors, so each task resolves to a fixed commit, and they include codebases larger than the common public suites reach: Chromium at roughly 35 million lines, Firefox at 20 million, LLVM at 15 million, the Android platform frameworks at 12 million, and the Linux kernel, JDK, and LibreOffice at comparable magnitudes. The two arms differ only in code access. The baseline arm holds the repositories locally and uses built-in file tools; the retrieval arm reaches the same code through a retrieval server. Chapter 3 argued that a benchmark supports a capability claim only when its workload matches the deployment. This suite was built for multi-repository work in large codebases, and the results below estimate performance on that workload, not on single-repository issue resolution.

That interval treated all 370 task-level differences as independent. The tasks came from 46 anchor repositories and were also grouped into 20 suites, but the resampling represented neither structure. The interval therefore likely understates uncertainty and should not be read as a valid bound on it.

Chapter 1's rule about choosing the resampling unit applies directly, but promoting the unit from task to repository or suite is not sufficient by itself. Repositories and suites are two groupings over the same tasks. Whether they are nested or crossed determines whether one clustered unit is adequate or whether the analysis requires hierarchical or multiway resampling.

The retrieval measures showed that the tool had become substantially better at placing relevant repository evidence in front of the agent. The end-to-end score showed that the system as a whole changed little. The instruments estimated different events.

That disagreement changed the engineering question. Had I retained only the final reward, I would have treated retrieval as a weak intervention and looked elsewhere. The stage measurements showed that retrieval improved while much of the additional evidence failed to become correct edits. The next investigation therefore belonged downstream, in context assembly, evidence use, generation, or verification.

The reverse pattern would require another explanation. An end-to-end gain without a retrieval gain could come from a model change, a prompt difference, or another uncontrolled part of the pipeline.

Repository retrieval belongs to a causal chain. A task becomes one or more queries, each query searches an index, and ranked results enter a context budget. A model interprets that context before execution and verification turn its output into a scored result. Pass rate records only the state at the end of the chain and cannot identify which transition produced it.

\hypertarget{score-retrieval-and-generation-separately}{%
\section{Score retrieval and generation separately}\label{score-retrieval-and-generation-separately}}

Chapter 2 used ablations to isolate a component by removing it while holding the surrounding system fixed. Stage-level scoring applies the same attribution discipline within a run. Measure whether the retriever surfaced the evidence the task required, where that evidence appeared, whether it survived the context cutoff, and whether the generator used it. Report those measurements beside final task completion because each describes a different condition for success.

The first condition is \textbf{availability}. The relevant file, symbol, documentation block, or prior decision must exist in the searchable corpus and its index. A retriever cannot return a file omitted by an indexing rule, a generated tree excluded as noise, or a recent change still waiting for an asynchronous update. Treating those cases as ranking failures directs the investigation toward query tuning when the actual defect is corpus coverage or freshness.

The second condition is \textbf{retrieval}. Given an indexed target, the query must return it.

The third is \textbf{placement}. A relevant item can appear too low in the ranking to survive a fixed result depth or token budget.

The fourth is \textbf{contextual sufficiency}. A method signature may be relevant but insufficient when the implementation, callers, or type definition determine the correct edit.

The fifth is \textbf{use}. The model may receive sufficient evidence and still ignore it, misinterpret it, or produce an edit that fails for an unrelated reason.

An end-to-end failure is compatible with failure at any one of these stages, or at several simultaneously. A tool may return the correct method at position twelve, while the context builder retains only ten results and the generator invents an interface from the remaining snippets. The final failure does not reveal whether query formation, ranking, truncation, or generation should change. The trace can, but only when the evaluation records the boundaries between those stages.

\begin{figure}[htbp]
\centering
\includegraphics{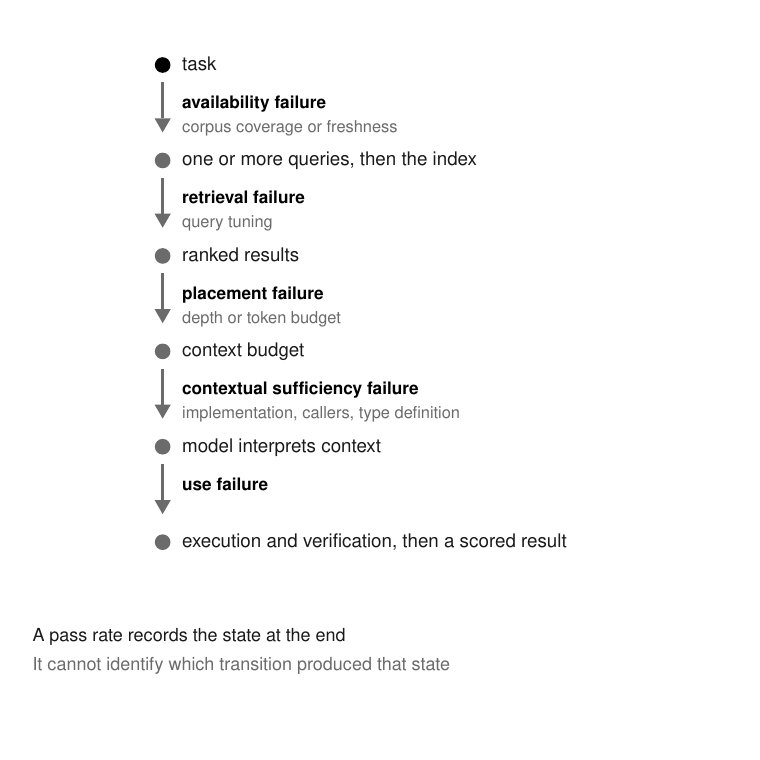}
\caption{The retrieval chain separates availability, retrieval, placement, contextual sufficiency, and use failures across the transitions from task formulation to scored execution and verification.}
\end{figure}

Pass rate captures only the final state. It cannot reveal which transition failed or which condition governed the result.

A passing task is equally ambiguous. A generator may solve a familiar edit from model knowledge while the retriever returns irrelevant material. A broad retriever may place the answer somewhere in a large dump that exceeds the model's useful attention and still receive credit because the final edit happens to work. Chapter 6's return-everything case achieved recall 1.0 by construction. Correctness at the end of a run does not certify the evidence path that preceded it.

\Needspace{5\baselineskip}

Retain a per-task record containing at least:

\begin{itemize}
\tightlist
\item
  the identifiers of the judged relevant items;
\item
  their positions in the returned list;
\item
  the subset that entered model context; and
\item
  the final task outcome.
\end{itemize}

When the system exposes citations, tool references, or another credible evidence-use signal, also record which items the generator appeared to use. A copy of retrieved text in the prompt establishes exposure, not use.

Repository-level code-completion research provides a clean example of this decomposition. RepoBench, from Liu et al.~(\href{https://arxiv.org/abs/2306.03091}{2023}), evaluates retrieval, completion with supplied context, and the combined pipeline as separate conditions. The design distinguishes a system that cannot retrieve useful cross-file context from one that receives the context and cannot complete the code. A better ranker addresses the first failure and leaves the second untouched.

Memory evaluations expose a similar split through different measurements. MemConflict, from Tao et al.~(\href{https://arxiv.org/abs/2605.20926}{2026}), combined final-answer scoring with observations of whether the required memory was absent, ranked too low, retrieved but unused, or used incorrectly. Its systems sometimes answered correctly even though the best reported conflict-recognition score reached only 0.2501.

Memory Circuit Analysis, from Mao et al.~(\href{https://arxiv.org/abs/2605.03354}{2026}), reports a stage-level diagnostic that attributes silent failures to extraction, retention, or retrieval. These results support stage localization, but their memory workloads do not reproduce every constraint of repository-scale software work.

Three newer studies provide directional support for the same measurement shape. SkillEvolBench, from Lei et al.~(\href{https://arxiv.org/abs/2605.24117}{2026}), and EngramaBench, from Acuna (\href{https://arxiv.org/abs/2604.21229}{2026}), examine whether stored or evolved material remains available and useful while also reporting final answers. Wolff and Bennati (\href{https://arxiv.org/abs/2601.07978}{2026}) separate what a memory system retrieves from what the answering model produces. I use these studies to establish direction, not as a numerical basis for repository-retrieval policy.

Agent Retrieval Bench provides a direct repository-scale instrument for the upstream stage. Qin and Xie (\href{https://arxiv.org/abs/2607.24882}{2026}) built 427 samples across 25 repositories, including positive retrieval tasks, natural no-gold cases, and counterfactual wrong-repository controls. No retrieval family dominated all measures, and logged agent trajectories missed every gold file on 27 to 35 percent of samples. A controlled seed pilot found that retrieval-derived initial context raised file F1 while reducing later exploration relative to random non-gold context. These are strong benchmark and pilot results under the released design; they do not select one retrieval architecture for every repository.

\hypertarget{retrieval-metrics}{%
\section{Retrieval metrics}\label{retrieval-metrics}}

When a task requires one item, its \(\mathrm{Recall}@k\) value is binary: the item either appears among the first \(k\) results or it does not. When several items are required, \(\mathrm{Recall}@k\) is the fraction found within those results.

The aggregate may be computed as the mean of per-task recall values or as the pooled fraction across all required items. These are different estimands when tasks require unequal numbers of items, so the report should name which one it uses. It must also state what counts as relevant.

A file-level measure asks whether the needed file appeared. A chunk-level measure may require the particular definition, implementation, or surrounding block. Those targets answer different questions and should not share one unlabeled score.

The \textbf{rank} of a needed item is its position in the ordered results, with position one returned first. Items at positions one and ten both count as retrieved at \(k = 10\), although the first consumes less context and is less vulnerable to truncation. \(\mathrm{Recall}@k\) does not preserve that difference.

When several items are required, retain each position or report a declared aggregate. One rank value rarely describes the full set.

Retrieval depth is an experimental variable. Sweep \(k\) across a range and plot recall against depth. The \textbf{recall plateau} is the region in which the curve flattens and additional results recover little more required evidence. Fixing \(k\) before inspecting the curve can distort the comparison.

A shallow cutoff can make a viable retriever appear incapable. A deep cutoff can conceal poor ranking because a disordered list still contains the target somewhere among many returned items. The plateau does not determine the operating point by itself. Latency, context consumption, and distractors may justify stopping earlier. It shows the recall the ranker can reach and the marginal number of returned items required to reach it.

Multiple retrieval channels produce ordered lists whose numeric scores may have incompatible meanings. BM25, a lexical ranking function based on term occurrence and document statistics, can produce values well above one. Cosine similarity between embedding vectors usually occupies a bounded interval. Adding or averaging their raw scores gives arbitrary influence to whichever implementation emits larger numbers.

\textbf{Reciprocal-rank fusion} combines lists by position instead. Each item receives a contribution such as \(1/(c+r)\) from each channel, where \(r\) is its rank and \(c\) is a fixed damping constant. The fused order reflects agreement and placement across lists without assuming comparable score scales.

Each metric remains incomplete. \(\mathrm{Recall}@10\) does not show whether ten falls before or after the recall plateau. Mean rank excludes or obscures tasks in which the target never appeared unless missing results receive an explicit convention. A fused list may preserve coverage when one channel fails, but that failure remains visible only if the original channel results are retained.

Stage-level evaluation requires relevance judgments. In a curated benchmark, maintainers may identify the file, symbol, span, or dependency set needed for each task. In a production repository, relevance may be plural and conditional. One correct patch may rely on an interface declaration and a caller. Another may use the implementation and a test. Declaring only one path relevant can penalize an equally valid evidence route.

When the evaluation warrants the expense, construct judgments from the accepted solution, dependency structure, and expert review. Retain graded labels when an item can be useful without being sufficient. A class declaration might be relevant, while the concrete override is necessary for sufficiency. Whatever schema is used, the record should state what qualified an item and which valid solution paths the reviewers considered.

The cost limits where this decomposition can be applied. Repository-scale relevance judgments require people who understand both the task and the codebase, and repository changes can invalidate them. Sampling controls that expense. Annotate a stratified subset, preserve the adjudication record, and use it to diagnose the larger evaluation without pretending that unlabeled tasks have measured retrieval quality.

A proxy based only on file overlap with one reference patch is cheaper and should be reported as that proxy.

\Needspace{5\baselineskip}

Five terms help describe retrieved context:

\begin{itemize}
\tightlist
\item
  \textbf{Relevance} asks whether an item bears on the task.
\item
  \textbf{Sufficiency} asks whether the returned set contains enough evidence to act correctly.
\item
  \textbf{Isolation} asks whether the evidence is separated from material likely to mislead the model.
\item
  \textbf{Economy} measures the context consumed to convey the evidence.
\item
  \textbf{Provenance} records where the item came from and which version it represents.
\end{itemize}

This vocabulary comes from a position paper by Vishnyakova (\href{https://arxiv.org/abs/2603.09619}{2026}) and is not itself a measurement protocol. Each property requires an operational definition tied to the evaluated system before it can enter a score.

Evidence use is harder to measure than exposure. Model-generated citations may be incomplete or retrospective. Removing one retrieved item and rerunning the task is an ablation, but model variability may change the result for reasons unrelated to that item. Attention weights do not establish causal use.

\Needspace{5\baselineskip}

Combine the available signals and state what each supports:

\begin{itemize}
\tightlist
\item
  Exposure logs show what the model received.
\item
  Citations show what the model claimed to use.
\item
  Controlled removal across repeated runs estimates whether an item changed the outcome.
\end{itemize}

This decomposition also changes how verbose retrieval should be interpreted. Returning more material generally raises the chance that relevant evidence appears, but it can reduce economy, isolation, and sometimes final performance. Precision and recall should therefore be reported with context-token use and completion score.

A system that improves recall by filling the context window has made a different tradeoff from one that promotes the same evidence into its first few results.

Stage scoring is well established in benchmarked retrieval pipelines and less established in end-to-end agent evaluation. Agents reformulate queries, inspect results, open additional files, and change direction during a run. One static top-\(k\) list does not represent that trajectory.

\Needspace{5\baselineskip}

Record each retrieval event with:

\begin{itemize}
\tightlist
\item
  the query;
\item
  retrieval channel;
\item
  index version;
\item
  returned results and ranks;
\item
  the subset selected into context; and
\item
  the task state when retrieval occurred.
\end{itemize}

The evaluation can then ask whether the required evidence ever appeared, whether it appeared before the decision that needed it, and whether a later query repaired an earlier miss.

This record supports engineering decisions that the final score cannot. A missing indexed target, weak recall before the plateau, strong eventual recall with poor early ranks, and a relevant item excluded from context each identify a different stage. The categories are not perfectly independent, but they make the next ablation narrower and more informative.

\hypertarget{recover-the-evidence-a-single-lane-misses}{%
\section{Recover the evidence a single lane misses}\label{recover-the-evidence-a-single-lane-misses}}

My literature-review pipeline once ran with its semantic search lane down while lexical search continued returning plausible results. Nothing in the completed reviews revealed that one channel was missing. After I restored the lane and repeated the searches, the combined system surfaced 64 additional papers across nine queries for one review and 42 across seven queries for another. Manual curation retained 12 and 7 of them.

This case illustrates the danger of a single-lane failure: the output can remain coherent. Lexical search still returns documents containing the query terms, so operators see populated result lists rather than an outage. The missing papers become missing claims, qualifications, and related work downstream. The example does not estimate how many sources a semantic lane will recover in general. It shows why both lane health and lane contribution should stay observable even when another channel still appears to work.

Each retrieval lane has characteristic blind spots. Lexical systems rank through shared tokens. They are strong when the query contains a rare identifier, exact error string, configuration key, function name, or the spelling used in code. They weaken when a task describes behavior using language different from the implementation. A request to ``stop duplicate jobs after a worker restart'' may depend on code named \texttt{claim\_epoch}, \texttt{lease\_owner}, or \texttt{dedupe\_key}, even though none of those terms appears in the request.

CodeSearchNet, from Husain et al.~(\href{https://arxiv.org/abs/1909.09436}{2019}), framed this as a vocabulary gap between natural-language intent and source code. Repository agents face a wider form of the same problem because their queries combine prose, symbols, stack traces, and partial hypotheses. An embedding lane can connect semantically related text with little token overlap. It may retrieve a lease-recovery routine for the duplicate-job request even when the task never uses the routine's identifiers.

The software-engineering venue audit found two reviews that sharpen this stage boundary. Rahman et al.~(\href{https://doi.org/10.1145/3607179}{2023}) screened 2,970 candidates and synthesized 70 primary studies on automated query reformulation for source-code search. Sun et al.~(\href{https://doi.org/10.1145/3656341}{2024}) organized 68 code-search studies around query understanding, code understanding, and query-code matching. Their taxonomies support recording query formation as its own stage rather than attributing every miss to ranking. Neither review tests the hybrid rank-fusion protocol developed here.

The complement runs in the other direction. Semantic retrieval may place conceptually similar utilities near one another while missing the exact declaration that governs an edit. A query containing \texttt{ERR\_REPLAY\_DIVERGED}, a database column name, or a generated type gives lexical search unusually discriminating evidence. Mapping those characters into a semantic neighborhood can discard the rarity that made them useful.

The strongest code-specific evidence for retaining both lanes comes from an industrial study. Across 1,669 internal WeChat repositories and 26 language models, Yang et al.~(\href{https://arxiv.org/abs/2507.18515}{2025}) found that combined lexical and semantic retrieval performed best for closed-source code completion. The channels recovered complementary evidence rather than acting as interchangeable versions of one ranking function.

That result does not establish the same ordering for every repository agent or newer embedding model. The workload was industrial, dominated by large C++ codebases and completion-style tasks. The experimental design transfers more readily than the result: run each lane independently on the operated corpus and measure which relevant items each uniquely recovers.

Production reports and late-interaction research point in the same direction without settling the question. A production account of agent memory from Trautmann and Sutter (\href{https://blog.cloudflare.com/introducing-agent-memory/}{2026}) describes parallel channels fused by rank as a deployed default. Adaptive memory retrieval, from Yan et al.~(\href{https://arxiv.org/abs/2603.16496}{2026}), and late-interaction retrieval, from Khattab and Zaharia (\href{https://arxiv.org/abs/2004.12832}{2020}), support retaining complementary representations rather than any specific fusion policy. None provides code-specific evidence for the effect of fusion.

Execute the channels separately and preserve their original result lists. The lexical lane records its query, tokenizer, filters, and index version. The semantic lane records its query, embedding model, filters, and vector-index version. Per-lane \(\mathrm{Recall}@k\) and target ranks show what each channel contributes. The fused list is a separate stage with its own recall, ranks, context consumption, and downstream completion score.

Rank fusion is necessary because channel scores do not share a unit. Suppose lexical search returns an exact symbol with a BM25 score of 18, while semantic search returns a related file with cosine similarity 0.84. The difference between 18 and 0.84 says nothing about their relative relevance. Multiplying the cosine score by 100 would reverse the implicit weighting without changing the semantic order. Any sum of raw scores therefore embeds an arbitrary scaling decision.

\textbf{Reciprocal-rank fusion} combines lists through position rather than score magnitude. An item receives a contribution such as \(1/(c+r)\) from each lane, where \(r\) is its rank and \(c\) is a fixed damping constant. An item ranked first by one lane and fifth by another receives support from both. An exact identifier ranked first lexically but absent semantically can remain competitive. A conceptual match ranked moderately in both lanes can rise above results supported by only one.

The damping constant controls how quickly contributions fall with rank. It belongs in the recorded configuration and should be selected before inspecting completion outcomes. Choosing it afterward creates another researcher degree of freedom of the kind Chapter 1 describes.

Fusion does not make agreement equivalent to relevance. Two lanes may share an ingestion error, retrieve the same stale version, or favor a heavily duplicated utility. Duplicate chunks can occupy several positions and create the appearance of independent support. Deduplicate stable item identities before or during fusion, and retain provenance so that several representations of one source do not count as separate evidence.

Filters also belong in each lane's record. Restrictions by language, repository area, branch, generated-code status, or modification time can improve isolation. A mistaken filter makes relevant evidence unreachable. When lanes apply different filters, their measured retrieval quality includes a corpus-selection difference as well as a ranking difference. A useful comparison either holds the eligible corpus fixed or declares corpus selection part of the lane being evaluated.

Versioned corpora create a particular risk for semantic retrieval. An obsolete implementation may be extremely close to the query and structurally similar to current code while prescribing an invalid interface. Such a result becomes harmful when the generator treats similarity as authority. Record source revision and freshness with every result so the freshness gate in Chapter 13 can reject evidence whose semantic closeness exceeds its authority.

Every added lane also adds cost. It requires index construction, storage, query compute, operational monitoring, and another source of distractors. Parallel execution can reduce wall-clock latency relative to sequential search, but it does not remove compute or maintenance cost. A lane belongs in the stack when it recovers useful evidence on the operated workload or supplies resilience the system needs.

Demand may be asymmetric. In one corpus served by my systems, I observed 7,993 keyword calls and 2,449 semantic calls across site traffic and benchmark runs. Those counts do not measure answer quality and reflect the interfaces and users exposed by that system. They do show that replacing lexical retrieval with semantic retrieval would have ignored most expressed demand, even though the semantic lane remained necessary for vocabulary gaps.

\Needspace{5\baselineskip}

Evaluate an added lane through a paired ablation. The same tasks run with:

\begin{enumerate}
\def\labelenumi{\arabic{enumi}.}
\tightlist
\item
  the existing lane;
\item
  the added lane alone; and
\item
  fused results under the same retrieval and context budgets.
\end{enumerate}

Because outcomes are paired by task, the analysis uses per-task differences rather than comparing aggregate recall values alone. Sweep retrieval depth for each original list and the fused list because fusion can move the point where recall plateaus. The per-task record identifies targets found only lexically, only semantically, by both lanes, or by neither.

Identifier-heavy tasks need their own stratum. Aggregate recall can conceal a fusion policy that helps prose queries while demoting exact symbols. Identify these tasks from task construction and repository artifacts before inspecting outcomes. Error-string lookup, symbol navigation, configuration lookup, and natural-language behavior queries may benefit from different mixtures. The strata should remain measurements of the workload rather than becoming routing rules justified by a small result.

The companion catalog contains more elaborate retrieval paths: explicit recovery across lexical gaps, iterative re-querying when a draft exposes missing context, retrieval gates based on expected benefit, investment in whichever end of the retrieval-and-generation pipeline is limiting, and a cheap-first funnel followed by reranking. Each is useful only when the operated system exhibits a failure that calls for it. None repairs an unobserved channel outage or makes raw scores commensurate.

The operational test is whether fusion recovers relevant evidence that the current lane systematically misses without consuming more context than the generator can use. Inspect the recovered items, their ranks, and the tasks whose outcomes they change. A lane that contributes only duplicates and stale neighbors has added cost without adding usable evidence, regardless of its semantic sophistication.

\hypertarget{preserve-code-structure-inside-the-retrieval-unit}{%
\section{Preserve code structure inside the retrieval unit}\label{preserve-code-structure-inside-the-retrieval-unit}}

Replacing fixed line windows with syntax-aware chunks increased \(\mathrm{Recall}@5\) by 4.3 percentage points on RepoEval, a repository-level retrieval and completion benchmark, in the cAST study from Zhang et al.~(\href{https://arxiv.org/abs/2506.15655}{2025}). The same study reported a 2.67-percentage-point \(\mathrm{Pass}@1\) gain on SWE-bench generation across languages.

These are modest end-to-end changes. They are still large enough to investigate at the chunking-policy layer because the intervention changes neither the model nor the information present in the repository.

The chunking policy defines the units a retriever can rank and a context builder can select. An index rarely stores an entire repository as one document. It divides files into spans, attaches identifiers and metadata, and represents each span for lexical or semantic search. Those spans become the objects over which \(\mathrm{Recall}@k\) is measured. A relevant function cannot rank as one coherent item when the chunker separates its signature, body, and surrounding contract into unrelated records.

Fixed windows choose boundaries by line or token count. They are simple, deterministic, and applicable to any text format. Their weakness follows from ignoring code structure. A 100-line window may end after a function signature and place its body in the next chunk. Another may combine the end of one class, several imports, and the beginning of an unrelated helper merely because those lines are adjacent.

Overlap reduces some boundary damage without eliminating it. Repeating twenty lines between windows may preserve a short definition, but it also duplicates imports, comments, and boilerplate across many indexed records. Long methods still split. Repeated spans can crowd result lists and distort fusion unless the system deduplicates them. Overlap trades index size and redundancy for fewer severed boundaries.

An abstract syntax tree, or AST, represents source code as nested syntactic constructs produced by a parser. A class contains methods, a method contains statements and expressions, and a conditional contains branches. Syntax-aware chunking uses these structures as boundary constraints while still respecting a size budget. It does not require every AST node to become a separate retrieval unit.

The process begins with a file-level tree and a maximum chunk size expressed in tokens or another measure tied to the context system. When a node fits, the chunker retains it as a candidate. When it is too large, the chunker descends into its children and repeats the test. Small adjacent siblings are merged while the combined representation remains within budget. The resulting units tend to preserve complete functions, methods, classes, or coherent statement groups rather than arbitrary line intervals.

Comments, decorators, attributes, and signatures require explicit attachment rules. A parser may represent them as siblings or metadata even though a programmer treats them as part of the following declaration. Losing a decorator can remove authorization or routing behavior. Losing a leading comment can remove a constraint not expressed in code. Test these attachment rules by language, because the generic tree shape may not represent the unit a programmer needs.

Imports and type definitions introduce another choice. Copying all imports into every chunk improves local interpretability at a high token cost. Keeping imports only in a file header preserves the original structure but can leave a retrieved method with unresolved names. A practical design records stable file and symbol metadata on every chunk and allows the context assembler to retrieve a small companion span when import or type resolution is needed. Sufficiency and economy determine the policy. The tree alone does not.

The concept is language-general, but parsing is not. Every supported language needs a parser compatible with the repository's syntax version, including extensions, generated forms, and incomplete files. Parse failures require an explicit fallback, such as fixed windows marked with a failure flag. Silently dropping an unparseable file creates a corpus-coverage defect that later resembles a retrieval miss.

Incremental indexing adds state-management requirements. When a file changes, the system must parse the new version, retire old chunk identities, write the replacements, and update references without exposing an empty or mixed version to queries. Identities based only on line offsets churn whenever lines are inserted near the top of a file. Symbol-qualified identities survive some edits, but overloaded, anonymous, or generated constructs still need versioned disambiguation. Every chunk should record the source revision from which it was derived.

Syntax is only a partial account of meaning. A large function may contain several responsibilities that deserve separate units. A small method may be unintelligible without a protocol implemented across nearby methods. Macros, reflection, generated bindings, configuration, and build files may carry semantics an AST does not represent. Syntax-aware boundaries improve structural coherence without discovering every semantic dependency.

The reported 4.3-percentage-point retrieval gain should be interpreted at that layer. It shows that changed chunk boundaries moved relevant code into the first five results on the studied tasks. It does not show that structural chunking repairs a missing repository, stale index, incorrect query, or retriever unsuited to code. A chunker can shape only the items that enter the index and only the evidence the ranking system can recognize.

The 2.67-percentage-point generation gain is smaller and farther downstream. Better retrieval must survive context selection and then alter the model's output. The attenuation is consistent with the stage chain developed earlier, although the study does not identify one cause for it. Some newly retrieved chunks may be redundant, appear below the usable cutoff, or fail to change generation.

\Needspace{5\baselineskip}

Compare chunking policies on identical repository revisions and tasks. The fixed-window and syntax-aware indexes use the same:

\begin{itemize}
\tightlist
\item
  eligible files;
\item
  retrieval methods;
\item
  queries;
\item
  filters; and
\item
  top-\(k\) sweeps.
\end{itemize}

Relevance judgments identify both the required symbol and acceptable supporting neighbors. Then compare chunk recall, target rank, returned tokens, duplicated content, parse-failure coverage, and final task completion.

Chunk recall requires a declared unit. If the target is a method, a chunk containing one line from that method should not count as fully relevant. The chunk must contain the task-specific evidence named in the judgment, such as the complete signature and relevant branch. Partial matches can be reported separately when they help explain a failure. File-level recall remains useful for corpus and coarse-retrieval diagnosis, but it cannot establish that the returned chunk is interpretable.

Chunk size needs its own sweep. Very small syntax-aware chunks improve isolation but can remove relationships required to understand state changes. Very large chunks preserve local structure while consuming more context and reducing the number of distinct candidates the model can inspect. Sweep size together with retrieval depth, because the two controls interact. Ten 200-token chunks and ten 1,500-token chunks impose different costs even though both are reported as \(\mathrm{Recall}@10\).

The implementation should also preserve ordering and provenance. When several chunks from one file enter context, source order can help the model reconstruct control flow. Every chunk should identify its file, symbol path, source revision, and line range so a failed completion can be traced back to the exact indexed representation.

Structural chunking is justified when its measured benefit covers parser maintenance and indexing complexity for the supported languages. A repository dominated by languages with reliable parsers and long structured files is a strong candidate. A heterogeneous corpus containing templates, notebooks, generated fragments, and proprietary languages may require mixed policies. The appropriate retrieval unit follows the workload and the structure available within it.

The final comparison returns to decomposed scoring. Measure whether syntax-aware chunks recover the required evidence and where they rank, then whether those chunks improve completion under the same context budget. Better chunk recall with unchanged completion remains a useful diagnostic result. It shows that the chunker repaired its stage and that the remaining limitation lies elsewhere in the path.

\hypertarget{run-the-retrieval-protocol-on-the-operated-workload}{%
\section{Run the retrieval protocol on the operated workload}\label{run-the-retrieval-protocol-on-the-operated-workload}}

Begin with the stack that serves real tasks. The first version can use a small sample from one query class, sized to the available annotation budget and stratified across the repositories or task shapes relevant to the decision. Judge the evidence each task requires and instrument the production retrieval path. A retrieval-depth sweep on that stratum can answer one question: whether the current top-\(k\) cutoff should change for that query class.

\Needspace{5\baselineskip}

For each task, identify the file, symbol, or evidence set that a competent solution could use. Record:

\begin{itemize}
\tightlist
\item
  every retrieval query;
\item
  the index revision;
\item
  returned item identities and ranks;
\item
  which items entered model context; and
\item
  the final task outcome.
\end{itemize}

When the system exposes citations or another credible evidence-use signal, preserve it. Do not treat context exposure alone as proof that the model used the evidence.

Compute file-level and chunk-level recall separately. Sweep retrieval depth because a cutoff such as ten is usually a tool default rather than a measured operating point. Plot recall against depth, identify where additional results recover little new evidence, and record latency and context-token use at each point.

Select the operating point under the workload's context and cost constraints. Retain the full sweep with the result so another reader can see what the chosen cutoff excluded.

Separate tasks by the evidence they require. Exact symbols, error strings, natural-language behavior descriptions, cross-file dependencies, and version-sensitive questions exercise different retrieval paths. Define those strata before comparing systems. A gain concentrated in one query class may justify routing or an additional lane even when its aggregate effect is small.

If the production stack has only one retrieval lane, add the missing lexical or semantic lane as a controlled arm. Run each lane independently and preserve its original ranking. Fuse ranks rather than raw scores because the score scales are not commensurate.

\Needspace{5\baselineskip}

Measure:

\begin{itemize}
\tightlist
\item
  which relevant items each lane uniquely recovers;
\item
  which stale or duplicated items each adds;
\item
  how fusion changes target ranks;
\item
  how much context the fused list consumes; and
\item
  whether completion changes under a fixed context budget.
\end{itemize}

Keep lane health visible in production so a populated result list cannot conceal the failure of one channel.

When reliable parsers exist, build a second index using syntax-boundary chunks. Hold corpus eligibility, repository revision, task set, queries, and retriever fixed. Sweep chunk size and retrieval depth together, and report parse failures and fallback coverage.

\Needspace{5\baselineskip}

Compare the syntax-aware and existing chunkers on:

\begin{itemize}
\tightlist
\item
  target rank;
\item
  file and chunk recall;
\item
  returned tokens;
\item
  duplicated content;
\item
  parse and indexing coverage; and
\item
  final task completion.
\end{itemize}

The measured difference, including no meaningful difference, determines whether parser maintenance and indexing complexity belong in that repository.

Interpret the resulting table by stage. Missing targets indicate a corpus-coverage, indexing, or freshness problem. Targets still absent when the recall curve plateaus indicate a query or retrieval-lane problem. Targets retrieved but ranked beyond the context cutoff indicate a ranking, fusion, or context-selection problem.

When sufficient evidence reaches the model but completion remains unchanged, move the next ablation downstream. Investigate evidence use, generation, execution, or verification rather than continuing to modify retrieval.

This decomposition narrows the next intervention without treating any one retrieval metric as a measure of the whole system.

The retrieval depths, fusion constants, and chunk sizes in this chapter are not recommended defaults. Each is a parameter whose effect must be measured under the repository distribution, task mix, model, and context budget in operation.

Chapter 13 turns to the architecture around these measurements: cheap retrieval funnels, typed indexes, and the conditions under which apparently relevant context should be rejected as stale.

\section*{Sources and evidence}

\textbf{Score retrieval and generation separately}

\begin{itemize}
\tightlist
\item
  Strong evidence: Liu, T., et al.~(2023), ``RepoBench: Benchmarking Repository-Level Code Auto-Completion Systems,'' ICLR 2024, arXiv:2306.03091. (R/C/P decomposition localizes failure to retrieval vs completion vs pipeline stage.)
\item
  Strong evidence: MemConflict (Tao et al.~2026), arXiv:2605.20926. (Black-box + white-box memory evaluation; localizes missing vs low-ranked vs retrieved-but-unused; best reported conflict-recognition score 0.2501.)
\item
  Strong evidence: Memory Circuit Analysis (Mao et al.~2026), arXiv:2605.03354. (Stage-level diagnostic localizing a silent failure to the responsible memory operation; extraction/retention/retrieval fail silently behind fluent answers.)
\item
  Directional evidence: SkillEvolBench (Lei 2026), arXiv:2605.24117.
\item
  Directional evidence: EngramaBench (Acuna 2026), arXiv:2604.21229.
\item
  Directional evidence: Cost-and-Accuracy study (Wolff \& Bennati 2026), arXiv:2601.07978.
\item
  Corroborating case: Vishnyakova, O. (2026), arXiv:2603.09619. (Names the five context-quality criteria; position paper.)
\item
  Strong evidence for the benchmark measurements and controlled seed pilot: Qin, B., and Xie, Y. (2026), ``Agent Retrieval Bench: Evaluating Repository Context Retrieval for Coding Agents,'' arXiv:2607.24882. The benchmark supports stage-level evaluation rather than a universal retrieval-family ranking.
\end{itemize}

\textbf{Hybrid retrieval fused on ranks}

\begin{itemize}
\tightlist
\item
  Strong evidence: Yang, Z., et al.~(2025), ``A Deep Dive into Retrieval-Augmented Generation for Code Completion: Experience on WeChat,'' arXiv:2507.18515 (industrial study).
\item
  Directional evidence for the vocabulary-gap claim: CodeSearchNet (Husain 2019), arXiv:1909.09436. The study establishes the mismatch between natural-language queries and code vocabulary. It does not test the chapter's rank-fusion protocol; Yang et al.~supply the direct comparison of combined lexical and semantic retrieval.
\item
  Directional evidence (one record, two accounts): Yan et al.~(2026), ``AdaMem: Adaptive User-Centric Memory for Long-Horizon Dialogue Agents'', arXiv:2603.16496, whose own retrieval route is semantic retrieval with conditional graph expansion rather than rank fusion, so it supports retaining a complementary semantic lane only; and Trautmann \& Sutter (2026), {``Agents that remember: introducing Agent Memory''}, Cloudflare, a production account carrying the shipped default of parallel channels fused with reciprocal-rank fusion.
\item
  Directional evidence: ColBERT (Khattab and Zaharia 2020), arXiv:2004.12832.
\item
  Directional evidence: Rahman, M. M., et al.~(2023), ``A Systematic Review of Automated Query Reformulations in Source Code Search,'' \emph{ACM Transactions on Software Engineering and Methodology}, \href{https://doi.org/10.1145/3607179}{DOI:10.1145/3607179}. The review supports treating query formation as a distinct retrieval stage, not the chapter's fusion policy.
\item
  Directional evidence: Sun, W., et al.~(2024), ``A Survey of Source Code Search: A 3-Dimensional Perspective,'' \emph{ACM Transactions on Software Engineering and Methodology}, \href{https://doi.org/10.1145/3656341}{DOI:10.1145/3656341}. The survey separates query, code, and matching components; transfer to repository agents remains directional.
\end{itemize}

\textbf{Chunk on AST boundaries}

\begin{itemize}
\tightlist
\item
  Strong evidence: Zhang, Y., et al.~(2025), ``cAST: Enhancing Code Retrieval-Augmented Generation with Structural Chunking via Abstract Syntax Tree,'' arXiv:2506.15655.
\item
  Corroboration for this entry: none on record. The other two entries in this chapter include author-system cases as illustrations rather than independent sources.
\end{itemize}

\chapter{Localization funnels, repository indexes, and freshness checks}
\label{ch13-localization-funnels-repository-indexes-freshness-checks}
\begin{quote}
\textbf{Evidence profile.} 6 strong \(\cdot\) 5 directional \(\cdot\) 1 corroborating evidence items across 3 developed practices (\erca{078}, \erca{084}, \erca{174}).

\textbf{Chapter claim.} Evidence without revision identity is stale, not current.
\end{quote}

In one evaluation of my memory system, run across four seeds, the task required passing a recalled value exactly as a tool argument. Retrieval by stable memory identity scored 1.0. Retrieval by token similarity returned superseded values and scored 0.0.

The similarity lane had not malfunctioned. It returned real, highly ranked records that described a state the system had already left behind. This author-system case carries no evidentiary weight, but it separates two questions that a retrieval score can collapse into one: whether a result ranks well, and whether it still describes the repository or system the worker is acting on.

Repository work presents three related architectural decisions. Some tasks benefit from a funnel that separates coarse file selection from fine edit selection. Some require typed structural relationships because the issue language and the change site share no useful vocabulary. Every retriever, index, and cache also describes a particular repository state, which may expire while its answers remain fluent and specific.

Each decision carries a different cost. A funnel adds handoffs, and an early omission propagates through every later stage. A typed index requires construction, language coverage, storage, query tools, and continuing maintenance. A freshness gate can reject useful evidence when its identity checks are too coarse. All three are therefore architecture choices with observable failure modes rather than retrieval features to add by default.

\hypertarget{stage-localization-before-repair}{%
\section{Stage localization before repair}\label{stage-localization-before-repair}}

Hierarchical localization determines where each comparison occurs and what evidence crosses the boundary between stages. Xia et al.~(\href{https://arxiv.org/abs/2407.01489}{2024}) removed open-ended tool-use autonomy from repository repair and retained three fixed phases: hierarchical localization, repair, and patch validation. On SWE-bench Lite, a curated subset of SWE-bench, that pipeline outperformed the autonomous agents evaluated alongside it while costing substantially less.

Each phase received bounded input, produced a bounded artifact, and handed that artifact to the next phase. The result showed that staged narrowing could recover much of the value then attributed to open-ended agency. Later systems built on stronger models moved ahead of that fixed pipeline, which limits the conclusion. The transferable design principle is to separate narrowing decisions and validate the transition between them; neither the original pipeline nor a general claim against agency follows from the result.

Repository structure may identify the correct region without identifying the function or statement that owns a failure. In a disconnect case, search may return a response writer, several transport adapters, and a lifecycle callback. The state transition responsible for the malformed response may sit behind that callback in another package.

\begin{figure}[htbp]
\centering
\includegraphics{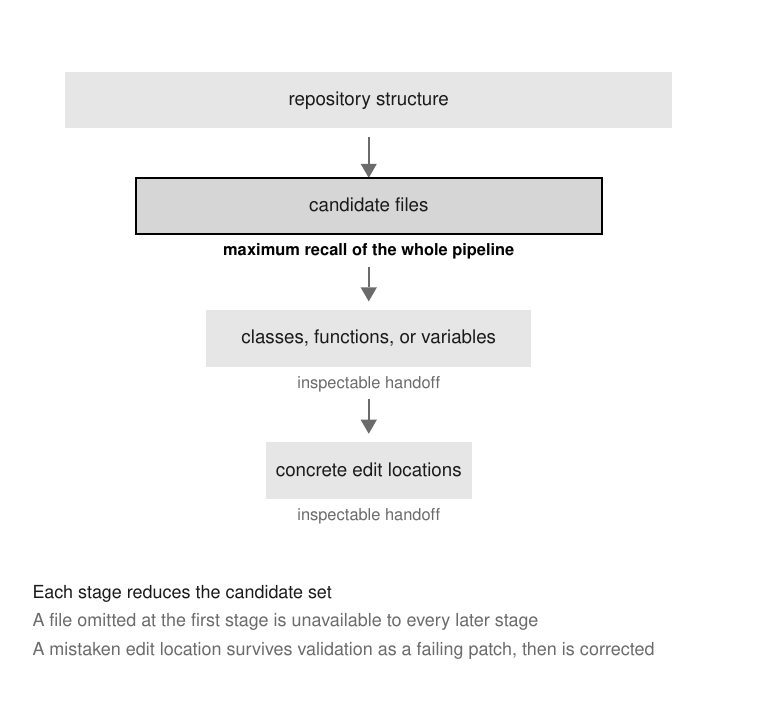}
\caption{The localization funnel progressively narrows repository evidence, while candidate file selection caps recall because any omitted file remains unavailable to every later inspection and edit stage.}
\label{fig:ch13-localization-funnel}
\end{figure}

The evidence needed to select the package differs from the evidence needed to select the function and edit location. A localization funnel separates those comparisons, narrowing repository structure to candidate files, then to candidate symbols, then to concrete edit locations, and finally to patch validation. Figure \ref{fig:ch13-localization-funnel} shows the sequence. Each stage should reduce the candidate set and leave an inspectable handoff.

In the disconnect case, the first stage narrows the repository to the transport and lifecycle packages. The file handoff retains the response writer and callback and records why each survived. Symbol inspection follows the callback to the function that changes response state. The final stage proposes a guard at that assignment, and validation exercises the disconnect path. If an intermediate check shows that the state owner was omitted, the funnel widens before repair begins.

The stages have asymmetric consequences. Validation can sometimes correct a mistaken edit location after a patch fails. A file omitted during the first stage is unavailable to every later stage, however capable the repair model may be. The maximum recall of the complete pipeline is therefore capped by file selection.

Inspectable handoffs make downstream errors visible, but validation can repair a poor edit choice only when the necessary file survived initial selection.

Sepidband et al.~(\href{https://arxiv.org/abs/2604.05481}{2026}) measured this asymmetry across 500 SWE-bench Verified instances and 61 context configurations. Relative to the no-file condition, adding file-level context produced 15 to 17 times as much repair improvement as later localization refinements.

The same experiment found that successful repairs clustered around approximately six to ten relevant files for the model family studied. That range is useful as a diagnostic band, not as a quota for other models, repositories, or task distributions. Both findings come from one benchmark and one sweep of context configurations.

The practical implication is to measure file selection before adding more reasoning downstream. Chapter 12's decomposed scoring makes that diagnosis possible. When file \(\mathrm{Recall}@k\) is low, a stronger patch generator is being evaluated on a candidate set that often excludes the answer. When file recall is already high and repair still fails, additional localization machinery may move cost into a stage that is no longer limiting the system.

\Needspace{5\baselineskip}

The funnel also constrains what context should cross each boundary. A file-selection stage should not silently pass its complete search transcript downstream, because doing so recreates the mixed-granularity problem inside a larger prompt. Its handoff should identify:

\begin{itemize}
\tightlist
\item
  the files retained;
\item
  the evidence for retaining them;
\item
  the candidates rejected; and
\item
  any uncertainty that remains.
\end{itemize}

The next stage can then inspect those files for structural candidates without inheriting every abandoned hypothesis.

Inspectable handoffs also create validation points. Before symbol inspection begins, a file set can be checked against import neighborhoods, ownership boundaries, or another retrieval lane. Before concrete lines are proposed, the symbol set can be checked for definitions, references, and state ownership. These checks need not be performed by an autonomous agent. They need only expose a miss while widening remains possible.

Skipping a stage may appear cheaper because it removes a model call or retrieval pass. Xia et al.~reported the opposite. When their pipeline moved directly from file selection to edit locations, both cost and performance worsened because the remaining stage had to search too much source at once. Removing a stage increased the work inside the stage that remained. Pipeline length is therefore a poor proxy for pipeline cost.

Chang et al.~(\href{https://arxiv.org/abs/2502.15292}{2025}) trained separate models for file, function, and statement localization and found value in treating each level as a distinct representation and discrimination problem. That support is directional and does not determine the implementation. A fixed prompt, cheap classifier, retrieval operation, or stronger general model may be appropriate at a particular stage depending on workload, accuracy, and latency.

The funnel fits issue-resolution work because those tasks often decompose into localization, repair, and validation. It is less convincing for exploratory refactoring, architecture discovery, or changes whose scope becomes visible only after editing begins. Those workloads require an explicit return path that can widen the candidate set after a failed hypothesis. Without one, a rigid pipeline continues refining the wrong corpus.

Later agent systems that surpassed the fixed pipeline also limit any claim that autonomy itself was unnecessary. Stronger models can search, revise, and use tools while preserving hierarchical localization. The evidence supports measuring the fixed narrowing architecture first and then asking which remaining failures autonomy repairs. It does not support freezing the pipeline around one generation of models.

The companion material covers narrower implementation choices: structural retrieval as a lane beside text search, stable structural anchors injected into context, and query-time graph walks when stored expansion would become too large. Those choices follow the architectural decision developed here. Preserve granularity, validate each handoff, and measure file selection as the stage capable of foreclosing every later success.

\hypertarget{build-a-typed-index-only-when-someone-will-maintain-it}{%
\section{Build a typed index only when someone will maintain it}\label{build-a-typed-index-only-when-someone-will-maintain-it}}

Consider an issue reporting that cancelling one export leaves later exports blocked. The change belongs in a generic state-transition helper reached through a queue consumer and shared lifecycle component. None of those artifacts uses the language of exports or cancellation.

Exact search can find the user-facing entry point. Similarity search can find code that resembles the reported symptom. Neither query expresses the dependency path from the entry point to the component that owns the blocked state. A typed index can answer that path question, but the decision to build one begins with maintenance. If no named owner accepts the continuing work, the projected retrieval gain has no durable system behind it.

\Needspace{5\baselineskip}

That work includes:

\begin{itemize}
\tightlist
\item
  parsing each supported language;
\item
  assigning stable identities to entities;
\item
  recording typed relationships;
\item
  handling generated code and partial parses;
\item
  publishing updates without exposing incomplete state;
\item
  versioning the schema and query surface; and
\item
  testing correspondence between index results and repository state.
\end{itemize}

Each new language or artifact type extends the obligation. Index construction is a continuing product surface even when users encounter it only through search results.

Scale is what makes that surface unavoidable rather than optional. Potvin and Levenberg (\href{https://doi.org/10.1145/2854146}{2016}) describe one organization holding billions of lines in a single repository and the purpose-built navigation, search, and dependency tooling its scale required. This is a practice report from one company with no comparison against alternatives, and it predates coding agents, so it establishes that the obligation exists at that scale rather than which index answers it.

The cheaper alternative is often a generated repository map or skeleton placed near the front of the model's context. It can list directories, important files, top-level symbols, and ownership boundaries. Navigation then relies on inexpensive operations such as exact search, opening definitions, finding references, and listing callers.

A repository map cannot answer arbitrary multi-hop questions. It is, however, easy to regenerate and inspect. For a modest repository, low change rate, or occasional task, that trade may be preferable to maintaining a graph service.

A typed index becomes plausible when the workload repeatedly asks structural questions that the map cannot express. The representation parses directories, files, classes, functions, and linked development artifacts into typed nodes. In the export example, invocation and reference edges connect the queue consumer to the lifecycle component and then to the state helper. A broader index may also record containment, imports, inheritance, definitions, ownership, and links to issues or tests.

This representation is a \textbf{typed knowledge graph}. It stores entities under declared kinds and records how each pair of connected entities relates. The types constrain traversal and make a returned path inspectable. They also create schema, parser, identity, and migration obligations that a similarity index does not have.

The export question can then become a bounded traversal. It starts at the export handler named by the issue, follows invocation edges into the queue consumer, follows references into the lifecycle component, and returns the functions that write the blocked state. The issue and the final helper need no shared token. The graph supplies a path that can be inspected edge by edge.

A structural path can also be more compact than the alternatives. Instead of serializing the repository or opening every file encountered during exploration, the retriever can return a small subgraph containing the entry point, intermediate calls, state owner, and relationship types. The model receives both the candidates and an explanation of why they were selected.

Published results show why teams consider paying this cost. Chen et al.~(\href{https://arxiv.org/abs/2503.09089}{2025}) reported 92.7 percent file-level localization accuracy for LocAgent, with performance declining at finer granularities. That pattern supports using the graph to protect file localization while measuring symbol and line localization separately. The figure describes localization, not repair, and should not be read as an issue-resolution rate.

LocAgent's fine-tuned 7B and 32B models approached the reported Claude 3.5 localization accuracy at approximately 86 percent lower cost. Its ablations also showed that the gain depended on the complete search and traversal toolset around the graph, not on the stored representation alone.

Yang et al.~(\href{https://arxiv.org/abs/2503.21710v1}{2025}) provide more direct but weaker evidence for multi-hop demand. Among bugs their repository-aware graph successfully localized, 69.7 percent required traversal across several relationships. That figure comes from the first paper revision, and headline results changed later. It also comes from SWE-bench Lite. The result establishes direction within one benchmark, not the prevalence of multi-hop demand in an arbitrary production repository.

The same study capped repair input at the top 20 functions. That cap is part of the architecture. A traversal system can continue finding plausible neighbors after useful evidence has been exhausted. Passing every discovered node to the generator allows additional candidates to interfere with repair. A graph can therefore improve localization while degrading generation when no boundary separates the two stages.

Ma et al.~(\href{https://arxiv.org/abs/2406.01422}{2024}) measured this tension in LingmaAgent across a larger exploration budget. Its issue-resolution rate rose from 16.0 percent with no exploration iterations to 21.3 percent at 600 iterations. Exploration continued to uncover useful code across that range.

The patch-application rate, which measured whether generated diffs applied cleanly, peaked at 200 iterations and then declined through 600. Clean application was necessary but not sufficient for resolution. Fewer patches could apply while more tasks were resolved if the patches that did apply were more often correct. The result suggests that additional exploration continued to improve localization while making the downstream generation problem harder.

Both measurements come from one benchmark configuration. The paper reports the 200-iteration location of the application-rate peak but not the curve's values, so the peak is useful only as a prompt to measure the local stopping point.

A traversal budget must therefore be evaluated against both localization and patch application. It may stop after a fixed number of expansions, when no new relationship type appears, or when the candidate set reaches the repair budget. The literature does not identify one stopping rule that dominates across repositories.

\Needspace{5\baselineskip}

A defensible local rule requires recording:

\begin{itemize}
\tightlist
\item
  each expansion;
\item
  the relationship followed;
\item
  candidates admitted and pruned;
\item
  the downstream candidate limit; and
\item
  the point at which further exploration stopped changing repair outcomes.
\end{itemize}

On SWE-bench Lite, the same study reported an 18.5 percent relative improvement over SWE-agent. That result applies to the benchmark configuration. Its production report used different operating measures: 16.9 percent of in-house issues were resolved automatically and 43.3 percent after human intervention. Automatic resolution, assisted resolution, and benchmark performance describe different systems and should not be combined.

Two additional experiments extend the evidence without removing its limits. Ouyang et al.~(\href{https://arxiv.org/abs/2410.14684}{2024}) added RepoGraph, a line-level definition and reference graph, as a plug-in and reported gains in both pipeline-style and agent-style frameworks. CodexGraph, from Liu et al.~(\href{https://arxiv.org/abs/2408.03910}{2024}), provides directional support for graph queries when similarity retrieval has weak multi-hop recall. Together they support a specialist structural lane that can attach to different orchestration styles.

Fine-grained localization remains difficult. LocAgent's accuracy declined below the file level, and serializing a graph as text can erase distinctions explicit in storage. Edge direction, relationship type, entity scope, and alternative paths can collapse into an ordering the model misreads. Extending the graph below functions may help line-level work, but it also increases node count, serialization cost, and the number of nearly equivalent candidates.

Query construction adds another failure surface. A valid graph can return an irrelevant answer because the model selected the wrong starting entity, followed the wrong relationship, or stopped one hop too early. Bounded operations such as \texttt{find\ definitions}, \texttt{list\ callers}, and \texttt{follow\ imports} are easier to validate than a general graph query generated in one attempt. Simple entity questions should use those cheaper operations, reserving the graph for structural questions that require it.

The representation returned to the model must also be measured. An index may emit paths, source excerpts, signatures, summaries, edge lists, or a combination. Each format changes token cost and the evidence available to the generator. A compact summary may omit the exact call that proves relevance. Raw source may conceal the path that made it relevant. Serialization is therefore part of the retrieval design rather than a presentation detail.

My architecture-analysis toolkit provides one author-system example. Its typed symbol graph measures change propagation by counting files reachable from a changed entity within a bounded depth. The same representation can support impact analysis through relationships that lexical similarity does not encode. This is narrative corroboration and supplies no evidence about agent repair rates.

The index owner must also own correspondence with the repository. Incremental parsers fail, branches diverge, generated artifacts retain obsolete identities, and linked issue data changes independently of code. An index that is not checked against repository state after those events silently answers questions about an earlier repository.

\Needspace{5\baselineskip}

Every returned path should therefore carry enough identity to establish:

\begin{itemize}
\tightlist
\item
  the repository and branch;
\item
  the source revision;
\item
  the index and schema versions;
\item
  the entity identities;
\item
  the freshness status; and
\item
  any parse or ingestion failures affecting the path.
\end{itemize}

Readers unwilling to fund that admission and maintenance path should choose a generated repository map and fixed navigation operations explicitly. Readers willing to fund it should cap candidate sets, budget traversal, validate query construction, test serialization, and measure the graph as one component of the surrounding retrieval system.

The maintenance cost is justified only while a named owner can show which repository state each answer describes. An external code-intelligence service changes who performs the parsing, indexing, and publication work. It does not change who owns correspondence, since the operator still has to know which repository state an answer describes and still absorbs the cost when that answer is wrong. Adoption is a decision about which of the obligations listed above transfer and which remain local, and the ones that remain local are the ones this section has already named.

A shared index across repositories also creates an authority question. Chapter 8 treats the read boundary as part of containment, because anything an identity can read can leave through model output. An index built over many repositories concentrates the material that per-repository access control was meant to separate. If it serves results without evaluating the caller's permissions, an agent reads code its operator cannot open, and the rule Chapter 17 states for the agent is broken by the retrieval layer instead. The local test issues one query under two identities with different repository access and compares the result sets, including the paths and snippets carried in the returned evidence. Identical answers mean the index is a read path around the permission model, and its results are a disclosure surface whatever its ranking quality.

\hypertarget{admit-only-evidence-tied-to-the-current-state}{%
\section{Admit only evidence tied to the current state}\label{admit-only-evidence-tied-to-the-current-state}}

Freshness is one instance of the state-consistency problem Chapter 7 states for the whole factory. A derived index, cache, or repository map describes a particular version of the world, and contract I7 makes retrieved evidence valid only for the state it observed. Before that evidence may influence a mutation against another version, the consumer must know which version it describes. The retrieval system is therefore not a special case with its own freshness rules. It is one more producer of versioned records inside a system whose other components, workers, verifiers, and schedulers, already carry the same obligation.

That framing also sharpens the observability question. What code exists is only part of it; the operable question is which state the agent, the scheduler, the verifier, and the publisher each observed when they acted. The mechanisms in this section, identity checks at the retrieval boundary, atomic generation publication, and periodic reconciliation, exist to make that question answerable for the retrieval component.

Weng et al.~(\href{https://arxiv.org/abs/2605.14478}{2026}) changed Python helper signatures in 17 curated examples and compared stale with current retrieved snippets. This small diagnostic study is the only literature support for the practice developed here. It establishes the direction of the failure, but one form of staleness tested on two models supplies neither a general effect size nor a safe refresh cadence.

Under stale-only retrieval, Qwen2.5-Coder-7B-Instruct produced 15 outputs incompatible with the current helper signature, and gpt-4.1-mini produced 13. Both models received the same 17 examples. Under current-only retrieval, neither produced an incompatible output.

These paired results support a narrow architectural rule: retrieved code should enter model context only when the system can tie it to the repository state the worker is allowed to modify. Rejecting evidence whose state identity does not match is contract I7 enforced at admission, and it follows contract I6's discipline: a record that contradicts the current state triggers reconciliation, not a guess about which version is right.

A stale result is an active hazard rather than an ordinary retrieval miss. It offers a concrete and plausible description of an API that no longer exists. High rank cannot compensate for evidence drawn from the wrong state.

The contrast with no retrieval makes the failure sharper. Without retrieved context, the models tended to fail without calling the obsolete signature. With stale context, they generated executable-looking code against the wrong contract. Retrieval converted uncertainty into a specific but incompatible implementation.

Adding current evidence reduced that harm even when stale snippets remained present. Across the tested model and retrieval conditions, adding valid current snippets to stale ones lowered the rate of current-state-incompatible outputs by 47 to 65 percentage points relative to stale-only retrieval over the same 17 examples. Permuting result order produced no significant difference. In this diagnostic, the presence of current evidence mattered more than whether it appeared first.

The paired comparison corresponds to the binary-outcome design in Chapter 1. Both conditions used the same 17 examples, each outcome was compatible or incompatible, and McNemar's test analyzed the discordant pairs.

The exact two-sided values were \(6.1 \times 10^{-5}\) for Qwen2.5-Coder-7B-Instruct and \(2.4 \times 10^{-4}\) for gpt-4.1-mini. Under the null hypothesis of no difference between retrieval conditions, each value gives the probability of observing discordance at least as extreme as the measured result.

Those values establish a difference within the curated sample. They do not estimate the effect size in another repository.

The study does not show that stale retrieval will fail at similar rates elsewhere, that every form of drift is equally harmful, or that one refresh interval is safe. Helper-signature changes are unusually direct because stale code presents a callable contract. Changes to behavior, configuration, ownership, or surrounding assumptions may fail differently and may be harder to observe.

\Needspace{5\baselineskip}

A freshness gate needs two identities:

\begin{itemize}
\tightlist
\item
  the repository state from which the retrieval artifact was built; and
\item
  the repository state the worker is allowed to edit.
\end{itemize}

A commit identity can represent a clean tree. When relevant files contain uncommitted changes, the commit alone is insufficient because it describes a state different from the worker's current view. The gate then needs content identities for the affected files or another representation of the working state. The commit-plus-working-tree pair also names where authority actually lives. The mutable target contract I2 protects is the working tree the worker can change, not the commit behind it, so the identity the gate compares against must describe that target, not an ancestor of it.

Every retrieval response should carry its indexed state identity so the caller can compare it with the working state before exposing any snippet to the model.

The comparison belongs at the retrieval boundary. Checking only when indexing begins leaves a race in which files change during construction or after a long-running query starts. A safe builder reads from a fixed snapshot, constructs the new index privately, and atomically publishes the completed generation together with its state identity. Queries then observe one complete generation rather than a mixture of old and new records. Atomic publication with a state identity in every response is contract I7 in mechanism form: each answer names the version it is bound to, and no reader can observe a version that never coherently existed.

Suppose an index was built from a clean commit and a worker later edits a helper without committing it. A query returns the old signature together with the index commit. The gate checks the in-scope working file, detects that its content no longer matches the indexed state, and withholds the snippet before generation begins.

Live exact search can supply the current helper while the richer index rebuilds. The builder snapshots the edited state, publishes a completed generation with the new identity, and repeats the original query. The caller admits the new result only after its identity matches the worker's state.

Concurrent work makes the choice of identity important. Two agents may share a base commit while holding different uncommitted edits. A commit-level check can therefore admit evidence stale for both working trees. Hashing every file in a large repository on every query may be too expensive.

\Needspace{5\baselineskip}

A practical design can:

\begin{itemize}
\tightlist
\item
  pin the base commit;
\item
  track content identities for files covered by each response;
\item
  invalidate affected entries from filesystem events; and
\item
  reconcile periodically against the actual working tree.
\end{itemize}

The system should measure both the coverage of that check and the race window it leaves open.

Caches require the same treatment. A query cache keyed only by search text can return an obsolete result after the underlying index refreshes. Repository-state identity must participate in the cache key, or the cache entry must be invalidated when a new index generation publishes. Otherwise the freshness gate protects the index while a faster layer bypasses it.

Retries also need a state transition. If a query fails because its index generation is stale, retrying against the same generation cannot improve the result. The retry must wait for or request a generation built from an accepted state and then repeat the query against that generation.

A timeout can bound the wait, but it should end in an explicit freshness failure rather than silently returning stale evidence.

Refusal has an operational cost. Live exact search may remain current while a structural or semantic index rebuilds. The system can therefore fall back to lower-capability navigation that is tied to the working state. When no current source is available, a visible miss that preserves uncertainty is safer than a fluent answer built from an obsolete API. The generator can request more evidence or fail explicitly rather than act on a false contract.

The literature reviewed for this chapter supplies no universal refresh cadence. Repository change rate, index-build duration, query latency, and the consequence of stale evidence can inform a local policy, but they do not yield an evidence-backed interval.

\Needspace{5\baselineskip}

Event-driven refresh can shorten the stale window after ordinary edits, although missed events and parser failures still require reconciliation. Scheduled rebuilding can provide that reconciliation, but its period should follow measured drift and recovery cost. Periodic reconciliation after missed events is contract I6 applied to the index's own records, and contract I10 governs the reconciler itself: it runs with authority over the index and is production code, tested under the same invariants as the normal ingestion path. Useful retrieval metrics include:

\begin{itemize}
\tightlist
\item
  the age of admitted results;
\item
  the frequency of state mismatches;
\item
  the duration of rebuild windows;
\item
  the rate of freshness refusals; and
\item
  the success and cost of fallback navigation.
\end{itemize}

My session-snapshot system illustrates a strict version of this gate. It pins cryptographic hashes for all in-scope files together with the repository commit. Any drift marks the snapshot stale, and the system regenerates it rather than branching from it. Remotely sourced repository knowledge is excluded because the system cannot bind it to the same content check.

My published literature-review index provides the counterexample. One surface advertised a paper count that disagreed with the linked page, and synchronized copies diverged further. The index continued to appear authoritative after it stopped describing its own contents. This contrary case exposes the maintenance cost of the prescription: a freshness policy that no one operates becomes another stale artifact.

Freshness does not determine authorization. Permission policy requires separate identities, rules, and audit records. The evidence here supports refusing obsolete repository states. It does not establish who may retrieve a current one.

\hypertarget{choose-the-next-retrieval-intervention}{%
\section{Choose the next retrieval intervention}\label{choose-the-next-retrieval-intervention}}

Begin with Chapter 12's decomposed measurements and determine which stage is limiting the operated system. Inspect file-level retrieval before changing the agent architecture.

File \(\mathrm{Recall}@k\) asks whether the repair stage receives the necessary files at all. When those files rarely appear, improve file selection and staged localization before adding more open-ended reasoning. A stronger repair model cannot act on evidence the retrieval path excluded.

\Needspace{5\baselineskip}

Next, decide whether repeated structural questions justify a typed index. Before approving the build:

\begin{itemize}
\tightlist
\item
  name the owner;
\item
  define the supported languages and relationship types;
\item
  specify how repository correspondence will be maintained;
\item
  cap the candidates the index can send downstream;
\item
  define its publication and rollback process; and
\item
  record how freshness failures will be detected.
\end{itemize}

When no one owns that lifecycle, choose the cheaper architecture explicitly: a generated repository map near the front of context, together with exact search, definition lookup, reference search, and bounded caller navigation.

Every index, snapshot, and cache should carry a commit or content identity. Compare that identity with the worker's current state at retrieval time. Record mismatches as an operating metric, and refuse the result or fall back to current navigation when the identities differ.

A freshness refusal is visible and recoverable. A stale result can silently become the basis of a plausible patch.

Chapter 14 turns from which evidence the system retrieves to how much of that evidence the model can use.

\section*{Sources and evidence}

\textbf{Stage localization before repair}

\begin{itemize}
\tightlist
\item
  Strong evidence: Xia, C. S., et al.~(2024), ``Agentless: Demystifying LLM-based Software Engineering Agents,'' arXiv:2407.01489.
\item
  Directional evidence: Chang, J., et al.~(2025), ``Bridging Bug Localization and Issue Fixing: A Hierarchical Localization Framework Leveraging Large Language Models,'' arXiv:2502.15292, which names its framework BugCerberus. (Per-hierarchy-level specialization.)
\item
  Strong evidence: Sepidband, M., Viet Pham, H., Hemmati, H. (2026), ``On the Role of Fault Localization Context for LLM-Based Program Repair,'' arXiv:2604.05481.
\item
  Corroboration: none on record.
\end{itemize}

\textbf{Index the repository as a knowledge graph}

\begin{itemize}
\tightlist
\item
  Strong evidence: Chen, Z., et al.~(2025), ``LocAgent: Graph-Guided LLM Agents for Code Localization,'' arXiv:2503.09089.
\item
  Directional evidence: Yang, B., et al.~(2025), ``Enhancing Repository-Level Software Repair via Repository-Aware Knowledge Graphs'' (KGCompass), arXiv:2503.21710. (The 69.7 percent multi-hop figure is carried from v1; later versions report different headline values, so the inline link pins v1.)
\item
  Strong evidence: Ma, Y., et al.~(2024), ``Alibaba LingmaAgent: Improving Automated Issue Resolution via Comprehensive Repository Exploration,'' arXiv:2406.01422.
\item
  Directional evidence: Liu, X., et al.~(2024), ``CodexGraph: Bridging Large Language Models and Code Repositories via Code Graph Databases,'' ICLR 2025, arXiv:2408.03910.
\item
  Strong evidence: Ouyang, S., et al.~(2024), ``RepoGraph: Enhancing AI Software Engineering with Repository-level Code Graph,'' arXiv:2410.14684.
\item
  Strong evidence: Liu, J., et al.~(2026), ``AOCI: Symbolic-Semantic Indexing for Practical Repository-Scale Code Understanding with LLMs,'' arXiv:2605.02421.
\item
  Directional evidence: Tao, W., et al.~(2024), ``MAGIS: LLM-Based Multi-Agent Framework for GitHub Issue Resolution,'' arXiv:2403.17927.
\item
  Corroborating case: Potvin, R., Levenberg, J. (2016), ``Why Google Stores Billions of Lines of Code in a Single Repository,'' \emph{Communications of the ACM} 59(7), \href{https://doi.org/10.1145/2854146}{DOI:10.1145/2854146}. Documents the scale at which one organization's repository required purpose-built navigation and dependency tooling. A single-company practice report predating coding agents, with no comparison against alternatives.
\end{itemize}

\textbf{Gate retrieval on freshness}

\begin{itemize}
\tightlist
\item
  Directional evidence: Weng, H., et al.~(2026), ``When Retrieval Hurts Code Completion: A Diagnostic Study of Stale Repository Context,'' arXiv:2605.14478.
\end{itemize}

\chapter{Usable context budgets, consolidated-spec restarts, and file-based tool output}
\label{ch14-usable-context-budgets-spec-restarts-file-output}
\begin{quote}
\textbf{Evidence profile.} 6 strong \(\cdot\) 2 directional \(\cdot\) 0 corroborating \(\cdot\) 2 null or conflicting evidence items across 4 developed practices (\erca{072}, \erca{088}, \erca{143}, \erca{178}).

\textbf{Chapter claim.} Advertised capacity is not usable context.
\end{quote}

Guidance I had placed in a system prompt remained inside the context window twenty turns later but appeared nowhere in the model's stated reasoning. Across 1,705 visible thinking blocks in 199 traces from my trace-diagnostics corpus, that instruction appeared zero times.

Visible reasoning is an incomplete observation of model state, so absence from those blocks does not prove that the instruction had no influence. It does show that material can remain inside the context window without remaining visibly active in the run. Benchmark evidence exposes the same gap through measured performance.

Only about half of the 17 models evaluated by Hsieh et al.~(\href{https://arxiv.org/abs/2404.06654}{2024}) maintained satisfactory performance at 32,000 tokens, even though every model advertised a context window at least that large. Their RULER benchmark combined 13 task types, including retrieval and aggregation.

Rando et al.~(\href{https://arxiv.org/abs/2505.07897}{2025}) found a sharper decline on realistic repository issues as supplied context increased. On LongSWE-Bench, Claude 3.5 Sonnet fell from 29 percent at 32,000 tokens to 3 percent at 256,000 tokens, the model's maximum context in that study. On the separate LongCodeQA task, Qwen2.5-14B-Instruct scored 70.2 percent at 512,000 tokens and 40.0 percent at one million.

The measurement problem is clearer than the prescription. I ranked ten candidate context-assembly practices three times: by how much each would change an engineering decision, how much of the problem it covered, and how teachable it was. The three rankings agreed on only one practice. Five practices appeared in exactly one ranking, and three appeared in none. I take that spread as evidence that the field has not converged. Every recommendation in this chapter is therefore either measured locally or identified as a convention.

The advertised window is a capacity limit. The usable window is a property measured on the work the model must perform.

Once that working limit is known, three assembly decisions follow. A run that has lost its governing specification should restart from a consolidated statement of the settled requirements. Large tool outputs should remain outside the active window in retrievable artifacts. A standing repository context file should be compared with no standing file before its permanent token cost is accepted.

\hypertarget{measure-the-context-that-remains-useful}{%
\section{Measure the context that remains useful}\label{measure-the-context-that-remains-useful}}

\textbf{Effective context length} here means the largest assembled context at which a model still meets a specified reliability threshold on a defined workload. The value belongs to a model, configuration, workload, and engineering threshold. A model may accept 128,000 tokens and retrieve one fact at that length while having a much shorter effective context for a code change that depends on relationships among several files.

Measurement begins with task shape. A production coding agent may need to retrieve a declaration from one file, connect it to a caller in another, incorporate a test failure from a third, and preserve a constraint supplied near the beginning of the run. A single hidden fact embedded in filler tests only one part of that work.

\Needspace{5\baselineskip}

The qualification workload should represent the operations the deployed system actually performs, including:

\begin{itemize}
\tightlist
\item
  retrieval across several documents;
\item
  aggregation across several pieces of evidence;
\item
  retention of early instructions;
\item
  reasoning over long code; and
\item
  tool-using trajectories whose context grows over time.
\end{itemize}

The benchmark record shows why these distinctions matter. RULER found that advertised capacity often exceeded the length at which performance remained satisfactory. Leng et al.~(\href{https://arxiv.org/abs/2411.03538}{2024}) evaluated 20 models from 2,000 to 128,000 tokens on retrieval-augmented generation. Accuracy saturated at a model-specific context size, around 64,000 tokens for most frontier models in that study. Beyond that point, model families exhibited different combinations of refusals, repetition, and neglected instructions.

Code work depends more heavily on relationships among units than simple fact retrieval does. Li et al.~(\href{https://arxiv.org/abs/2503.04359}{2025}) introduced LONGCODEU, covering nine models and eight long-code understanding tasks, and found substantial degradation beyond 32,000 tokens despite advertised windows of 128,000 to one million tokens. Understanding relationships among code units was the weakest area.

LongCodeBench evaluated issue answering and bug fixing rather than document retrieval. Its model-task pairs also covered different maximum lengths, so the reported declines belong to those pairs and should not be compared as one curve.

These studies do not identify a universal ceiling. Their results depend on the model generation, prompts, tasks, and scoring rules. The durable finding is the repeated gap between capacity and competence, together with a protocol for measuring it. The knee can move with a new model and may differ between retrieval and code modification even within the same release.

Each study finds degradation within the tested or advertised window. Differences in tasks and metrics prevent direct comparison; the practical response is the local context-size sweep below.

Qualify a model using context-size strata around the range you might deploy. A first pass can include a small condition, two intermediate conditions, and one near the advertised limit. The exact token values depend on the model and workload.

Each stratum should contain the same task families and, when possible, the same underlying tasks with controlled additions to context. This separates the effect of context length from a change in task difficulty.

For repository work, each item begins from a fixed repository state and a known evidence set. The smallest arm receives the minimum material needed to solve the task. Larger arms add realistic neighboring files, test output, prior discussion, generated summaries, or tool schemas in a recorded order.

The model receives the same task, permissions, and tool surface in every arm. Pin the model version, decoding settings, harness version, prompts, and evaluator so that context size remains the deliberate change.

Each stratum is a measured comparison, so Chapter 1's requirements apply. Tasks should be paired across context sizes wherever possible, and every stratum needs repeated independent runs. One run at each length cannot distinguish a context-length effect from ordinary run-to-run variation.

\Needspace{5\baselineskip}

Four outcomes should remain separate:

\begin{itemize}
\tightlist
\item
  \textbf{Retrieval accuracy} records whether the required evidence was found.
\item
  \textbf{Task correctness} records whether the answer or code change was right.
\item
  \textbf{Reliability} records how often the result recurred across repeated runs.
\item
  \textbf{Cost} records tokens, model calls, and tool use.
\end{itemize}

Latency and usability may matter operationally, but neither can substitute for correctness. A longer prompt can reduce retrieval calls while lowering the probability of a correct patch.

Plot outcome distributions by context stratum, and reject thresholds based on a single passing run. The \textbf{saturation point} is the region where adding context stops improving the relevant outcome. The operational knee may occur earlier, where the marginal gain no longer justifies added cost or variance.

A decline may be gradual, abrupt, or isolated to one task family. Those shapes support different operating limits.

Suppose retrieval accuracy improves through the intermediate conditions and then flattens, while code-change success begins declining earlier. Retrieval alone would permit a larger budget. The coding workload would not. Set the production limit from the weakest task that must remain reliable, then place retrieval and harness caps below its measured knee.

The margin accounts for token-estimation error, task variation, and fixed context that accumulates during execution. Its size follows consequence. A workflow with cheap retries and deterministic verification may operate close to the knee. A production change with costly failure and weak verification requires more distance.

The final record should state the threshold, uncertainty, and consequence that justified the margin so the decision can be revisited when the system changes.

Several mechanisms can produce decline. Additional context introduces competing versions of facts, similarly named symbols, and relationships the model must reconstruct. Position affects whether early instructions and evidence influence later actions. Repetition can induce repetition, while conflicting material can cause refusal or stale instruction following. The model may also spend more of its output budget restating or reconciling context rather than acting.

\Needspace{5\baselineskip}

Token count alone therefore does not diagnose the failure. Retain traces from every stratum and annotate the first consequential divergence. Useful classes include:

\begin{itemize}
\tightlist
\item
  retrieval failure;
\item
  incorrect symbol relationships;
\item
  forgotten constraints;
\item
  repetition;
\item
  stale or conflicting evidence; and
\item
  harness truncation.
\end{itemize}

The curve identifies where performance weakens. The trace identifies what to change.

This distinction prevents an apparatus failure from being mistaken for a model limit. A harness may silently clip messages at its own cap. An evaluator that sees only the final answer may classify an instruction-following failure as missing knowledge. Larger contexts may also contain more stale repository material, confounding length with freshness.

Preserve the exact assembled prompt, record token allocation by source, and retain every truncation or omission decision with the run.

Context also grows during tool use. A starting prompt well below the limit may cross the measured knee after searches, test logs, edits, and correction attempts accumulate. Qualification should therefore include both static prompts and representative trajectories, or replays of observed context growth.

Measure the initial assembly and the maximum context reached before compaction or restart. A limit applied only to retrieval at time zero does not govern later accumulation.

Every consequential change to the model, workload, or harness requires requalification. A new model may move the knee or change the dominant failure mode. A harness may add standing instructions that consume the safety margin before useful work begins. Preserve the old strata, add cases shaped by the change, and compare the curves.

An unchanged average can conceal a shorter effective context offset by stronger short-context performance. When noise obscures the knee, Chapter 1's power analysis determines whether to add tasks or narrow the claim.

Synthetic probes remain useful because they isolate positional recall, aggregation, and distractor effects. Their thresholds transfer only approximately to repository work. Realistic tasks entangle retrieval, reasoning, tools, and verification, making mechanism attribution harder but deployment relevance greater.

Synthetic probes explain a failure; repository-shaped trials set the operating limit. Neither establishes the other's threshold without measurement.

\Needspace{5\baselineskip}

The final configuration should specify:

\begin{itemize}
\tightlist
\item
  the maximum initial assembly;
\item
  per-source retrieval limits;
\item
  a reserve for tool output and correction;
\item
  the action taken when that reserve is exhausted; and
\item
  the workload and model version that justified each value.
\end{itemize}

A run can remain inside this measured budget and still fail when its specification has become fragmented or superseded.

\hypertarget{restart-from-the-current-specification}{%
\section{Restart from the current specification}\label{restart-from-the-current-specification}}

A model reads the first version of a requirement, resolves an ambiguity, and begins implementing that interpretation. Three turns later, the user corrects the assumption. The correction appears in the transcript, but the plan, file selection, and partial implementation still encode the earlier choice. Each later instruction asks the model to repair work whose structure continues reproducing the misunderstanding.

This can look like stubbornness or weak reasoning. A more useful explanation is commitment under an incrementally revealed specification.

The current task no longer exists in one authoritative place. It is distributed across the initial request, later qualifications, rejected proposals, tool observations, and the model's own accounts of what was decided. Before choosing another action, the system must distinguish accepted decisions from abandoned ones.

Laban et al.~(\href{https://arxiv.org/abs/2505.06120}{2025}) simulated more than 200,000 conversations across 15 models and compared instructions delivered together with instructions spread across several turns. Sharding the instructions produced an average performance decline of 39 percent relative to the single-turn condition. Models showed only slightly lower aptitude but much greater unreliability. They made early assumptions, committed to them, and often failed to recover when later turns supplied missing constraints.

The study used simulated conversational generation tasks. It did not evaluate coding agents operating tools, receiving compiler feedback, or modifying persistent files. Tests and environment errors may correct some assumptions while partial implementations entrench others. The 39 percent result is evidence for the commitment mechanism, not an effect-size estimate for coding work.

The value is also an average across models and tasks rather than a per-conversation quantity. It could represent a few catastrophic failures or smaller degradation across many conversations. Those distributions would justify different interventions.

The trace corpus described at the beginning of this chapter illustrates a related problem without strengthening the causal evidence. Guidance delivered once, twenty turns earlier, remained inside the window but appeared in none of the visible reasoning. Initial delivery is not evidence that a long-running agent still governs its work by the same instruction.

The operational response is consolidation. When requirements emerge through discussion, rewrite the accepted decisions, constraints, definitions, interfaces, and unresolved questions into one coherent specification. The document identifies itself as authoritative and explicitly supersedes earlier proposals on the same subjects.

It represents the current agreement rather than the chronology through which that agreement emerged.

\Needspace{5\baselineskip}

Consolidation must remain selective because a transcript contains several kinds of state:

\begin{itemize}
\tightlist
\item
  Settled requirements belong in the specification.
\item
  Open questions remain explicitly unresolved.
\item
  Test failures, command output, changed files, and deployment state belong in an execution record.
\item
  Rationale stays with a decision when removing it would invite accidental reversal.
\end{itemize}

The distinction between specification and observation determines what a restart may discard.

Suppose a run has established that an API must remain backward compatible, selected a migration sequence, edited several files, and discovered that a test fixture relies on an undocumented field. The compatibility requirement and migration sequence belong in the consolidated specification. The changed-file list, diff, test command, failure output, and fixture discovery belong in the execution handoff.

Combining both into undifferentiated prose makes it difficult to tell which statements govern future work and which report what happened.

\Needspace{5\baselineskip}

Restart when the run's behavior shows that appended corrections no longer alter its governing interpretation. Indicators include:

\begin{itemize}
\tightlist
\item
  repeatedly proposing an interface that was already rejected;
\item
  continuing to edit files selected under an obsolete plan;
\item
  explaining a current failure through a superseded assumption; or
\item
  spending several turns reconciling conflicting summaries.
\end{itemize}

A growing token count may increase the risk, but it does not determine the boundary. A short run can become incoherent after one consequential misunderstanding. A long run can remain stable when its decisions are consistently restated.

The restart package contains three artifacts.

First, the \textbf{consolidated specification} contains settled requirements and genuinely unresolved questions.

Second, the \textbf{execution handoff} records the repository state, completed changes, current failures, verification already performed, and raw artifacts available for inspection.

Third, the \textbf{starting instruction} identifies the authoritative sources and states the next action.

The package should be built from inspectable state. Use the current diff, test results, issue record, and relevant files as evidence. When a prior claim conflicts with the repository, the repository state governs, and the discrepancy remains visible. A consolidation built from an inaccurate summary produces a cleaner version of the same error.

Restarting has costs. It loses conversational nuance, discarded alternatives, and tacit knowledge about how the work reached its current form. It may repeat exploration or freeze a provisional design too early.

Consolidate only settled decisions. When two designs remain plausible, the specification records both, the evidence supporting each, and the unresolved choice. Consolidation becomes harmful when it silently turns ambiguity into authority.

Environment feedback also resists conversion into static requirements. A failed test may be obsolete, flaky, misconfigured, or decisive. A compiler error is an observation produced against one repository state. The restarted run needs the raw output or a retrievable reference, the command that produced it, and the state against which it ran.

A polished sentence may erase the uncertainty the next investigation needs to resolve.

Versioning prevents authoritative specifications from competing. Every consolidation receives an identity, names the prior version it supersedes, and remains visible to the run. Concurrent workers submit changes through one owner or merge process because requirements that govern all workers require an ordering rule.

A failing run should not decide alone that its interpretation is accurate enough to preserve. The restart boundary belongs to the control plane and should incorporate the trace, repository state, verification results, and user decisions. The model may draft the consolidation, but an inspectable comparison should show what it retained, changed, omitted, or marked unresolved.

High-consequence requirements warrant human review before the new run treats the specification as authoritative.

The choice is between continuing from entangled state and paying to reconstruct clean state. Appending another correction is inexpensive when the run has incorporated earlier corrections and its working interpretation remains coherent. Restarting becomes cheaper when every new turn must fight decisions embedded in the existing trajectory.

The 39 percent result helps explain why repeated correction messages can fail even when a model can solve the same task from a complete specification. When that pattern appears, restart from a consolidated statement of the settled requirements and carry execution evidence separately as raw observations rather than conclusions.

\hypertarget{move-bulk-output-out-of-the-active-context}{%
\section{Move bulk output out of the active context}\label{move-bulk-output-out-of-the-active-context}}

In my agent fleet, a snapshot of the shared skill catalog once traveled between sessions through an environment variable. As sessions inherited, transformed, and re-emitted it, the snapshot drifted. Moving the same context surface into a fingerprinted file made each session's copy atomic and allowed every reload to validate its identity.

The incident illustrates how a named artifact can make ownership, validation, and retrieval explicit. Model-performance effects require a separate comparison.

The evidence for file-backed tool output is limited to two directional practitioner accounts and no strong studies. It counts as a convention because it is a simple way to control raw-output growth. Long terminal sessions, search results, test logs, tool responses, and pre-compaction history move into session-scoped artifacts. The active context retains a concise description, a stable pointer, and enough metadata for the agent to decide whether loading more is worth the cost.

This changes where state lives. Without externalization, a tool response exists only in the conversation history. Truncation can remove it, while summarization replaces it with an interpretation produced before anyone knows which details will later matter.

\Needspace{5\baselineskip}

With an external artifact, the raw output belongs to the run rather than to the current context window. The active context can retain an index entry containing:

\begin{itemize}
\tightlist
\item
  the command or tool call;
\item
  the time and completion status;
\item
  the byte or token size;
\item
  a content fingerprint;
\item
  the artifact location; and
\item
  whether the content is complete or truncated.
\end{itemize}

Retrieval then becomes incremental. An agent can inspect the final lines of a failed build, search for an error code, or load a bounded range around a match. After a context reset, the replacement run can query the original output instead of relying on a lossy summary. Bulk evidence remains available without occupying the working window on every turn.

The pointer should expose the cost before loading the content. My memory system follows this pattern. An index operation reports the expected token cost, and a separate content operation loads the requested material. Injected lessons can cite their evidence while leaving the raw trace unloaded. The agent may choose a short excerpt when the complete artifact would consume too much of the working window.

Cursor's engineering blog (\href{https://cursor.com/blog/dynamic-context-discovery}{2026}) reports writing long tool output to files that the agent can read back, reducing forced summaries near the context limit. The Hightouch team, in an interview published by Amplify Partners (\href{https://www.amplifypartners.com/blog-posts/how-hightouch-built-their-long-running-agent-harness}{2026}), describes buffering large tool results similarly and reports that the design kept the active window focused on reasoning rather than raw output.

Hightouch also reported that allowing the model to choose when to buffer performed better for its system than a coded decision tree. These accounts provide implementation direction rather than a controlled comparison. Model-selected buffering remains probabilistic, and another model or prompt may make worse choices.

\Needspace{5\baselineskip}

A safe interface therefore places deterministic controls around that choice. The harness owns:

\begin{itemize}
\tightlist
\item
  maximum inline output size;
\item
  artifact-write error handling;
\item
  retention limits;
\item
  access controls;
\item
  permitted read ranges; and
\item
  the behavior when preservation fails.
\end{itemize}

The model may choose among tail, search, range read, and full load within those limits. A failed write returns an explicit failure. The system must not claim that output was preserved when no artifact exists.

Pointers must remain valid across restarts. A path tied only to one temporary process may disappear before a replacement worker can use it. Associate each artifact with a run identity and tool-call identity, preserve ordering, and record completeness. Concurrent calls write distinct artifacts and publish their pointers atomically so one worker cannot overwrite another's evidence.

Security policy follows the data. Terminal output may contain credentials, proprietary source, personal information, or production identifiers. Removing it from the prompt may reduce incidental exposure during later turns, but it also creates a retained artifact that requires permissions, redaction, encryption where appropriate, and a deletion policy.

A pointer must not allow the model to read outside the session's authorized scope. Retention and pointer validity are one decision: deleting an artifact also invalidates every pointer to it in an older transcript.

Files are an implementation choice rather than the essential abstraction. An object store, content-addressed artifact service, or trace database may offer stronger retention and concurrency guarantees. Files are useful early because existing tools can search and range-read them, their failure modes are visible, and they avoid committing to a more elaborate service before the access patterns are understood.

The durable requirement is retrievable raw evidence behind a stable, costed pointer.

Externalization addresses only one source of context growth. Plans, speculative explanations, corrections, and abandoned approaches still accumulate in the active window. A run can write every log to disk and still lose its specification. Compaction and restart from a consolidated specification remain separate controls.

The convention also introduces retrieval failures. An agent may never follow the pointer, search for the wrong term, or load an excerpt that omits the decisive line. An index may point to deleted content. A summary may misdescribe an artifact and discourage inspection.

Retain the raw output, validate pointer resolution, expose search and range operations, and record which portions the agent actually loaded.

\Needspace{5\baselineskip}

The first tests are operational:

\begin{enumerate}
\def\labelenumi{\arabic{enumi}.}
\tightlist
\item
  Can a replacement run recover a required detail after context reset?
\item
  Do concurrent calls preserve separate outputs?
\item
  Does every excerpt resolve to its original artifact?
\item
  Are truncation and failed writes visible?
\item
  Do expired artifacts invalidate their pointers explicitly?
\end{enumerate}

Only after those mechanics work does task-level evaluation ask whether the convention lowers cost or improves completion on workloads with large outputs.

That comparison inherits the resolution problem from Chapter 1. A difference smaller than run-to-run variation can still be measured, but doing so requires enough paired runs. Size the evaluation around the smallest cost or completion change that would alter the engineering decision, and interpret the interval around the paired difference rather than one pair of totals.

The current justification is inspectable preservation and ownership. Improvement in task outcomes remains unmeasured.

\hypertarget{make-standing-context-justify-its-permanent-cost}{%
\section{Make standing context justify its permanent cost}\label{make-standing-context-justify-its-permanent-cost}}

Gloaguen et al.~(\href{https://arxiv.org/abs/2602.11988}{2026}) tested whether repository context files already used by coding teams improved task success. Across the evaluated agents and models, they did not. The files increased inference cost by roughly 20 percent, even though the agents followed their instructions.

This is a null result from one study. It should reverse the assumption that a standing context file is beneficial by default, without becoming a universal verdict against such files.

Khatri (\href{https://arxiv.org/abs/2607.27250}{2026}) ran a second controlled ablation across two agents, 17 real tasks from three repositories, and 288 evaluated runs. Equivalence tests bounded any correctness effect to no more than roughly 10 to 15 percentage points and found no measurable improvement. This null result strengthens the case for a no-file baseline while remaining limited by the agents, repositories, tasks, and equivalence bounds studied.

Instruction adherence and task completion measure different effects. A file can cause an agent to run a command, follow a naming rule, or avoid a directory while leaving the final task score unchanged. It can also consume context, provoke additional tool calls, or direct attention toward an overview the model did not need.

Evidence that the file was read and followed establishes treatment delivery; task and cost outcomes determine benefit.

The null result must be interpreted through the power analysis from Chapter 1. Failure to detect improvement may mean that the average effect is small, that positive and negative effects cancel across the task mix, or that the experiment could not resolve the effect it sought.

The measured cost remains part of the result, as does the evidence that the files changed agent behavior. Repositories, instruction categories, models, and safety outcomes may produce a different balance.

The local decision requires a no-file baseline. The same representative tasks and repository states run under the same model, prompts, tools, harness, and evaluator with and without the standing file. The comparison is paired by task, and repeated runs remain necessary when model variation is material.

\Needspace{5\baselineskip}

Before executing the comparison, record:

\begin{itemize}
\tightlist
\item
  the primary task-success metric;
\item
  cost measures;
\item
  model and harness versions;
\item
  decoding settings;
\item
  the context-file revision;
\item
  the task-set version; and
\item
  the decision threshold.
\end{itemize}

This prevents an unfavorable result from being rescued after the fact by whichever secondary measure happened to move. Diagnostic measures such as instruction adherence, tool calls, latency, and failure class remain useful for explaining the primary result.

Inference cost includes more than the file's input tokens. An instruction may trigger additional searches, builds, reviews, or explanations. A concise repository-specific command may instead prevent failed exploration and lower total cost despite adding permanent context.

Record input and output tokens, model calls, tool execution, and cost per completed task. File length alone is incomplete. A reported 20 percent aggregate difference is also one estimate and deserves the same repeated measurement as a completion score.

\Needspace{5\baselineskip}

The task sample must exercise the claims the file makes. A suite of isolated function edits says little about migration rules, generated code, deployment checks, or concurrent changes. Include:

\begin{itemize}
\tightlist
\item
  ordinary repository work;
\item
  tasks that encounter unusual local constraints; and
\item
  tasks where violating a rule would produce a consequential failure.
\end{itemize}

Expected use determines the mix.

An all-or-nothing comparison may hide which content helps or harms. When the task budget permits, ablate instruction categories separately. One arm may retain only build and test commands, another only failure-prevention rules, and another the architectural overview.

The categories can interact, so the decomposition is imperfect. It is still more informative than concluding that the entire file either works or does not.

The Gloaguen et al.~result gives narrow failure-prevention rules a plausible role because agents followed the instructions even when aggregate task success did not improve. Permanent inclusion still requires a rule-specific outcome and cost test.

A rule such as ``do not regenerate this checked-in directory manually'' names a concrete failure and can be tested on tasks that encounter that directory. A page describing the repository architecture may duplicate information that search can retrieve only when needed.

Permanent inclusion therefore deserves a high bar. The default file should focus on unusual constraints and failure-prevention rules that a capable agent is likely to miss and cannot cheaply discover at the moment of need. Common language conventions, framework summaries, and directory tours usually belong behind on-demand retrieval instead.

One of my repositories uses its context file as a failure-mode ledger. Each prohibition names the incident class it prevents. That design connects a rule to a repository-specific need and allows its removal when the hazard disappears. This author-system example illustrates maintenance practice but provides no measured completion benefit.

Actual context files are broader. Chatlatanagulchai et al.~(\href{https://arxiv.org/abs/2511.12884}{2025}) analyzed 2,303 files from 1,925 repositories. Implementation details appeared in 69.9 percent, architecture information in 67.7 percent, and build or execution guidance in 62.3 percent. Security and performance guidance each appeared in 14.5 percent.

Those percentages describe what maintainers wrote, not what improved outcomes. The relative absence of security and performance guidance is a useful review prompt because coding agents can execute commands and modify production-relevant code; it is not a reason to add generic text without a concrete local failure mode.

\Needspace{5\baselineskip}

A file retained after evaluation should be maintained like configuration. Every instruction needs:

\begin{itemize}
\tightlist
\item
  an owner;
\item
  a reason for existing;
\item
  a failure or outcome it is meant to affect;
\item
  a review trigger; and
\item
  a deletion condition.
\end{itemize}

Changes should arrive as small diffs so reviewers can see which behavior is intended to change. Large generated rewrites make stale claims and contradictions harder to detect.

Chatlatanagulchai et al.~also found that context files often evolve through frequent small additions. That observation is compatible with configuration-like maintenance but says nothing about the quality of the changes. A new line may document a real failure, duplicate an existing rule, or preserve an obsolete workaround. Change frequency is not correctness.

Staleness is especially dangerous because the file is authoritative by placement. An obsolete build command can cause the agent to modify generated output or skip a required check. An architectural overview decays as files move. A prohibition tied to a removed failure mode can continue shaping every run after its justification disappears.

I have seen this drift in my own agent-facing documents. A review-checker file contained 35 rules while its header and wrapping skill still claimed 29. A skill-library README contained three incompatible counts. Neither document had a mechanism capable of detecting the discrepancy.

These cases do not measure task impact. They show that documents consumed by agents can become internally inconsistent when their claims are not checked.

Some claims are mechanically verifiable. Counts can be recomputed, paths validated, commands exercised in a clean environment, and generated sections tied to source fingerprints. Another of my knowledge maps stores a source hash so that changes to the material it describes make drift observable.

Semantic claims still require review, but deterministic checks reduce the surface that depends on human memory.

Ownership should be specific enough to trigger review when the build system, repository structure, release process, or protected failure mode changes. Context assembly should also preserve identity and precedence for requirements repeated across system prompts, repository files, task descriptions, and tool documentation. A pre-run check can detect contradictory instructions before the agent must decide which source governs.

The comparison with no file should be interpreted by instruction category and failure consequence as well as by aggregate completion. A file with no measurable mean benefit may still prevent a rare destructive event that the task suite lacks power to observe. That claim needs a specific threat model, a compliance test, and an explicit cost decision.

A general completion benchmark cannot validate or dismiss safety value when its scoring rule omits the relevant failure.

Correctness, reliability, cost, and usability remain separate columns in the decision record. Retain the no-file condition after the first evaluation, because a new model, harness change, or substantial file revision may change both adherence and cost. The earlier result belongs to the configuration that produced it.

When the comparison is inconclusive, adherence cannot substitute for the missing outcome. Report the interval the evaluation supports, inspect whether the sample exercised the file's claims, and decide whether a larger test is justified.

A file may remain provisionally for a concrete safety reason, but the exception and its permanent cost should be explicit. An underpowered null does not prove uselessness, just as instruction compliance does not prove benefit.

The best standing file will often become shorter after evaluation. Rules without a concrete purpose, duplicated repository descriptions, stale counts, and material better served through on-demand retrieval can be removed.

What remains is a reviewed configuration surface whose permanent token cost is tied to an observed failure or measured outcome. That is narrower than claiming that repository context files help, and it is a claim the repository can test.

\hypertarget{set-the-context-operating-limits}{%
\section{Set the context operating limits}\label{set-the-context-operating-limits}}

The companion catalog contains narrower techniques for systems that need them. It covers placing load-bearing evidence near a context edge, testing semantic rather than verbatim recall, compacting at task milestones, pricing compression by completed work rather than token reduction, minimizing always-loaded context, and auditing the assembled prompt as an inspectable artifact.

Each technique changes one part of the context path. None removes the need to measure the complete path on the operated workload.

Begin with one production-shaped task family and run it across several context sizes. Include relevant multi-document evidence, realistic distractors, and the relationships the agent must reconstruct. Measure correctness, reliability, and cost separately. Inspect the traces around the decline and place the harness limit below the measured knee for the weakest task that must remain reliable.

Record the model, workload, margin, and changes that require requalification.

\Needspace{5\baselineskip}

Next, identify one long-running task whose governing specification has become fragmented. Stop appending correction messages. Consolidate the accepted decisions, constraints, and unresolved questions into one authoritative specification. Inspect the repository state and verification evidence, then restart from a package containing:

\begin{itemize}
\tightlist
\item
  the consolidated specification;
\item
  an execution handoff containing raw state and observations; and
\item
  a clear next action.
\end{itemize}

The new run should be able to distinguish what it must obey from what it must investigate.

Move large tool outputs into session-scoped artifacts and retain costed pointers in the active context. Verify recovery after reset, separate preservation of concurrent calls, explicit handling of failed writes, and traceability from every excerpt to its raw source. Treat this as an operational convention until a local paired comparison demonstrates a cost or completion effect.

Finally, compare the repository context file with no standing file on the same tasks and repository states. Record task success, reliability, instruction adherence, total inference and tool cost, and the exact file revision. The roughly 20 percent increase reported in one study is a reason to measure, not a transferable estimate.

The paired comparison determines whether the file justifies its permanent cost. Category-level ablations determine which instructions remain.

\Needspace{5\baselineskip}

Together, these controls establish:

\begin{itemize}
\tightlist
\item
  how much context the system admits;
\item
  when a run restarts from a clean specification;
\item
  where bulk evidence lives; and
\item
  which instructions deserve permanent residence in every run.
\end{itemize}

Chapter 15 turns to what survives between sessions and in what form.

\section*{Sources and evidence}

\textbf{Budget to measured effective context}

\begin{itemize}
\tightlist
\item
  Strong evidence: Hsieh, C.-P., et al.~(2024), ``RULER: What's the Real Context Size of Your Long-Context Language Models?'' COLM 2024, arXiv:2404.06654.
\item
  Strong evidence: Leng, Q., et al.~(2024), ``Long Context RAG Performance of Large Language Models,'' Databricks Mosaic Research, arXiv:2411.03538.
\item
  Strong evidence: Li, J., et al.~(2025), ``LONGCODEU: Benchmarking Long-Context Language Models on Long Code Understanding,'' arXiv:2503.04359.
\item
  Strong evidence: Rando, S., et al.~(2025), ``LongCodeBench: Evaluating Coding LLMs at 1M Context Windows,'' arXiv:2505.07897.
\item
  Corroboration: none on record.
\end{itemize}

\textbf{Consolidate the specification and restart lost runs}

\begin{itemize}
\tightlist
\item
  Strong evidence: Laban, P., et al.~(2025), ``LLMs Get Lost In Multi-Turn Conversation,'' arXiv:2505.06120.
\item
  Corroboration, narrative only: the author's trace-diagnostics corpus showed that guidance delivered in the system prompt twenty turns earlier was absent from the visible reasoning: 0\% across 1,705 thinking blocks in 199 traces. The literature result, not this case, carries the claim.
\end{itemize}

\textbf{Persist transient context as files}

\begin{itemize}
\tightlist
\item
  Directional evidence: \href{https://cursor.com/blog/dynamic-context-discovery}{``Dynamic Context Discovery in Coding Agents''}, Cursor engineering blog, 2026-01-07.
\item
  Directional evidence: \href{https://www.amplifypartners.com/blog-posts/how-hightouch-built-their-long-running-agent-harness}{``How Hightouch built their long-running agent harness''}, Amplify Partners engineering interview, 2026-01-20.
\item
  Corroboration, narrative only: the author's memory system exposes token cost before loading content, and the author's agent fleet stopped a drifting skill-catalog snapshot by moving it to a fingerprinted, session-atomic file validated on reload.
\end{itemize}

\textbf{Measure context files and maintain them like configuration}

\begin{itemize}
\tightlist
\item
  Null or conflicting result: Gloaguen, T., Mündler, N., Müller, M., Raychev, V., Vechev, M. (2026), ``Evaluating AGENTS.md: Are Repository-Level Context Files Helpful for Coding Agents?'' arXiv:2602.11988.
\item
  Null or conflicting result: Khatri, P. (2026), ``Do Context Files Help Coding Agents? A Two-Agent Ablation Study on Real Repositories,'' arXiv:2607.27250. Across 288 runs, equivalence tests bounded correctness effects to no more than roughly 10 to 15 percentage points in the studied conditions.
\item
  Strong evidence: Chatlatanagulchai, W., et al.~(2025), ``Agent READMEs: An Empirical Study of Context Files for Agentic Coding,'' arXiv:2511.12884.
\item
  Corroboration, narrative only: the author's systems include a context file maintained as a failure-mode ledger and a knowledge map with a machine-checkable source hash; contrary cases include a review-checker whose live rule count disagreed with its header and wrapper, and a skill-library README with three inconsistent counts.
\end{itemize}

\chapter{Cross-session memory, raw traces, and compaction policies}
\label{ch15-cross-session-memory-raw-traces-compaction}
\begin{quote}
\textbf{Evidence profile.} 2 strong \(\cdot\) 4 directional \(\cdot\) 1 corroborating evidence items across 3 developed practices (\erca{081}, \erca{116}, \erca{118}).

\textbf{Chapter claim.} Preserve raw events; rebuild derived memory.
\end{quote}

One of my memory systems rebuilds its entire derived layer whenever the schema version changes. Nothing in that layer is migrated in place. Three append-only tables containing lessons, memory events, and provenance events cross the rebuild boundary through mechanical export and import. Everything else is regenerated from the work record that produced it.

The arrangement is deliberately inconvenient. I have no comparative evidence that it outperforms a system that rewrites its memory incrementally.

The reason for accepting that inconvenience has been measured elsewhere. Zhang (\href{https://arxiv.org/abs/2605.12978}{2026}) studied memories that language models update continuously and found that they degrade. A production account of Slack's context management describes the same general pattern. The Agentic Context Engineering synthesis (\href{https://arxiv.org/abs/2510.04618}{2025}) names two degradation modes: context collapse and brevity bias.

Each rewriting pass optimizes a representation already produced by earlier optimization. Deletion and distortion can therefore accumulate without any explicit destructive operation. Routine maintenance gradually damages the evidence.

The complete chronological record of a session's inputs, actions, outputs, and state changes is the source record for cross-session memory, the raw, or episodic, trace. Every representation derived from that trace must remain rebuildable. Observed query and task failures, rather than anticipated ones, determine when the storage and retrieval layers deserve additional complexity.

Seven items support the chapter's three design entries; two are strong, four are syntheses rather than controlled experiments, and one is a practitioner account. The contested storage recommendation has no strong supporting item. The evidence defines architectures and measurement decisions, but supplies no defensible retention period, retrieval threshold, or compression ratio.

\hypertarget{keep-the-source-record-rebuildable}{%
\section{Keep the source record rebuildable}\label{keep-the-source-record-rebuildable}}

Continuous rewriting can leave the latest memory without evidence that earlier versions contained. Support for this mechanism comes from one source study retained as strong by the evidence audit, one directional synthesis, and the Slack production account described above. The directional synthesis, Agentic Context Engineering (2025), reports no quantitative result and therefore supports only the direction of the practice. No controlled comparison of storage architectures is available.

Separate the immutable source record from every summary, profile, or extracted fact derived from it. Material removed during an earlier rewrite cannot be reconstructed from the latest version. Another pass may produce a better summary of what remains, but it has no observation from which to recover what disappeared.

The system has crossed a recovery boundary without recording that it did so. Treating each rewrite as the new authoritative state hides the loss behind a successful update.

An immutable raw layer changes that failure mode. The recorder appends original session events and does not ask a model to revise them. A separate process reads those events and produces summaries, profiles, or extracted facts. That output is the \textbf{distillate}: a derived representation constructed from the source record for a stated retrieval purpose.

When the distillate omits relevant material, the omission becomes a retrieval or consolidation defect that can be repaired from preserved evidence.

\Needspace{5\baselineskip}

The architecture has two ownership rules:

\begin{itemize}
\tightlist
\item
  The recorder owns fidelity to what happened, including ordering, identity, and state references.
\item
  The distiller owns one interpretation of that record for a declared retrieval purpose.
\end{itemize}

A summary may state what the system believed about a user. The trace records which event produced that belief, when it arrived, and which later event contradicted it. Combining those roles in one mutable record turns an inference into history.

\Needspace{5\baselineskip}

Rebuildability follows from this separation. Every derived item carries provenance to:

\begin{itemize}
\tightlist
\item
  the source events that produced it;
\item
  the version of the extraction or summarization rule;
\item
  the schema under which it was stored; and
\item
  any source material excluded by policy.
\end{itemize}

When the schema changes, the system builds a new derived store from the raw record. It does not ask a model to translate each old conclusion into a new one in place. Parallel rebuilds also allow an operator to compare the old and new distillates before switching readers to the new version.

\begin{figure}[htbp]
\centering
\includegraphics{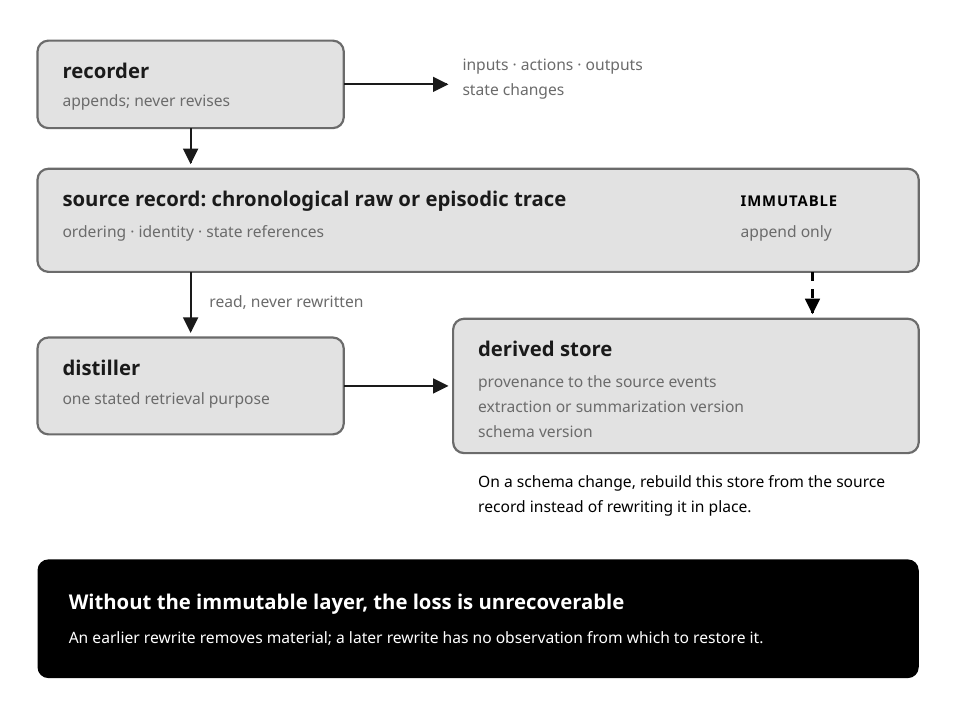}
\caption{The recorder appends an immutable source record; the distiller reads it for one stated retrieval purpose and writes a derived store whose items carry provenance, extraction version, and schema version. A schema change rebuilds that store from the record, because material removed by an in-place rewrite has no observation left from which a later rewrite could restore it.}
\end{figure}

The underlying artifact is the replayable trace from Chapter 10, used for another purpose. For recovery, the trace reconstructs execution after a crash. For memory, it reconstructs a representation after that representation proves incomplete or wrong.

The requirements overlap, but memory adds retrieval patterns, retention policy, user correction, supersession, and deletion obligations that execution replay alone does not resolve.

\Needspace{5\baselineskip}

Consider a coding agent that learns three facts during a repository investigation:

\begin{enumerate}
\def\labelenumi{\arabic{enumi}.}
\tightlist
\item
  An interface was renamed.
\item
  A compatibility wrapper remains for one release.
\item
  One service still calls the wrapper during rollback.
\end{enumerate}

A concise summary may preserve the rename and wrapper while omitting the rollback dependency as a low-frequency detail.

\Needspace{5\baselineskip}

Weeks later, an incident task asks whether the wrapper can be removed. A query against the summary returns a confident but incomplete answer. When the raw trace remains available, the system can:

\begin{itemize}
\tightlist
\item
  retrieve the original observation;
\item
  rebuild the derived facts under a schema that represents rollback dependencies; or
\item
  state that the current distillate does not cover the question.
\end{itemize}

Without the trace, the only available repair is another model-generated inference over already reduced evidence.

Continuous rewriting also collapses time. Suppose one session records Priya as the owner of a service, and a later session records that ownership moved to Luis. A later task asks who approved a decision between those dates. A profile optimized for the current owner may erase Priya entirely.

Preserving both events allows the derived layer to represent succession, validity intervals, or uncertainty over time.

The companion catalog develops this temporal design separately. Work on temporal knowledge-graph memory from Kim et al.~(\href{https://arxiv.org/abs/2408.05861}{2024}) contributes directional support. The recommendation to record two time axes, one for when a claim was true and another for when the system learned it, is practitioner-weighted rather than a measured result of that work.

The raw event sequence supplies the observations. The derived schema still decides whether a claim is current, obsolete, disputed, or historically valid.

Immutability does not imply public access or indefinite retention. Raw traces may contain credentials, personal information, proprietary code, and conclusions later shown to be false. Their value as recovery material increases the harm of unauthorized access.

\Needspace{5\baselineskip}

The raw layer is therefore sensitive source material:

\begin{itemize}
\tightlist
\item
  writers receive append-only authority;
\item
  readers cross an audited access boundary;
\item
  retained data is encrypted where appropriate;
\item
  sensitive fields follow explicit redaction rules; and
\item
  every event belongs to a stated retention policy.
\end{itemize}

Deletion must follow provenance in the reverse direction. If a user or policy requires removal of a raw event, every summary, profile, embedding, cache entry, and graph edge derived from it becomes suspect. Removing only the transcript leaves the more convenient copies intact.

\Needspace{5\baselineskip}

The system therefore needs a mechanical path to:

\begin{enumerate}
\def\labelenumi{\arabic{enumi}.}
\tightlist
\item
  identify affected derivatives;
\item
  invalidate them;
\item
  remove unauthorized cached representations; and
\item
  rebuild the remaining store from authorized source material.
\end{enumerate}

Forgetting is part of the architecture, and it is the least-measured part of the memory stack in the available evidence. A system may demonstrate successful retrieval of retained facts without testing whether deleted or superseded facts still reach generation through a summary, cache, embedding, or profile.

Deletion propagation and stale-claim retrieval should therefore be measured separately from task completion. A task can succeed while a deleted fact continues shaping the answer.

Retention also has a boundary that the recovery argument cannot override. Some systems cannot lawfully or safely preserve complete traces. Others should not preserve them because the likely harm exceeds their diagnostic value.

\Needspace{5\baselineskip}

The raw layer may therefore require:

\begin{itemize}
\tightlist
\item
  field-level redaction;
\item
  shorter retention windows;
\item
  content-addressed references to separately governed artifacts;
\item
  deliberate noncollection; or
\item
  irreversible deletion of selected event classes.
\end{itemize}

The design requires each loss to be explicit and testable, without requiring universal retention. A record already compressed by a model cannot serve as the raw recovery boundary.

\textbf{Companion catalog note (limited support).} Persisting the repository regions an agent has already explored may reduce repeated discovery across sessions. The idea rests on one directional source from Pan et al.~(\href{https://arxiv.org/abs/2507.19942}{2025}). Whether it reduces redundant retrieval without preserving stale interpretations requires local measurement.

\textbf{Companion catalog note (limited support).} Durable project state survived crashes and context loss in the long-horizon engineering system evaluated by Chen (\href{https://arxiv.org/abs/2604.13018}{2026}). This provides directional support for handing artifact references between agents. No controlled comparison with transcript-summary handoff is available, so any fidelity advantage remains an inference. The receiving agent must still verify artifact identity, version, and permissions.

\hypertarget{add-memory-infrastructure-only-for-measured-retrieval-failures}{%
\section{Add memory infrastructure only for measured retrieval failures}\label{add-memory-infrastructure-only-for-measured-retrieval-failures}}

At an architecture review for a new agent project, the proposed design already included a vector database and a knowledge graph. The query log was empty. Nobody had yet asked the system a question, but the retrieval architecture had already been chosen.

The case against that default is directional, practitioner-weighted, and contested. It draws on three source syntheses and no strong evidence item. Menschikov (\href{https://arxiv.org/abs/2506.17001}{2025}) compared knowledge-graph memory with lighter substrates in a personal-AI setting. Wolff and Bennati (\href{https://arxiv.org/abs/2601.07978}{2026}) compared cost and accuracy, although the evidence audit classifies the result as directional and records no quantitative finding behind it. Yang (\href{https://arxiv.org/abs/2602.05665}{2026}) surveyed graph-based agent memory and serves as a counterweight to the position taken here rather than direct support for it.

A practitioner argument that knowledge graphs are the wrong abstraction for agent memory accompanies those sources without a study behind it. Together they establish a presumption in favor of simpler infrastructure. They do not establish that relational storage is best across workloads.

\Needspace{5\baselineskip}

Begin with the corpus the agent actually records:

\begin{itemize}
\tightlist
\item
  session events;
\item
  tool observations;
\item
  decisions;
\item
  outcomes; and
\item
  provenance connecting derived facts to those events.
\end{itemize}

A relational database can store these records with stable identities, explicit time fields, transactional deletion, and access controls. Full-text search provides an inspectable first retrieval path. The operator can see the query, matched terms, filters, and returned records without first interpreting a learned similarity score or model-generated relation.

\Needspace{5\baselineskip}

This starting point is deliberately modest. It handles:

\begin{itemize}
\tightlist
\item
  exact identifiers;
\item
  quoted phrases;
\item
  error messages;
\item
  names and paths;
\item
  structured filters;
\item
  time-bounded queries; and
\item
  lexical combinations over those fields.
\end{itemize}

It also exposes its failures clearly.

Suppose a user asks for ``the outage caused by stale ownership data,'' while the trace says that ``the on-call mapping lagged the transfer.'' Lexical retrieval may miss the relationship. Repeated misses of this kind provide evidence for a semantic lane.

\Needspace{5\baselineskip}

Vector retrieval addresses lexical mismatch by comparing learned representations of the query and stored text. It may be justified when:

\begin{itemize}
\tightlist
\item
  users repeatedly describe the same event in different language;
\item
  terminology varies across teams;
\item
  names change over time; or
\item
  the useful passage shares a concept but few tokens with the query.
\end{itemize}

It also introduces an embedding model whose version, chunking policy, and source selection affect the representation. The additional index needs its own update, provenance, access-control, and deletion path. Assess those costs against observed lexical failures rather than anticipated ones.

A knowledge graph answers a different class of question. It represents entities and relationships, allowing traversal such as:

\begin{quote}
Which services depend on a library owned by a team whose runbook changed after an incident?
\end{quote}

When relationships are the retrieval target and multi-hop questions recur, the schema and maintenance burden may be justified. Similarity search is not a reliable substitute for explicit traversal when the answer depends on a sequence of named relations.

\Needspace{5\baselineskip}

The graph obtains those relations through an extraction pipeline. A model or parser reads an event or summary, identifies entities, resolves identities, and emits edges. Each step can:

\begin{itemize}
\tightlist
\item
  invent an entity;
\item
  merge distinct entities;
\item
  split one entity into several;
\item
  assign an unsupported relationship; or
\item
  attach a relationship to the wrong time interval.
\end{itemize}

The graph can then return a mechanically valid traversal over semantically incorrect edges. Structured presentation may make the answer appear better supported than the underlying evidence warrants.

High-degree entities create another scaling problem. Common concepts such as production, a central platform team, or a widely used library can become hubs connected to much of the store. Traversal through those hubs expands quickly and returns weakly related paths.

\Needspace{5\baselineskip}

Operating the graph then requires:

\begin{itemize}
\tightlist
\item
  relation constraints;
\item
  temporal filters;
\item
  traversal budgets;
\item
  edge-confidence policy;
\item
  identity reconciliation; and
\item
  pruning or correction of weak relationships.
\end{itemize}

Graph construction becomes an ongoing information-quality program rather than a one-time schema decision.

\Needspace{5\baselineskip}

The two additions address different failures:

\begin{itemize}
\tightlist
\item
  A vector index addresses recurring lexical mismatch.
\item
  A graph addresses recurring relational traversal.
\end{itemize}

A workload may need both, but combining them compounds provenance, synchronization, access-control, deletion, and rebuild obligations. The architecture record should name which representative queries each component answers and which observed failure made the simpler system inadequate.

The decision resembles Chapter 13's repository-index choice, but the corpus differs. Chapter 13 addressed indexing a repository whose entities already exist in source code. Here the corpus is the agent's own experience, and extraction may create entities and relationships absent from the original event.

That difference raises the cost of opaque derivation errors.

Claims of local deployment need the same scrutiny. A framework may store its database on a workstation while sending extraction or embedding requests to a remote service. Another may require a cloud model key for its default path even though the storage package is open source.

\Needspace{5\baselineskip}

Inspect the effective dependency graph, including:

\begin{itemize}
\tightlist
\item
  model calls;
\item
  embedding services;
\item
  telemetry;
\item
  background synchronization;
\item
  credential flows; and
\item
  remote administration.
\end{itemize}

The location of the database alone does not establish that the memory system is local.

Governance may determine the preferred substrate before retrieval quality does. A relational event store may fit existing backup, residency, audit, row-level access, and deletion controls. A managed vector service may satisfy those obligations with less operational effort elsewhere. A graph can expose sensitive relationships that no individual source event revealed.

Storage cost alone cannot compare systems that create different privacy and accountability surfaces.

The evaluation should keep correctness, performance, and governance separate. A full-text query that returns no relevant event has a retrieval-correctness failure even if it finishes in milliseconds. A graph query that finds the correct path but expands too slowly has a performance failure. A semantically useful vector result that cannot be traced to an authorized source has a governance failure.

Calling all three ``memory quality'' conceals which component must change.

\Needspace{5\baselineskip}

Before adding infrastructure, construct a small query set containing:

\begin{itemize}
\tightlist
\item
  ordinary exact lookups;
\item
  paraphrased requests;
\item
  time-bounded questions;
\item
  deletion and supersession checks;
\item
  provenance checks; and
\item
  the relational questions motivating the proposed graph.
\end{itemize}

\Needspace{5\baselineskip}

For each query, record:

\begin{itemize}
\tightlist
\item
  whether the relevant evidence was retrieved;
\item
  which irrelevant records displaced it;
\item
  how the result traces to source events;
\item
  latency and cost;
\item
  whether access controls held; and
\item
  whether deleted or superseded material appeared.
\end{itemize}

The set comes from real tasks and failures and is written before candidate stores are compared. Choosing examples after seeing which store handled them would favor that store, just as Chapter 2's task-exclusion rules must be fixed before outcomes are inspected.

The decision remains reversible while the raw source layer stays independent. A vector index can be regenerated from authorized trace segments. A graph can be rebuilt under a revised extraction schema and compared with its predecessor. If either becomes necessary, it joins the derived layer without replacing the source record.

The recovery boundary remains intact while retrieval becomes more specialized.

A simple store can become a false economy. Repeated lexical misses can force users to restate questions, cause agents to proceed without prior evidence, and consume review time. Repeated manual joins can conceal relationships the workload depends on. The presumption for simplicity expires when measured query failures and operational constraints show that the initial path cannot support the work.

The opposite failure is easier to overlook because the system appears sophisticated. A graph built for anticipated questions can make entity reconciliation, schema maintenance, and edge pruning the dominant memory workload before any production query uses traversal. A vector lane can retain embeddings derived from source text that has since been deleted or superseded.

Those maintenance costs belong in the same decision record as the retrieval gains.

Scheduled consolidation into typed, pruned records can reduce query-time work once the types correspond to recurring retrieval needs. Scheduling changes when interpretation happens. It does not change the evidence boundary. Every consolidated item still requires source provenance and a rebuild path.

Running consolidation on a timer before the types have stabilized merely moves speculative extraction into a batch process.

Product and decision context may also deserve separation from repository search. Dillon (\href{https://arxiv.org/abs/2605.08112}{2026}) measured coding-agent compliance when recorded product, design, and engineering decisions were available. That evidence is directional. Extending the idea to runbooks, ownership changes, rejected designs, and incident lessons is an inference, although those records answer questions repository structure cannot.

Such material should retain explicit source and time semantics. Code search should remain scoped to the repository rather than becoming an undifferentiated memory store.

\textbf{Companion catalog note (limited support).} Memory portability favors exportable events, explicit schemas, and provenance another system can audit. The support from Munirathinam (\href{https://arxiv.org/abs/2606.01138}{2026}) is anecdotal. A portability claim should therefore be tested by exporting and importing an actual corpus. A feature statement in product documentation is not a portability test.

\hypertarget{let-observed-failures-determine-what-compaction-preserves}{%
\section{Let observed failures determine what compaction preserves}\label{let-observed-failures-determine-what-compaction-preserves}}

A reviewer opens a failed trace annotated under Chapter 11's protocol and finds that the first upstream error followed a condensed-history update. \textbf{Compaction} replaces working history with a shorter representation so a run can remain inside its context budget. When a fixed summarization prompt is treated as routine maintenance, its failure profile remains unmeasured.

One policy may preserve goals while dropping constraints. Another may retain decisions while obscuring the tool observations that supported them. A third may preserve every apparent dependency and fill the effective context window with obsolete branches.

The evidence for treating compaction as a tunable component rests on one strong study, from Kang et al.~(\href{https://arxiv.org/abs/2510.00615}{2025}), covering app, office, and question-answering agents. It includes no coding-agent benchmark, so transfer to repository work remains an inference. The study supports treating each compaction policy as a candidate component whose omissions, distortions, and distractions should be measured through downstream tasks rather than assumed away.

Chapter 11 asks the reader to build a corpus of roughly one hundred diverse traces, cluster them into five to ten failure classes, and assign the first upstream failure in each run. Those counts are unmeasured starting recommendations. For compaction failures, add a context label only when the preserved trace supports it.

\Needspace{5\baselineskip}

The label records whether necessary evidence was:

\begin{itemize}
\tightlist
\item
  omitted;
\item
  retained with altered meaning; or
\item
  present but obscured by irrelevant history.
\end{itemize}

It also identifies the compaction-policy version and points to the source events that should have survived or been removed.

This label is an attribution judgment rather than a direct measurement. Chapter 11 requires a named human to approve any attribution that changes remediation or the interpretation of an evaluation. Revising the compaction policy is such a decision.

The label also prevents a common attribution error. A task may fail after compaction even when compaction played no causal role. The repository may have changed, a tool may have timed out, or the agent may have ignored evidence that remained available. Revise the policy only when the trace supports a counterfactual claim: a different representation of the same prior history could plausibly have changed the relevant decision.

\Needspace{5\baselineskip}

The initial policies are therefore hypotheses. Before a local failure corpus exists, only structural rules are available, such as preserving:

\begin{itemize}
\tightlist
\item
  the current goal;
\item
  unresolved constraints;
\item
  artifact identities;
\item
  decision provenance;
\item
  contradictory observations; and
\item
  uncertainty about external state.
\end{itemize}

Their relative value for a particular workload cannot be known in advance. The early system should preserve the raw trace, record which policy produced each condensed history, and avoid presenting an untested prompt as optimized.

Once failures accumulate, change one policy element at a time. A revision might preserve the exact observation supporting an unresolved dependency, distinguish superseded claims from deleted claims, or remove completed exploratory branches after their artifacts have been recorded.

Then run paired repeats on the same tasks using the discipline from Chapter 1. The comparison measures task correctness and the context consumed. Token count remains a resource measure rather than the primary outcome.

A comparison between two compaction policies inherits the run-to-run variation Chapter 1 describes. One run under each policy supplies one draw from each condition. A difference that remains smaller than the measured spread should not justify replacing the incumbent.

The compatibility-wrapper example provides a concrete policy test. The raw trace records that one service still calls the wrapper during rollback. The compacted history drops that observation. The agent recommends removing the wrapper, and the reviewer assigns missing context as the first upstream cause of the error.

Revise one rule so that unresolved rollback dependencies retain their supporting observations, then rebuild the condensed history from the same raw trace. Paired runs on the same removal task test whether the agent now preserves the wrapper and uses the retained evidence.

Accept the revision only when task success improves without introducing distraction failures in neighboring cases. Otherwise the rule remains a hypothesis.

Kang et al.~reported peak-token reductions of 26 to 54 percent alongside improved task success across three app, office, and question-answering benchmarks. They also reported gains of up to 46 percent for small-model agents after distracting context was removed. Under those experimental conditions, less context coincided with better task performance.

Those figures do not establish an expected token reduction or success gain for coding agents.

The study covers app, office, and question-answering benchmarks rather than coding agents. Transfer to repository work remains an inference.

A plausible mechanism is the effort required to discriminate among records. A model presented with every prior attempt must determine which constraints remain active, which observations are stale, and which abandoned branches still matter. Irrelevant history competes with useful evidence, while repetition can make an obsolete plan appear current. Removing that material can improve correctness while reducing inference cost.

Compression creates the inverse failure. A short history may omit a rare constraint because it appeared once or consumed few tokens. Frequency is a poor proxy for importance when one low-salience fact determines whether a destructive operation is allowed. The failure corpus reveals which details became decisive in real investigations, allowing the policy to preserve evidence according to function rather than prominence.

\Needspace{5\baselineskip}

Task correctness remains too coarse for diagnosis. Inspect whether the compacted context:

\begin{itemize}
\tightlist
\item
  retained the decisive observation;
\item
  preserved its qualifications;
\item
  distinguished current from superseded information; and
\item
  removed material that no longer affected the task.
\end{itemize}

A successful run can conceal damaged memory when the model guesses correctly or recovers through another tool call. A failed run can contain a sound compacted history that the model then misuses. Keeping those cases separate prevents the compactor from absorbing every downstream error.

Policy versions must remain replayable. The same raw trace should produce the same condensed representation under a named policy and model configuration. When nondeterminism prevents exact reproduction, the rebuild should record that variation explicitly.

Retries also need a declared rule. The system records whether a retry reuses the existing compacted state or regenerates it from the raw trace. Otherwise a successful retry can conceal that the two attempts received different histories.

Once a policy stabilizes across new failures and held-aside traces from the same workload, its examples may support training or distilling a smaller compaction model. That step changes latency and cost while introducing another source of semantic drift.

The smaller model's outputs and downstream task results should therefore be compared with the accepted policy. A route back to the larger compactor should remain available for cases outside the smaller model's validated range.

The tuning loop cannot begin on the first day because no local failure evidence exists. Borrowed rules can provide an initial safety hypothesis, while preserved raw traces make mistakes recoverable by allowing the derived representation to be rebuilt.

Optimization begins only after the traces show which omissions, distortions, and distractions caused consequential errors. This dependence on Chapter 11's failure corpus is why compaction tuning follows the broader retrieval and context architecture rather than preceding it.

\hypertarget{build-memory-in-reversible-stages}{%
\section{Build memory in reversible stages}\label{build-memory-in-reversible-stages}}

Storage and retention decisions take effect from the first session. Compaction-policy optimization waits for the failure corpus from Chapter 11. The sequence below separates controls required immediately from those that depend on observed failures.

From the first run, I preserve every permitted raw trace in inexpensive immutable storage. Every summary, profile, extracted fact, embedding, and graph edge is marked as derived. Provenance, schema versions, and policy versions are recorded before the derived layer becomes operationally important.

Retention limits and deletion propagation are tested as part of the architecture. A recovery boundary that violates its governance boundary cannot remain in service.

Begin retrieval with a relational event store and full-text search. This is a presumption drawn from contested, directional evidence rather than a measured conclusion. Local query failures determine whether the system needs more.

\Needspace{5\baselineskip}

Before adding another retrieval component, write down the representative queries and observed failures that would justify it:

\begin{itemize}
\tightlist
\item
  recurring lexical mismatch may justify a vector lane;
\item
  recurring relational traversal may justify a graph.
\end{itemize}

Operating cost, remote dependencies, provenance, deletion behavior, and access control belong in the same decision.

Initial compaction rules remain explicit hypotheses, and every condensed history stays reproducible from the raw trace. Once the Chapter 11 corpus contains failures attributed to context omission, distortion, or distraction, revise the policy against those cases and remeasure task correctness through paired runs.

Token reduction is useful for capacity planning. It cannot substitute for evidence that the agent made better decisions.

Review capacity is finite and does not increase automatically with fleet size. Part V turns to the people who remain accountable for this work and to the interfaces, review policies, and escalation paths built around their limited attention.

\section*{Sources and evidence}

\textbf{Preserve raw traces and distill separately}

\begin{itemize}
\tightlist
\item
  Strong evidence for the degradation mechanism: ``Useful Memories Become Faulty When Continuously Updated by LLMs'' (Zhang 2026), arXiv:2605.12978. The study measures degradation under repeated LLM memory updates; it does not compare immutable-source and rebuildable-distillate architectures.
\item
  Corroborating case: De Simone, S. (2026), ``How Slack Manages Context in Long-Running Multi-agent Systems,'' \href{https://www.infoq.com/news/2026/04/slack-agent-context-management/}{InfoQ}, April 28, 2026. Slack's engineers replaced accumulated chat logs with three structured memory channels and a credibility-scored review step. The account supports operational plausibility without extending the strong result.
\item
  Directional evidence: Agentic Context Engineering, arXiv:2510.04618.
\item
  Corroboration (narrative only): the author's memory system rebuilds its derived layer on schema changes and mechanically carries three append-only event tables across rebuilds.
\end{itemize}

\textbf{Use a light store by default}

\begin{itemize}
\tightlist
\item
  Directional evidence: PersonalAI KG comparison (Menschikov), arXiv:2506.17001, with a contested practitioner thread named in the same synthesis.
\item
  Directional evidence: Cost-and-Accuracy study (Wolff \& Bennati 2026), arXiv:2601.07978.
\item
  Directional evidence (boundary citation): Graph-based Agent Memory survey (Yang 2026), arXiv:2602.05665. Cited as the counterweight, not support.
\item
  Corroboration: none on record.
\end{itemize}

\textbf{Optimize compaction from failures}

\begin{itemize}
\tightlist
\item
  Strong evidence for the measured failure-driven compression comparison: Kang, M., et al.~(2025), ``ACON: Optimizing Context Compression for Long-horizon LLM Agents,'' arXiv:2510.00615 (Microsoft). The study tests failure-guided compression on app, office, and question-answering benchmarks; transfer to coding agents remains directional.
\item
  Corroboration: none on record.
\end{itemize}

\textbf{Companion-catalog records named inline}

These are not part of this chapter's three developed practices. Each is named in the prose so that its source and evidence grouping remain visible.

\begin{itemize}
\tightlist
\item
  Directional evidence: Temporal KG memory (Kim et al.), arXiv:2408.05861, carried by the companion record on modeling time explicitly. Supports the two-time-axes aside only.
\item
  Directional evidence: Prometheus (Pan, H., et al.~2025), arXiv:2507.19942, carried by the companion record on persisting explored context. No figure quoted here.
\item
  Directional evidence: ``Toward Autonomous Long-Horizon Engineering for ML Research'' (Chen et al.~2026), arXiv:2604.13018, carried by the companion record on durable artifact handoff.
\item
  Corroborating case: memorywire (Munirathinam 2026), arXiv:2606.01138, carried by the companion record on memory portability.
\item
  Directional evidence: Product context and coding-agent decision compliance (Dillon 2026), arXiv:2605.08112, carried by the companion record on the tribal-knowledge substrate. Its compliance figure is not used in the prose.
\end{itemize}

\part{Human review and accountability engineering}
\gdef\currentparttitle{Part V: Human review}

\chapter{Efficient verification interfaces and risk-based human escalation}
\label{ch16-verification-interfaces-risk-based-escalation}
\begin{quote}
\textbf{Evidence profile.} 3 strong \(\cdot\) 4 directional \(\cdot\) 0 corroborating evidence items across 3 developed practices (\erca{148}, \erca{153}, \erca{157}).

\textbf{Chapter claim.} Verification must be cheaper than uncritical acceptance.
\end{quote}

On my site, every agent output arrives as a schema-validated Git diff. The agent cannot write to the production database or publish a page. The architecture places the proposed state transition in an artifact that I can inspect, execute, reject, or revise before it acquires production authority.

It provides no evidence that the proposed change is correct. A diff that is easy to approve can look much like one that has been verified.

A reviewer who skims that diff is making a cost-benefit decision. Reading every changed branch, reconstructing the intended behavior, running the relevant tests, and comparing the result with the task costs far more than clicking approve. When the expected consequence of a missed defect feels remote, shallow review becomes the rational response even for someone who intends to be careful.

Vasconcelos et al.~(\href{https://arxiv.org/abs/2212.06823}{2023}) tested this account across five experiments. They studied \textbf{overreliance}, the failure in which a person accepts an incorrect AI output. Overreliance changed when the economics of checking changed. Participants engaged more when verification required less effort or when the cost of an error became more salient. Explanations reduced overreliance when they altered that calculation. They did not act as a general antidote to misplaced trust.

Software-engineering studies locate the same problem closer to the operated tools. Johnson et al.~(\href{https://doi.org/10.1145/3772370}{2026}) derived a trust framework from 18 practitioner interviews and tested its factors in a 368-response Microsoft survey. Tufano et al.~(\href{https://doi.org/10.1109/TSE.2023.3348172}{2024}) manually inspected 2,291 predictions from code-review automation techniques, identifying task-specific success and failure classes along with dataset defects. These studies support showing reviewers the conditions under which a tool succeeds and the evidence behind a particular output. They do not test the patch-gate interfaces proposed in this chapter, so that transfer remains directional.

\Needspace{5\baselineskip}

This changes the engineering question. Telling reviewers to be more careful treats attention as a personality trait. A review system should instead account for:

\begin{itemize}
\tightlist
\item
  the cost of obtaining evidence;
\item
  the point at which a person can accept the output; and
\item
  the finite supply of human judgment.
\end{itemize}

These are properties of the workflow and can be inspected, measured, and changed.

Review capacity does not expand because an agent fleet can produce more patches. Practitioner reports already describe systems generating substantial code volume behind retained human gates. InfoQ reported (\href{https://www.infoq.com/news/2026/03/stripe-autonomous-coding-agents/}{2026}) that Stripe's Minions system produces more than 1,300 production pull requests each week.

That number says nothing about review depth, escaped defects, or whether the human gate changed any verdict. It does illustrate the queueing problem. Generation can scale faster than the people authorized to accept its consequences.

\Needspace{5\baselineskip}

This mismatch is three connected design problems:

\begin{enumerate}
\def\labelenumi{\arabic{enumi}.}
\tightlist
\item
  The review surface must make correctness claims inexpensive to challenge.
\item
  Friction must interrupt rapid acceptance without taxing every interaction equally.
\item
  Automated monitoring must direct scarce human attention toward cases that warrant judgment while leaving the verdict with a named person.
\end{enumerate}

Each intervention changes the cost or allocation of review. None creates correctness.

Nearly all of the empirical oversight evidence used here predates long-horizon agents. Researchers have rarely tested whether findings about reviewing suggestions transfer to autonomous actions, although the transfer is often assumed.

A code completion that waits for a programmer to accept one function creates a different control problem from an agent that edits several files, runs commands, revises its plan, and opens a proposed patch after a long trajectory. The older evidence remains useful for identifying mechanisms. It does not settle how those mechanisms behave as autonomy, duration, and action space increase.

The practical standard is therefore narrower than ``keep a human in the loop.'' A workflow can include a person who sees too little, arrives too late, lacks an affordable way to check the work, or has no authority to stop it.

\Needspace{5\baselineskip}

The useful questions are concrete:

\begin{itemize}
\tightlist
\item
  What evidence reaches the reviewer?
\item
  At which decision is independent judgment required?
\item
  Which cases consume scarce review capacity?
\item
  Who owns the final verdict?
\item
  What happens after that person decides?
\end{itemize}

\hypertarget{make-correctness-cheap-to-challenge}{%
\section{Make correctness cheap to challenge}\label{make-correctness-cheap-to-challenge}}

An explanation describes why an output might be right. A verification interface gives the reviewer a practical way to discover that it is wrong.

Tests, executable examples, type checks, continuous-integration gates, sandboxed runs, and evidence placed beside the claim all serve the second purpose. Their value comes from the failures they can expose.

\Needspace{5\baselineskip}

Suppose an agent changes a retry loop. A prose rationale may say that the new backoff prevents overload while preserving responsiveness. A useful review surface instead exposes:

\begin{itemize}
\tightlist
\item
  the changed state machine;
\item
  the maximum attempt count;
\item
  the errors classified as retryable;
\item
  cancellation behavior;
\item
  the idempotency boundary; and
\item
  a test that advances a fake clock through failure and recovery.
\end{itemize}

\Needspace{5\baselineskip}

The reviewer can then challenge specific claims:

\begin{itemize}
\tightlist
\item
  Do retries stop?
\item
  Does a non-retryable error escape immediately?
\item
  Is cancellation honored?
\item
  Can two workers duplicate the side effect?
\item
  Does recovery preserve the intended state?
\end{itemize}

The explanation may help locate those claims. The executable evidence performs the check.

The same distinction applies to evidence presentation. Deciding what belongs in the review packet and locating the decisive material within an admitted artifact are separate problems. A system may correctly include logs, traces, generated files, and dependency changes while burying the relevant event among thousands of lines.

Curation helps only when the omitted material remains reachable and the curation rule can itself be inspected. Otherwise the interface reduces apparent complexity by hiding the evidence required to falsify the output.

Fok and Weld (\href{https://arxiv.org/abs/2305.07722}{2024}) synthesized research on explainable AI and reached a compatible but directional conclusion. Explanations rarely produced \textbf{complementary performance}, in which the human-AI pair outperforms either one alone, unless the explanation helped the person verify the answer.

The literature spans tasks with different error costs, expertise requirements, and verification opportunities. It does not establish that one explanation format will improve software review. It does support a sharper design test:

\begin{quote}
After seeing the explanation, what can the reviewer now check that was previously unavailable or too expensive?
\end{quote}

Software is unusually favorable under that test because many claims admit partial mechanical oracles. A compiler can reject a type error. A property test can generate cases a reviewer would not enumerate. Deterministic replay can expose an ordering defect. A sandbox can record which files and network endpoints an action touched.

None proves complete correctness. Each moves a claim away from something the reviewer must read and trust and toward something the reviewer can execute and challenge.

That advantage has strict limits. A passing test suite shows only that the tested behavior passed under the tested conditions. Generated tests can repeat the implementation's mistaken assumptions. Type systems encode some properties and omit others. Sandboxes constrain effects without establishing intent.

A green continuous-integration result can itself become a trust signal that accelerates approval even when the changed risk lies outside the tested surface.

A large diff with weak tests can therefore make verification much more expensive than acceptance. The reviewer must reconstruct behavior across files, infer unstated requirements, and judge interactions that no executable check covers. Adding a longer reasoning trace increases reading cost without creating an oracle. Under those conditions, the cost-benefit model predicts shallow engagement.

Optimizing first for a complete agent explanation is therefore the wrong starting point. Begin with the claims whose failure matters and ask what artifact could disconfirm each one.

\Needspace{5\baselineskip}

For example:

\begin{itemize}
\tightlist
\item
  A database migration may require forward and rollback rehearsal against representative data.
\item
  An authorization change may require a matrix of identities and denied actions executed as integration tests.
\item
  A concurrent worker may require a trace of ownership, ordering, retries, and idempotency keys under an injected failure.
\item
  A dependency upgrade may require compatibility checks against affected consumers rather than only the changed package.
\end{itemize}

The check also needs a sound scope. A changed-file hook that formats or tests only affected modules may shorten feedback time, but its dependency model must be correct. When a shared schema changes, the affected set may extend far beyond the edited directory.

Cheap verification produced through unsound scoping is an instrumentation artifact. The dashboard becomes greener because the check stopped observing where the failure could occur.

Displayed confidence does not repair that defect. A confidence signal should be calibrated before it reaches a reviewer because an uncalibrated percentage adds persuasive precision without dependable information. Calibration is a prerequisite for presentation, not a substitute for checking the claim underneath it.

\Needspace{5\baselineskip}

The reviewer still needs to know:

\begin{itemize}
\tightlist
\item
  which event the probability describes;
\item
  which population produced the estimate;
\item
  which threshold generated the displayed category; and
\item
  whether the current case resembles that population.
\end{itemize}

The verification cost also includes more than time. A check may require scarce domain expertise, access to production-like state, coordination with another owner, or authority to run a destructive rehearsal safely. A one-command test can be computationally cheap and organizationally expensive.

Sarkar et al.~(\href{https://arxiv.org/abs/2208.06213}{2022}) advance a related hypothesis that treats verification as the dominant cost of agent-produced software. Their support is anecdotal synthesis, and they do not quantify the magnitude, so the hypothesis is useful for inspecting a workflow rather than as a measured law.

The more defensible claim is conditional: as generation becomes cheaper, verification becomes a larger share of the work unless the cost of establishing correctness falls with it. Capacity plans that count generated patches while omitting review evidence measure supply while ignoring the constrained resource.

The design objective is not to make every check effortless. Some claims have no cheap oracle. The objective is to avoid charging people for work the system can perform mechanically and to make the remaining uncertainty visible.

An explanation without a practical check is context. It may orient the reviewer, but it should not be recorded as evidence that the change is safe.

\hypertarget{interrupt-acceptance-where-overreliance-occurs}{%
\section{Interrupt acceptance where overreliance occurs}\label{interrupt-acceptance-where-overreliance-occurs}}

Cheap verification does not ensure that anyone performs it. A reviewer may adopt a global heuristic such as ``this agent is usually right'' and apply it to the next output without examining the particulars. The interface must then interrupt the act of acceptance rather than add ceremony to every earlier step.

\Needspace{5\baselineskip}

A \textbf{cognitive forcing function} requires deliberate judgment at a point where a person might otherwise rely on a fast heuristic. In an AI review workflow, it might:

\begin{itemize}
\tightlist
\item
  require an independent answer before revealing the agent's answer;
\item
  delay the reveal at a consequential checkpoint;
\item
  ask the reviewer to name the expected failure mode before showing the test result;
\item
  require a reason tied to evidence before accepting a high-risk change; or
\item
  require the reviewer to identify which invariant was checked.
\end{itemize}

The intervention repays its cost only when it changes the decision process at the point where overreliance occurs.

Buçinca, Malaya, and Gajos (\href{https://arxiv.org/abs/2102.09692}{2021}) tested cognitive forcing functions in a controlled experiment with 199 participants. The interventions reduced overreliance where explanations did not, but they did not eliminate it. The most effective interventions were also rated least favorably.

Benefits accrued disproportionately to participants with high \textbf{Need for Cognition}, a measured tendency to engage in effortful thought. Friction changed behavior, but it imposed a real usability cost and did not affect everyone equally.

That unpopularity is not merely an implementation defect. Effective friction blocks the shortest path through the task, so users experience its mechanism as inconvenience. A team that evaluates the gate only through satisfaction scores will select against the property that makes it work. A team that ignores usability will invite workarounds, perfunctory responses, and pressure to remove the control.

\Needspace{5\baselineskip}

The useful comparison has three parts:

\begin{enumerate}
\def\labelenumi{\arabic{enumi}.}
\tightlist
\item
  the additional defects detected at the targeted decision;
\item
  the delay and cognitive load imposed there; and
\item
  the behavior that emerges after repeated use.
\end{enumerate}

A forcing function may improve an experimental average while teaching experienced reviewers to enter meaningless text. It may also move acceptance into a less observable channel. The control must be evaluated as part of the workflow it changes.

Barke et al.~(\href{https://arxiv.org/abs/2206.15000}{2023}) conducted a grounded-theory study of twenty programmers that suggests where to target the intervention. They observed two recurring interaction modes.

In \textbf{acceleration mode}, the programmer believed the plan was already understood and used the model to move faster. Validation tended to be shallow.

In \textbf{exploration mode}, the programmer used suggestions to discover an approach and already compared, revised, and rejected alternatives more actively.

This distinction supports a better allocation rule than uniform friction. Requiring an independent judgment during acceleration may interrupt a reflexive accept loop. Applying the same ceremony during exploration may add cost to work that already contains deliberation.

Uniform friction is administratively simple. It spends reviewer tolerance without regard to the behavior it is intended to change.

Barke et al.~studied suggestion-level interaction rather than autonomous agents producing multi-file diffs after long trajectories. Acceleration and exploration may coexist within one agent run. A programmer may explore the architecture, delegate the implementation, and then accelerate through review because the overall plan feels familiar.

The distinction is therefore a useful hypothesis for instrumentation, not a validated classification for autonomous-agent review.

\Needspace{5\baselineskip}

A practical system needs observable proxies for the decision context without pretending to read the reviewer's mind. Candidate signals include:

\begin{itemize}
\tightlist
\item
  changes to high-impact file classes;
\item
  a previously approved implementation plan;
\item
  repeated acceptance from the same agent;
\item
  a large generated diff;
\item
  a short interval between presentation and approval;
\item
  weak or missing verification evidence; and
\item
  an irreversible downstream action.
\end{itemize}

Each proxy can be wrong. Diff size is not risk. Speed is not negligence. Familiarity may reflect genuine expertise. The signals should determine where to test an intervention, not establish that a reviewer was careless.

\Needspace{5\baselineskip}

Targeting therefore requires its own evaluation. Record:

\begin{itemize}
\tightlist
\item
  where the forcing function appeared;
\item
  whether the reviewer changed or rejected the output;
\item
  how long the decision took;
\item
  which defects later appeared;
\item
  whether defects concentrated in bypassed cases; and
\item
  whether users routed work around the gate.
\end{itemize}

Segmenting results by reviewer experience and task class may reveal useful differences, but each segment adds another comparison. Chapter 1's multiplicity rules apply before a subgroup result is treated as real.

An aggregate change in acceptance rate cannot show whether the intervention improved judgment or merely annoyed one group into leaving the workflow.

Time pressure deserves separate treatment. Rosbach et al.~(\href{https://arxiv.org/abs/2411.00998}{2024}) studied computational pathology and found that time pressure increased the severity of automation-bias errors while leaving their frequency unchanged.

That setting does not establish the same effect in software review. It does identify a possible design error: adding a forcing function while retaining an impossible service deadline may reduce the opportunity for reflection on the cases whose consequences are greatest. Protecting a checkpoint from time pressure may matter as much as adding the checkpoint.

The decision prompt should solicit evidence tied to the action it authorizes. ``Have you reviewed this?'' can be answered with a checkbox.

A prompt such as:

\begin{quote}
Which authorization boundary changed, and which denied case demonstrates that it remains closed?
\end{quote}

requires a claim tied to an observable check. The answer can still be wrong, but shallow assent becomes harder to reconcile with the authorized action.

\Needspace{5\baselineskip}

Forcing functions also need explicit escape paths. An emergency may justify bypassing an independent pre-commit judgment, but the bypass should retain:

\begin{itemize}
\tightlist
\item
  the responsible owner;
\item
  the reason;
\item
  the scope of the exception;
\item
  the evidence available at the time; and
\item
  the required follow-up state.
\end{itemize}

A gate with no exception can become unsafe when delay carries greater cost. A gate whose exception leaves no durable record becomes optional whenever pressure rises.

Chapter 17 examines how to determine whether documented gates continue to affect execution. The narrower conclusion here is that friction is unpopular by construction. Its existence and actual use cannot be inferred from the workflow diagram.

\hypertarget{allocate-review-where-it-can-change-the-outcome}{%
\section{Allocate review where it can change the outcome}\label{allocate-review-where-it-can-change-the-outcome}}

Kale et al.~(\href{https://arxiv.org/abs/2508.19461}{2025}) reported that targeted escalation improved true-positive rate over blanket oversight at a false-positive rate of 0.01, with a gain of approximately 15 percent. Blanket oversight does not guarantee greater detection at fixed human capacity. Spreading attention across every case reduces the scrutiny available to each one, while targeting concentrates the same attention on fewer cases. The study's hierarchical and sequential monitor scaffolds also allowed weaker models to oversee stronger agents.

This result is unusually relevant because it measured the allocation of oversight rather than only asking whether a judge could classify isolated outputs. Its setting was a covert-misbehavior arena rather than production software review. The operating point should not be treated as a deployment guarantee, and a gain measured at a 0.01 false-positive rate need not persist at another threshold.

The allocation problem is broader than that setting. When every agent action receives the same depth of human review, review burden grows directly with action volume. A fleet can then increase output until reviewers skim, queues stall, or teams weaken the gate. Blanket oversight preserves the appearance of equal scrutiny while reducing the scrutiny available per case.

\Needspace{5\baselineskip}

Targeted escalation changes the unit of human work. Automated monitors inspect the full stream, while people receive the cases selected for judgment. An escalation route must connect each flag to:

\begin{itemize}
\tightlist
\item
  a named decision owner;
\item
  the evidence supporting the flag;
\item
  an available verdict;
\item
  the state paused by the escalation; and
\item
  the event that resumes execution.
\end{itemize}

Without those elements, a flag is only a notification. It can remain unread, move among teams, or leave execution suspended indefinitely.

\Needspace{5\baselineskip}

A monitor scaffold should assign each detection problem to the least complex mechanism capable of resolving it:

\begin{itemize}
\tightlist
\item
  Deterministic rules handle exact violations.
\item
  A model judge applies a fixed rubric to contextual cases.
\item
  An agent judge examines live traces when the decision depends on sequence, changing state, or tool interaction.
\item
  Human reviewers calibrate the layers and own the final verdict.
\end{itemize}

Kale et al.~measured the value of monitoring scaffolds. The particular ordering above is a design convention drawn from one practitioner account and has no independent measurement behind it.

A deterministic rule can establish that an agent wrote outside an allowed directory, attempted a forbidden operation, or omitted a required approval token. Asking a model to infer an exact violation adds variance without adding useful judgment.

A model judge becomes appropriate when the rubric requires contextual interpretation, such as whether a migration plan includes a credible rollback for the affected data path. An agent judge is appropriate only when no single artifact contains the evidence, such as whether a sequence of individually permitted actions collectively bypassed a control.

\begin{figure}[htbp]
\centering
\includegraphics{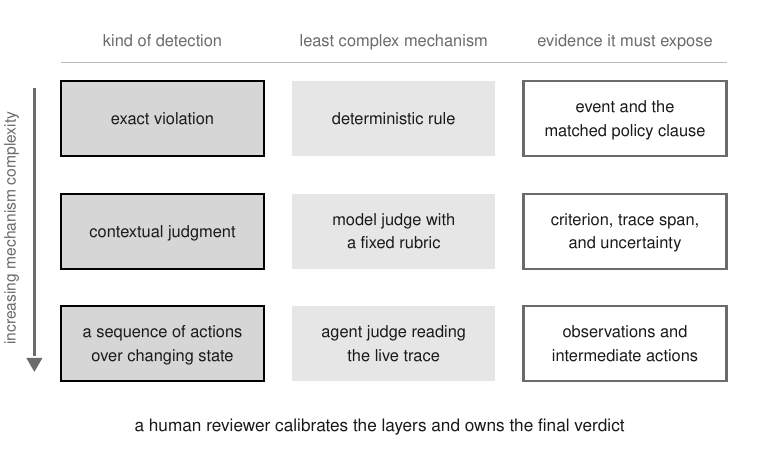}
\caption{Each kind of detection routes to the least complex mechanism capable of checking it, and every layer must expose the evidence the next one needs.}
\label{fig:ch16-detection-ladder}
\end{figure}

Each layer must expose its evidence to the next, as Figure \ref{fig:ch16-detection-ladder} sets out.

A bare risk score forces the reviewer either to repeat the discovery work or to accept the classifier's authority without an affordable check.

Calibration follows the Chapter 5 protocol. Select the operating point on an expert-labeled, stratified, held-out set representing consequential case types, including rare failures. Report true-positive rate at the chosen false-positive rate and preserve the confusion matrix by stratum.

Both rates are estimates from finite labeled samples. A stratum with few cases produces wide uncertainty. Re-run the evaluation when the agent, rubric, tools, environment, or case mix changes. Overall accuracy is especially misleading when most actions are benign and false positives determine whether the human queue remains usable.

The queue budget is part of the operating point. A false-positive rate of 0.01 produces one expected false alarm across 100 benign actions and 10,000 across one million. When the flagged event is rare, false alarms may outnumber correct flags even at a seemingly low false-positive rate. This is the prevalence effect Chapter 5 develops for graders.

The arithmetic is linear, but its operational consequence depends on prevalence: even a low false-positive rate can overwhelm the review queue when benign actions dominate. The approximately 15 percent true-positive-rate gain comes from one covert-misbehavior comparison at one operating point.

\Needspace{5\baselineskip}

Escalation is therefore a budgeted online decision. As cases arrive, the system must decide which uncertainty deserves scarce human attention. The decision may consider:

\begin{itemize}
\tightlist
\item
  expected consequence;
\item
  reversibility;
\item
  novelty;
\item
  disagreement among monitors;
\item
  evidence quality; and
\item
  availability of a qualified owner.
\end{itemize}

These factors should not disappear into an unexplained semantic risk score. They should define the evidence presented to the judge and the explicit constraints governing the queue.

My orchestration fleet uses a related division of labor. Mechanical work proceeds autonomously, while judgments the system should not make unilaterally enter a structured morning decision record. The system does not silently resolve them. The design reserves interruption for decisions that require a person.

I have not measured maintainer time, review depth, or escaped defects, so the attention benefit remains a design claim rather than a demonstrated result.

\Needspace{5\baselineskip}

The associated wake-up contract makes the route more specific. The fleet should wake me for suspected wrong calls and decisions about scope or trust. It should not wake me for passing runs or routine plumbing. Each escalation must:

\begin{itemize}
\tightlist
\item
  name the outcome at stake;
\item
  present a question answerable in one line;
\item
  list the available options;
\item
  state the recommendation; and
\item
  explain why the system cannot decide.
\end{itemize}

The success criterion is behavioral: engineers should not learn to ignore the messages.

This separates notification quality from verdict authority. A monitor may decide that a case satisfies an escalation rule. It should not convert that classification into the substantive decision under review. The named owner must be able to reject the work, request evidence, narrow its scope, or authorize resumption. Naming someone who lacks those powers assigns accountability without control.

The resumption state also requires explicit design because asynchronous review creates races. If an agent continues acting while escalation is pending, rejection may arrive after dependent effects have occurred. If execution stops globally, one ambiguous case can block unrelated low-risk work.

\Needspace{5\baselineskip}

The route should state:

\begin{itemize}
\tightlist
\item
  whether the action, trajectory, repository, or fleet pauses;
\item
  which state becomes durable while it waits;
\item
  which later actions are prohibited;
\item
  which decision token releases the pause; and
\item
  how stale approvals are rejected.
\end{itemize}

Checkpoint placement should follow observed failure paths. If an action becomes irreversible at deployment, a checkpoint after deployment can only document the result. If failures begin when a plan silently expands scope, waiting for the final diff makes intervention expensive.

The narrow companion rule is to place human checkpoints where the evidence shows that intervention can still change the outcome.

Human expertise is another boundary. A reviewer who handles only rare anomalies may lose familiarity with ordinary system behavior, while automation absorbs the routine work that previously maintained that familiarity. Bainbridge (\href{https://doi.org/10.1016/0005-1098(83)90046-8}{1983}) described this problem in industrial process control as an ``irony of automation.'' Its application to software-agent oversight is analogical rather than measured.

The design question remains relevant: how will the person expected to intervene retain the knowledge required to recognize and correct an abnormal state?

Xie et al.~(\href{https://arxiv.org/abs/2408.01703}{2024}) propose rendering agent actions as live state so a reviewer can inspect a trajectory before it terminates. The support is one prototype-scale result and does not establish improved oversight. The proposal is useful when delayed reconstruction repeatedly fails, provided the live interface preserves the durable trace rather than replacing it.

Monitoring can also alter the behavior being measured. An agent aware of evaluation may adapt to the evaluation frame. This is a design caution rather than a measured result from Kale et al., whose reported finding was that monitor scaffolding mattered more than monitor awareness.

My evaluation harnesses therefore use de-framed prompts that omit evaluation vocabulary. Removing those cues is a precaution against one measurement effect. It does not prove that the resulting behavior matches unobserved production behavior.

Judge drift creates another measurement problem. Deterministic rules drift when policy or event schemas change. Model judges drift when the model, rubric interpretation, or evaluated distribution changes. Agent judges can drift more severely because their decisions depend on longer sequences of observations and actions.

A monitor calibrated last quarter may preserve the same interface while changing which cases reach people.

\Needspace{5\baselineskip}

Layering redistributes detection work but does not answer who evaluates the evaluators. Human owners need:

\begin{itemize}
\tightlist
\item
  adjudication of flagged cases to estimate false alarms;
\item
  sampled review of unflagged cases to estimate misses;
\item
  versioned records connecting each verdict to its monitor configuration; and
\item
  targeted samples after relevant system changes.
\end{itemize}

The unflagged sample cannot contain only easy negatives. It must include strata in which the monitor is expected to fail. When a stratified sample is used, its estimates must be weighted back to the deployed case mix before they describe production performance.

These audits do not require human review of every action. They require enough independent labels to determine whether the chosen operating point still holds. When performance degrades, the response may be recalibration, a narrower deployment boundary, a new deterministic check, a revised rubric, or greater human review.

The correct response depends on the failure mechanism. ``Use a stronger judge'' is not a diagnosis.

\hypertarget{build-the-review-and-escalation-path}{%
\section{Build the review and escalation path}\label{build-the-review-and-escalation-path}}

Start with one consequential agent workflow. Identify the decision that transfers authority from proposed work to an external effect, such as merge, deployment, publication, data modification, customer contact, or a change to another system.

Name the person or role that owns that decision. When no owner can be named, escalation cannot end in an accountable verdict.

List the claims the reviewer must establish before acceptance. Attach a falsification artifact to each claim where one is available: a test, typed interface, deterministic replay, structured trace, sandbox result, policy match, or executable example.

Mark the claims that still require expert judgment. Do not count an explanation as evidence unless it makes one of those checks possible.

Measure the review path before adding friction. Record how long evidence takes to obtain, which tools and permissions it requires, where reviewers stop checking, which failures remain invisible after a green run, and how often approval occurs without accessing the available evidence.

The objective is to identify where the system charges a person for mechanical work or presents confidence without a reachable basis. No universal review-time threshold applies.

Place a forcing function at the acceptance decision for cases in which shallow validation is both plausible and consequential. Require an independent judgment or claim-specific response before revealing the agent's recommendation.

Preserve an emergency bypass with a named owner, reason, defined scope, the evidence available at the time, and a required follow-up state.

Evaluate detection, delay, bypass frequency, and workarounds together. Satisfaction alone will favor the controls that interfere least, not necessarily those that improve judgment.

\Needspace{5\baselineskip}

Build the monitor scaffold from exact checks outward:

\begin{enumerate}
\def\labelenumi{\arabic{enumi}.}
\tightlist
\item
  Use deterministic rules for exact policy violations.
\item
  Use a model judge with a versioned rubric for contextual classification.
\item
  Use an agent judge only when sequence and changing state make it necessary.
\item
  Keep the final verdict with a named person.
\end{enumerate}

Every layer should return the evidence required to challenge its conclusion.

Calibrate the complete stack on an expert-labeled, stratified, held-out set. Choose and record the operating point, then translate its rates into expected queue volume using the deployed action volume and event prevalence. A percentage without the expected number of human cases does not define an operating policy.

Define the escalation route as part of deployment. Every flag should identify the decision owner, outcome at stake, supporting evidence, available options, recommended action, reason automation cannot decide, state currently paused, and event or token that resumes execution.

Test the route by processing a real rejection through it. When the named owner cannot stop, revise, or narrow the action, assign authority before assigning accountability.

After deployment, sample both flagged and unflagged cases. Compare human verdicts with monitor decisions by risk stratum and monitor version. Recalibrate after material changes to agents, tools, rubrics, or workload. Examine whether knowledge of monitoring changes behavior.

Finally, budget practice for the people expected to handle rare failures. A name in an escalation table does not preserve operational skill.

\section*{Sources and evidence}

\textbf{Building verification surfaces so verifying costs less than blind acceptance}

\begin{itemize}
\tightlist
\item
  Strong evidence: Vasconcelos, H., Jörke, M., Grunde-McLaughlin, M., Gerstenberg, T., Bernstein, M., Krishna, R. (2022/2023). Explanations Can Reduce Overreliance on AI Systems During Decision-Making. PACM HCI 7(CSCW1). arXiv:2212.06823. Five experiments treat overreliance as a cost-benefit choice and find greater engagement when verification is cheaper or stakes are higher.
\item
  Directional evidence: Fok, R., Weld, D.S. (2023/2024). In Search of Verifiability: Explanations Rarely Enable Complementary Performance in AI-Advised Decision Making. AI Magazine 45(3). arXiv:2305.07722. Across the XAI-reliance literature, explanations help only insofar as they enable verification.
\item
  Directional evidence: Johnson, B., et al.~(2026). ``Facilitating Trust in AI-assisted Software Tools.'' \emph{ACM Transactions on Software Engineering and Methodology}. \href{https://doi.org/10.1145/3772370}{DOI:10.1145/3772370}. Eighteen interviews and a 368-response survey identify factors shaping trust in software tools; autonomous patch-gate transfer remains untested.
\item
  Directional evidence: Tufano, M., et al.~(2024). ``Code Review Automation: Strengths and Weaknesses of the State of the Art.'' \emph{IEEE Transactions on Software Engineering}. \href{https://doi.org/10.1109/TSE.2023.3348172}{DOI:10.1109/TSE.2023.3348172}. Manual analysis of 2,291 predictions exposes capability boundaries and dataset defects; it does not evaluate this chapter's interface design.
\item
  Corroboration (illustration only): the author's website, ``Two retrieval systems write this site'' (2026-06-10), describes schema-validated agent output landing as a Git diff that serves as the review gate. This author-system case does not count as independent external evidence.
\end{itemize}

\textbf{Putting the forcing function at the accept decision, aimed at acceleration mode}

\begin{itemize}
\tightlist
\item
  Strong evidence: Buçinca, Z., Malaya, M.B., Gajos, K.Z. (2021). To Trust or to Think: Cognitive Forcing Functions Can Reduce Overreliance on AI in AI-assisted Decision-making. PACM HCI 5(CSCW1), 188. arXiv:2102.09692. Cognitive forcing functions reduced overreliance where explanations failed (N=199).
\item
  Directional evidence: Barke, S., James, M.B., Polikarpova, N. (2022/2023). Grounded Copilot: How Programmers Interact with Code-Generating Models. PACMPL 7(OOPSLA1). arXiv:2206.15000. Grounded theory from 20 programmers supports the suggestion-level acceleration/exploration split. Transfer to autonomous-agent diffs is untested. This directional item is the only source for the chapter's targeting rule.
\item
  Corroboration: none on record.
\end{itemize}

\textbf{Escalating only monitor-flagged cases to a human who owns the verdict, with a scaffolded monitor}

\begin{itemize}
\tightlist
\item
  Strong evidence: Kale, N., et al.~(2025). Reliable Weak-to-Strong Monitoring of LLM Agents. arXiv:2508.19461. Targeted escalation adds approximately 15\% TPR at FPR=0.01. Hybrid hierarchical-sequential scaffolds let weaker models monitor stronger agents. The paper reports that scaffolding mattered more than monitor awareness; the awareness caution in this chapter is my design caution, not a result from this paper. The setting is a covert-misbehavior arena, not production code review. This is the sole admitted evidence item for the whole grouping.
\item
  Corroboration (illustration only): the author's website, ``Running my own agent city on Slack'' (2026-06-16), describes mechanical work proceeding autonomously while judgment calls accumulate in a decision ledger for a human. The author's WF2026 field notes describe a three-layer judge stack reported in conference talks. Neither account counts as independent external evidence, and the reviewed source set contains no direct study of that scaffold ordering.
\end{itemize}

\textbf{Companion material carried in-text but not developed as a practice}

\begin{itemize}
\tightlist
\item
  Strong evidence: Rosbach, E., Ganz, J., Ammeling, J., Riener, A., Aubreville, M. (2024). Automation Bias in AI-Assisted Medical Decision-Making under Time Pressure in Computational Pathology. arXiv:2411.00998. This is the basis for the time-pressure paragraph. The setting is computational pathology, not software review.
\item
  Directional evidence: Bainbridge, L. (1983). Ironies of Automation. Automatica 19(6), 775-779. \url{https://doi.org/10.1016/0005-1098(83)90046-8}. This is the single source behind the skill-retention paragraph. The evidence is pre-AI and industrial. Transfer to code review is analogical.
\item
  Corroborating case: Sarkar, A., et al.~(2022). What is it like to program with artificial intelligence? PPIG 2022. arXiv:2208.06213. This is the basis for the companion hypothesis that verification is the dominant cost. The magnitudes are not quantified.
\item
  Directional evidence: Xie, L., Zheng, C., Xia, H., Qu, H., Zhu-Tian, C. (2024). WaitGPT: Monitoring and Steering Conversational LLM Agent in Data Analysis with On-the-Fly Code Visualization. UIST 2024. arXiv:2408.01703. This prototype-scale single item is the basis for the live-state proposal and the thinnest support in the chapter. Transfer to arbitrary code agents requires an operation-level abstraction that may not exist.
\item
  Directional evidence: DeepRare (Zhao 2026); Manager Agent (Masters 2025, arXiv:2510.02557); NIST AI RMF.
\item
  Directional evidence: Bounded Autonomy for Enterprise AI (Sohail 2026, arXiv:2604.14723).
\item
  Directional evidence: AMBIPOM human-LLM co-planning (He 2026, arXiv:2605.23023).
\item
  Directional evidence: Token Budgets (Khan 2026, arXiv:2606.04056). Together, the four synthesis items support only the direction of the checkpoint-placement claim. None is strong.
\item
  Corroboration (illustration only): author-system cases concerning a repository-scale benchmark and a background-agent review pipeline illustrate checkpoint placement. They do not count as independent external evidence.
\end{itemize}

\textbf{Opening scene}

\begin{itemize}
\tightlist
\item
  Corroborating evidence: InfoQ, ``Stripe Engineers Deploy Minions, Autonomous Agents Producing Thousands of Pull Requests Weekly,'' 2026-03-20, \url{https://www.infoq.com/news/2026/03/stripe-autonomous-coding-agents/}. Reports over 1,300 production pull requests per week with human review retained. Carried as scale context; it reports no measure of review depth or defect escape.
\end{itemize}

\chapter{Autonomy calibration, provenance, effective gates, and accountability}
\label{ch17-autonomy-provenance-gates-accountability}
\begin{quote}
\textbf{Evidence profile.} 2 strong \(\cdot\) 5 directional \(\cdot\) 0 corroborating evidence items across 4 developed practices (\erca{145}, \erca{149}, \erca{154}, \erca{165}).

\textbf{Chapter claim.} A gate that cannot change execution records assent and nothing more.
\end{quote}

I built a human-approval queue to hold decisions an agent was not permitted to make. The architecture appeared conservative: the agent could prepare an action, but execution stopped at a human-only boundary.

An audit found paths around that boundary. Some scripts failed open when a required component was missing, and command construction reached execution without the validation the gate was supposed to enforce. The system could report that approval existed while allowing an action to bypass the person. The conservative appearance of the apparatus had concealed a material difference between policy and execution.

The chapter's four entries rest on seven evidence items, two of them strong. The accountability literature represented here consists largely of surveys, frameworks, and position papers, with one practitioner account and little production-grounded measurement. The practices are therefore testable defaults rather than universal policy; the entries on gate effectiveness and alignment between accountability and control have no strong supporting item.

The failed queue illustrates the broader problem. An autonomy policy constrains nothing unless the system enforces the boundary it names. A provenance label changes nothing unless it changes what a reviewer can learn or do. A human gate controls nothing unless the person can alter the execution path. An accountability assignment prevents nothing unless the named person has authority over the outcome.

These are separate design objects, but they fail in the same way when the operational path differs from the policy description.

Agent systems acquire more kinds of authority as they grow. One may restart a service, propose a migration, merge documentation, rotate a credential, or delete a branch. Calling the system ``autonomous'' collapses those different powers into one word. That word says nothing about reversibility, blast radius, available evidence, ownership of state, or the ability to interrupt execution.

\Needspace{5\baselineskip}

The useful unit is the transfer of control for one action class. Each transfer has:

\begin{itemize}
\tightlist
\item
  an initiator;
\item
  a proposed action;
\item
  an approving or executing party;
\item
  an artifact supporting the decision; and
\item
  a recorded outcome.
\end{itemize}

Once those elements are explicit, autonomy can be calibrated from observations, provenance can travel with the artifact, and the gate can be tested as an executable control. Accountability can then be assigned to a role that actually holds authority.

\hypertarget{widen-authority-one-action-class-at-a-time}{%
\section{Widen authority one action class at a time}\label{widen-authority-one-action-class-at-a-time}}

An autonomy ladder widens control one rung at a time for a defined action class. Each class keeps its own approval, modification, and outcome record. A \textbf{promotion threshold} is the criterion for moving that action to a wider rung.

One practitioner article, Priya C (\href{https://devops.com/when-should-a-devops-agent-act-without-human-approval/}{2026}), proposes roughly 95 percent unmodified approvals as a starting policy. It reports no measurement supporting that cutoff. The number has no claim to universality.

A team should instead define the performance difference that would justify wider authority, gather enough observations to detect that difference, and retain the underlying counts. The protocol here rests on one directional practitioner source and one strong controlled study concerning reviewer comprehension.

The ladder needs separate rails because action classes expose different risks. A service restart changes transient process state and can often be reversed with another restart. A credential rotation changes distributed configuration, invalidates clients, and may lock operators out of recovery. A branch deletion changes repository state. A proposed text edit changes only an artifact that still requires merge.

\begin{figure}[htbp]
\centering
\includegraphics{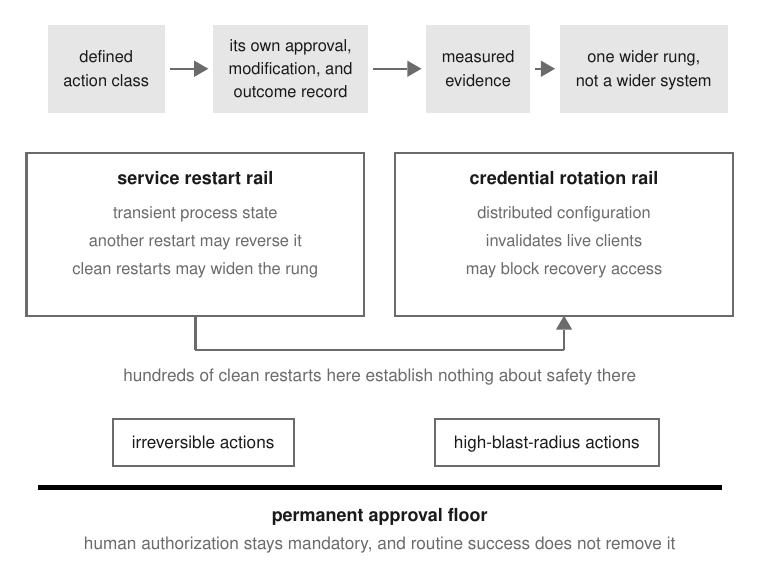}
\caption{Authority widens one rung at a time within a defined action class. Separate rails transfer no evidence between them, and the approval floor beneath them is permanent.}
\label{fig:ch17-authority-rails}
\end{figure}

Hundreds of clean service restarts establish nothing about whether the same system can rotate credentials safely. Figure \ref{fig:ch17-authority-rails} shows the separation and the floor beneath it.

My operator policy makes this separation explicit. Service restarts are pre-approved. Destructive operations such as recursive deletion, force-push, and branch deletion require renewed confirmation every time. The policy is durable configuration, so a new session inherits the same boundary. This illustrates the shape of an action-specific policy; it is not evidence that these classifications suit another system.

\Needspace{5\baselineskip}

A transfer-of-control record documents every decision to let the system act, require approval, or return control to a person. For each action class, it should preserve:

\begin{itemize}
\tightlist
\item
  the proposed action;
\item
  the autonomy rung in force;
\item
  the reviewer's decision;
\item
  any human modification;
\item
  the observed outcome; and
\item
  later reversal or incident evidence.
\end{itemize}

Outcome correctness requires an action-specific definition. For a restart, ``the command exited zero'' is insufficient if the service never became healthy. For a patch, ``merged'' is insufficient if production failure caused an immediate revert.

Approval measures are proportions over repeated observations:

\Needspace*{10\baselineskip}
\begin{align*}
  \text{approval rate} &= \frac{\text{approved proposals}}{\text{reviewed proposals}} \\[4pt]
  \text{unmodified approval rate} &= \frac{\text{proposals approved without human change}}{\text{reviewed proposals}} \\[4pt]
  \text{modification rate} &= \frac{\text{proposals changed by a human}}{\text{reviewed proposals}}
\end{align*}

The denominators must refer to the same action class and autonomy rung. Combining restarts with schema changes can produce a stable aggregate while both underlying rates move in opposite directions. Excluding rejected or timed-out proposals makes the system look more acceptable by removing the transfers that failed. Recording only executed actions introduces survivorship bias because abandoned proposals disappear before the outcome record is built.

Part I established why an observed proportion is not the underlying property. Nine unmodified approvals among ten proposals and ninety among one hundred both produce 90 percent, but the first estimate is much less precise. The interval method depends on the analysis plan; the operational conclusion does not. A small-\(n\) percentage should not transfer authority merely because it crossed a displayed threshold.

The sampling unit also matters. A series of transfers reviewed by one person measures the reviewer-system pair as much as the system itself. A track record collected under one model version, prompt set, or harness revision describes that configuration. Changing any of them reopens the question the record was intended to settle.

Promotion also requires a power analysis. The team first defines the smallest change that would reverse the decision, such as the smallest increase in post-action failures that would make a wider rung unacceptable. It then estimates how many observations are needed to detect that change at the chosen error rates.

Rare, high-consequence failures may require more observations than the deployment can plausibly accumulate. In that case, the action cannot earn wider autonomy from its local record, even when every observed outcome has been clean.

Approval alone omits information captured by modification. A reviewer may approve every database query only after correcting its time window, tenant filter, or target environment. Approval is then 100 percent, even though autonomous execution would repeatedly reproduce a material error.

The modification record should retain what changed. Correcting punctuation and changing a production target have different implications. Modifications can be classified for analysis when the categories and inter-rater reliability procedure are fixed before the rates are examined.

Outcome correctness is a third measure because approval can reflect reviewer behavior rather than system quality. Reviewers under queue pressure may approve familiar actions quickly or reserve comments for severe defects. A rising unmodified approval rate combined with a rising rollback rate argues against promotion. Delayed outcomes must also remain attached to the original transfer, or failures discovered after the review window will be counted as successes.

Chen et al.~(\href{https://arxiv.org/abs/2507.08149}{2025}) complicate a capability-only account of automation. In their controlled study, limited understanding of agent behavior constrained willingness to adopt higher automation even as capability increased. Capability alone did not determine the limit.

\Needspace{5\baselineskip}

The verification surface is therefore part of the autonomy mechanism. A reviewer needs:

\begin{itemize}
\tightlist
\item
  a behavior summary;
\item
  the relevant artifact or diff;
\item
  the intended state transition;
\item
  the observed state transition; and
\item
  the evidence connecting the proposed action to the stated objective.
\end{itemize}

When those artifacts do not support comprehension, even an accurate system may fail to earn legitimate authority.

Comprehension should be tested at the rung under consideration. Ask reviewers to predict affected resources, identify the rollback path, and locate the evidence behind the system's claim before revealing the answer. Record where their understanding diverges from execution. No universal quiz score follows from this design.

The companion entry \texttt{require-comprehension-before-merge} makes the explain-back check explicit. An independent checker should evaluate the generating system's work.

Some action classes remain above a permanent approval floor, below which human authorization is never removed. Irreversible and high-blast-radius actions belong there when the consequence of a mistaken transfer exceeds what the observed record can justify. The floor is not a trust ramp and does not disappear after routine successes.

Least privilege should still constrain the executing identity so that a mistaken approval cannot authorize effects beyond the operation being reviewed.

Most teams should keep most consequential actions below full autonomy today because the supporting evidence remains thin and action-specific. The companion entry \texttt{set-autonomy-defaults-per-task-type} can initialize a policy table, but it cannot create a track record or replace the ladder.

\Needspace{5\baselineskip}

The executable decision is narrower:

\begin{enumerate}
\def\labelenumi{\arabic{enumi}.}
\tightlist
\item
  Define one action class.
\item
  Define a correct outcome and the material difference that would reverse the policy.
\item
  Record every proposed transfer, including rejection and timeout.
\item
  Estimate uncertainty.
\item
  Test reviewer comprehension.
\item
  Widen authority by one rung only when the observations support it.
\end{enumerate}

Two companion entries sit beside this practice. \texttt{measure-oversight-with-decomposed-metrics} separates overreliance from underreliance and classifies review interactions before counting coverage. The evidence for \texttt{attach-reliance-disclaimers} supports only its inclusion as a marked design option.

\hypertarget{put-provenance-where-review-happens}{%
\section{Put provenance where review happens}\label{put-provenance-where-review-happens}}

Tang et al.~(\href{https://arxiv.org/abs/2405.16081}{2024}) observed 28 developers in a laboratory study and found that they were unreliable at recognizing machine-generated code without assistance. When told that the code was generated, participants searched and verified more, and their repair performance improved. Cognitive workload also increased.

This study provides the strong evidence for provenance disclosure in this chapter. It used short code fragments under laboratory conditions.

A provenance label records how an artifact was produced and presents that record where the artifact is reviewed. It may be a pull-request label, a structured commit trailer, or an editor marker.

Recognition is the wrong task on which to spend reviewer attention. Machine-generated code has no stable visual signature. Asking reviewers to infer its origin consumes attention before they inspect its behavior and produces selective scrutiny: obvious generations receive extra review while plausible generations pass as ordinary work.

Attaching provenance at the review surface makes origin an input to the decision rather than a detection task.

The study suggests a specific mechanism. Disclosure caused participants to search and verify more, giving them additional evidence for repair. The label did not establish that the code was defective and performed no verification itself. Its value depended on the behavior it produced in a reviewer who had the tools and time to investigate.

A common misuse ignores that dependence. A warning badge can become a substitute for testing, as though declaring machine authorship discharged the maintainer's responsibility. It can also become a weak liability transfer in which the system announces risk while giving the reviewer no practical way to inspect it.

\Needspace{5\baselineskip}

A useful label connects to:

\begin{itemize}
\tightlist
\item
  the generation context;
\item
  the proposed diff or artifact;
\item
  the tests and verification results;
\item
  later human modifications; and
\item
  the role answerable for integration.
\end{itemize}

None of those artifacts is thereby proven correct.

The disclosure boundary needs a defined object. One pull request may contain human-written scaffolding, generated implementation, generated tests, and later human repairs. A request-level label is simple but loses that composition. A marker on every line is more precise but noisy and dependent on editor and diff support. Commit trailers provide a durable repository unit, although squashing, copying, and cherry-picking can detach the trailer from the code it described.

There is no universal granularity. The team should identify the review decision the label is meant to influence and choose the smallest unit whose provenance survives the repository workflow.

\Needspace{5\baselineskip}

Test that survival through:

\begin{itemize}
\tightlist
\item
  rebases;
\item
  cherry-picks;
\item
  squash merges;
\item
  file moves;
\item
  copied patches; and
\item
  partial adoption of generated work.
\end{itemize}

A provenance control that disappears during the ordinary merge path creates a confident but incomplete history.

My authorship-measurement project illustrates why provenance counts must remain narrow. I measured a 14.5 percent trailer-signed floor, a lower bound on visible trailer-marked authorship, because every detected trailer was treated as a true positive. The figure says nothing about total agent contribution and cannot be averaged with an inferred share.

A preregistered replication could not estimate the share of agent-authored code beyond commits carrying the trailer. I therefore discard the earlier exploratory range and report only the directly observed trailer-marked floor.

Provenance is also distinct from answerability. In my maintainer practice, an adoption pull request can preserve a contributor's commits while stating:

\Needspace*{5\baselineskip}
\begin{record}
\begin{recordbody}
Supersedes #X
Credit: original fix by @X (commit preserved with original authorship)
\end{recordbody}
\end{record}

The original author remains attached to the work. The maintainer who adopts and submits it becomes answerable for the integration decision. Both facts survive in the artifact.

The format establishes attribution rather than correctness. It removes the need for a later reviewer to infer who produced the change and who accepted responsibility for putting it forward.

Persistent provenance also supports incident reconstruction. A reviewer investigating a regression can ask whether the failure arose during generation, human modification, integration, or a later environmental change. That distinction is harder when provenance exists only in an editor and disappears before merge.

The companion entry \texttt{record-steps-in-hash-chained-ledger} describes one tamper-evident history mechanism. The argument here does not depend on that implementation.

Tang et al.'s measured workload cost limits the recommendation. Increased cognitive effort may be acceptable when a marker directs attention to occasional generated work. In a codebase where nearly every change carries the label, constant exposure may tax reviewers and lose salience.

Large diffs, persistent queues, and experienced teams may respond differently over time. Local evaluation should therefore measure reviewer behavior and repair, not label coverage alone.

\Needspace{5\baselineskip}

Useful observations include whether reviewers:

\begin{itemize}
\tightlist
\item
  open referenced files;
\item
  inspect or run tests;
\item
  search documentation;
\item
  change the proposed code;
\item
  detect seeded defects; and
\item
  report excessive workload.
\end{itemize}

Compare equivalent review artifacts with and without disclosure under the actual tools and queue constraints. Using the same artifacts removes change difficulty from the comparison, as Chapter 1 recommends for paired designs.

\Needspace{5\baselineskip}

The experimental unit is the reviewer-artifact pair, not the artifact alone. The design must state:

\begin{itemize}
\tightlist
\item
  how reviewers are assigned to conditions;
\item
  whether any reviewer sees the same artifact twice;
\item
  how condition order is counterbalanced;
\item
  which defects count; and
\item
  how cognitive workload is measured.
\end{itemize}

The study provides no production threshold to copy.

If disclosure increases activity without improving repair, the marker may be producing ritual rather than evidence. If repair improves while queue delay or abandonment rises sharply, the team has found a real tradeoff between scrutiny and capacity. Either result is more useful than a policy requiring an ``AI-generated'' badge everywhere.

The relevant question is whether provenance changes verification in the deployed review environment.

The companion entry \texttt{write-an-agent-contribution-policy} covers repository rules for disclosure, attribution, and accepted contribution paths. A policy can standardize the wire format. It cannot make the marker effective. Effectiveness remains an observed relationship among the label, verification surface, reviewer behavior, and repair outcome.

\hypertarget{prove-that-a-gate-can-change-execution}{%
\section{Prove that a gate can change execution}\label{prove-that-a-gate-can-change-execution}}

Human gates have only directional support in this chapter and no strong evidence item. Sterz et al.~(\href{https://arxiv.org/abs/2404.04059}{2024}) propose an interdisciplinary framework for effective human oversight but provide no empirical validation of a universal test. Green (\href{https://arxiv.org/abs/2109.05067}{2022}) compared 41 government oversight policies with findings from human-computer interaction research and concluded that the human functions prescribed by those policies were generally not performable.

These sources justify an audit structure and a burden of proof. Whether a particular gate works remains unresolved until the deploying team tests it.

In a workflow diagram, the word ``gate'' implies causal control. In execution, an approval step may only record that someone clicked before an action the person and the click could not alter.

\textbf{Compliance theater} is a control whose visible form satisfies the policy description while its execution path does not mitigate the named risk. A \textbf{fail-open gate} stops enforcing its boundary when a dependency, validation step, or error path fails.

\begin{figure}[htbp]
\centering
\includegraphics[width=1\textwidth,height=\textheight,keepaspectratio]{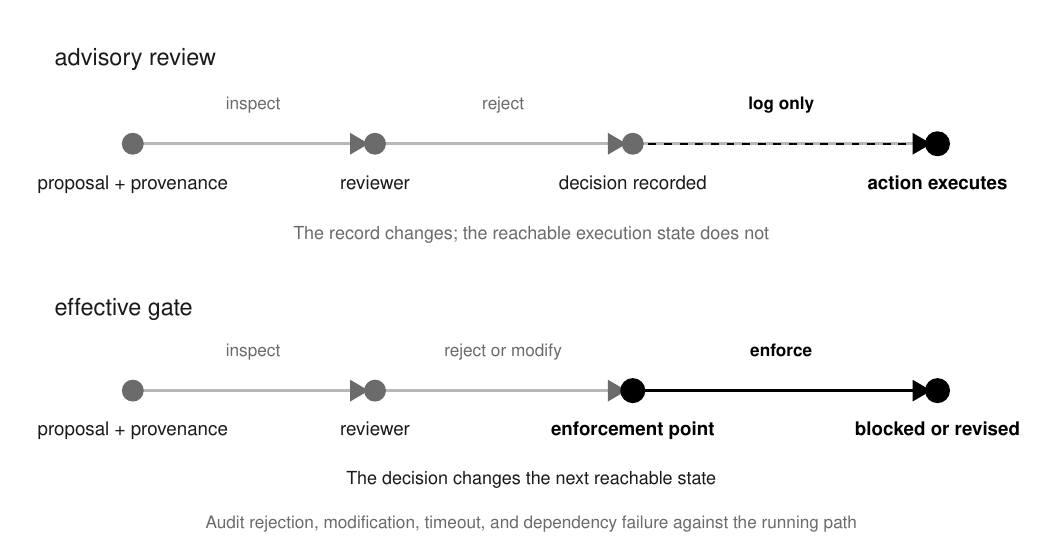}
\caption{An advisory review can record rejection while leaving execution reachable. An effective gate connects the review decision to an enforcement point, so rejection or modification changes the next reachable state.}
\end{figure}

\Needspace{5\baselineskip}

The interdisciplinary framework identifies four conditions for effective oversight:

\begin{itemize}
\tightlist
\item
  \textbf{Causal power:} the person can stop or change the consequential action.
\item
  \textbf{Epistemic access:} the person can obtain the evidence needed to understand the decision before it takes effect.
\item
  \textbf{Self-control:} the person can exercise judgment rather than follow a compelled path.
\item
  \textbf{Fitting intentions:} the person intends and is prepared to perform the assigned oversight function.
\end{itemize}

Evaluate those conditions against an execution trace rather than the policy document. For every named gate, follow the proposed action through approval, mutation, dispatch, and durable state change. Inject the failures the gate claims to handle. Observe whether the reviewer receives the required evidence, whether rejection blocks every path, and whether a modified action requires renewed approval.

\begin{longtable}[]{@{}
  >{\raggedright\arraybackslash}p{(\columnwidth - 6\tabcolsep) * \real{0.2500}}
  >{\raggedright\arraybackslash}p{(\columnwidth - 6\tabcolsep) * \real{0.2500}}
  >{\raggedright\arraybackslash}p{(\columnwidth - 6\tabcolsep) * \real{0.2500}}
  >{\raggedright\arraybackslash}p{(\columnwidth - 6\tabcolsep) * \real{0.2500}}@{}}
\caption{Conditions under which a human gate can exercise effective control.}\tabularnewline
\toprule\noalign{}
\begin{minipage}[b]{\linewidth}\raggedright
Condition
\end{minipage} & \begin{minipage}[b]{\linewidth}\raggedright
Question for the running gate
\end{minipage} & \begin{minipage}[b]{\linewidth}\raggedright
Observation to collect
\end{minipage} & \begin{minipage}[b]{\linewidth}\raggedright
Failure implication
\end{minipage} \\
\midrule\noalign{}
\endfirsthead
\toprule\noalign{}
\begin{minipage}[b]{\linewidth}\raggedright
Condition
\end{minipage} & \begin{minipage}[b]{\linewidth}\raggedright
Question for the running gate
\end{minipage} & \begin{minipage}[b]{\linewidth}\raggedright
Observation to collect
\end{minipage} & \begin{minipage}[b]{\linewidth}\raggedright
Failure implication
\end{minipage} \\
\midrule\noalign{}
\endhead
\bottomrule\noalign{}
\endlastfoot
Causal power & Can the person stop or alter this exact action? & Rejection, cancellation, edit, and timeout traces & The click records assent but does not control execution \\
Epistemic access & Can the person inspect the relevant inputs, artifact, target, and consequences before deciding? & Missing context, inaccessible logs, and reconstruction errors & Approval rests on incomplete state \\
Self-control & Can the person decide without a forced default or impossible queue constraint? & Default acceptance, time pressure, and override use & The workflow substitutes acquiescence for judgment \\
Fitting intentions & Is the role expected and prepared to inspect the named risk? & Review actions, escalation behavior, and interviews & The role is ceremonial or aimed at another risk \\
\end{longtable}

The acceptance criteria must be system-specific. One team may require a deployment approver to cancel a queued release before the first production mutation and to see the exact artifact digest being deployed. Another may require a migration reviewer to revise the plan while withholding direct execution authority.

Name the claimed oversight function, build a trace capable of falsifying it, and decide in advance what observation would establish control.

Removing a dependency from a gate is itself a state change, so this failure test belongs in a contained environment. Chapter 8 supplies that boundary. Auditing a production gate by allowing its negative result to mutate production converts a test into an incident.

My approval queue failed this audit. Missing components allowed scripts to bypass the path that should have waited for a person. Unvalidated command construction weakened the connection between the reviewed proposal and the executed command. The existence of the queue did not establish causal power.

The failure is useful contrary evidence precisely because the queue was designed to preserve human-only decisions.

Another of my systems used a reproduction-before-mutation hook intended to require a reproducible failure before code could change. When a required binary was missing, the hook stopped enforcing the gate. The project documented the fail-open behavior rather than counting installation of the hook as coverage.

The case does not estimate how often gates fail this way. It shows how one environmental dependency can erase a governance boundary.

Mechanical and attentional gates fail differently.

A \textbf{mechanical gate} uses execution state to block an action until a condition holds, such as a verified signature or explicit \texttt{-\/-apply} flag. It may enforce a condition that does not establish the property for which the gate receives credit.

An \textbf{attentional gate} presents evidence to a person and requires a decision, such as review of a diff and its supporting checks. It may obtain a click without scrutiny.

Agent memory is neither. A remembered instruction has no independent causal path to enforcement.

My approval-gate configuration contains seven action-class rows. Each identifies the gate as mechanical or attentional, names an owner, and avoids reliance on agent memory. Mutating commands dry-run by default and require \texttt{-\/-apply}, placing a mechanical boundary between inspection and state change.

This is a design example, not evidence that seven rows, that flag name, or those classifications improve outcomes elsewhere.

More elaborate gates require the same audit. In pull request 1558 of my orchestration system, the review surface recorded:

\Needspace*{5\baselineskip}
\begin{record}
\begin{recordbody}
Latest review attempt: 6
Quality score: 950/1000, threshold 850
\end{recordbody}
\end{record}

The attempt count and score are verified properties of that artifact. The gate was intended to concentrate human attention after repeated model review, and degradation of the gate was treated as an incident.

The claimed attention saving remains unmeasured because maintainer time, scrutiny depth, and escaped-defect rate were not recorded.

Optimizer and model-review gates depend on the same distinction. A threshold may mechanically block execution while providing no assurance that its score tracks the failure a person cares about. The companion entries \texttt{treat-objective-weights-as-reviewed-policy} and \texttt{gate-optimizer-output-behind-human-review} address those policy surfaces.

Their presence does not remove the need to test whether rejection changes system state and whether the reviewer can interrogate the score.

Green's analysis places the burden of proof on the institution deploying the automation. The team should justify why automation is appropriate for the action and demonstrate that reviewers can perform the function assigned to them.

When a reviewer cannot see the relevant evidence or cannot stop the action, attaching their name legitimizes the decision without reducing its risk. Accountability then moves toward the person with the least practical control.

That negative conclusion is bounded. Code review can offer unusually strong epistemic access because a reviewer can inspect an exact diff, run tests, examine history, and delay merge. Government-benefit decisions, live vehicle intervention, and repository changes do not create the same verification problem.

The gate must be tested in its own domain. These sources establish no universal conclusion about either human incapacity or human benefit.

Chapter 8 also showed why a control cannot be accepted merely because the harness reports that it ran. The governance extension asks whether a person can independently inspect the relevant state and interrupt the path that changes it. An override button without an independent signal provides nominal authority without usable control.

Three companion entries sit beside this practice: \texttt{separate-control-from-oversight}, \texttt{write-an-oversight-card}, and \texttt{keep-override-authority-real}. Each rests on one evidence item and can help distinguish control before action, observation after action, stated reviewer duties, and override paths.

None establishes that a particular gate works. That requires a trace showing that the reviewer had the evidence, authority, and execution path needed to change the outcome.

\hypertarget{assign-responsibility-only-where-control-exists}{%
\section{Assign responsibility only where control exists}\label{assign-responsibility-only-where-control-exists}}

Cavalcante Siebert et al.~(\href{https://arxiv.org/abs/2112.01298}{2023}) provide a conceptual bridge from responsibility theory to engineering practice: responsibility should be commensurate with the ability and authority to control an outcome. Suryana et al.~(2025) operationalized related questions in another domain through interviews with 103 users of partially automated driving systems, identifying gaps between expected and actual behavior and inconsistent adherence to operating protocols.

These are two directional literature items, and this practice has no strong supporting evidence item. Organizational-accountability research for agent systems is thin and recent. The sources support a defensible audit method. They provide no compliance threshold and no estimate of failure rates in software delivery.

After an incident, organizations often assign responsibility to the engineer who approved the action, monitored the system, or happened to be on call. That assignment may be administratively clear while remaining causally false. When the person lacked authority to halt the action, access to the relevant state, tools to change it, or time to intervene, responsibility has followed proximity rather than control.

The resulting responsibility gap produces blame without creating a path that could have prevented the failure.

\Needspace{5\baselineskip}

The control check begins before an incident. For every role named as accountable, identify the exact state transitions that role can:

\begin{itemize}
\tightlist
\item
  authorize;
\item
  alter;
\item
  stop; or
\item
  reverse.
\end{itemize}

Test those powers using the role's deployed identity rather than an administrator's demonstration. Record the information available at decision time, the tools through which intervention occurs, and the interval during which intervention remains effective.

Authority, access, tooling, and time are separate requirements. An engineer may have permission to cancel a deployment while lacking access to the tenant, model trace, or configuration needed to know that cancellation is warranted. A rollback command may exist while an irreversible external action commits before the reviewer has time to evaluate the evidence.

A role has control only when those elements combine into an effective intervention path.

\Needspace{5\baselineskip}

When responsibility exceeds control, the organization has two coherent options:

\begin{enumerate}
\def\labelenumi{\arabic{enumi}.}
\tightlist
\item
  Repair the role by granting the required authority, access, tooling, and time, subject to least privilege.
\item
  Move responsibility to the role that already holds those controls.
\end{enumerate}

Leaving the responsibility label in place while operational control remains elsewhere preserves the gap by design.

The check should include the identity that executes the action. In my enterprise data-access service, mutations are unavailable, requests use the caller's identity, and the audit log records that person. The agent cannot silently accumulate authority beyond what its operator already holds.

This design aligns execution and identity more closely than a shared privileged service account. Access justification and caller understanding remain separate checks.

\Needspace{5\baselineskip}

Identity passthrough also carries a cost. It improves attribution and limits silent privilege expansion, but it can reduce availability when the caller lacks a permission that a service account previously supplied. The system should first determine which authority the action genuinely requires, then measure:

\begin{itemize}
\tightlist
\item
  authorization failures;
\item
  escalation requests;
\item
  escalation completion time;
\item
  unauthorized access attempts; and
\item
  cases in which the agent requests authority broader than the task requires.
\end{itemize}

Provenance answers a neighboring but different question. It records who or what contributed to an artifact and how that contribution moved through the workflow. Accountability identifies the person or institution answerable for the decision to accept and deploy it.

The adoption pull-request format shown earlier preserves original authorship while transferring answerability for integration to the maintainer. Conflating the two roles would either erase contribution history or blame a contributor for a deployment decision they did not control.

The literature on meaningful human control offers two additional concepts: \textbf{tracking} and \textbf{tracing}.

Tracking asks whether system behavior follows the relevant and justified reasons or expectations of the people affected by it. Tracing asks whether a capable and aware person can be connected to the decision and its control path.

These concepts become operational only after the team translates them into observable behavior for its own system.

Suryana et al.~applied tracking and tracing through interviews with 103 users of Tesla Autopilot and Full Self-Driving features. The interviews revealed expectation-reality gaps and inconsistent adherence to operating protocols. This evidence comes from driving rather than software delivery; its transferable contribution is the interview method for revealing an operational control model different from the one described in design documents.

\Needspace{5\baselineskip}

For an agent that prepares production changes, a tracking interview might ask:

\begin{itemize}
\tightlist
\item
  What do users believe the agent may modify?
\item
  Which constraints do they expect it to preserve?
\item
  What evidence do they expect to stop execution?
\item
  When do they believe human approval is required?
\item
  What do they believe happens after rejection?
\end{itemize}

Those answers should be compared with policy and with actual execution traces.

\Needspace{5\baselineskip}

A tracing interview might ask:

\begin{itemize}
\tightlist
\item
  Who can explain this action?
\item
  Who could have changed or stopped it at each stage?
\item
  Which identity executed it?
\item
  Who could reverse it?
\item
  Where should a user escalate when behavior diverges?
\item
  What happens if that person is unavailable?
\end{itemize}

Compare the answers across operators, reviewers, system owners, and affected users. Inconsistent answers are evidence that the operational responsibility model is not shared.

The audit identifies distinct failures rather than producing one certification score. A shared credential may erase the responsible identity even when the system tracks user expectations well. A complete decision history may coexist with systematic misunderstanding about when the system will act. Combining both into one ``meaningful control'' score would hide the repair each problem requires.

\Needspace{5\baselineskip}

Operational definitions must therefore remain local. A deployment system may define a traceable decision through:

\begin{itemize}
\tightlist
\item
  the artifact digest;
\item
  approver identity;
\item
  execution identity;
\item
  target environment;
\item
  authorization token; and
\item
  rollback authority.
\end{itemize}

\Needspace{5\baselineskip}

A data-access agent may require:

\begin{itemize}
\tightlist
\item
  caller identity;
\item
  query purpose;
\item
  source authorization;
\item
  returned-field records; and
\item
  a durable record of disclosure.
\end{itemize}

A partially automated vehicle has different timing, embodiment, and safety constraints. The questions about authority, awareness, and intervention transfer across domains. The measured rates and exact controls do not.

Collective work complicates responsibility without removing it. Models propose actions, tools execute them, reviewers approve them, platform teams configure permissions, and managers allocate review capacity. The companion entry \texttt{rebuild-collective-accountability} addresses how those contributions form an institutional account, while \texttt{scale-review-capacity-to-experience} addresses review planning by contributor experience.

Neither should diffuse answerability so widely that no role owns a state transition it can actually stop.

\hypertarget{audit-the-path-from-policy-to-execution}{%
\section{Audit the path from policy to execution}\label{audit-the-path-from-policy-to-execution}}

Run four checks against one defined action class and one real repository workflow. The first asks whether the autonomy rung the system claims is the rung it operates at. The second follows a provenance marker attached at generation and asks whether it still identifies responsibility after the repository has rebased, squashed, and moved the code. The third asks whether a gate that reports a decision can change what executes. The fourth separates the people who hold control from the people the record names.

Run them in that order. A provenance marker is worth testing only once the action class is defined, and responsibility can be assigned only after the gate is known to bind.

\hypertarget{test-the-autonomy-transfer}{%
\section{Test the autonomy transfer}\label{test-the-autonomy-transfer}}

Define the action class, current autonomy rung, correct outcome, material difference that would justify wider authority, and failures that would prevent promotion.

Record every proposed transfer, including rejected, modified, timed-out, abandoned, and executed proposals. Do not combine action classes or autonomy rungs, and do not build the record from executed actions alone.

Calculate uncertainty around approval, modification, and outcome rates. Test whether reviewers understand the affected state, evidence, and rollback path. Widen authority by one rung only when those observations support the change.

\hypertarget{test-whether-provenance-survives-review-and-integration}{%
\section{Test whether provenance survives review and integration}\label{test-whether-provenance-survives-review-and-integration}}

Choose the smallest provenance unit that can survive the repository workflow while still representing the review decision it is intended to affect.

Exercise that marker through rebases, cherry-picks, squash merges, file moves, copied patches, partial adoption, and later human modification.

Verify that a reviewer can distinguish generation, human revision, and integration responsibility after each transformation.

Keep the count semantics narrow. The 14.5 percent trailer-signed floor described earlier is a lower bound on visible trailer-marked authorship. It must not be combined with an inferred contribution share.

\hypertarget{test-the-gate-as-a-failure-experiment}{%
\section{Test the gate as a failure experiment}\label{test-the-gate-as-a-failure-experiment}}

Run the gate audit in a contained environment. Name the action, gate, and decision owner, then capture the complete execution trace.

Remove or fail each gate dependency in turn. Exercise approval, rejection, modification, cancellation, timeout, missing validation, missing binaries or services, stale approval tokens, and attempts to execute through alternate paths.

Inspect the evidence available to the reviewer and verify the durable state produced by each decision. A rejected action must fail to execute through every path. A modified action must receive renewed review when the modification changes the authorized state transition.

Interview the assigned reviewer about the risk they believe they are checking, then compare that account with the actual execution path. Repair or remove a gate that cannot demonstrate causal power and epistemic access. Decide explicitly how queue pressure, defaults, incentives, and emergency procedures will preserve self-control and fitting intentions.

\hypertarget{audit-accountability-in-two-passes}{%
\section{Audit accountability in two passes}\label{audit-accountability-in-two-passes}}

\hypertarget{pass-one-verify-operational-control}{%
\paragraph{Pass one: verify operational control}\label{pass-one-verify-operational-control}}

For each role named as accountable, select a real action and test the role's deployed permissions.

Exercise whether the role can reject the action, alter its scope, stop execution, inspect the relevant evidence, initiate recovery, and reverse the result where reversal is part of the claimed control.

Record the state transitions outside that role's authority, the information it cannot access, the tools it lacks, and the point after which intervention becomes ineffective.

Where control is missing, either repair the role within least-privilege constraints or reassign responsibility to the role that holds the necessary authority.

\hypertarget{pass-two-compare-the-stated-and-experienced-control-models}{%
\paragraph{Pass two: compare the stated and experienced control models}\label{pass-two-compare-the-stated-and-experienced-control-models}}

Interview actual users, operators, reviewers, and owners using tracking and tracing questions. Compare their justified expectations with policy and execution traces.

Record expectation mismatches separately from protocol deviations, untraceable actions, reviewers who cannot explain an approval, responsibility assigned without intervention power, ineffective escalation routes, and delays that allowed an action to become irreversible before review.

Repeat the first pass whenever permissions, action classes, execution paths, or recovery mechanisms change. Repeat the second when the user population, operating protocol, interface, or escalation policy changes.

Measure the findings that carry operational weight: untraceable actions, failed stops, unauthorized execution paths, expectation mismatches, dead-end escalations, stale approvals, late interventions, and time lost before someone with effective control becomes involved.

Do not adopt a pass rate borrowed from another system. Decide in advance which failures invalidate the accountability claim and preserve the identities, traces, interviews, and execution evidence required to reproduce that decision.

\section*{Sources and evidence}

Support is thin across the chapter: four developed practices carry seven evidence items, of which two are strong and five directional; six are scholarly sources and one is a practitioner source. \texttt{audit-human-gates-for-effectiveness} and \texttt{align-accountability-with-actual-control} carry no strong evidence item.

\textbf{graduate-autonomy-per-action-track-record}

\begin{itemize}
\tightlist
\item
  Directional evidence: ``When Should a DevOps Agent Act Without Human Approval?'', Bala Priya C, DevOps.com, 2026-05-11, \href{https://devops.com/when-should-a-devops-agent-act-without-human-approval/}{article}. Supports action-type autonomy promotion at roughly 95\% unmodified approval, permanent approval floors for irreversible or high-blast-radius actions, and decision-ready gates with timeout plans. The protocol comes from one practitioner article; the threshold is stated, not measured.
\item
  Strong evidence: Chen, V., Talwalkar, A., Brennan, R., Neubig, G. (2025). Code with Me or for Me? How Increasing AI Automation Transforms Developer Workflows. arXiv:2507.08149. In the controlled study, users' failure to understand agent behavior, not capability, limited broader adoption.
\item
  Corroboration (narrative only): a demoted author-distilled field note about a 2025 lights-off agent factory failure, the operator policy, and an essay describing per-action autonomy with a batched morning decision ledger. The field note is not independent support.
\end{itemize}

\textbf{label-ai-provenance}

\begin{itemize}
\tightlist
\item
  Strong evidence: Tang, N., Chen, M., Ning, Z., Bansal, A., Huang, Y., McMillan, C., Li, T. J.-J. (2024). A Study on Developer Behaviors for Validating and Repairing LLM-Generated Code Using Eye Tracking and IDE Actions. arXiv:2405.16081. Supports improved repair performance and greater verification effort with provenance disclosure; recognition without disclosure was unreliable. Limited by a small laboratory sample and code chunks of tens of lines; the attention cost may not scale to constant exposure.
\item
  Corroboration: none on record.
\end{itemize}

\textbf{audit-human-gates-for-effectiveness}

\begin{itemize}
\tightlist
\item
  Directional evidence: Sterz, S., et al.~(2024). On the Quest for Effectiveness in Human Oversight: Interdisciplinary Perspectives. ACM FAccT 2024. arXiv:2404.04059. Supports four conditions for effective oversight persons; gates failing any condition are compliance theater. The framework is motivated by EU AI Act Article 14, not empirically validated.
\item
  Directional evidence: Green, B. (2022). The Flaws of Policies Requiring Human Oversight of Government Algorithms. Computer Law \& Security Review 45, 105681. arXiv:2109.05067. Compares 41 policies with the human-computer interaction record and shifts the burden of proof to the deploying institution. The negative claim transfers only where reviewers lack real verification surfaces; code review differs in verifiability.
\item
  Corroboration (narrative only, contrary): the human-approval queue audit found fail-open scripts and unvalidated command construction, while the reproduction-before-mutation hook stops enforcing when a required binary is absent.
\end{itemize}

\textbf{align-accountability-with-actual-control}

\begin{itemize}
\tightlist
\item
  Directional evidence: Cavalcante Siebert, L., et al.~(2023). Meaningful human control: actionable properties for AI system development. AI and Ethics 3, 241-255. arXiv:2112.01298. Supports responsibility commensurate with ability and authority to control, the framework's third actionable property.
\item
  Directional evidence: Suryana, L. E., Nordhoff, S., Calvert, S., Zgonnikov, A., van Arem, B. (2025). Meaningful human control of partially automated driving systems: Insights from interviews with Tesla users. Transportation Research Part F 113, 213-236. Applies tracking and tracing criteria to 103 users to localize expectation-reality gaps and inconsistent protocol adherence. The method requires case-specific operationalization, yields failure localization rather than a compliance score, and is evidenced here in the driving domain. The record carries no arXiv identifier; the published article is doi:10.1016/j.trf.2025.04.026.
\end{itemize}

\part{Research agenda: work allocation and cost engineering}
\gdef\currentparttitle{Part VI: Research agenda}

\chapter{Agent topology selection and dynamic task allocation}
\label{ch18-agent-topology-dynamic-task-allocation}
\begin{quote}
\textbf{Evidence profile.} 6 strong \(\cdot\) 16 directional \(\cdot\) 1 corroborating \(\cdot\) 0 null or conflicting evidence items across 4 developed practices (\erca{044}, \erca{093}, \erca{101}, \erca{107}). One additional source establishes historical lineage.

\textbf{Chapter claim.} Coordination must justify its cost against a live single-agent baseline.
\end{quote}

In one CodeProbe rerun (\href{https://github.com/sjarmak/codeprobe}{public repository}), I changed an agent's preamble to correct a retrieval behavior I had already diagnosed. The patch reduced cost while leaving the motivating failure intact.

On the affected task family, cost fell by 28 percent, while the primary reward difference was +0.0048 with a t statistic of 0.27, indistinguishable from noise in that comparison. The wall-clock prediction failed in both direction and magnitude. I had predicted a 40 percent reduction, but elapsed time increased by 3.9 percent. The agent still accepted a false-negative retrieval result and constructed its answer around missing evidence.

The patch changed the wrong variable. The preamble instructed the agent to synthesize for coverage, but the execution path still treated an empty tool result as authoritative. That coupling belonged to the harness rather than to the sentence-level wording of the prompt.

Prompt changes are cheap, visible, and easy to isolate in an experiment. Structural changes alter component boundaries: which worker retrieves evidence, which worker interprets it, what crosses the handoff, and which component can reject the result. They are more expensive to implement and harder to evaluate because several causal paths may move together. They are also sometimes the only intervention aimed at the point where the failure is produced.

\hypertarget{when-does-a-persistent-failure-require-structural-repair}{%
\section{When does a persistent failure require structural repair?}\label{when-does-a-persistent-failure-require-structural-repair}}

A failure class does not become persistent merely because it occurred more than once. A class counts as persistent only after a paired comparison across model versions, with repeated trials, shows that it survives. One successful run on the newer model proves no more than one failed run on the older one.

Chapter 1's variance discipline still applies when the proposed treatment is a model upgrade. Analyze paired per-item differences rather than comparing two aggregate scores.

Once persistence is established, Chapter 11's taxonomy and first-upstream-failure method locate the earliest supported cause. That location constrains the repair. If retrieval produced an incomplete candidate set, adding a stronger instruction for the final writer to cite sources attacks the last visible symptom. The relevant repair belongs where the candidate set is formed, checked, or handed off.

The companion catalog states the broader rule as repairing the supported root cause rather than the final visible error. Its evidence is thin, so it serves here only as a diagnostic reminder.

Cemri et al.~(\href{https://arxiv.org/abs/2503.13657}{2025}) developed MAST, a taxonomy of failures in multi-agent language-model systems. They reported limited gains from prompt improvements on persistent classes and stronger results from interventions such as redesigned verification topology and modular roles while retaining the same underlying model.

The evaluated systems included benchmark frameworks such as AG2 and ChatDev rather than production deployments. The findings therefore establish plausible structural interventions within those frameworks, not portable recipes for arbitrary agent systems.

Kim et al.~(\href{https://arxiv.org/abs/2602.09937}{2026}) provide a stronger cross-version result in another domain. Across the capability tiers evaluated in OpenRCA, dominant failure modes in cloud root-cause analysis remained present. Prompt engineering did not eliminate communication failures, while richer communication protocols reduced them by as much as 15 percentage points.

That is evidence that an architectural failure can survive a more capable model. It does not establish a coding-agent failure rate or show that prompt engineering is generally ineffective.

The OpenRCA result separates two failure mechanisms. Protocol enrichment addressed communication failures, whereas hallucinated interpretation required another intervention. A structured handoff cannot repair a worker that invents the meaning of accurate evidence, and a stronger verifier cannot recover evidence no upstream component retrieved.

Oskooei et al.~(\href{https://arxiv.org/abs/2608.01507}{2026}) measured one such boundary on SWE-QA. Semantic search answered 65.2 percent of repository questions correctly, compared with 46.2 percent for planner-to-subagent search, at less than half the cost per correct answer. Handoff loss accounted for 41.8 percent of the subagent system's failures. This is strong evidence for the compared configurations on SWE-QA and directional evidence for topology selection beyond read-only repository questions.

``Change the structure'' is therefore not a remedy. It is an instruction to revisit the causal boundary that produced the observed class.

A \textbf{multi-agent system} consists of several model-driven workers coordinating on one task. Its \textbf{topology} determines which workers exchange observations, where decisions converge, and which component owns shared state. Topology is distinct from implementation. Two programs can use the same queue library while exposing different communication structures, and two runtimes can implement the same topology.

Role specialization is one possible structural intervention. It assigns workers different responsibilities, evidence access, and acceptance conditions instead of asking each worker to perform the complete task.

In the opening retrieval case, specialization could separate candidate discovery from evidence adjudication. The retrieval worker would return candidates together with explicit coverage information. An adjudicator would decide whether an empty result supported a negative conclusion before the writer could treat absence of results as absence of evidence.

The specialization is justified only because it creates a boundary at which the false-negative claim becomes visible and rejectable. Adding personas without changing evidence or authority does nothing.

Structured handoffs can address the same coupling:

\Needspace*{8\baselineskip}
\begin{record}
\begin{recordbody}
query: exact request issued to the retriever
candidates: identifiers returned
coverage_checks: alternate queries and their results
claim: sender's interpretation
status: supported | incomplete | failed
\end{recordbody}
\end{record}

The schema does not make the interpretation correct. It keeps evidence, interpretation, and completion status separate. An incomplete search can no longer become indistinguishable from a supported negative result unless another component explicitly makes that conversion.

During failure, the status must remain \texttt{failed} or \texttt{incomplete}. Converting it to an empty success recreates the original defect behind a cleaner interface.

\Needspace{5\baselineskip}

The structural boundary may also be a verification loop rather than another worker. One agent can produce an artifact, run a deterministic check, inspect the failure, and revise. An independent reviewer becomes useful when:

\begin{itemize}
\tightlist
\item
  the producer is poorly positioned to detect its own mistake;
\item
  the reviewer needs independent evidence access;
\item
  verification policy should remain isolated from generation; or
\item
  the consequence warrants a separate decision owner.
\end{itemize}

Role specialization describes distinct responsibilities, not necessarily distinct model instances.

The first structural question is therefore:

\begin{quote}
Which decision requires an independent state or authority boundary?
\end{quote}

The number of agents follows from that answer. Sometimes no additional worker is needed. Sometimes one verifier is enough. Five workers sharing the same incomplete evidence can amplify confidence without changing the failure path.

Accounts grouped under ``harness engineering'' describe teams accumulating capability in execution boundaries, verification loops, repository instructions, rollback paths, and approval controls rather than in prompts alone. Böckeler (\href{https://martinfowler.com/articles/exploring-gen-ai/harness-engineering.html}{2026}) analyzes vendor-reported practice from the OpenAI Codex team. Jain (\href{https://www.reddit.com/r/devops/comments/1touxz4/}{2026}) writes from practitioner experience rather than a controlled evaluation.

Neither source supplies enough verification detail to estimate an effect. They describe where practitioners report making repairs. Cemri et al.~and Kim et al.~carry the evidentiary claim that some persistent classes responded to structural intervention.

Lin et al.~(\href{https://arxiv.org/abs/2604.25850}{2026}) add preprint evidence pointing the same way. Across ten harness iterations, Terminal-Bench 2 pass@1 rose from 69.7 percent to 77.0 percent with the model held fixed, and their ablations locate the gains in tools, middleware, and memory rather than in the system prompt alone. That is a benchmark result under one team's iteration process. I take from it the direction, that structural elements of the harness carried measurable capability, and I do not transfer the magnitude to production systems.

My trace data adds a caution about what prompt evidence can establish. Across 1,705 visible thinking blocks from 199 traces, agents never named the two dominant tools in their written reasoning. Sixty-five explicit intentions to find references produced no use of the purpose-built reference tool.

Zero mentions did not imply zero prompt influence. In a later 17-run experiment, deliberation-style preambles changed tool selection on matching tasks while scores remained 4.0 versus 4.0.

These observations separate two effects. Prompt steering may improve selection efficiency without improving task success. The target retrieval failure in my rerun still survived. Another wording change was therefore not the primary repair. The system needed a boundary that checked retrieval coverage before synthesis trusted the result.

\Needspace{5\baselineskip}

Structural repair carries costs. Splitting work introduces:

\begin{itemize}
\tightlist
\item
  handoff failures;
\item
  additional tokens and latency;
\item
  more state identities;
\item
  more opportunities for stale or conflicting evidence;
\item
  false rejection by verifiers; and
\item
  loss of context at specialist boundaries.
\end{itemize}

A protocol may omit the field later needed for diagnosis, while a specialist may lose information a generalist would have retained. Representing topology in configuration may make revision easier, but that companion proposal has thin support and an unmeasured maintenance burden.

\Needspace{5\baselineskip}

The repair criterion is causal and comparative:

\begin{enumerate}
\def\labelenumi{\arabic{enumi}.}
\tightlist
\item
  Change the smallest boundary capable of interrupting the observed failure path.
\item
  Preserve the failure evidence through that boundary.
\item
  Replay the same cases under paired repeated trials.
\item
  Measure whether the target class moved.
\item
  Inspect which new classes appeared.
\end{enumerate}

If the class does not change, the new structure added coordination without repairing the cause.

\hypertarget{when-does-fan-out-beat-the-live-single-agent-baseline}{%
\section{When does fan-out beat the live single-agent baseline?}\label{when-does-fan-out-beat-the-live-single-agent-baseline}}

Every proposal to add debate, delegation, or parallel workers needs a live single-agent control.

\textbf{Fan-out} assigns one task, or parts of it, to several workers whose outputs must later be selected or combined. \textbf{Delegation} transfers responsibility for a defined unit of work to another worker. Neither mechanism deserves credit merely because the configuration makes it available.

The golden set from Chapter 4 supplies the comparison. Replay the current single-agent workflow and the proposed team on the same versioned tasks and executable oracles. Record the promotion rule before the first run. Use the same number of repeated trials in both arms and score reliability through \(\mathrm{pass}^{k}\) where that metric matches the deployment semantics.

The multi-agent treatment advances only when it clears both the success requirement and the relevant cost limit. Chapter 2's Pareto discipline rejects a quality gain whose operating cost makes the workflow unusable.

The gate does not presume that fan-out will fail. Work with independent branches, high error costs, or reasoning demands beyond one reliable agent trajectory may clear it. Those properties make fan-out plausible. They do not remove the need to measure whether coordination and aggregation consume the expected benefit.

Chun et al.~(\href{https://arxiv.org/abs/2503.12029}{2025}) compared debate configurations with task baselines and reported the inference cost of the debate variants themselves. They did not compare debate with a single-model baseline on cost and identified that comparison as an open question. Debate therefore receives no presumed efficiency benefit.

Li et al.~(\href{https://arxiv.org/abs/2606.00655}{2026}) provide a controlled cost result from SIMAS. Their multi-agent configurations used roughly 15 times as many tokens, returns were non-monotonic as agents were added, and debate sometimes underperformed self-correction.

These are strong results within the tested tasks and configurations. They do not supply a universal exchange rate for another worker. Token premiums depend on prompt length, number of rounds, context sharing, model mix, and duplicated work.

A strong source synthesis across several studies supports the general need for a gate, but no standardized experiment settles it. Other directional syntheses cover heterogeneous and homogeneous teams, consensus systems, and scaling networks: Tian et al.~(\href{https://arxiv.org/abs/2509.23537}{2025}), Kumar et al.~(\href{https://arxiv.org/abs/2604.13120}{2026}), Bertalani\v{c} et al.~(\href{https://arxiv.org/abs/2605.00914}{2026}), and Qian et al.~(\href{https://arxiv.org/abs/2406.07155}{2024}).

Because those syntheses overlap in their underlying sources and were not retained as strong controlled evidence, I use them only for direction. The operational decision still comes from reproducing the comparison on the target workload.

Aggregation belongs inside the treatment definition. Chapter 5 described an exploratory plurality-vote oracle gap of up to 32.3 percentage points from Bertalani\v{c} et al.~A correct answer existed among the candidates, but the vote selected another answer. That establishes a failure mode rather than its frequency elsewhere.

\Needspace{5\baselineskip}

It is enough to separate two outcomes:

\begin{itemize}
\tightlist
\item
  at least one worker produced a correct candidate; and
\item
  the system returned the correct candidate.
\end{itemize}

Suppose five workers review a code change. Three repeat the same shallow diagnosis, one identifies a real race condition, and one reports an unrelated issue. Plurality voting selects the shallow diagnosis even though the candidate set contains the critical finding.

\Needspace{5\baselineskip}

A synthesis worker may recover the minority report, but it can also suppress it when all findings arrive as undifferentiated prose. The evaluation must therefore:

\begin{itemize}
\tightlist
\item
  score the final system output;
\item
  retain candidate-level results;
\item
  record the aggregation decision; and
\item
  report whether the aggregator discarded an oracle-correct candidate.
\end{itemize}

The same requirement applies to debate. Later turns can make workers converge by exposing them to one early answer even when independent reasoning never resolves the problem. If independent judgment is the reason for adding reviewers, the treatment must preserve that independence long enough to test it.

Blind review lanes, in which reviewers do not see the producing worker's identity or conclusion, are a thinly supported companion option. They reduce one obvious coupling path but are not an established remedy for correlated errors.

\Needspace{5\baselineskip}

Cost measurement must include the full topology:

\begin{itemize}
\tightlist
\item
  planner input and output;
\item
  worker context and generation;
\item
  inter-worker messages;
\item
  retries;
\item
  failed branches;
\item
  synthesis;
\item
  verification; and
\item
  the final response.
\end{itemize}

Measure latency separately from total computation. Parallel execution may reduce wall-clock time while increasing tokens and monetary cost. That trade may be justified for a high-consequence incident and unacceptable for routine maintenance.

Chapter 19 develops routing and cost engineering in more detail. The minimum requirement here is to choose the cost axis that matters, measure it end to end, and record the acceptable increase before seeing the treatment result. A ceiling selected afterward can always be adjusted to preserve the favored architecture.

One validity check comes before any outcome comparison: the multi-agent mechanism must actually execute.

In my repository-scale benchmark, I configured a lean-subagent treatment across four arms and roughly 200 trials, including investigation tasks, and observed zero subagent spawns. The result showed that the treatment never entered the causal path. It did not show that subagents were ineffective.

Configuration is not behavior. A system may expose a delegation tool while the planner never invokes it. It may dispatch workers and ignore their responses. It may ask several workers questions that do not meaningfully partition the task. All can be labeled ``multi-agent'' while operating as an effectively single-agent system.

\Needspace{5\baselineskip}

Instrument the treatment at three levels:

\begin{enumerate}
\def\labelenumi{\arabic{enumi}.}
\tightlist
\item
  \textbf{Configuration:} which topology and tools were available.
\item
  \textbf{Execution:} delegations, messages, worker completions, retries, and aggregation decisions.
\item
  \textbf{Outcome:} how those events contributed to the final result and cost.
\end{enumerate}

A run enters the multi-agent comparison only when the trace shows that the defining coordination behavior occurred.

That eligibility rule must be declared before execution. Writing it afterward selects runs based on observed behavior or outcome. Declared in advance, it distinguishes noncompliance from architectural failure.

\Needspace{5\baselineskip}

The distinction guides diagnosis:

\begin{itemize}
\tightlist
\item
  No delegation means the planner or tool boundary failed to activate the treatment.
\item
  Useful branches discarded by aggregation indicate a convergence failure.
\item
  Identical branch errors from shared context indicate a common upstream dependency.
\item
  Divergent correct candidates with a wrong final answer indicate a selection failure.
\item
  Expensive duplicate work with no gain indicates a decomposition failure.
\end{itemize}

The single-agent baseline must remain live as models change. A topology that once improved performance may become unnecessary when a stronger model absorbs the relevant capability. It may also become more valuable when the stronger model becomes a better planner while specialists remain inexpensive.

Replaying both configurations for each consequential release turns that shift into a measured comparison. Without the single-agent arm, improving team scores may conceal that the simpler system improved faster.

A compact promotion record can contain:

\Needspace*{31\baselineskip}
\begin{record}
\begin{recordbody}
golden_set_version: repository-maintenance-2026-07

single_agent:
  all_k_success:
  total_tokens:
  elapsed_time:
  failures_by_class:

multi_agent:
  all_k_success:
  total_tokens:
  elapsed_time:
  failures_by_class:

mechanism_check:
  delegations_observed:
  branches_consumed:
  aggregator_used:

aggregation_check:
  correct_candidates_retained_or_discarded:

promotion_rule:
  success_floor:
  cost_ceiling:

decision:
  promote | retain baseline | redesign treatment
\end{recordbody}
\end{record}

This gate answers whether to divide the work. Topology selection answers how. A multi-agent design that cannot clear the gate should return to structural diagnosis. The next experiment may require a narrower specialist boundary, a different aggregator, or no additional worker.

\hypertarget{which-topology-fits-the-task-and-contains-its-faults}{%
\section{Which topology fits the task and contains its faults?}\label{which-topology-fits-the-task-and-contains-its-faults}}

Jia et al.~(\href{https://arxiv.org/abs/2602.19843}{2026}) injected synthetic faults into three multi-agent architectures in MAS-FIRE. Closed-loop architectures neutralized more than 40 percent of faults that caused the linear pipeline to collapse, and stronger foundation models did not uniformly produce greater robustness.

The faults were synthetic and the comparison covered three architectures. The result does not establish a general preference for closed loops. It does show that message routes and feedback paths can determine whether a local error remains local or becomes a system failure.

A \textbf{pipeline} passes work through a fixed sequence of stages. It fits tasks that decompose into stable ordered transformations: retrieve candidate files, extract the relevant regions, generate a patch, run the tests. Its advantage is inspectability. State moves in one direction, and each transition has a defined producer and consumer. Its weakness is the same property. An early omission can flow through every later stage without any component returning to the source.

Pipelines also compress meaning at each handoff. If retrieval emits only file excerpts, the patch stage may never learn that coverage was uncertain. A status field can preserve that uncertainty, but it does not create a return route. When downstream evidence can invalidate an upstream decision, the task no longer has a purely linear structure even if its implementation remains a sequence of functions.

An \textbf{orchestrator-worker topology} assigns one component responsibility for planning, dispatch, and integration while workers perform bounded tasks. It fits work whose decomposition depends on the initial state but whose branches can be evaluated separately.

A repository migration may require the orchestrator to identify affected packages, send independent compatibility checks to workers, and combine the findings into an ordered plan.

The orchestrator owns global task state. Workers should receive enough evidence for their assignments without inheriting the complete history by default. Their outputs return through a defined convergence point where conflicts and missing work can be detected.

Failure of the orchestrator has a larger blast radius than failure of one worker. Durable state, stable identities, and replay at that boundary therefore matter more to reliability than adding another worker.

A \textbf{hierarchy} or explicit leadership structure extends this arrangement when a flat team would exchange too many messages or when subproblems require their own coordination. A lead assigns work to subordinate groups and accepts bounded summaries rather than allowing every worker to communicate with every other worker.

This can reduce message growth. It also introduces information loss and concentrates authority. A hierarchy repays its cost only when the lead measurably reduces communication and integration burden without suppressing critical evidence. Titles assigned to the agents establish nothing.

A \textbf{blackboard} is shared evolving state that workers read and update. It fits tasks whose next useful action depends on discoveries made by any participant, such as an incident investigation in which logs, hypotheses, eliminated causes, and requested checks change throughout the run.

The blackboard externalizes state that would otherwise be scattered across pairwise conversations. It also creates concurrency obligations. Updates require identities, versions, conflict rules, provenance, and a distinction between an untested hypothesis and an established observation.

\Needspace{5\baselineskip}

Consider a security review of a large change:

\begin{itemize}
\tightlist
\item
  A pipeline fits fixed analyzers followed by one synthesis step.
\item
  An orchestrator-worker design fits a planner dividing the diff by attack surface and assigning bounded reviews.
\item
  A hierarchy becomes plausible when package leads coordinate their own specialist groups.
\item
  A blackboard fits an exploratory investigation in which one discovery should redirect every other reviewer.
\end{itemize}

These topologies can be combined. An orchestrator may dispatch pipelines. A hierarchy may use a blackboard for cross-group evidence while keeping control messages within leadership paths.

The relevant architectural properties are ownership of each state transition and the participants able to observe, challenge, or revise it; a configuration label alone does not establish either property.

Zhao et al.~(\href{https://arxiv.org/abs/2605.26178}{2026}) provide strong experimental evidence from ATOM that topology is a first-order variable and that adapting it to task difficulty can outperform one average design. Their evaluation used five-agent teams with two open models across MMLU, GSM8K, HumanEval, AQuA, MultiArith, and SVAMP.

Those are short-form question-answering, mathematics, and code benchmarks. The study supports difficulty-conditioned orchestration within that scope.

Topology-to-task matching remains unvalidated for long-horizon repository work, where files persist, tools have side effects, branches remain partially complete, tests are expensive, and concurrent edits conflict. The topology menu is therefore an engineering hypothesis to test rather than a benchmark conclusion to import.

Two directional research lines expand the vocabulary without resolving the transfer. Huang and Zhou (\href{https://arxiv.org/abs/2605.13850}{2026}) organize agent design patterns along two dimensions. Shang (\href{https://arxiv.org/abs/2604.08206}{2026}) relates blackboard coordination to a global workspace in Theater of Mind. Both provide exploratory support without a quantitative claim here.

An active-blackboard variant that schedules contributors for diversity remains a thinly supported companion idea rather than an established practice.

Fault propagation provides a practical topology test.

In a pipeline, a retrieval error may contaminate every downstream artifact. In an orchestrator-worker system, the same error may remain local when the orchestrator compares workers against independent evidence, or become global when it broadcasts one corrupted premise to all of them.

In a hierarchy, a lead's inaccurate summary can erase correct subordinate findings. On a blackboard, an unsupported claim can spread through repeated reads unless provenance and status remain attached.

\begin{figure}[tbp]
\centering
\includegraphics{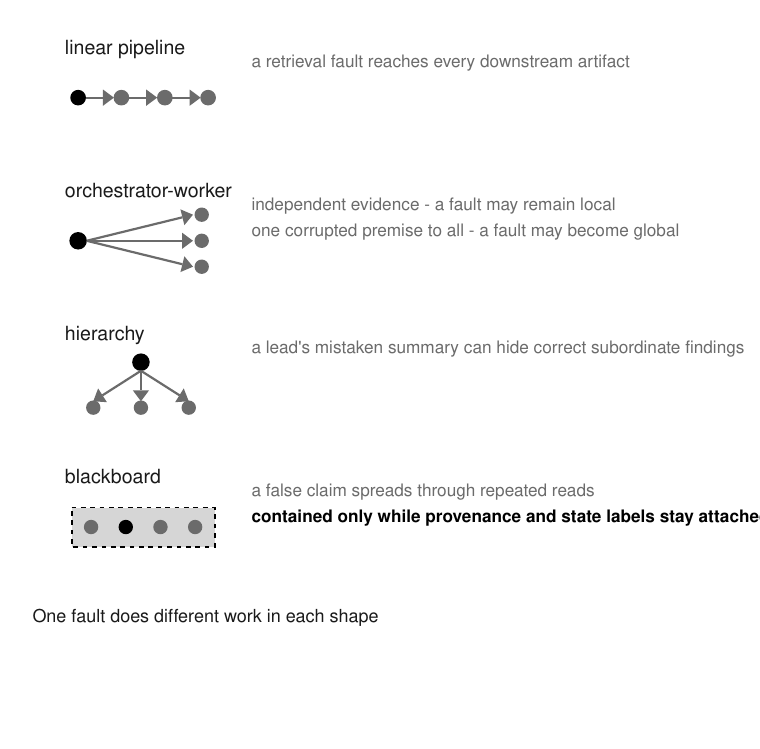}
\caption{Fault propagation depends on collaboration topology, ranging from downstream contamination in pipelines to contained local errors when shared systems preserve provenance and state labels.}
\end{figure}

Containment is conditional. A shared corrupted premise, an inaccurate lead summary, or repeated reads of an unqualified claim can turn a local error into a global one.

\Needspace{5\baselineskip}

Observation routes are therefore part of correctness. Record:

\begin{itemize}
\tightlist
\item
  which workers see raw tool output;
\item
  which receive only summaries;
\item
  whether a downstream component can request new upstream evidence;
\item
  whether workers see others' tentative conclusions before forming their own;
\item
  where contradictory results converge; and
\item
  which component can mark shared state disputed or invalid.
\end{itemize}

These properties govern error correlation, recovery, and attribution even when every worker uses the same model and prompt.

Runtime oversight can represent the topology as an interaction graph whose nodes are workers and artifacts and whose edges are messages or state transitions. That representation is a thinly supported companion idea and has not been validated as a monitoring standard. It remains a useful inspection tool because it exposes unreachable reviewers, unexpected broadcasts, and missing convergence paths.

Topology should be tested through faults, not inferred from diagrams.

\Needspace{5\baselineskip}

For a pipeline:

\begin{itemize}
\tightlist
\item
  remove or corrupt one stage output;
\item
  verify that downstream work stops or marks the state incomplete;
\item
  test whether the pipeline can widen or restart upstream.
\end{itemize}

\Needspace{5\baselineskip}

For an orchestrator-worker system:

\begin{itemize}
\tightlist
\item
  delay one worker;
\item
  inject contradictory findings;
\item
  corrupt one worker's input;
\item
  verify that the orchestrator distinguishes missing, late, and conflicting evidence.
\end{itemize}

\Needspace{5\baselineskip}

For a hierarchy:

\begin{itemize}
\tightlist
\item
  insert a critical minority finding at a subordinate level;
\item
  verify that it survives summarization;
\item
  test whether the lead can request source evidence.
\end{itemize}

\Needspace{5\baselineskip}

For a blackboard:

\begin{itemize}
\tightlist
\item
  submit conflicting updates;
\item
  verify that provenance and ordering remain attached;
\item
  preserve unresolved status rather than allowing last-write-wins to manufacture agreement.
\end{itemize}

The strongest evidence here, from Jia et al., concerns synthetic injected faults. Naturally occurring production faults lie outside that result, and the injected fault distribution determines what the comparison can show.

The reported closed-loop advantage supplies a reason to run local fault tests. It does not remove their necessity.

\Needspace{5\baselineskip}

The acceptable topology is the simplest one that:

\begin{itemize}
\tightlist
\item
  preserves the evidence the task requires;
\item
  contains representative faults;
\item
  exposes incomplete and disputed state;
\item
  keeps aggregation failures observable; and
\item
  clears the live single-agent gate on the target workload.
\end{itemize}

\hypertarget{which-work-is-eligible-to-dispatch}{%
\section{Which work is eligible to dispatch?}\label{which-work-is-eligible-to-dispatch}}

In a narrative reconstruction from my workflow harness, six planned work items initially appeared ready to run together. A pre-dispatch overlap check showed that four touched the same adapter file and its test. The schedule therefore became two waves. Three items could run in parallel, including two disjoint changes and one low-risk item from the colliding group. The remaining three ran sequentially in increasing risk order.

The scheduler selected a maximum independent set of nonconflicting work. Maximizing the number of simultaneous workers was not the objective.

The work-item identities, overlap matrix, and dispatch log from that sweep were not preserved, so the example illustrates the scheduling decision without providing an auditable result. The evidence for dynamic task graphs is also directional. The example cannot strengthen it.

A \textbf{dependency graph} represents units of work as nodes and prerequisites or conflicts as edges. It becomes a \textbf{task graph} when the runtime uses those records to decide which nodes are eligible for dispatch. A node becomes ready only when its prerequisites have completed successfully and its declared resources do not conflict with another running node. Independent work need not wait behind an unrelated serial step.

Execution can also change which work exists. A repository investigation may begin with one node that locates the affected adapter, then create package-specific repair nodes only after the adapter is identified. A failed compatibility check may add a migration node and block integration. A fixed schedule must either predict those branches in advance or leave workers idle while a central planner reconstructs dependencies from messages.

Dynamic dispatch does not mean unconstrained autonomy. The runtime performs a mechanical transition from \texttt{blocked} to \texttt{ready} when explicit conditions become true. Semantic decisions, such as whether a test failure warrants a new task, still require a model or person to propose the node and its dependencies. The scheduler validates that record and executes it under policy.

Yu et al.~(\href{https://arxiv.org/abs/2503.07675}{2025}) provide the only direct systems report attached to this practice, at directional strength. DynTaskMAS contains no controlled comparison between dynamic and fixed scheduling. The study compared parallel with serial execution on one seven-agent travel-planning workload. It did not compare a dynamic schedule with a fixed parallel schedule or isolate the task graph from the asynchronous execution engine.

The reported improvement therefore shows only that parallel execution outperformed serial execution in that setting. It does not show that dynamic scheduling outperformed a well-designed fixed schedule.

Rose et al.~(\href{https://arxiv.org/abs/2605.15132}{2026}) in APWA, Luo et al.~(\href{https://arxiv.org/abs/2502.13965}{2025}) in Autellix, and Masters et al.~(\href{https://arxiv.org/abs/2510.02557}{2025}) in the Manager Agent research challenge provide separate directional examples of related planning and scheduling architectures. None replicates a dynamic-versus-fixed comparison.

No controlled result in the cited evidence establishes that a dynamic dependency graph improves repository work relative to a fixed schedule.

The narrower systems argument applies when dependencies vary during execution. If independent branches exist, their prerequisites are observable, and their durations are uncertain, an explicit graph can expose safe concurrency without forcing unrelated work through one serial queue. If the workflow is short, fixed, and inexpensive, a graph scheduler may add persistence and recovery complexity without creating useful parallelism.

Dependency accuracy is the central risk. A missing edge can dispatch two workers that edit the same file, migrate the same schema, or invalidate each other's assumptions. A false edge serializes work that could safely overlap.

\Needspace{5\baselineskip}

File overlap is only one conflict signal. Work may also conflict through:

\begin{itemize}
\tightlist
\item
  generated artifacts;
\item
  shared test fixtures;
\item
  database schemas;
\item
  deployment environments;
\item
  locks or leases;
\item
  external services;
\item
  mutable caches; or
\item
  logical invariants spanning disjoint files.
\end{itemize}

A task graph is not enough for repository work. Its edges record which work items the planner believes depend on one another. Dispatch also needs a versioned view of the code those items may affect. Each dependency or conflict claim should identify the repository, branch, source revision, index version, and affected entity. Each running attempt should publish the same information for its in-flight artifact. Otherwise two tasks can appear independent in the work ledger while changing disjoint files that implement the same interface, or while an unmerged change on another host has already invalidated one worker's base. Code understanding is an observability input for the factory control plane, not only context supplied to the coding agent. This is a proposed control-plane requirement rather than a measured performance result. Test it by retaining predicted conflict edges and comparing them with observed rebase failures, overlapping verification failures, stale-base rejections, and missed migration targets.

Each node therefore needs enough state to support both scheduling and recovery:

\Needspace*{16\baselineskip}
\begin{record}
\begin{recordbody}
node_id: stable logical identity
inputs: immutable artifact versions
repositories: the repositories the node may affect
input_revisions: branch and revision per repository
code_index_version: the index generation conflict claims were computed from
affected_entities: files, symbols, or interfaces the node claims to touch
depends_on: prerequisite nodes that must succeed
conflicts_with: resources or nodes that cannot overlap
owner: currently assigned worker
attempt: retry identity
in_flight_artifacts: published state of the running attempt's artifact
status: blocked | ready | running | succeeded | failed
outputs: versioned artifacts and supporting evidence
\end{recordbody}
\end{record}

The attempt identity prevents a late response from an expired worker from overwriting a successful retry. Immutable input versions reveal when a ready node was planned against stale state. Versioned outputs allow downstream nodes to distinguish attempts that produced different artifacts.

These requirements come from the execution topology. Drawing the graph supplies none of them.

Mokhov, Mitchell, and Peyton Jones (\href{https://www.microsoft.com/en-us/research/wp-content/uploads/2018/03/build-systems-final.pdf}{2018}) make the same separation precise for build systems in Build Systems à la Carte. The dependency graph is not the complete execution semantics. How tasks are scheduled, whether dependencies are declared statically or discovered during execution, what triggers a rebuild, and what trace information persists across runs are separate design choices, and existing build systems occupy different points in that space. The lesson I transfer is structural: an explicit graph permits reasoning about safe parallelism only when the recorded dependencies are correct, and the scheduling and rebuild policies deserve their own decisions rather than riding in unexamined with the graph. The analogy has a limit. A coding factory operates on mutable worktrees and issues external effects, conditions that ordinary build DAGs constrain more tightly through hermetic inputs and declared outputs, so the decomposition names the design axes without supplying the fault model.

In my workflow harness, notification-driven dispatch ends the coordinating turn after work is assigned and resumes it when completion events arrive. This avoids a polling loop that consumes attention while no state has changed. Mechanical parallel-set detection checks dependency layers, blocking edges, and overlapping file paths.

Those rules illustrate the mechanism. They do not provide comparative performance evidence.

A per-step time limit contains a slow or failed worker only when expiration produces and propagates a real failure. Treating a timeout as success destroys the evidence of failure and may release downstream nodes without their required inputs. The schedule then appears healthy because it erased the event needed to diagnose it.

Dispatch should fail closed. When a worker exceeds its time limit, the attempt fails visibly, dependent nodes remain blocked, and the retry, escalation, or cancellation policy runs.

Retries require the same discipline. A retry creates a new attempt for the same logical node rather than a second independent task. The scheduler must either establish that repeated execution is idempotent or isolate side effects before trying again. Otherwise a timeout can produce duplicate comments, partial writes, or two workers claiming the same resource.

\hypertarget{a-lease-is-not-write-authority}{%
\section{A lease is not write authority}\label{a-lease-is-not-write-authority}}

The owner field in the node record is commonly implemented with a lease: a worker holds a time-bounded claim on the work and renews it while making progress, and when renewals stop, the control plane may reassign. That reading is correct, and it is the only question a lease answers. A lease resolves a liveness and allocation question: after the timeout, the control plane is permitted to give the work to someone else. It does not establish that the previous worker is dead. A process paused under memory pressure resumes after its lease has expired. A network partition heals after reassignment has already happened. A mutation issued before expiry arrives at its target after it.

A lease does not guarantee a single writer. Safety comes from a separate mechanism at the boundary being protected, described with the \texttt{work\_id}, \texttt{ownership\_epoch}, and \texttt{attempt\_id} identities from Chapter 7. This is the mechanism for which the ownership epoch exists: a monotonic generation of write authority, incremented each time the control plane reassigns the work. The attempt field in the node record above is the attempt identity; the epoch completes the picture the earlier paragraph on late responses began. Preventing an expired worker's late response from overwriting a successful retry is exactly the fencing problem, stated at the scheduler. The general form states it at every protected boundary.

The rule is contract I2 from Chapter 7: only the current ownership generation may commit a mutation at a protected boundary, and a superseded worker must be rejected even if it is still running. Every mutation to a protected shared target carries the current \texttt{ownership\_epoch}, and the target itself, or an authoritative mutation gateway in front of it, rejects any older epoch. Contract I3 follows at the ledger: a durable state transition validates generation and attempt identity rather than trusting the scheduler's belief about who is active.

The worked sequence in the figure below makes the mechanism concrete.

\begin{figure}[htbp]
\centering
\includegraphics[width=1\textwidth,height=\textheight,keepaspectratio]{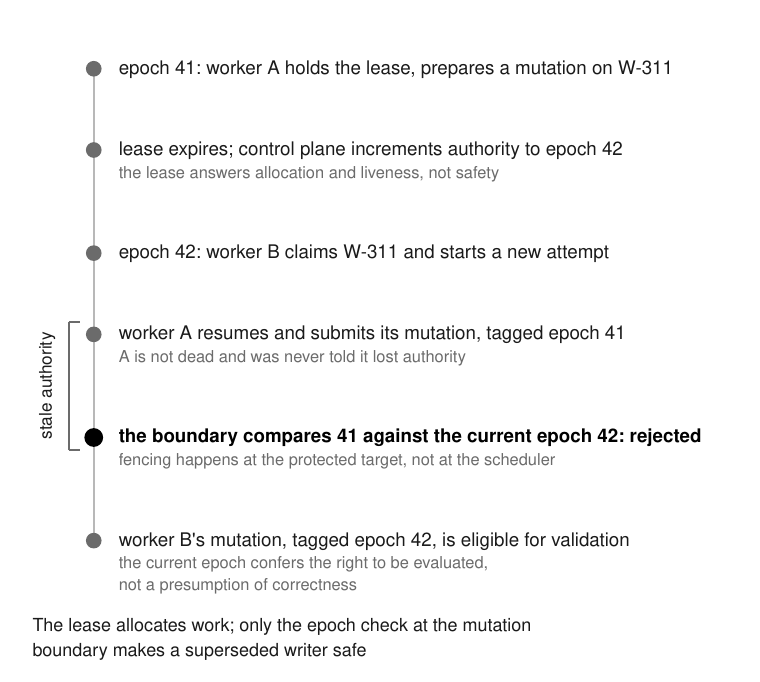}
\caption{Epoch fencing on work W-311. Worker A is not dead and behaves correctly by its own lights, having never been told it lost authority; only the epoch check at the protected target stops its stale mutation. Worker B's mutation, carrying the current epoch, is eligible for normal validation, not automatically accepted: holding the current epoch confers the right to be evaluated, not a presumption of correctness.}
\end{figure}

The check must occur at the mutable boundary whose integrity matters: the branch head, the work ledger, the external-effect adapter. Validating only at the scheduler before dispatch is insufficient, because the scheduler's belief about which workers are alive is precisely the information the failure invalidates. By the time worker A's late mutation arrives, the scheduler's dispatch-time check has already passed.

The precedent is old. Burrows (\href{https://www.usenix.org/conference/osdi-06/chubby-lock-service-loosely-coupled-distributed-systems}{2006}) describes sequencers in Chubby: a lock holder obtains a generation number it passes along with its requests, and servers receiving those requests validate the number rather than trusting that the lock is still held. This is historical lineage, and I transfer the principle, not any parameter of Chubby's deployment. The stale-authority problem the sequencer solves is the same problem the paused worker A presents.

Where should the epochs and lease state themselves live? Hunt et al.~(\href{https://www.usenix.org/conference/usenix-atc-10/zookeeper-wait-free-coordination-internet-scale-systems}{2010}) built ZooKeeper on the observation that coordination state belongs in a service designed to survive individual process failure rather than in the memory of any process it coordinates. I take that narrow principle and nothing more. It does not follow that every factory needs ZooKeeper, or a consensus protocol at all; a single durable database that records epochs transactionally can enforce I2 for a modest fleet. The requirement is that the record of who currently holds authority survives the failure of the workers it governs.

Testing this machinery requires perturbing the controller's view of the world, not just the workers. The scheduler and any reconciling component need tests under stale work status, intermediate artifact states, transitions the controller never observed, late completion arriving after reassignment, and controller restart during active work. Sun et al.~(\href{https://www.usenix.org/system/files/osdi22-sun.pdf}{2022}) built Sieve on exactly this method for Kubernetes controllers: perturb what the controller observes, then compare resulting cluster state against an unperturbed run. They found real bugs in mature controllers whose logic was correct under an accurate view. I transfer the method, at directional strength, not the tooling. Chapter 10's fault-injection protocol supplies the harness discipline; the epoch-41 sequence above is precisely the kind of scenario such a test suite should replay on every scheduler change.

\Needspace{5\baselineskip}

Dynamic graphs are most useful when they expose decisions that were previously implicit. The runtime should be able to explain:

\begin{itemize}
\tightlist
\item
  why a node remains blocked;
\item
  which successful output released it;
\item
  which conflict prevents concurrent execution;
\item
  which attempt currently owns it;
\item
  which input version it was planned against; and
\item
  why a failure created, removed, or blocked another node.
\end{itemize}

If those answers cannot be reconstructed, the graph has become another hidden coordinator rather than an observable allocation mechanism.

\hypertarget{what-evidence-should-promote-a-topology}{%
\section{What evidence should promote a topology?}\label{what-evidence-should-promote-a-topology}}

Chapter 1 defines the paired comparison, Chapter 2 the baseline and cost coordinates, Chapter 10 the fault-injection method, and Chapter 11 the first-upstream-failure label. A topology experiment should reuse those protocols rather than restating them.

Begin with one recurring failure class and draw the smallest boundary at which it could have become observable and stoppable. Change one structural element at that boundary: a verification loop, specialist role, handoff, communication route, aggregation rule, or dependency-aware dispatcher. Preserve the live single-agent workflow as the cheapest credible baseline.

\Needspace{5\baselineskip}

The promotion record adds three topology-specific conditions:

\begin{itemize}
\tightlist
\item
  \textbf{Mechanism entry:} a delegated result was consumed, independent nodes actually overlapped, or the proposed aggregation rule selected among real candidates.
\item
  \textbf{Fault containment:} an omitted artifact, delayed worker, contradictory finding, late completion, or dependency error remained visible and did not release downstream work as complete.
\item
  \textbf{Coordination return:} any gain in the target failure class cleared the predeclared success, reliability, latency, review-burden, and total-cost limits.
\end{itemize}

Run the baseline and treatment on identical task versions and preserve the complete configuration identity. Promote only if all three conditions hold. Retain the baseline when added structure does not repay its coordination cost; redesign when the mechanism executed but left the target failure path intact.

The record should remain intelligible without the architecture's label. ``Hierarchical,'' ``dynamic,'' and ``multi-agent'' describe arrangements; mechanism entry, containment, and measured return decide whether the arrangement earned adoption.

\section*{Sources and evidence}

\begin{itemize}
\tightlist
\item
  Bertalani\v{c}, et al.~2026. \emph{The Cost of Consensus}. arXiv:2605.00914. Exploratory evidence on aggregation failure; Chapter 5 carries the reported oracle-gap result.
\item
  Böckeler, Birgitta. 2026. ``Harness Engineering.'' MartinFowler.com, February 17. Practitioner account based partly on vendor-reported Codex-team practice.
\item
  Burrows, Mike. 2006. ``The Chubby Lock Service for Loosely-Coupled Distributed Systems.'' OSDI. Historical lineage for generation-validated authority; sequencers let servers reject requests carrying a stale lock generation. Principle transferred, not parameters.
\item
  Cemri, M., et al.~2025. ``Why Do Multi-Agent LLM Systems Fail?'' arXiv:2503.13657. MAST failure taxonomy and benchmark-framework interventions.
\item
  Chun, Yong Jin, et al.~2025. ``Enhancing LLM Performance Through Debate: An Empirical Study on Multi-Agent Debate for Coding Tasks,'' arXiv:2503.12029. Audit-retained strong evidence on debate performance against task baselines; reports inference cost across debate variants and leaves the single-model cost comparison open. The v1 record carried the title ``Is Multi-Agent Debate (MAD) the Silver Bullet?''; this entry cites the current version.
\item
  Huang, Jia, and Joey Tianyi Zhou. 2026. ``A Two-Dimensional Framework for AI Agent Design Patterns: Cognitive Function and Execution Topology.'' arXiv:2605.13850. Directional taxonomy of topology choices.
\item
  Hunt, Patrick, et al.~2010. ``ZooKeeper: Wait-Free Coordination for Internet-Scale Systems.'' USENIX ATC. Directional; cited only for the principle that coordination state should survive individual process failure, not as a requirement that factories adopt consensus systems.
\item
  Jia, J., et al.~2026. ``MAS-FIRE: Fault Injection and Reliability Evaluation for LLM-Based Multi-Agent Systems.'' arXiv:2602.19843. Synthetic fault injection across three architectures.
\item
  Jain, Prateek. 2026. ``Harness Engineering: The New DevOps Layer for AI Agents.'' r/devops, May 27. Anecdotal practitioner account.
\item
  Kim, T., et al.~2026. ``Why Do AI Agents Systematically Fail at Cloud Root Cause Analysis?'' arXiv:2602.09937. OpenRCA evidence on failure persistence across capability tiers and protocol intervention.
\item
  Lin, et al.~2026. \emph{Agentic Harness Engineering}. arXiv:2604.25850. Preprint; ten harness iterations raised Terminal-Bench 2 pass@1 from 69.7 to 77.0 percent, with ablations locating gains in tools, middleware, and memory. Direction transferred, not magnitude.
\item
  Oskooei, A. R., et al.~2026. ``Deep Agentic Search for Repository-Level Code Question Answering: An Empirical Study.'' arXiv:2608.01507. Strong evidence for the SWE-QA comparison and its measured handoff failures; broader topology transfer is directional.
\item
  Kumar, et al.~2026. \emph{AgentForge}. arXiv:2604.13120. Directional evidence on multi-agent design.
\item
  Li, et al.~2026. \emph{SIMAS}. arXiv:2606.00655. Audit-retained strong evidence on token cost, non-monotonic scaling, and debate versus self-correction.
\item
  Luo, et al.~2025. \emph{Autellix}. arXiv:2502.13965. Directional scheduling and orchestration example.
\item
  Masters, et al.~2025. \emph{Orchestrating Human-AI Teams: The Manager Agent as a Unifying Research Challenge}. arXiv:2510.02557. Directional planning and scheduling example.
\item
  Mokhov, Andrey, Neil Mitchell, and Simon Peyton Jones. 2018. ``Build Systems à la Carte.'' ICFP. Directional; separates the dependency graph from scheduling, dependency discovery, rebuild decisions, and trace persistence. The mutable-worktree, external-effect setting of a coding factory limits the analogy.
\item
  Qian, et al.~2024. \emph{MacNet}. arXiv:2406.07155. Directional evidence on network scaling.
\item
  Rose, Evan, et al.~2026. \emph{APWA: A Distributed Architecture for Parallelizable Agentic Workflows}. arXiv:2605.15132. Directional orchestration example.
\item
  Shang, Wenlong. 2026. \emph{``Theater of Mind'' for LLMs: A Cognitive Architecture Based on Global Workspace Theory}. arXiv:2604.08206. Directional blackboard and global-workspace design; single-author architecture proposal, not a controlled measurement.
\item
  Sun, Xudong, et al.~2022. ``Sieve: Automatic Reliability Testing for Cluster Management Controllers.'' OSDI. Directional; perturbed-view testing of reconciliation controllers, transferred as method for scheduler and reconciler tests, not as tooling.
\item
  Tian, et al.~2025. ``Beyond the Strongest LLM.'' arXiv:2509.23537. Directional evidence on multi-agent teams.
\item
  Yu, Junwei, Yepeng Ding, and Hiroyuki Sato. 2025. \emph{DynTaskMAS}. arXiv:2503.07675. Directional parallel-versus-serial result on a seven-agent travel-planning workload; no controlled dynamic-versus-fixed comparison.
\item
  Zhao, et al.~2026. \emph{ATOM}. arXiv:2605.26178. Audit-retained strong evidence on difficulty-conditioned topology across six short-form benchmarks.
\end{itemize}

\textbf{Author-system illustration cited inline}

\begin{itemize}
\tightlist
\item
  Not an evidence item: CodeProbe, the author's task-mining evaluation tool, \href{https://github.com/sjarmak/codeprobe}{public repository}. Named inline for the preamble rerun described in the opening, which is narrative illustration.
\end{itemize}

\chapter{Cost-aware fleet scheduling and model routing}
\label{ch19-cost-aware-fleet-scheduling-model-routing}
\begin{quote}
\textbf{Evidence profile.} 2 strong \(\cdot\) 17 directional \(\cdot\) 2 corroborating \(\cdot\) 0 null or conflicting evidence items across 4 developed practices (\erca{171}, \erca{187}, \erca{191}, \erca{206}). Four of the directional items are inference-serving systems results, admitted for the serving-conditions section and measured outside any agent fleet.

\textbf{Chapter claim.} Re-decide from observed state, then ship the best feasible incumbent on time.
\end{quote}

In an eleven-week replay of my fleet ledger, 1,286 work items crossed 22 execution pools. Age-only first-come scheduling produced 6.84 hours of priority-weighted flow time; a hybrid rule, plain priority, and a four-feature weighted index each produced 6.70 hours, within 0.1 percent of one another. I had fixed a 15 percent promotion requirement before the replay, so none of the more elaborate policies shipped.

A fleet can increase throughput while also increasing the cost of each accepted result. Its own traces must distinguish the two. Concurrent execution, together with a willingness to attempt work a human team might have left queued, exposes more tasks to model inference, verification, review, and recovery. Throughput alone therefore does not establish an economic gain.

The evidence is largely transferred from other fields. One strong software-engineering result concerns cheap baselines; observatory scheduling, compute-cluster management, multi-tenant resource control, search-based software engineering, budget-constrained bandit theory, and router calibration supply directional mechanisms. The telescope-scheduling lineage that opens this chapter is one transfer lineage among several, alongside cluster management and multi-tenant resource control, rather than the dominant systems foundation. My fleet replay remains a narrative illustration.

\Needspace{5\baselineskip}

These systems schedule telescopes and compute clusters, not software-agent fleets. The transferable design is narrower:

\begin{itemize}
\tightlist
\item
  make new decisions from current state;
\item
  limit how long the decision process may run;
\item
  compare sophisticated policies with tuned cheap alternatives; and
\item
  preserve enough evidence to detect when the policy is worse.
\end{itemize}

Their scheduling cadence, thresholds, cost ratios, and performance constants do not transfer.

Allocation policy is easy to mistake for a local configuration detail. In an agent fleet, it determines which model receives a request, which queued task moves first, whether running work remains fixed, and when an obsolete plan is discarded. These choices connect inference cost to elapsed time, reviewer demand, and the probability that completed work will survive evaluation.

A cheaper model call can trigger enough repair and review to cost more per accepted result. A mathematically better schedule can become operationally worse if computing it delays the work beyond its deadline. Scheduling and routing are therefore measured decisions rather than static configuration.

Production traces also show that coding-agent traffic differs from chatbot traffic. Liu et al.~(\href{https://arxiv.org/abs/2608.00101}{2026}) sampled GitHub Copilot activity from 3.2 million users, 13 million sessions, 761 million model calls, and 95 trillion tokens. KV-cache hit rates averaged 90 percent within a turn and 55 percent across turn boundaries; a lightweight predictor captured 86 to 90 percent of total user idle time. The study strongly characterizes one production workload and directionally supports measuring turn boundaries and idle periods in capacity policy.

Those rates belong to one product's traffic and do not transfer to another coding agent. The shape of the workload is what does carry. An agent session accumulates inference state that later calls in the same session can reuse, and much less of it survives the boundary between turns. Cache behavior of that kind sets how much concurrent work an inference endpoint can carry and what each call costs, which is why the serving conditions behind a model endpoint get a section of their own later in this chapter. That constrains how allocation is measured. An experiment that samples model calls as interchangeable independent requests destroys the reuse structure it is supposed to be measuring, so replays and shadow runs must preserve turn and session boundaries.

\hypertarget{what-allocation-decision-is-the-system-making}{%
\section{What allocation decision is the system making?}\label{what-allocation-decision-is-the-system-making}}

\textbf{Cheapest-sufficient routing} selects, for each request, the least expensive worker predicted to satisfy that request's declared requirement. The worker may be a model tier, a model-plus-tool configuration, or another execution pool with a known price.

The important word is \emph{sufficient}. The router does not minimize the invoice in isolation, and it does not assign every task class through a fixed hand-written rule. It estimates expected performance and expected cost for the current request, then chooses the least expensive option predicted to satisfy the stated tradeoff.

\Needspace{5\baselineskip}

This definition leaves two deployment decisions open:

\begin{itemize}
\tightlist
\item
  how performance is predicted; and
\item
  how sufficiency is expressed.
\end{itemize}

Both affect calibration and evaluation.

\textbf{Shipping the incumbent at the deadline} places a hard time limit on schedule computation and executes the best feasible plan available when that limit expires. A feasible plan respects constraints that cannot be violated, such as capacity, hard deadlines, and work already in flight. The \textbf{incumbent} is the best feasible plan found so far.

Additional solve time may find a better plan or establish that the incumbent is optimal. The schedule has operational value only while it still describes the fleet's current state. A solve deadline makes decision latency an explicit policy rather than an accidental result.

\Needspace{5\baselineskip}

The \textbf{optimality gap} records how far the shipped plan may remain from the best possible solution, using the strongest bound available when the deadline expires. Preserving that gap distinguishes two operationally different outcomes:

\begin{itemize}
\tightlist
\item
  a plan known to be close to optimal; and
\item
  a merely feasible plan found before search made much progress.
\end{itemize}

Operations may reasonably execute either. Evaluation should not treat them as equivalent.

\begin{figure}[htbp]
\centering
\includegraphics{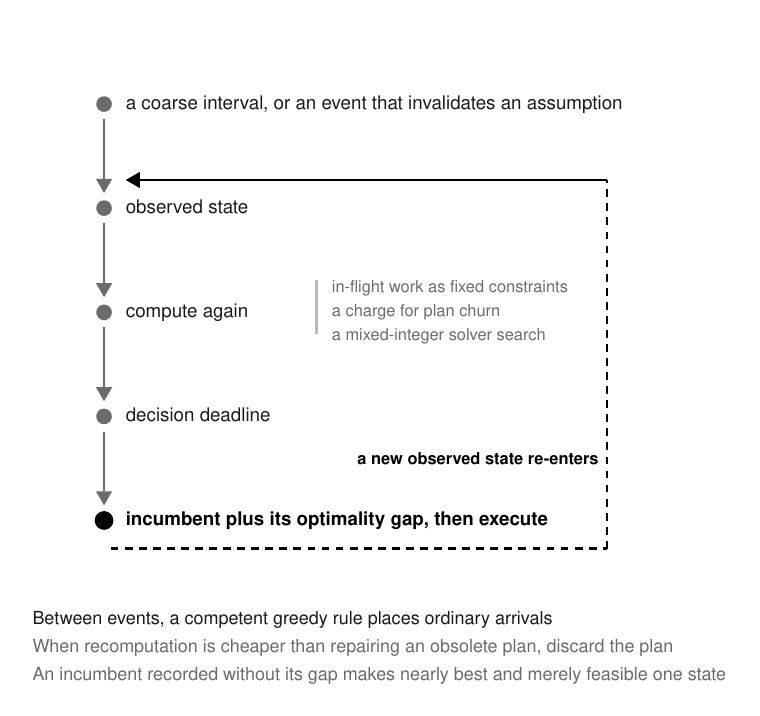}
\caption{The re-decision loop recomputes from observed events under fixed work and churn costs, then executes a deadline capped incumbent while retaining its optimality gap.}
\end{figure}

Ordinary arrivals may use a cheap greedy rule between re-decision events. The system discards the existing plan only when the cost of recomputing is lower than the expected cost of executing or repairing an obsolete one.

\Needspace{5\baselineskip}

These terms identify the decisions that must be made before selecting an algorithm:

\begin{itemize}
\tightlist
\item
  What counts as sufficient performance?
\item
  How long may schedule computation run?
\item
  Which constraints are inviolable?
\item
  How is running work represented?
\item
  What reassignment cost counts as plan churn?
\item
  Which measure of remaining solution uncertainty must be retained?
\end{itemize}

The companion catalog contains algorithmic patterns. The narrower question here is whether these decisions lower accepted-result cost or improve flow on the fleet's own workload.

\hypertarget{when-should-the-scheduler-recompute-and-when-should-it-stop}{%
\section{When should the scheduler recompute, and when should it stop?}\label{when-should-the-scheduler-recompute-and-when-should-it-stop}}

An observatory scheduler can spend the night improving a plan that clouds have already invalidated, or compute another imperfect plan from the sky that remains observable. Bellm et al.~(\href{https://arxiv.org/abs/1905.02209}{2019}) describe the Zwicky Transient Facility taking the second approach in production since 2018. When changing conditions invalidate the schedule, it resolves the nightly integer program again rather than enumerating every weather scenario in advance.

Three directional observatory-scheduling sources support transferring this decision architecture to agent fleets. None provides a strong result for software fleets. The recommendation is therefore limited to choosing and measuring a re-decision policy. The appropriate cadence remains a local deployment decision.

The schedule should be treated as the current output of a control loop. At a coarse interval, or after an event invalidates an important assumption, the scheduler computes again from observed state. Between those points, a competent greedy rule can place ordinary arrivals without paying for a complete solve.

The system should abandon an obsolete plan when recomputation costs less than repairing or completing that plan. This comparison must include decision latency. An excellent schedule returned after the queue, capacity, or deadlines have changed is a schedule for a fleet that no longer exists.

\Needspace{5\baselineskip}

Current state includes more than queued work. Running tasks may:

\begin{itemize}
\tightlist
\item
  consume scarce model or execution capacity;
\item
  hold repository leases;
\item
  occupy reviewer slots;
\item
  have produced costly intermediate state; or
\item
  contain work another worker cannot reconstruct cheaply.
\end{itemize}

A re-solve should represent those commitments explicitly. Treating work in flight as fixed is a conservative default unless preemption has been separately justified.

The objective should also charge for \textbf{plan churn}: avoidable reassignment or resequencing that forces workers to discard useful preparation or partially completed work. Cadence, invalidation triggers, preemption, and churn penalties belong to the scheduling design rather than being added after the objective is chosen.

When the solve deadline arrives, execute the incumbent and retain its optimality gap. This follows from the mismatch between proof time and operational time. A solver may find a feasible plan quickly but take much longer to establish how close it is to the best plan. Meanwhile, new work arrives and the state used to construct the plan ages.

\Needspace{5\baselineskip}

The deadline prevents optimization itself from becoming another queue. The stored gap supports later questions:

\begin{itemize}
\tightlist
\item
  Did large gaps coincide with poorer flow?
\item
  Would more solve time have changed the executed assignments?
\item
  Did a tuned greedy policy perform just as well?
\item
  Did solve latency consume the predicted scheduling benefit?
\end{itemize}

Parazin et al.~(\href{https://arxiv.org/abs/2203.00013}{2022}) compared a mixed-integer scheduler for gravitational-wave follow-up with a tuned greedy heuristic. Neither won consistently. Under a 500-second cap, each performed better on different skymaps, and only 37 of 97 well-localized skymaps improved under the reported optimization comparison. Across 951 simulated events, a hybrid improved detection efficiency by roughly 3 to 11 percent. The authors also reported that reducing the cap to 100 seconds would have truncated only 64 additional schedules without a substantial efficiency loss.

\Needspace{5\baselineskip}

These are directional results from gravitational-wave follow-up rather than operating parameters for an agent fleet. They support evaluating:

\begin{itemize}
\tightlist
\item
  the solve deadline;
\item
  the retained gap;
\item
  the greedy baseline; and
\item
  a hybrid policy.
\end{itemize}

They argue against assuming that either the optimizer or the heuristic will win in advance.

Naghib et al.~(\href{https://arxiv.org/abs/1810.04815}{2019}) describe another observatory design in which a feature-based, memoryless policy chooses again from current state at every step. \emph{Memoryless} here does not mean that durable state is discarded. It means that the next scheduling action does not depend on preserving a brittle sequence of prior planned actions.

Interruptions are absorbed by observing the new state and making another decision. The evidence comes from a framework and simulation, so it supports a plausible architecture rather than a measured advantage for agent execution.

The fleet-wide and worker-local decisions should remain separate. A global allocator determines how scarce model tiers, reviewer capacity, or execution pools are divided among classes of work. A worker-local dispatcher determines which eligible item a particular worker takes next.

Combining both choices in one opaque score makes it difficult to determine whether a poor outcome arose from resource allocation or queue ordering. It can also allow local dispatch to undo a capacity decision the global layer just made.

The companion entry \texttt{formalize-work-as-constraints-and-solve-globally} proposes representing requests in one constraint language so an allocator can optimize across them. Its lineage reaches Hubble scheduling in 1990. The architecture is useful only when that representation captures the constraints operations actually enforce.

A nominally global solution built from missing deadlines, false independence assumptions, or omitted affinity constraints can be less coherent than separate local rules. The companion aside \texttt{audit-the-allocation-layer} therefore asks which policy actually controls execution, planning, and optimization. Its observatory analogy has no software-fleet measurement.

Probe costs belong in the same accounting. The thinly supported companion entry \texttt{probe-only-above-an-uncertainty-threshold} asks whether a dry run, canary, or classifier call reduces enough uncertainty to justify its latency and inference cost. Its theoretical constant comes from adversarial single-machine scheduling and does not transfer.

The local question is operational:

\begin{quote}
For which uncertain requests does a probe change the selected worker or schedule often enough to repay its cost?
\end{quote}

When logs cannot answer that question, probing remains another unmeasured allocation policy.

The evidence boundary should remain visible. The source survey does not cover much of the major USENIX and ACM cluster-scheduling literature. Agent runtime variance is also partly endogenous. Repeated tool failures, unstable environments, and unnecessary retries should be repaired at their source before the scheduler is asked to absorb them.

Nor does the cited evidence establish a general prohibition on preempting long-running agent sessions. Keeping running work fixed during a re-solve is a conservative representation of existing commitments rather than proof that preemption is always wrong.

Choose the cadence and solve deadline through a short pilot. Retain the incumbent, optimality gap, solve time, assignment decisions, and resulting execution trace at every re-decision. Compare the complete policy with a tuned greedy baseline.

The companion entry \texttt{pilot-to-pick-the-algorithm-then-commit-the-budget} states the same logic: spend a small amount distinguishing among methods, then commit the operating budget to the one that performs better.

\Needspace{5\baselineskip}

The design becomes auditable when the following are explicit:

\begin{itemize}
\tightlist
\item
  the re-decision trigger;
\item
  the solve deadline;
\item
  treatment of in-flight work;
\item
  the churn cost;
\item
  the simple baseline; and
\item
  the metrics used to compare them.
\end{itemize}

Its value remains an empirical question until the fleet's traces answer it.

\hypertarget{admission-backpressure-and-recovery-capacity}{%
\section{Admission, backpressure, and recovery capacity}\label{admission-backpressure-and-recovery-capacity}}

In the eleven-week replay described later in this chapter, the scheduling effect concentrated entirely in one contended pool, where eligible work regularly exceeded available capacity; the uncontended pool showed almost no difference across policies. That retained observation locates where overload policy matters: not everywhere, but at the resources where demand exceeds capacity, and most acutely when capacity has just fallen. A queue records demand; it does not create capacity. Retry policy can increase demand precisely when capacity has fallen. A reliable factory therefore needs overload policy above individual queues, distinct from the re-decision and routing questions elsewhere in this chapter: those decide which work runs where, and overload policy decides how much work is allowed to exist inside the system at all.

The failure mode this section guards against is mechanical. A dependency slows, tasks time out, retries multiply the request rate against the slowed dependency, queues grow, work waits long enough that its callers retry the whole task, and the fleet spends its capacity on demand that its own policies manufactured. Nothing in that sequence requires a model to misbehave. The Google SRE chapters on \href{https://sre.google/sre-book/addressing-cascading-failures/}{cascading failures} and \href{https://sre.google/sre-book/handling-overload/}{handling overload} describe this pattern across production services; I cite them as practitioner guidance rather than controlled evidence, and none of their capacity or threshold figures transfer to an agent fleet. The figure below shows the amplification loop and the policies that bound it. Overload policy has three responsibilities: bound total admitted demand, prevent one ownership or workload domain from consuming the fleet, and release retry and recovery traffic at a controlled rate.

\begin{figure}[htbp]
\centering
\includegraphics[width=1\textwidth,height=\textheight,keepaspectratio]{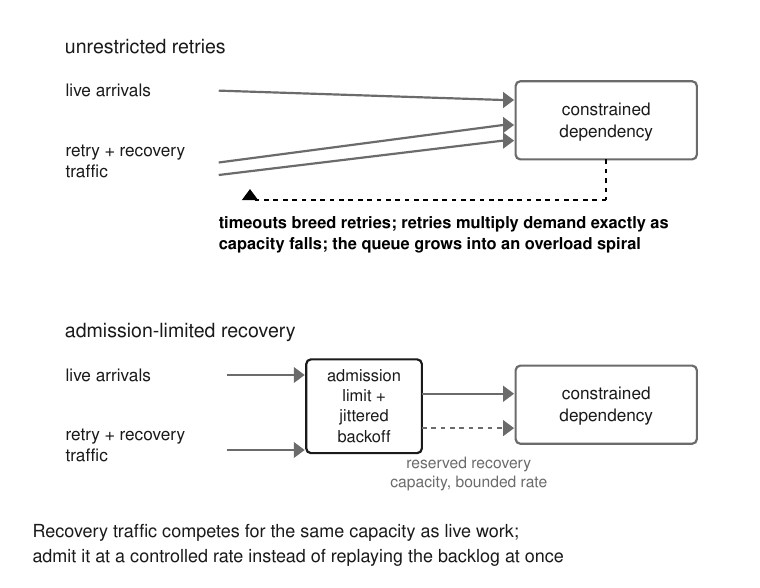}
\caption{Live arrivals and retry-driven recovery traffic compete for the same constrained dependency; unrestricted retries amplify load exactly when capacity has fallen, while admission limits, jitter, and reserved recovery capacity bound the interference.}
\end{figure}

\textbf{Bounding admitted demand.} The first policy decision is whether new work enters the system at all. Verma et al.~(\href{https://research.google/pubs/large-scale-cluster-management-at-google-with-borg/}{2015}) describe Borg holding submitted jobs in a pending state until admission, packing admitted work against declared resource requirements, isolating tenants, and treating overcommitment as an explicit cluster policy rather than an accident of scheduling. The transfer is directional: an agent fleet benefits from the same separation between accepting work and running it, and from making overcommitment a decision with an owner. Borg's packing algorithms, machine-share targets, and utilization figures do not transfer as parameters.

Admission requires bounded queue growth. An unbounded queue converts overload into unbounded latency, and by contract I9 admissible work cannot starve invisibly. A bounded queue forces the honest alternatives into view: reject explicitly, shed load by declared class, or wait with a visible position and age. Explicit rejection at admission is cheaper than the same rejection discovered hours later as a timeout, because the caller learns immediately and the system spends nothing executing work it will abandon. When a downstream stage saturates, the signal must propagate upstream as backpressure rather than accumulate in an intermediate queue; when slowing is not enough, the system sheds load by explicit policy, rejecting the lowest declared class first and telling the caller. Silent degradation, where work is accepted and then quietly never runs, violates I9 directly.

\textbf{Isolating demand domains.} Aggregate capacity limits are not enough, because a single hot tenant or repository can fill a shared queue and starve unrelated work while the aggregate limit is still respected. Partitioning queues by tenant, repository, or work class bounds the damage one demand source can do to another. Mace et al.~(\href{https://www.usenix.org/conference/nsdi15/technical-sessions/presentation/mace}{2015}) built Retro around per-tenant resource monitoring and central control points that throttle at the granularity of the tenant, and they observed that maintenance traffic itself can become the overload source. Both observations transfer directionally. In an agent fleet, the analogue of maintenance traffic is recovery, reconciliation, and retry work, and it competes for the same models, runners, and reviewers as live work.

\textbf{Concurrency limits by constrained resource.} Concurrency limits belong on the resource that is actually constrained, not on an arbitrary global worker count. Model-tier tokens per minute, repository leases, reviewer attention, and external API rate limits are separate constraints with separate saturation points. A single fleet-wide concurrency number either wastes the uncontended resources or oversubscribes the contended one. The contended-pool observation from my replay applies here in reverse: limits matter where eligible work exceeds capacity, and a limit on an uncontended resource is dead configuration.

\textbf{Retry budgets, backoff, and jitter.} Retries are the mechanism by which demand rises as capacity falls, so they need a budget, not just a count. A per-attempt retry cap still permits a retry storm when many attempts fail together; a budget bounds the fraction of total traffic that retries may consume. Individual retries should back off exponentially, and backoff must carry jitter, because deterministic backoff synchronizes the failed population into retry waves that arrive together and re-saturate the recovering dependency. The AWS Builders' Library articles on \href{https://aws.amazon.com/builders-library/timeouts-retries-and-backoff-with-jitter/}{timeouts, retries, and backoff with jitter} and on \href{https://aws.amazon.com/builders-library/making-retries-safe-with-idempotent-APIs/}{making retries safe with idempotent APIs} state both mechanisms clearly; they are practitioner guidance, and their specific multipliers and cap values are their systems' parameters, not this book's. The idempotency requirement is not optional decoration here. A retry is a new attempt at the same logical work (contract I4), and a retried external effect is safe only under the effect-identity contracts of I5. A retry policy layered over non-idempotent effects converts overload into duplicated external commits.

\textbf{Recovery capacity.} Recovery and reconciliation need their own capacity policy, so that backlog cannot consume every worker needed to recover it. After a worker loss or a dependency outage, the system holds a backlog of interrupted attempts, orphaned leases, and unreconciled external effects. Recovering those requires the same constrained resources as live work. If the backlog is admitted at full priority into an undifferentiated queue, recovery work and live work starve each other, and the reconciliation that would shrink the backlog never gets scheduled. Contract I10 requires that recovery preserve the same obligations as normal execution and be tested as production code; that includes its capacity allocation. A recovery path that only works when the fleet is idle has not been tested under the one condition where it runs. Pilot-execution work by Li, Cai, and Lou (\href{https://www.usenix.org/conference/nsdi26/presentation/li-zhenyu}{2026}) adds the sharper directional warning that recovery actions themselves cause severe failures, which is a second reason to bound the rate at which recovery traffic is released rather than replaying the entire backlog at once. No universal fraction of capacity is prescribed here for recovery. That is a measured deployment decision, and the experiment below is how to measure it.

The global-allocation versus worker-local-dispatch separation defended earlier in this chapter has a precedent in this literature. Schwarzkopf et al.~(\href{https://research.google/pubs/omega-flexible-scalable-schedulers-for-large-compute-clusters/}{2013}) built Omega as parallel schedulers operating over shared cluster state with optimistic concurrency control, precisely because a single monolithic scheduler conflated policies with different cadences and stakes. Hindman et al.~(\href{https://www.usenix.org/conference/nsdi11/mesos-platform-fine-grained-resource-sharing-data-center}{2011}) reached a related two-level split in Mesos, with a central allocator offering resources and frameworks deciding what to run on them. Both are directional support for keeping capacity allocation and dispatch as separately auditable layers; neither system's conflict rates, offer semantics, or scale figures describe an agent fleet.

\Needspace{5\baselineskip}

The replay protocol later in this chapter tests dispatch policy under normal arrivals. Overload policy needs its own experiment, because its interesting behavior only appears when capacity falls or backlog is released. I add a recovery-capacity experiment to the protocol:

\begin{enumerate}
\def\labelenumi{\arabic{enumi}.}
\tightlist
\item
  Establish the normal arrival rate and the capacity of each constrained resource from the ledger.
\item
  Inject a worker loss or a dependency failure of recorded duration.
\item
  Release a controlled backlog and retry population against the recovering system.
\item
  Compare unrestricted recovery traffic with a bounded-recovery policy that reserves declared capacity for live work and admits recovery work at a controlled rate.
\item
  Measure, by declared class: live-work tail latency, recovery completion time, queue depth, retry count, request rate against the recovering dependency, starvation against the existing guard, useful throughput, and duplicate external effects.
\end{enumerate}

The comparison should show where each policy fails, not which one wins in general. Unrestricted recovery typically minimizes recovery completion time while damaging live-work tails and re-saturating the dependency; an over-tight bound protects live work while the backlog ages past usefulness. The right division is a property of the fleet's own workload, which is why this section gives an experiment rather than a universal recovery-capacity fraction.

\hypertarget{which-worker-is-cheapest-and-sufficient}{%
\section{Which worker is cheapest and sufficient?}\label{which-worker-is-cheapest-and-sufficient}}

Cheapest-sufficient routing has support from three directional literature items and one recent benchmark-bound controlled result. Cayci, Eryilmaz, and Srikant (\href{https://arxiv.org/abs/2003.00365}{2020}) developed budget-constrained online learning under broad cost and reward distributions. Somerstep et al.~(\href{https://arxiv.org/abs/2502.03261}{2025}) analyzed a calibrated static router. Li (\href{https://arxiv.org/abs/2502.02743}{2025}) reported preference-conditioned dynamic routing in a single-author preprint.

None evaluates a production software-agent fleet. This section therefore defines the routing decision and the measurements required to test it. It does not establish that a learned router will lower the cost of a particular fleet.

My fleet orchestrator illustrates the starting point. Each execution pool has a model assigned through a static configuration string. There is no request-level performance estimate, cost model, deadline, or lookahead.

The static table is predictable and legible. It also supplies the fallback when a selective router lacks evidence. Its limitation is that dispatch cannot respond to variation in request difficulty.

\Needspace{5\baselineskip}

A cheapest-sufficient router replaces the static assignment with two estimates for every candidate worker on the current request:

\begin{itemize}
\tightlist
\item
  expected task performance; and
\item
  expected total cost.
\end{itemize}

It then selects the least expensive worker predicted to satisfy the declared requirement.

Bhola, Krishnan, and NS (\href{https://arxiv.org/abs/2608.04804}{2026}) evaluated a scout-and-route configuration on 266 Python tasks from SWE-bench Pro. It solved 159 tasks, compared with 158 for the best single model, at about one fifth of the reported total cost per solve. A no-router ablation that always selected the cheapest fixer but retained the verified handoff tied the routed system, locating the measured gain in the handoff rather than the routing decision. The result strongly supports that benchmark-specific configuration and its null routing ablation; generalized cheapest-sufficient routing remains directional.

Somerstep et al.~analyze this plug-in structure in CARROT and argue that its direct routing overhead is negligible because selection requires evaluating the two predictors for each available model. The useful behavior still depends on calibration: predicted performance must correspond to observed performance on the deployment distribution.

A mathematically correct selector cannot compensate for a predictor that assigns unjustified confidence to unfamiliar work.

\Needspace{5\baselineskip}

The sufficiency requirement must be operational rather than rhetorical. ``Good enough'' might mean:

\begin{itemize}
\tightlist
\item
  passing an executable verifier;
\item
  remaining below a defect-probability threshold;
\item
  satisfying a calibrated review rubric;
\item
  meeting a deadline;
\item
  preserving a reliability requirement; or
\item
  satisfying a caller-selected cost-quality tradeoff.
\end{itemize}

Each defines a different routing problem.

The roughly 15-fold token premium reported for some fan-out configurations in Chapter 18 makes this choice economically consequential. Token count is still not the final objective. A cheap model that triggers repeated repair, verification, and review can cost more per accepted result than a stronger model would have cost initially.

When sufficient outcome data accumulates, the router can update through a budget-constrained bandit, such as Thompson sampling, with the algorithmic details left to the companion catalog. Cayci, Eryilmaz, and Srikant establish an asymptotic regret guarantee under moment conditions that permit heavy-tailed, correlated, and even negative cost-reward pairs.

Those properties resemble agent dispatch more closely than a setting with bounded independent rewards. The guarantee does not tell an operator whether the reward represents useful software work, and it provides limited guidance during the early period when production exploration is most costly.

Online updating addresses a real limitation of static routing. Models and workloads change. A model revision can change which tier is sufficient. A new repository or task family can move requests outside the router's calibration population.

Li conditions routing on a user-selected performance-cost preference, allowing one learned policy to represent several tradeoffs at inference time. The evidence is benchmark-based and comes from a single-author preprint. It supports testing preference conditioning. Calibration under delayed software outcomes remains unmeasured.

\Needspace{5\baselineskip}

The reward definition is the controlling measurement choice. Software signals arrive at different times and answer different questions:

\begin{itemize}
\tightlist
\item
  tests may be immediate but incomplete;
\item
  review may identify defects later;
\item
  rework or rollback may occur after acceptance;
\item
  user rejection may arrive after the router has recorded success; and
\item
  review effort may erase the apparent savings of the selected model.
\end{itemize}

Combining those signals into one reward is a construct-validity decision outside the routing mathematics. The reward definition should be versioned so that policies trained under different meanings of success are not silently compared.

\Needspace{5\baselineskip}

A learned policy should remain advisory until its recommendations correlate with an outcome measured outside the router and relevant to accepted work. In advisory mode, record:

\begin{itemize}
\tightlist
\item
  the requested route;
\item
  the route actually taken;
\item
  predicted performance;
\item
  predicted cost;
\item
  uncertainty;
\item
  the routing reason; and
\item
  the eventual outcome.
\end{itemize}

The thinly supported companion entry \texttt{log-scheduling-decisions-for-selection-effect-modeling} adds an important field: why the request received that route. Without the selection reason, later analysis cannot distinguish model quality from the policy that sent different work to each model.

When observations are sparse, fall back to the static cost-performance table. That floor is the current configuration and can be measured directly. The learned router should take control only where its errors can be estimated with adequate density.

Rare task families, new models, and new reward definitions should remain on the static floor until the router exposes enough evidence and uncertainty to justify a different choice.

\Needspace{5\baselineskip}

Calibration can fail before deployment. Benchmark-derived routing data may encode:

\begin{itemize}
\tightlist
\item
  regular prompt formats;
\item
  benchmark-specific scoring;
\item
  an unrealistically fixed model menu;
\item
  task distributions unlike production; or
\item
  missing repair and review costs.
\end{itemize}

CARROT assumes a static model pool. Production pools change as prices move, versions are retired, rate limits vary, and tool compatibility excludes some models. Each change requires recalibration or creates a region in which the router is extrapolating.

Online exploration creates another cost. A learning policy must sometimes choose an uncertain worker to gather information, and the routed request is real production work. A budget constraint reveals the cost but does not decide who may bear it.

\Needspace{5\baselineskip}

Deployment therefore needs:

\begin{itemize}
\tightlist
\item
  an exploration budget;
\item
  exclusions for high-consequence work;
\item
  a stop condition tied to routing error;
\item
  a static fallback; and
\item
  a record of which production requests were used for learning.
\end{itemize}

An asymptotic guarantee does not excuse poor performance during the learning period.

The companion pattern \texttt{cost-cascade-routing} provides another response to uncertainty. Begin with a cheaper route and escalate when calibrated confidence is insufficient. Chang (\href{https://arxiv.org/abs/2604.12262}{2026}) reports that confidence-gated cascades often outperformed always-on strong-model fan-out in both cost and quality.

That result supports the catalog pointer rather than a universal deployment rule. A cascade changes latency and couples its stages through the escalation signal. It must be evaluated as one complete route, with cost measured across every call.

A routing decision record should include:

\Needspace*{15\baselineskip}
\begin{record}
\begin{recordbody}
sufficiency_requirement:
reward_version:
calibration_population:
model_pool_version:
predicted_performance:
predicted_cost:
uncertainty:
routing_reason:
exploration_budget:
static_fallback:
eventual_outcome:
accepted_result_cost:
\end{recordbody}
\end{record}

Evaluation should report cost per accepted result and failure rates by request class, not only average token use.

Cheapest-sufficient routing is useful because it makes the economic policy explicit: what is being predicted, what requirement must be met, and which tradeoff is being selected. It cannot repair an unrepresentative calibration set, a reward that measures the wrong outcome, or a task whose true cost appears only after repair and review.

\hypertarget{under-what-serving-conditions-should-that-inference-run}{%
\section{Under what serving conditions should that inference run?}\label{under-what-serving-conditions-should-that-inference-run}}

Choosing a model settles one question and leaves another. Where does the call execute, and what state is that endpoint in when it arrives? A model endpoint is not an unbounded pool of identical workers. Behind it sit finite accelerators, finite device memory, finite memory bandwidth, a queue, a batching policy, and inference state that is either present or absent. Those conditions move latency, cost, and failure rate with no change in the model's capability, which puts them in the same decision record as price and predicted performance.

A model call runs in two phases. \textbf{Prefill} processes the input, which for a coding agent means system instructions, tool definitions, repository context, retrieved evidence, and the conversation so far. \textbf{Decode} then generates output tokens one at a time. The phases load the hardware differently and surface as different operational measures: time to first token is dominated by prefill and by whatever queueing precedes it, while generation latency scales with output length. A record that reports only total call latency cannot separate a long input from a slow generation, and the two have different remedies.

Decode avoids recomputing the attention state of every earlier token by retaining it, which is the key-value cache. The cache trades compute for memory, and that memory is what binds. Kwon et al.~(\href{https://arxiv.org/abs/2309.06180}{2023}) report that holding each request's cache in one contiguous block loses enough of it to fragmentation and duplication to limit how many requests fit at once, and that paging that memory instead, with sharing within and across requests, raises throughput by two to four times at the same latency. Two consequences follow at fleet scale. Cached state has a size, so a worker's remaining capacity is not described by its request count. Cached state also has a location, so it is present on one replica and absent on another.

Coding-agent sessions make both consequences larger than a short exchange would. A session carries a long and slowly changing prefix: system instructions, tool schemas, repository conventions, retrieved files, and an accumulating history of prior tool outputs. Reuse of a shared prefix across separate requests is a different mechanism from retaining state within one generation, and serving systems implement it explicitly. Zheng et al.~(\href{https://arxiv.org/abs/2312.07104}{2023}) retain previously computed prefixes in a radix tree and report up to 6.4 times higher throughput on workloads whose calls share structure, including agent control and multi-turn chat. Whether an agent workload has that structure is a property of the harness and its prompt construction, not of the model, and the realized gain depends on the serving implementation as well. Both are measurable on the operated fleet, and assuming them is how a routing experiment ends up crediting the wrong cause.

Two replicas exposing the same model are therefore not operationally equivalent at a given instant if one already holds the session's prefix. Routing inputs can then include model capability, price, queue depth, memory pressure, the request's deadline, and cache locality. This is the two-level structure the Mesos and Omega lineage already gives the fleet, one layer lower. A fleet-wide allocator decides what should run and against which requirement, and a worker-local dispatcher places the call where current serving conditions can meet it. None of this establishes that a cache-aware router is better than a simple one. It is another policy with its own cost and failure modes, and it earns promotion through the same fixed-arrival replay as any other.

Batching sits underneath both layers. A server that groups concurrent requests keeps the accelerator busy and raises tokens served per second; it also introduces queueing, and a request can wait behind work unrelated to it. That is this chapter's throughput-versus-cost distinction one level down. Maximizing tokens per second is not the same as minimizing time to an accepted result, and a fleet usually carries both kinds of demand. An interactive session spends most of its perceived quality on time to first token. An overnight migration, evaluation sweep, or background repair spends none of it there and cares about completion time and price. Those sit at different points on the batching tradeoff, and a serving policy can only honor the difference if the request declares which one it is.

Implementation-level optimizations move the envelope again. Paged cache management, speculative decoding (Leviathan, Kalman, and Matias \href{https://arxiv.org/abs/2211.17192}{2022}), and reduced-precision execution (Dettmers et al.~\href{https://arxiv.org/abs/2208.07339}{2022}) each change how much work an accelerator completes and at what latency. None is recommended here, and their reported gains are properties of the models and workloads they were measured on. They matter to a scheduler because a serving change can alter the resource envelope beneath the fleet with no change in the harness, which puts the serving configuration in capacity planning and in the version fields of any routing experiment.

\Needspace{8\baselineskip}

The reliability consequence is that inference conditions can produce what the fleet records as an agent failure:

\begin{itemize}
\tightlist
\item
  inference demand exceeds the serving capacity available to the fleet, through concurrency, cache pressure, or a shift in input length;
\item
  queueing rises, and with it time to first token and total call latency;
\item
  calls begin to exceed the harness timeout;
\item
  the harness retries, which adds fresh prefill work and, when the session's cached prefix has been evicted, recomputes it; and
\item
  the added demand deepens the saturation that produced the timeout.
\end{itemize}

Nothing in that sequence requires the model to have become less capable, and nothing in it is visible from the agent's trajectory alone. It is the retry-amplification shape from earlier in this chapter with an inference endpoint as the constrained dependency, and the same controls apply: bound admitted demand, jitter and cap retries, and reserve capacity so that recovery traffic does not compete with the load that displaced it. Pilot Execution's finding that recovery actions themselves cause severe failures has an instance here. So does Chapter 11's attribution discipline. A first-upstream-failure analysis that stops at ``the agent timed out'' has assigned the failure to the wrong component.

Telling that story from the record rather than from memory requires the call's own serving observations, which the allocation-ledger table later in this chapter lists alongside the rest. Managed providers expose different subsets of those fields, and several expose almost none of them. A field the provider does not report is not a field to estimate; it is a limit on the causal claims the operator may make from the record, and it should be stated as one. These observations also stay separate from the chapter's other measurements. Cached-input share is a cost and performance observation. It says nothing about whether the result was correct, and nothing about whether the work survived review. Collapsing the two into one efficiency ratio would hide the failure this section is about.

None of the serving work cited here was measured on a coding-agent fleet. It establishes that these mechanisms exist, that they bind on memory and locality rather than on request count, and that an operator can change them. Whether any of them is worth routing on in a particular fleet is a question for that fleet's replay.

\hypertarget{does-the-policy-survive-a-fixed-arrival-replay}{%
\section{Does the policy survive a fixed-arrival replay?}\label{does-the-policy-survive-a-fixed-arrival-replay}}

In one replay of my agent fleet, the interesting scheduling idea lost to a one-word configuration change. Across eleven weeks of recorded traffic, a four-feature weighted index performed within 0.1 percent of plain priority banding.

That result has narrative value only and is not primary evidence for this practice. The strong support comes from SWAY, the controlled software-engineering study by Chen et al.~(\href{https://arxiv.org/abs/1608.07617}{2016}), which shows why a proposed search method should have to beat a cheap baseline. Decima, from Mao et al.~(\href{https://arxiv.org/abs/1810.01963}{2018}), and the telescope-network work of Lampoudi, Saunders, and Eastman (\href{https://arxiv.org/abs/1503.07170}{2015}) provide directional support for learned scheduling and validation against known optima.

Fixed-arrival replay itself remains a reasoned transfer rather than a controlled result from an agent fleet.

An \textbf{arrival trace} is a time-ordered record of work entering the system, including the information available when each item arrived. A \textbf{dispatch policy} decides which eligible item receives a resource next. \textbf{First-come scheduling} orders eligible work by arrival time, subject to constraints that make an item impossible to run.

A \textbf{replay harness} re-executes the recorded arrivals under a candidate dispatch policy without changing production state. It differs from the single-run replay used for debugging in Chapter 10. That replay reconstructs one execution. This harness holds an entire workload sequence fixed while replacing the policy that allocates it.

\Needspace{5\baselineskip}

The trace needs, at minimum:

\begin{itemize}
\tightlist
\item
  arrival time;
\item
  priority;
\item
  resource eligibility;
\item
  claims and releases;
\item
  observed duration;
\item
  completion or failure outcome;
\item
  review verdict; and
\item
  work already running at each decision.
\end{itemize}

It also needs any constraint that affected feasibility. If the historical record omits capacity, cancellation, model compatibility, repository leases, or work already in flight, the replay may create choices the live scheduler never had.

Missing decision context is not neutral noise. It changes the feasible schedule.

Holding arrivals fixed creates a paired comparison. Every policy encounters the same demand in the same order. Differences in simulated flow can therefore be attributed to the policy and its interaction with the reconstructed state rather than to one arm receiving a busier week.

The replay is not causal in every respect. It does remove one large source of variation that often overwhelms scheduler comparisons.

\Needspace{5\baselineskip}

The first run should validate the harness rather than compare policies. Construct small cases whose best schedule is known, including:

\begin{itemize}
\tightlist
\item
  simultaneous arrivals;
\item
  a resource becoming unavailable;
\item
  a priority inversion;
\item
  incompatible execution pools;
\item
  a cancellation while work is queued; and
\item
  work already in flight when a new item arrives.
\end{itemize}

Confirm that the replay clock, eligibility logic, claim duration, completion events, and resource accounting reproduce the expected schedule.

Lampoudi, Saunders, and Eastman (2015) derived an integer-program formulation for the Las Cumbres Observatory network and reported computational validation at realistic problem sizes. The transferable practice is to test the scheduling kernel on instances with known answers before trusting its output on a historical trace.

\Needspace{5\baselineskip}

The second run establishes cheap baselines. Compare the proposed policy with:

\begin{itemize}
\tightlist
\item
  first-come scheduling;
\item
  a simple priority sort;
\item
  a tuned greedy rule; and
\item
  random oversampling.
\end{itemize}

Random oversampling generates several inexpensive randomized schedules and retains the best under the declared objective. Chen et al.~(2016) developed SWAY by sampling a large candidate population down to better solutions rather than evolving many generations. Across several software-engineering models, it was competitive with state-of-the-art evolutionary algorithms at substantially lower computational cost. The authors proposed it explicitly as a baseline for search-based software engineering.

The controlled result supports the discipline: an elaborate optimizer that cannot beat a cheap alternative has not earned adoption.

\Needspace{5\baselineskip}

The objective must be fixed before inspecting the policy results. Possible objectives include:

\begin{itemize}
\tightlist
\item
  mean completion time;
\item
  priority-weighted flow time;
\item
  deadline misses;
\item
  accepted-result cost;
\item
  reviewer delay; and
\item
  time spent blocked on a constrained resource.
\end{itemize}

These measures describe different failures. A lower fleet-wide mean can coexist with severe delay for one priority or request class.

\textbf{Starvation} is persistent denial or extreme delay to a class because other eligible work repeatedly moves ahead of it. Report outcomes by class and define a starvation guard before running the comparison. An aggregate gain is not acceptable when one class pays for it through unbounded delay.

The opening replay also moved P1 tail latency from 20.4 hours to 18.8 hours without triggering the starvation guard. Those figures explain what the protocol prevented me from building; they are not an expected gain for another fleet.

Capacity distribution explained more than the aggregate result. The visible improvement concentrated in one \textbf{contended pool}, where eligible work regularly exceeded available capacity. An \textbf{uncontended pool}, where capacity usually met demand, showed almost no difference across policies.

That is the expected shape of a genuine scheduling effect. When work rarely waits for a resource, queue order has little opportunity to change the outcome. A policy claiming large gains in an uncontended pool deserves scrutiny.

Decima provides a useful but bounded comparison. Mao et al.~(2018) trained a cluster scheduler under continuous stochastic arrivals and evaluated it on a 25-node Spark cluster. They reported at least a 21 percent improvement in average job-completion time over hand-tuned heuristics, with gains reaching approximately twofold under high load.

Those measurements support learned scheduling in that cluster environment. They do not evaluate fixed-arrival replay and do not establish an expected effect for agent work. The transferable experimental rule is narrower: preserve the arrival process when comparing policies and look for gains where resources are actually contested.

\Needspace{5\baselineskip}

Historical replay also preserves historical selection. Work absent from the ledger remains absent, including:

\begin{itemize}
\tightlist
\item
  requests users never submitted because the old system was slow;
\item
  tasks operators diverted manually;
\item
  work rejected before entering the queue; and
\item
  difficult requests routed to another worker group.
\end{itemize}

A route may therefore contain an artificially easy trace because the incumbent policy sent hard work elsewhere. The companion entry \texttt{log-scheduling-decisions-for-selection-effect-modeling} adds the reason each item entered a route, allowing later analysis to model selection. It cannot reconstruct demand that was never observed.

A live policy can also change future arrivals. Faster completion may create more demand. Priority treatment may change how users label requests. Model routing can change reviewer behavior and the kinds of tasks people submit.

\Needspace{5\baselineskip}

Historical replay cannot capture all such feedback. Use a \textbf{temporal holdout}, a later period excluded while policies are developed, to test whether the replay result survives workload drift. Then stage deployment through:

\begin{enumerate}
\def\labelenumi{\arabic{enumi}.}
\tightlist
\item
  shadow evaluation;
\item
  one-project or one-pool canary;
\item
  bounded live authority; and
\item
  fleet-wide rollout.
\end{enumerate}

Each stage should retain an external outcome measure and the starvation guard.

\Needspace{5\baselineskip}

A compact replay protocol is:

\begin{enumerate}
\def\labelenumi{\arabic{enumi}.}
\tightlist
\item
  Export an immutable arrival trace containing decisions, resources, claims, durations, outcomes, and review verdicts.
\item
  Build a discrete replay clock and validate the scheduler on constructed cases with known answers.
\item
  Record the objective, class-specific starvation guard, acceptance threshold, and cost accounting before comparing policies.
\item
  Replay first-come scheduling, simple priority, a tuned cheap baseline, random oversampling, and each candidate over the identical trace.
\item
  Report aggregate and per-class outcomes, decision cost, contention location, and optimality information where available.
\item
  Repeat the comparison on a temporal holdout.
\item
  Require shadow and canary evidence before expanding live allocation authority.
\item
  Run the recovery-capacity experiment from the admission and backpressure section: establish normal arrival rate and capacity, inject a worker loss or dependency failure, release a controlled backlog and retry population, and compare unrestricted recovery traffic with a bounded-recovery policy on live-work tail latency, recovery completion time, per-class queue depth, retry count, dependency request rate, starvation, useful throughput, and duplicate external effects.
\end{enumerate}

\Needspace{5\baselineskip}

This protocol is stricter than asking whether the queue feels better. It asks:

\begin{itemize}
\tightlist
\item
  whether the candidate changes a declared outcome on identical work;
\item
  whether a trivial policy captures the same gain;
\item
  whether one class pays for the improvement;
\item
  whether the effect occurs where capacity is constrained; and
\item
  whether it survives later traffic.
\end{itemize}

If the ledger already records the required state, the harness can be built without changing production scheduling. If it does not, the immediate scheduling task is instrumentation.

The earlier decisions in this chapter then become hypotheses rather than received settings. Replay several re-decision cadences and solve limits while preserving in-flight work. Measure flow, churn, solve latency, and the optimality gap retained at each limit.

Replay cheapest-sufficient routes from logged estimates when counterfactual outcomes are available. Otherwise run them in shadow. Measure accepted-result cost and calibration error by request class.

Build the trace, hold arrivals fixed, make the cheap policy compete, and let the fleet produce the evidence its allocation policy requires.

\hypertarget{what-must-be-recorded-before-optimization}{%
\section{What must be recorded before optimization?}\label{what-must-be-recorded-before-optimization}}

When the ledger does not preserve decision state, add fields in dependency order.

\begin{longtable}[]{@{}
  >{\raggedright\arraybackslash}p{(\columnwidth - 2\tabcolsep) * \real{0.5000}}
  >{\raggedright\arraybackslash}p{(\columnwidth - 2\tabcolsep) * \real{0.5000}}@{}}
\caption{Allocation-ledger fields in the order they become interpretable.}\tabularnewline
\toprule\noalign{}
\begin{minipage}[b]{\linewidth}\raggedright
Stage
\end{minipage} & \begin{minipage}[b]{\linewidth}\raggedright
Fields
\end{minipage} \\
\midrule\noalign{}
\endfirsthead
\toprule\noalign{}
\begin{minipage}[b]{\linewidth}\raggedright
Stage
\end{minipage} & \begin{minipage}[b]{\linewidth}\raggedright
Fields
\end{minipage} \\
\midrule\noalign{}
\endhead
\bottomrule\noalign{}
\endlastfoot
Observed events & Arrival time, priority, claim and release events, observed duration, completion or failure outcome, and review verdict. \\
Serving observations & Per call: endpoint or route selected, model and serving configuration version, input and output token counts, cached input share where the provider reports it, time to first token, total call latency, observed queueing, and retry count. Record which of these the provider does not expose. \\
Decision state & Available capacity, eligible execution pools, work already running, leases or locks, deadlines, model or tool compatibility, reviewer availability, and constraints that excluded an otherwise eligible item. \\
Policy estimates & Predicted cost, predicted task performance, routing uncertainty, expected duration, solver objective value, and optimality gap. \\
Code-estate state & Campaign identity, repository, base revision, host or workspace, affected entities, artifact version, code index version, conflict evidence, and publication state. \\
\end{longtable}

The observed fields reconstruct demand, occupancy, and accepted work. They also expose missing or contradictory histories before policy-specific estimates complicate the schema.

Record the route selected and the reason for that route in the same decision event. State and reason need a shared decision identity and timestamp. Without them, later analysis cannot determine which alternatives were feasible or distinguish worker performance from the policy that selected the work.

The code-estate fields carry the same versioned view of the code that Chapter 18's node record gives the task graph into the allocation ledger, so that a scheduling decision can later be checked against the repository state and conflict evidence it acted on; full cross-repository campaign semantics stay in the companion repository, because the current evidence does not support a developed practice claiming one campaign protocol is generally superior.

Estimates belong last because they are outputs of a policy rather than observations of what happened. Store the policy, model, feature, and reward versions that produced them. Recalibration should create a new estimate record rather than rewrite the historical one.

The first operating comparison can remain descriptive. Ask whether high-throughput intervals under the incumbent policy also have a higher accepted-result cost than lower-throughput intervals. Report both measures by request class and constrained resource.

That comparison cannot establish the counterfactual effect of a new router, the correct re-decision cadence or solver limit, an exploration guarantee, the effect of demand absent from the ledger, or the causal effect of throughput on accepted-result cost.

Those questions still require replay, shadow execution, or controlled rollout on the fleet's workload. None of the cited studies reports results from a software-agent fleet.

My replay supplies no default cadence, threshold, or cost ratio. Do not import an observatory schedule, a bandit guarantee, or my 15 percent promotion rule.

The useful outcome is an allocation record that reconstructs what arrived, what was eligible, which resources and alternatives existed, which policy selected the route and why, what the work cost, what outcome survived review, and which later events changed the interpretation of the result.

Until the ledger can answer those questions, choosing a more sophisticated policy is premature.

The repository artifact \href{https://github.com/sjarmak/engineering-reliable-coding-agents/blob/main/protocols/allocation-policy-replay.md}{\texttt{protocols/allocation-policy-replay.md}} defines the replay comparison, promotion rule, and retained decision record.

\section*{Sources and evidence}

\textbf{Re-decide cheaply and ship the incumbent}

\begin{itemize}
\tightlist
\item
  Directional evidence: Bellm, E. C., et al.~(2019), ``The Zwicky Transient Facility: Surveys and Scheduler,'' PASP 131, 068003, arXiv:1905.02209.
\item
  Directional evidence: Naghib, E., et al.~(2018/2019), ``A Framework for Telescope Schedulers: With Applications to the Large Synoptic Survey Telescope,'' AJ 157, 151 (2019), arXiv:1810.04815 (2018). Framework and simulation evidence.
\item
  Directional evidence: Parazin, B., et al.~(2022), ``Foraging with MUSHROOMS: A Mixed-integer Linear Programming Scheduler for Multimessenger Target of Opportunity Searches with the Zwicky Transient Facility,'' ApJ 935, 87, arXiv:2203.00013.
\item
  Strong evidence for production workload characterization, directional for capacity-policy transfer: Liu, B., et al.~(2026), ``Agentic Coding in the Wild: Characterizing GitHub Copilot Traces at Production Scale,'' arXiv:2608.00101.
\item
  No strong result and no software-fleet measurement support this entry; all three items are observatory scheduling, and the cadence, cap, and churn charge remain deployment decisions.
\item
  Corroboration (narrative only): the author's transfer notes on observatory scheduling restate the same ZTF and MUSHROOMS results already carried above as literature evidence.
\end{itemize}

\textbf{Admission, backpressure, and recovery capacity}

\begin{itemize}
\tightlist
\item
  Directional evidence: Verma, A., et al.~(2015), ``Large-scale cluster management at Google with Borg,'' EuroSys. Admission control, packing, isolation, and overcommitment as explicit cluster policy; mechanisms transfer, parameters do not.
\item
  Directional evidence: Schwarzkopf, M., et al.~(2013), ``Omega: flexible, scalable schedulers for large compute clusters,'' EuroSys. Parallel schedulers over shared state with optimistic concurrency; supports separating global allocation from worker-local dispatch.
\item
  Directional evidence: Mace, J., et al.~(2015), ``Retro: Targeted Resource Management in Multi-tenant Distributed Systems,'' NSDI. Per-tenant resource monitoring and central control points; maintenance and recovery traffic can itself become the overload source.
\item
  Directional evidence: Hindman, B., et al.~(2011), ``Mesos: A Platform for Fine-Grained Resource Sharing in the Data Center,'' NSDI. Two-level allocation between a central allocator and per-framework dispatch.
\item
  Directional evidence: Li, Z., Cai, Q., \& Lou, C. (2026), ``Pilot Execution,'' NSDI. Recovery actions themselves cause severe failures; supports bounding the release rate of recovery traffic.
\item
  Practitioner guidance: AWS Builders' Library, ``Timeouts, retries, and backoff with jitter'' and ``Making retries safe with idempotent APIs.'' Mechanism descriptions from production practice, not controlled experiments; their constants are their systems' parameters.
\item
  Practitioner guidance: Google SRE, ``Addressing Cascading Failures'' and ``Handling Overload.'' Same status.
\item
  No source in this group evaluates a software-agent fleet, and no recovery-capacity fraction is prescribed; the recovery-capacity experiment in the replay protocol is how a fleet measures its own division.
\end{itemize}

\textbf{Route each request to the cheapest sufficient model}

\begin{itemize}
\tightlist
\item
  Directional evidence: Cayci, S., Eryilmaz, A., \& Srikant, R. (2020), ``Budget-Constrained Bandits over General Cost and Reward Distributions,'' arXiv:2003.00365. Asymptotic guarantee under stated moment conditions.
\item
  Directional evidence: Somerstep, S., et al.~(2025), ``CARROT: A Cost Aware Rate Optimal Router,'' arXiv:2502.03261. Static model pool assumed.
\item
  Directional evidence: Li, Y. (2025), ``LLM Bandit: Cost-Efficient LLM Generation via Preference-Conditioned Dynamic Routing,'' arXiv:2502.02743. Single-author preprint, benchmark evidence.
\item
  Strong evidence for one benchmark-bound scout-and-fixer configuration and its no-router ablation: Bhola, I., Krishnan, A., and NS, M. (2026), ``Scrouting: Cost-Aware Routing of Coding Agents by Scouting the Repository First,'' arXiv:2608.04804. The handoff, rather than the router, carried the measured result.
\item
  Directional evidence: Kwon, W., et al.~(2023), ``Efficient Memory Management for Large Language Model Serving with PagedAttention,'' SOSP 2023, arXiv:2309.06180. Key-value cache memory, not request count, bounds concurrent serving capacity; the reported two-to-four-times throughput gain is a property of the evaluated systems and models.
\item
  Directional evidence: Zheng, L., et al.~(2023), ``SGLang: Efficient Execution of Structured Language Model Programs,'' arXiv:2312.07104. Cross-request prefix reuse makes two replicas of one model operationally unequal; the reported gains depend on workload structure and serving implementation.
\item
  Directional evidence: Leviathan, Y., Kalman, M., \& Matias, Y. (2022), ``Fast Inference from Transformers via Speculative Decoding,'' arXiv:2211.17192, and Dettmers, T., et al.~(2022), ``LLM.int8(): 8-bit Matrix Multiplication for Transformers at Scale,'' arXiv:2208.07339. Cited only as evidence that serving-level optimizations exist and change the resource envelope; no claim that either is appropriate for a given fleet.
\item
  The three routing-theory items remain directional, and no source evaluates a production software-agent fleet; generalized cheapest-sufficient routing therefore remains a decision to test rather than an established cost reduction. The four inference-serving items are systems results measured on serving benchmarks, not on agent workloads, and they support the existence and mechanics of the constraint rather than any routing policy over it.
\item
  Directional evidence for this developed practice: CascadeDebate (Chang 2026), arXiv:2604.12262. Its confidence-gated cascade comparison strongly supports the narrower companion-catalog claim in the tested setting; transfer to production software-agent routing remains directional.
\item
  Corroboration (narrative only): the author's transfer notes restate the budgeted-bandit formulation already carried above.
\end{itemize}

\textbf{Replay fixed arrival traces before changing policy}

\begin{itemize}
\tightlist
\item
  Strong evidence for the cheap-baseline comparison only: SWAY, Chen, J., et al.~(2016), ``Sampling as a Baseline Optimizer for Search-Based Software Engineering,'' arXiv:1608.07617. The study supports requiring a candidate optimizer to beat a cheap baseline; fixed-arrival replay is a transfer beyond its experiment.
\item
  Directional evidence: Decima, Mao, H., et al.~(2018), ``Learning Scheduling Algorithms for Data Processing Clusters,'' arXiv:1810.01963. Measures learned scheduling under stochastic arrivals, not fixed-arrival replay; the replay transfer is the synthesis's.
\item
  Directional evidence: Lampoudi, S., Saunders, E., \& Eastman, J. (2015), ``An Integer Linear Programming Solution to the Telescope Network Scheduling Problem,'' arXiv:1503.07170. Supports validating a scheduling kernel on instances with known optima.
\item
  Fixed-arrival replay itself is a reasoned transfer, not a directly controlled fleet result, and per-class starvation must be checked separately because a policy can improve the mean by starving one class.
\item
  Corroboration (illustration only): the author's eleven-week replay of 1,286 work items across 22 execution pools, and the author's shadow, canary, then fleet rollout design. These cases do not count as independent external evidence.
\end{itemize}

\gdef\currentparttitle{Closing}
\chapter{The evidence chain behind reliable agents}
\label{closing-evidence-chain}

The most useful result in my retrieval work was a disagreement between two instruments.

Three retrieval measures agreed that the system had become substantially better at placing relevant repository evidence in front of the agent. Across 370 paired tasks, however, the end-to-end reward moved by only +0.0349. Worse, the first interval I reported around that difference was too optimistic: it resampled tasks as though they were independent even though they came from 46 anchor repositories and 20 suites.

The retrieval result was real. I had measured the uncertainty around the downstream result incorrectly.

Understanding that experiment required two parts of this monograph. The stage decomposition from Part IV identified where the improvement occurred. The resampling rules from Part I determined whether the end-to-end difference had been measured adequately at all.

A coding agent's reliability depends on a chain of claims about the system. A strong model, high benchmark score, complete trace, sophisticated retrieval system, or gated deployment cannot establish reliability alone. Each claim gives the next layer something it is allowed to trust.

When an earlier claim is wrong, the failure rarely announces itself downstream. It arrives as a clean score, a confident grader verdict, a plausible trace, a relevant-looking context window, an approved change, or a successfully completed workflow.

The hard problem is deciding which claims about agent behavior deserve to survive contact with the next layer of the system.

\hypertarget{each-layer-determines-what-the-next-may-trust}{%
\section{Each layer determines what the next may trust}\label{each-layer-determines-what-the-next-may-trust}}

Part I sits beneath the rest of this monograph because every later practice is eventually accepted or rejected through a comparison.

One run is one draw from a stochastic system. Before a measured difference can describe the system rather than one favorable draw, the evaluation must characterize local variation, preserve the dependency structure of the sample, and be capable of resolving an effect large enough to matter.

Nineteen practices developed in later parts execute methods introduced in Part I. Removing the explicit evaluation machinery merely leaves those requirements unstated.

Part II turns measurements into operating verdicts.

A model grader, consensus vote, reward model, heuristic score, or proxy metric is an instrument. Each has an operating point, a distribution on which it was calibrated, and errors that can matter asymmetrically. Skalse et al.~(\href{https://arxiv.org/abs/2209.13085}{2022}) showed, in a bounded formal setting over stochastic policies, how restrictive the relationship between a useful reward and an unhacked optimization target can be. The practical lesson is broader than the theorem: a nonconstant proxy should not quietly become final authority merely because it is convenient to optimize.

A grading system therefore produces a verdict under stated conditions. It does not produce truth.

\Needspace{5\baselineskip}

Part III asks whether the record underneath those verdicts is evidence at all.

Four ordinary system failures can corrupt that record in ways no later statistical treatment repairs:

\begin{itemize}
\tightlist
\item
  One identity can reach both primary data and its recovery material.
\item
  A run can die without leaving a durable account of completed work.
\item
  A trace can record tool dispatch without distinguishing whether the external effect committed.
\item
  A quarantine or cleanup policy can remove failed runs before anyone examines them.
\end{itemize}

Even a complete record does not solve attribution automatically. Zhang et al.~(\href{https://arxiv.org/abs/2505.00212}{2025}) gave expert-annotated failure logs from 127 multi-agent systems to the strongest automated attribution methods they evaluated. The best method located the decisive step in only 14.2 percent of cases.

A trace can make an explanation possible, but it cannot manufacture causality.

\Needspace{5\baselineskip}

Part IV determines what evidence reached the model.

A failed implementation should not be assigned to reasoning until the evaluation can answer a more basic sequence of questions:

\begin{itemize}
\tightlist
\item
  Did the required evidence exist in the searchable corpus?
\item
  Was the relevant version present in the index?
\item
  Did retrieval return it?
\item
  Did ranking and context selection preserve it?
\item
  Did it describe the repository state the agent was actually modifying?
\end{itemize}

Stale context is the sharpest example because it can remain perfectly fluent after losing authority. The agent may reason correctly over an interface that no longer exists.

Part V treats human review as a finite system resource rather than an unlimited source of correctness.

Its allocation depends on monitors calibrated in Part II. Its interfaces determine how expensive it is to challenge a claim rather than accept one. Its gates depend on the authority boundaries and evidence preserved in Part III.

A gate that cannot change the execution path records assent. It does not provide control.

Part VI depends on artifacts produced by every earlier part. Routing and scheduling policies should be judged by replaying recorded arrivals, holding demand fixed, and forcing sophisticated policies to compete against cheap alternatives.

That requirement appears repeatedly in the literature because complexity has a habit of beating an absent baseline. Kapoor et al.~(\href{https://arxiv.org/abs/2407.01502}{2024}) found that simple model retries could match more elaborate agent architectures on a function-level coding benchmark at a fraction of their inference cost. Chen et al.~(\href{https://arxiv.org/abs/1608.07617}{2016}) made the same methodological point in search-based software engineering: an optimization method should first beat inexpensive random sampling.

The exact baseline changes, but the obligation does not.

The dependency chain also explains why some repairs are so expensive. Later machinery cannot reliably compensate for an earlier defect:

\begin{itemize}
\tightlist
\item
  More samples do not repair a task distribution that excludes production work.
\item
  More judges do not repair a rubric experts interpret inconsistently.
\item
  More agents do not repair a retrieval boundary that treats an empty result as authoritative.
\item
  More context does not repair a run that has lost its governing specification.
\item
  Better scheduling does not repair a work ledger that cannot identify which attempt owns the right to publish.
\item
  More observability does not repair an external effect whose identity was never recorded.
\end{itemize}

In each case, the added machinery is evaluated through the instrument that was already defective.

This repair asymmetry lets problems propagate forward cheaply while making downstream correction expensive.

\hypertarget{engineering-controls-can-precede-prevalence-evidence}{%
\section{Engineering controls can precede prevalence evidence}\label{engineering-controls-can-precede-prevalence-evidence}}

The evidence in this monograph is uneven, and that unevenness does not track operational consequence.

Strong-graded items make up 56 percent of the evidence in Part I but only 10 percent in Part III. Chapter 8 contains no strong-graded or direct scholarly evidence item even though its containment practices may be among the first controls needed by a system with production write access.

Within the 52-practice consequence-ranking subset, Spearman's rho between urgency rank and the presence of at least one strong evidence item is -0.004. In that selection, evidence strength and operational urgency are effectively unrelated.

Engineering sometimes has to move before prevalence has been estimated.

\texttt{Contain agent blast radius} ranked first by consequence in the catalog, yet its incident support consists of two corroborating practitioner reports: one involving a production workflow and another involving database deletion. Those reports establish possibility. They do not establish frequency.

I still recommend the control because its mechanism can be inspected directly. If one credential can alter both production state and the material required to recover that state, those resources share a failure domain. The authority boundary can be changed. A prohibited write can be exercised against a test resource. Ordinary and recovery authority can be separated. The control is reversible and its local effect is observable.

This is a mechanism-based justification, not a claim that the literature has proved a universal effect size.

Some practices in this monograph are supported by measured effects. Others are supported by a structural failure argument:

\begin{enumerate}
\def\labelenumi{\arabic{enumi}.}
\tightlist
\item
  The failure is possible under the current authority or state model.
\item
  Its consequence is material.
\item
  The proposed control changes the relevant mechanism rather than a proxy for it.
\item
  The change is reversible or bounded.
\item
  The resulting boundary can be exercised directly.
\end{enumerate}

The following are observations of local protections, not estimates of production frequency:

\begin{itemize}
\tightlist
\item
  A denied write.
\item
  A durable completion record that survives a worker crash.
\item
  A stale completion rejected by a newer ownership generation.
\item
  A replay in which the same recorded arrivals face two schedulers.
\end{itemize}

Each observation shows whether a claimed local protection exists.

The imperative headings in thin-evidence chapters should be read in this sense. They specify an engineering action and the observation that can falsify its local justification. They do not imply a universal prevalence estimate or effect size.

A mechanism that cannot be reduced to an executable observation has not yet been specified well enough to rely on.

Important gaps remain. Two studies would materially strengthen the record developed here:

\begin{itemize}
\tightlist
\item
  A controlled comparison of isolation and authority designs scored against observed incidents.
\item
  A recovery benchmark with a published fault menu, named kill placements, recurring failures, lost acknowledgements, and upgrades across runtime versions.
\end{itemize}

Part VI remains more transfer-heavy still. Its topology, admission-control, partitioning, and scheduling proposals are executable research questions for coding-agent fleets. They should not be mistaken for established production effects.

Useful engineering hypotheses need not masquerade as settled science. They need claims that state what kind of evidence supports them.

\hypertarget{start-smaller-than-this-book}{%
\section{Start smaller than this book}\label{start-smaller-than-this-book}}

A production system does not need every mechanism described in these chapters before it can improve.

It does need enough instrumentation to tell whether an improvement occurred and enough durable state to show what happened when something failed.

The literature does not provide universal values for repeat count, inter-rater agreement, context limits, refresh cadence, routing cost, retry count, or re-solve frequency. The values below are starting points for a new system, not transferred effect estimates.

\begin{longtable}[]{@{}
  >{\raggedright\arraybackslash}p{(\columnwidth - 4\tabcolsep) * \real{0.22}}
  >{\raggedright\arraybackslash}p{(\columnwidth - 4\tabcolsep) * \real{0.43}}
  >{\raggedright\arraybackslash}p{(\columnwidth - 4\tabcolsep) * \real{0.35}}@{}}
\caption{Personal starting points for a new system. These values are author defaults, not transferred effect estimates.}\tabularnewline
\toprule\noalign{}
\begin{minipage}[b]{\linewidth}\raggedright
Decision
\end{minipage} & \begin{minipage}[b]{\linewidth}\raggedright
Personal starting point
\end{minipage} & \begin{minipage}[b]{\linewidth}\raggedright
What changes it
\end{minipage} \\
\midrule\noalign{}
\endfirsthead
\endhead
\bottomrule\noalign{}
\endlastfoot
\textbf{Repeated evaluation} & Use \(k=3\) paired runs per item for inexpensive screening, but do not promote from that screen alone. Size release comparisons from a pilot and the smallest decision-relevant effect. & Higher variance, clustered tasks, or smaller meaningful effects require more runs. \\
\textbf{Reliability reporting} & Report both \(\mathrm{pass}@k\), where any attempt succeeds, and \(\mathrm{pass}^{k}\), where every attempt succeeds, with \(k\) printed beside the result. Begin with \(k=3\). & The deployed retry policy and the cost of intermittent failure determine the useful \(k\). \\
\textbf{Human-label agreement} & Treat Cohen's or Fleiss's kappa below 0.60 as a rubric-debugging trigger. A value above 0.60 is not a correctness gate. Promotion still depends on class-specific errors against expert adjudication. & Class imbalance, ambiguous labels, and high-consequence mistakes make the confusion matrix more informative than kappa alone. \\
\textbf{Promotion threshold} & Before execution, require either a three-percentage-point absolute success gain at no material cost increase or a ten-percent cost reduction at no success loss. Treat smaller effects as uncredited until a decision justifies them. & Task value, baseline success rate, and operating cost should replace these round numbers. \\
\textbf{Failure review} & Read twenty stratified failures after a material model, evaluation apparatus, or policy change. Preserve an abstain category when the trace cannot support attribution. & Rare high-consequence failure classes should be oversampled regardless of frequency. \\
\textbf{Context restart} & Restart or consolidate at 60 percent of the shortest context length at which the weakest required task shows material degradation in a local sweep. & New models, tools, instruction files, or task mixtures require another sweep. \\
\textbf{Freshness} & Require exact repository-revision identity for code-bearing retrieval and reject an index whose source revision is unknown. & A coarser content-addressed equivalence is acceptable only when the indexer can prove it. \\
\end{longtable}

The apparatus itself belongs in the cost model.

Each serious comparison should record inference and tool cost, elapsed time, labeling and adjudication hours, reviewer queue time, storage, and maintenance work. The denominator should be accepted work rather than model calls.

A control whose measurement cost remains invisible will eventually be bypassed, removed, or defended with stale evidence.

When capacity is constrained, remove complexity from the outside inward. Cut dynamic scheduling and learned routing before cutting the records needed to evaluate them. Cut multi-agent debate, specialist roles, and elaborate aggregation before cutting the acceptance check. Cut redundant model judges and retrieval lanes before cutting the ability to reconstruct which artifact was accepted. Reduce benchmark breadth before abandoning a small stratified set of consequential tasks.

Rich interfaces and secondary taxonomies can also go if their questions remain answerable from the durable record.

Four things should survive almost to the floor:

\begin{enumerate}
\def\labelenumi{\arabic{enumi}.}
\item
  \textbf{An executable acceptance check.}
\item
  \textbf{Separation between ordinary authority and recovery authority.}
\item
  \textbf{Versioned identity for the task, input state, configuration, attempt, and accepted artifact.}
\item
  \textbf{A paired comparison against the cheapest credible baseline.}
\end{enumerate}

A small team can operate those controls without reproducing the full apparatus described in this monograph.

This minimum viable reliability system can show whether the work passed and which state and configuration produced it. It can keep a failed worker from silently retaining authority and show whether a more complicated system beat the simpler one.

Everything added after that should have to earn its place.

\hypertarget{some-uncertainty-should-remain-visible}{%
\section{Some uncertainty should remain visible}\label{some-uncertainty-should-remain-visible}}

Several questions in this book remain research problems rather than missing configuration defaults.

No controlled result in the reviewed set establishes that a dynamic dependency graph outperforms a well-designed fixed schedule for repository work. Propagation of deletion through summaries, embeddings, caches, and graph edges remains poorly measured. The optimal recovery architecture for long-running coding agents is not settled. Neither are universal grader thresholds, context policies, retry counts, or fleet scheduling rules.

Many empirical conclusions here are also dated snapshots. Models, context windows, retrieval systems, agent runtimes, and costs will change. Benchmarks will become contaminated or obsolete.

The structural claims are harder to age out:

\begin{itemize}
\tightlist
\item
  One stochastic run is one draw.
\item
  A component that did not execute did not cause the observed result.
\item
  An instrument has to be calibrated for the decision it is being asked to make.
\item
  A trace can preserve evidence without proving causality.
\item
  An empty retrieval result is not proof that relevant evidence does not exist.
\item
  A stale artifact can remain fluent after losing authority.
\item
  An acknowledged workflow transition does not prove an external effect occurred exactly once.
\item
  A worker that has lost ownership must also lose the ability to make consequential mutations.
\item
  One identity that reaches production data and recovery material defines one failure domain.
\item
  Additional machinery should beat the cheapest credible alternative under the same workload before it earns adoption.
\end{itemize}

Those claims describe constraints on reasoning about systems, not preferences for today's agent stack. They are the part of this book I expect to last longest.

\hypertarget{the-system-around-the-model}{%
\section{The system around the model}\label{the-system-around-the-model}}

The easiest way to misunderstand a coding agent is to look only at the model.

A production result is jointly determined by the task presented to the system, the version of the world the system can see, the evidence retrieval makes available, the state retained across steps, the tools and permissions through which actions occur, the workflow that decides what may happen next, the acceptance path that decides what counts, and the evaluation that tells us whether any of those choices helped.

The model matters enormously, but it is only one component in the causal path.

A better model can fail inside a weak execution system; better retrieval can disappear inside a noisy end-to-end score; a complete trace can support the wrong attribution; a successful retry can duplicate an external effect; and an elaborate fleet scheduler can optimize a workload whose ownership semantics were never correct.

The practical consequence is a debugging order:

\begin{enumerate}
\def\labelenumi{\arabic{enumi}.}
\item
  Begin with the comparison you currently trust most.
\item
  Preserve the per-item result.
\item
  Ask whether repeated runs support the difference.
\item
  Check whether the verdict instrument agrees with expert adjudication on the errors that matter.
\item
  Confirm that the durable record can reconstruct the accepted attempt and artifact.
\item
  Verify that the evidence required by the task existed, was current, and reached the model.
\item
  Exercise the consequential boundary under failure.
\item
  Add machinery only after those checks pass.
\end{enumerate}

The six-step minimum pass is the shortest practical route through this dependency chain. It exposes the claims a system already depends on without prescribing a single agent architecture.

A reliable agent system preserves the evidence needed to understand a failure, prevents stale authority from surviving recovery, keeps uncertain measurements uncertain, and requires added complexity to improve accepted work.

The model can propose the work. The surrounding system determines what evidence the proposal used, what it could change, whether the result was accepted, whether the effect occurred, and what evidence survived.

\section*{Sources and evidence}

No evidence is introduced here. Each identifier below is carried by the chapter named, and the evidence grouping comes from that chapter's record.

\begin{itemize}
\tightlist
\item
  Strong evidence: Skalse, Howe, Krasheninnikov \& Krueger (2022). \emph{Defining and Characterizing Reward Hacking}. NeurIPS 2022. arXiv:2209.13085. Chapter 6, \texttt{layer-signals-beyond-single-proxy}.
\item
  Strong evidence: Zhang, S., et al.~(2025). \emph{Which Agent Causes Task Failures and When? On Automated Failure Attribution of LLM Multi-Agent Systems}. ICML 2025. arXiv:2505.00212. Chapter 11, \texttt{keep-humans-in-failure-attribution}.
\item
  Directional evidence: Kapoor, Stroebl, Siegel, Nadgir \& Narayanan (2024). \emph{AI Agents That Matter}. arXiv:2407.01502. Chapter 2, \texttt{report-cost-accuracy-pareto}.
\item
  Strong evidence: Chen, J., et al.~(2016). \emph{Sampling as a Baseline Optimizer for Search-Based Software Engineering}. arXiv:1608.07617. Chapter 19, \texttt{replay-traces-before-policy-changes}.
\end{itemize}

The part-level and chapter-level evidence shares restated above were recomputed from the companion catalog: Part I has 18 of 32 items grouped as strong, Part III has 3 of 30, Part IV has 20 of 42, and Part VI has 8 of 29. Chapter 8 has no strong or direct scholarly evidence item.

\clearpage
\appendix
\gdef\currentparttitle{Appendix}
\chapter{Practice catalog index}\label{appendix-practices}

Every reliability record in this edition carries a stable identifier of the form \texttt{ERCA-NNN}, after the initials of this book's title. The chapters name records by identifier rather than by title, because a title can be rewritten while the identifier must not be. This index resolves 206 reliability records: 193 gated practices, including 56 developed in the main chapters, plus 13 research leads. Every \texttt{ERCA-NNN} in the text links to its row here.

The \emph{chapter} column gives the chapter whose mechanism the record extends, including for records the chapters do not develop. The \emph{treatment} column separates records developed in the main text from compact companion entries and from leads, which are preserved for investigation rather than recommended. A dagger marks the 29 companion entries labeled limited support, where the available evidence does not support a recommendation.

This index carries identity and placement only. The action, mechanism, evidence, and boundary of each record live in the companion artifact, in \href{https://github.com/sjarmak/engineering-reliable-coding-agents/blob/main/companion/catalog.json}{\texttt{catalog.json}} and \href{https://sjarmak.ai/books/engineering-reliable-coding-agents/companion}{the companion site}.

\begingroup
\small
\setlength{\LTpre}{0.6\baselineskip}
\setlength{\LTpost}{0.6\baselineskip}
\begin{longtable}[]{@{}
  >{\raggedright\arraybackslash}p{0.145\linewidth}
  >{\raggedright\arraybackslash}p{0.585\linewidth}
  >{\raggedright\arraybackslash}p{0.05\linewidth}
  >{\raggedright\arraybackslash}p{0.135\linewidth}@{}}
\caption{Practice catalog index: identifier, practice, chapter, and treatment for all 206 records in this edition.}\tabularnewline
\toprule\noalign{}
\textbf{Identifier} & \textbf{Practice} & \textbf{Ch.} & \textbf{Treatment} \\
\midrule\noalign{}
\endfirsthead
\toprule\noalign{}
\textbf{Identifier} & \textbf{Practice} & \textbf{Ch.} & \textbf{Treatment} \\
\midrule\noalign{}
\endhead
\bottomrule\noalign{}
\endlastfoot
\raisebox{0pt}[0pt][0pt]{\hypertarget{erca-001}{}}\texttt{ERCA-001} & Strengthen weak test oracles before trusting pass rates & 3 & developed \\
\raisebox{0pt}[0pt][0pt]{\hypertarget{erca-002}{}}\texttt{ERCA-002} & Run memorization probes before attributing benchmark gains to capability & 3 & companion \\
\raisebox{0pt}[0pt][0pt]{\hypertarget{erca-003}{}}\texttt{ERCA-003} & Pair every public benchmark score with a matched private control and measure the gap & 3 & developed \\
\raisebox{0pt}[0pt][0pt]{\hypertarget{erca-004}{}}\texttt{ERCA-004} & Evaluate on post-cutoff tasks via temporal holdouts and live pipelines & 3 & developed \\
\raisebox{0pt}[0pt][0pt]{\hypertarget{erca-005}{}}\texttt{ERCA-005} & Measure train-test overlap with semantic matching, not just n-grams & 3 & companion \\
\raisebox{0pt}[0pt][0pt]{\hypertarget{erca-006}{}}\texttt{ERCA-006} & Measure contamination by estimated score benefit, not binary flags\,$\dagger$ & 3 & companion \\
\raisebox{0pt}[0pt][0pt]{\hypertarget{erca-007}{}}\texttt{ERCA-007} & Build fresh eval items under an anti-searchability annotation protocol & 3 & companion \\
\raisebox{0pt}[0pt][0pt]{\hypertarget{erca-008}{}}\texttt{ERCA-008} & Treat contamination as a (model x benchmark) cell, not a benchmark property & 3 & companion \\
\raisebox{0pt}[0pt][0pt]{\hypertarget{erca-009}{}}\texttt{ERCA-009} & Combine contamination detectors by threat model; treat negatives as weak evidence & 3 & companion \\
\raisebox{0pt}[0pt][0pt]{\hypertarget{erca-010}{}}\texttt{ERCA-010} & Run a construct-validity check before adopting or building any benchmark & 3 & companion \\
\raisebox{0pt}[0pt][0pt]{\hypertarget{erca-011}{}}\texttt{ERCA-011} & Build code benchmarks against the documented defect taxonomy\,$\dagger$ & 3 & companion \\
\raisebox{0pt}[0pt][0pt]{\hypertarget{erca-012}{}}\texttt{ERCA-012} & Evaluate agents jointly on cost and accuracy, with trivial baselines & 2 & developed \\
\raisebox{0pt}[0pt][0pt]{\hypertarget{erca-013}{}}\texttt{ERCA-013} & Give agent benchmarks real holdout sets, split by generalization level & 3 & companion \\
\raisebox{0pt}[0pt][0pt]{\hypertarget{erca-014}{}}\texttt{ERCA-014} & Pin the eval harness; publish prompts, code, and raw outputs & 3 & companion \\
\raisebox{0pt}[0pt][0pt]{\hypertarget{erca-015}{}}\texttt{ERCA-015} & Stress-test leaderboard rankings under problem and harness perturbations & 3 & companion \\
\raisebox{0pt}[0pt][0pt]{\hypertarget{erca-016}{}}\texttt{ERCA-016} & Report performance spread across prompt variants, not a single-format score & 1 & companion \\
\raisebox{0pt}[0pt][0pt]{\hypertarget{erca-017}{}}\texttt{ERCA-017} & Protect eval data at release: encryption, licensing, exclusion controls & 3 & companion \\
\raisebox{0pt}[0pt][0pt]{\hypertarget{erca-018}{}}\texttt{ERCA-018} & Purge label errors and invalid items before measuring reliability & 3 & companion \\
\raisebox{0pt}[0pt][0pt]{\hypertarget{erca-019}{}}\texttt{ERCA-019} & Treat leaderboard positions as gameable; audit the submission protocol & 3 & companion \\
\raisebox{0pt}[0pt][0pt]{\hypertarget{erca-020}{}}\texttt{ERCA-020} & Never report a single run; compare distributions over randomized repeats & 1 & developed \\
\raisebox{0pt}[0pt][0pt]{\hypertarget{erca-021}{}}\texttt{ERCA-021} & Audit your eval pipeline against the questionable-practices catalog\,$\dagger$ & 3 & companion \\
\raisebox{0pt}[0pt][0pt]{\hypertarget{erca-022}{}}\texttt{ERCA-022} & Search for an adequate existing benchmark before building one; upgrade, don't fork & 3 & companion \\
\raisebox{0pt}[0pt][0pt]{\hypertarget{erca-023}{}}\texttt{ERCA-023} & Report correlation-aware standard errors on every eval score & 1 & companion \\
\raisebox{0pt}[0pt][0pt]{\hypertarget{erca-024}{}}\texttt{ERCA-024} & Compare systems with paired tests matched to the metric & 1 & developed \\
\raisebox{0pt}[0pt][0pt]{\hypertarget{erca-025}{}}\texttt{ERCA-025} & Power-analyze the eval before running or trusting it & 1 & developed \\
\raisebox{0pt}[0pt][0pt]{\hypertarget{erca-026}{}}\texttt{ERCA-026} & Estimate pass@k with the unbiased combinatorial estimator & 1 & companion \\
\raisebox{0pt}[0pt][0pt]{\hypertarget{erca-027}{}}\texttt{ERCA-027} & Correct multi-benchmark victory claims with replicability analysis & 1 & companion \\
\raisebox{0pt}[0pt][0pt]{\hypertarget{erca-028}{}}\texttt{ERCA-028} & Shrink evals only with decision-impact validation & 1 & companion \\
\raisebox{0pt}[0pt][0pt]{\hypertarget{erca-029}{}}\texttt{ERCA-029} & Rank systems with a paired-comparison model, confidence intervals, and active sampling & 1 & companion \\
\raisebox{0pt}[0pt][0pt]{\hypertarget{erca-030}{}}\texttt{ERCA-030} & Profile a benchmark's noise before using it for decisions; engineer SNR up & 1 & companion \\
\raisebox{0pt}[0pt][0pt]{\hypertarget{erca-031}{}}\texttt{ERCA-031} & Declare the decoding configuration; test under both greedy and sampling & 1 & companion \\
\raisebox{0pt}[0pt][0pt]{\hypertarget{erca-032}{}}\texttt{ERCA-032} & Run agent benchmarks through a task-validity/outcome-validity checklist & 3 & companion \\
\raisebox{0pt}[0pt][0pt]{\hypertarget{erca-033}{}}\texttt{ERCA-033} & Design the experiment before running it: pilot, pre-commit, then confirm & 1 & companion \\
\raisebox{0pt}[0pt][0pt]{\hypertarget{erca-034}{}}\texttt{ERCA-034} & Report expected best performance as a function of tuning budget & 2 & companion \\
\raisebox{0pt}[0pt][0pt]{\hypertarget{erca-035}{}}\texttt{ERCA-035} & With a handful of runs, report IQM with stratified-bootstrap CIs and performance profiles & 1 & companion \\
\raisebox{0pt}[0pt][0pt]{\hypertarget{erca-036}{}}\texttt{ERCA-036} & Budget repeated sampling by coverage scaling laws and the verifier's false-positive ceiling & 2 & companion \\
\raisebox{0pt}[0pt][0pt]{\hypertarget{erca-037}{}}\texttt{ERCA-037} & Combine cheap judge labels with a small human-labeled set via valid inference & 5 & companion \\
\raisebox{0pt}[0pt][0pt]{\hypertarget{erca-038}{}}\texttt{ERCA-038} & Never gate on a single proxy metric & 6 & developed \\
\raisebox{0pt}[0pt][0pt]{\hypertarget{erca-039}{}}\texttt{ERCA-039} & Watch for proxy/true-signal divergence under optimization & 6 & developed \\
\raisebox{0pt}[0pt][0pt]{\hypertarget{erca-040}{}}\texttt{ERCA-040} & Treat mild gaming as a leading indicator & 6 & companion \\
\raisebox{0pt}[0pt][0pt]{\hypertarget{erca-041}{}}\texttt{ERCA-041} & Harden environments hardest where tasks are hardest & 6 & companion \\
\raisebox{0pt}[0pt][0pt]{\hypertarget{erca-042}{}}\texttt{ERCA-042} & Monitor the CoT, but never train against the monitor & 6 & developed \\
\raisebox{0pt}[0pt][0pt]{\hypertarget{erca-043}{}}\texttt{ERCA-043} & Add a monitor-independent effort signal & 6 & companion \\
\raisebox{0pt}[0pt][0pt]{\hypertarget{erca-044}{}}\texttt{ERCA-044} & Fix recurring failure classes structurally, not with prompt patches & 18 & developed \\
\raisebox{0pt}[0pt][0pt]{\hypertarget{erca-045}{}}\texttt{ERCA-045} & Derive the failure taxonomy from your own traces and target the mass & 11 & developed \\
\raisebox{0pt}[0pt][0pt]{\hypertarget{erca-046}{}}\texttt{ERCA-046} & Do not delegate failure attribution to a model & 11 & developed \\
\raisebox{0pt}[0pt][0pt]{\hypertarget{erca-047}{}}\texttt{ERCA-047} & Design traces for human audit, not LLM reading & 11 & developed \\
\raisebox{0pt}[0pt][0pt]{\hypertarget{erca-048}{}}\texttt{ERCA-048} & Repair the root cause, not the last step\,$\dagger$ & 18 & companion \\
\raisebox{0pt}[0pt][0pt]{\hypertarget{erca-049}{}}\texttt{ERCA-049} & Localize failures from a constraint-violation log, not the raw trace & 11 & companion \\
\raisebox{0pt}[0pt][0pt]{\hypertarget{erca-050}{}}\texttt{ERCA-050} & Use trajectory shape and early steps as an online failure signal & 11 & companion \\
\raisebox{0pt}[0pt][0pt]{\hypertarget{erca-051}{}}\texttt{ERCA-051} & Preempt environment errors before reasoning starts & 11 & companion \\
\raisebox{0pt}[0pt][0pt]{\hypertarget{erca-052}{}}\texttt{ERCA-052} & Fault-inject semantic failures with a realistic fault menu & 10 & companion \\
\raisebox{0pt}[0pt][0pt]{\hypertarget{erca-053}{}}\texttt{ERCA-053} & Attribute errors with hierarchical context plus aggregated independent judges\,$\dagger$ & 11 & companion \\
\raisebox{0pt}[0pt][0pt]{\hypertarget{erca-054}{}}\texttt{ERCA-054} & Differentially test agent vs environment before blaming the model & 11 & companion \\
\raisebox{0pt}[0pt][0pt]{\hypertarget{erca-055}{}}\texttt{ERCA-055} & Ensemble reward signals by weight averaging & 6 & companion \\
\raisebox{0pt}[0pt][0pt]{\hypertarget{erca-056}{}}\texttt{ERCA-056} & Evaluate in the feedback loop, not on static datasets & 6 & companion \\
\raisebox{0pt}[0pt][0pt]{\hypertarget{erca-057}{}}\texttt{ERCA-057} & Normalize trajectories into a canonical action taxonomy & 11 & companion \\
\raisebox{0pt}[0pt][0pt]{\hypertarget{erca-058}{}}\texttt{ERCA-058} & Benchmark any patch-correctness gate against random selection & 2 & companion \\
\raisebox{0pt}[0pt][0pt]{\hypertarget{erca-059}{}}\texttt{ERCA-059} & Oversee multi-agent execution as a graph, at node/edge/path levels\,$\dagger$ & 18 & companion \\
\raisebox{0pt}[0pt][0pt]{\hypertarget{erca-060}{}}\texttt{ERCA-060} & Test hack-detectors contrastively, with realistic base rates & 6 & companion \\
\raisebox{0pt}[0pt][0pt]{\hypertarget{erca-061}{}}\texttt{ERCA-061} & Gate self-correction on external feedback & 4 & developed \\
\raisebox{0pt}[0pt][0pt]{\hypertarget{erca-062}{}}\texttt{ERCA-062} & Use run-to-run consistency as a per-task risk signal\,$\dagger$ & 11 & companion \\
\raisebox{0pt}[0pt][0pt]{\hypertarget{erca-063}{}}\texttt{ERCA-063} & Separate submit rate from verified-resolve rate; reward abstention & 4 & companion \\
\raisebox{0pt}[0pt][0pt]{\hypertarget{erca-064}{}}\texttt{ERCA-064} & Assume tools fail silently; detect before you plan & 4 & companion \\
\raisebox{0pt}[0pt][0pt]{\hypertarget{erca-065}{}}\texttt{ERCA-065} & Cap turns dynamically, not generously & 4 & companion \\
\raisebox{0pt}[0pt][0pt]{\hypertarget{erca-066}{}}\texttt{ERCA-066} & Evaluate on your own workload, not public leaderboards & 3 & developed \\
\raisebox{0pt}[0pt][0pt]{\hypertarget{erca-067}{}}\texttt{ERCA-067} & Pair offline evals with online measurement, and calibrate the judge on live data & 3 & companion \\
\raisebox{0pt}[0pt][0pt]{\hypertarget{erca-068}{}}\texttt{ERCA-068} & Contain agent blast radius: least privilege, separated destructive capabilities, out-of-reach backups & 8 & developed \\
\raisebox{0pt}[0pt][0pt]{\hypertarget{erca-069}{}}\texttt{ERCA-069} & Never trust 'done': verify workspace state and budget for false self-reports & 8 & developed \\
\raisebox{0pt}[0pt][0pt]{\hypertarget{erca-070}{}}\texttt{ERCA-070} & Edge-load the load-bearing evidence & 14 & companion \\
\raisebox{0pt}[0pt][0pt]{\hypertarget{erca-071}{}}\texttt{ERCA-071} & Test semantic recall, not needle recall & 14 & companion \\
\raisebox{0pt}[0pt][0pt]{\hypertarget{erca-072}{}}\texttt{ERCA-072} & Budget context to measured, not advertised, length & 14 & developed \\
\raisebox{0pt}[0pt][0pt]{\hypertarget{erca-073}{}}\texttt{ERCA-073} & Never rely on in-context attention across a lexical gap & 12 & companion \\
\raisebox{0pt}[0pt][0pt]{\hypertarget{erca-074}{}}\texttt{ERCA-074} & Treat retrieval as a loop, not a preprocessing step & 12 & companion \\
\raisebox{0pt}[0pt][0pt]{\hypertarget{erca-075}{}}\texttt{ERCA-075} & Gate retrieval behind a will-this-help decision & 12 & companion \\
\raisebox{0pt}[0pt][0pt]{\hypertarget{erca-076}{}}\texttt{ERCA-076} & Score retrieval separately from final answers & 12 & developed \\
\raisebox{0pt}[0pt][0pt]{\hypertarget{erca-077}{}}\texttt{ERCA-077} & Fix both ends of the RAG pipeline & 12 & companion \\
\raisebox{0pt}[0pt][0pt]{\hypertarget{erca-078}{}}\texttt{ERCA-078} & Gate retrieved context on freshness and authorization & 13 & developed \\
\raisebox{0pt}[0pt][0pt]{\hypertarget{erca-079}{}}\texttt{ERCA-079} & Add a structural retrieval lane, not just token similarity & 13 & companion \\
\raisebox{0pt}[0pt][0pt]{\hypertarget{erca-080}{}}\texttt{ERCA-080} & Persist explored context across reasoning steps\,$\dagger$ & 15 & companion \\
\raisebox{0pt}[0pt][0pt]{\hypertarget{erca-081}{}}\texttt{ERCA-081} & Optimize compaction rules from the agent's own failures & 15 & developed \\
\raisebox{0pt}[0pt][0pt]{\hypertarget{erca-082}{}}\texttt{ERCA-082} & Make compaction an action the agent takes at milestones & 14 & companion \\
\raisebox{0pt}[0pt][0pt]{\hypertarget{erca-083}{}}\texttt{ERCA-083} & Run hybrid multi-channel retrieval fused on ranks & 12 & developed \\
\raisebox{0pt}[0pt][0pt]{\hypertarget{erca-084}{}}\texttt{ERCA-084} & Structure localization as a staged funnel & 13 & developed \\
\raisebox{0pt}[0pt][0pt]{\hypertarget{erca-085}{}}\texttt{ERCA-085} & Chunk code along AST boundaries, not line counts & 12 & developed \\
\raisebox{0pt}[0pt][0pt]{\hypertarget{erca-086}{}}\texttt{ERCA-086} & Give the agent a condensed map before it touches source & 13 & companion \\
\raisebox{0pt}[0pt][0pt]{\hypertarget{erca-087}{}}\texttt{ERCA-087} & Inject static-analysis facts as deterministic anchors & 13 & companion \\
\raisebox{0pt}[0pt][0pt]{\hypertarget{erca-088}{}}\texttt{ERCA-088} & Consolidate the spec; restart rather than patch a lost run & 14 & developed \\
\raisebox{0pt}[0pt][0pt]{\hypertarget{erca-089}{}}\texttt{ERCA-089} & Stage retrieval: cheap first pass, expensive rerank on the shortlist & 12 & companion \\
\raisebox{0pt}[0pt][0pt]{\hypertarget{erca-090}{}}\texttt{ERCA-090} & Schedule exploration by relevance and prune by graph distance & 13 & companion \\
\raisebox{0pt}[0pt][0pt]{\hypertarget{erca-091}{}}\texttt{ERCA-091} & Walk the call graph at query time when you cannot maintain an index & 13 & companion \\
\raisebox{0pt}[0pt][0pt]{\hypertarget{erca-092}{}}\texttt{ERCA-092} & Price context-compression transforms before adopting them & 14 & companion \\
\raisebox{0pt}[0pt][0pt]{\hypertarget{erca-093}{}}\texttt{ERCA-093} & Default to a single agent; gate fan-out on a measured baseline & 18 & developed \\
\raisebox{0pt}[0pt][0pt]{\hypertarget{erca-094}{}}\texttt{ERCA-094} & Ground evaluation in execution, not plausibility & 4 & developed \\
\raisebox{0pt}[0pt][0pt]{\hypertarget{erca-095}{}}\texttt{ERCA-095} & Replay a golden set every release, scored pass\textasciicircum{}k & 4 & developed \\
\raisebox{0pt}[0pt][0pt]{\hypertarget{erca-096}{}}\texttt{ERCA-096} & Calibrate LLM judges against expert human labels before scaling judgment & 5 & developed \\
\raisebox{0pt}[0pt][0pt]{\hypertarget{erca-097}{}}\texttt{ERCA-097} & Make every run a structured, replayable trace & 10 & developed \\
\raisebox{0pt}[0pt][0pt]{\hypertarget{erca-098}{}}\texttt{ERCA-098} & Treat side-effecting steps as transactions with receipts and compensators & 9 & companion \\
\raisebox{0pt}[0pt][0pt]{\hypertarget{erca-099}{}}\texttt{ERCA-099} & Hand off between agents via durable references, not chat summaries\,$\dagger$ & 15 & companion \\
\raisebox{0pt}[0pt][0pt]{\hypertarget{erca-100}{}}\texttt{ERCA-100} & Route cheap-first with uncertainty-gated escalation & 19 & companion \\
\raisebox{0pt}[0pt][0pt]{\hypertarget{erca-101}{}}\texttt{ERCA-101} & Schedule with dynamic task graphs, not static schedules & 18 & developed \\
\raisebox{0pt}[0pt][0pt]{\hypertarget{erca-102}{}}\texttt{ERCA-102} & Express agent roles in config and prompts, not code\,$\dagger$ & 18 & companion \\
\raisebox{0pt}[0pt][0pt]{\hypertarget{erca-103}{}}\texttt{ERCA-103} & Blind the review lanes and hide interim tallies\,$\dagger$ & 18 & companion \\
\raisebox{0pt}[0pt][0pt]{\hypertarget{erca-104}{}}\texttt{ERCA-104} & Treat orchestration topology as a security decision & 8 & companion \\
\raisebox{0pt}[0pt][0pt]{\hypertarget{erca-105}{}}\texttt{ERCA-105} & Defend against prompt injection in depth; control what the agent can do & 8 & developed \\
\raisebox{0pt}[0pt][0pt]{\hypertarget{erca-106}{}}\texttt{ERCA-106} & Place human checkpoints at measured failure points with a real escalation route & 16 & companion \\
\raisebox{0pt}[0pt][0pt]{\hypertarget{erca-107}{}}\texttt{ERCA-107} & Match orchestration topology to task shape & 18 & developed \\
\raisebox{0pt}[0pt][0pt]{\hypertarget{erca-108}{}}\texttt{ERCA-108} & Make blackboard writes active and build in a diversity check\,$\dagger$ & 18 & companion \\
\raisebox{0pt}[0pt][0pt]{\hypertarget{erca-109}{}}\texttt{ERCA-109} & Schedule consolidation in batches; type and prune the shared store & 15 & companion \\
\raisebox{0pt}[0pt][0pt]{\hypertarget{erca-110}{}}\texttt{ERCA-110} & Distribute guardrails; keep them stateful across sessions\,$\dagger$ & 8 & companion \\
\raisebox{0pt}[0pt][0pt]{\hypertarget{erca-111}{}}\texttt{ERCA-111} & Never treat consensus as verification & 5 & developed \\
\raisebox{0pt}[0pt][0pt]{\hypertarget{erca-112}{}}\texttt{ERCA-112} & Freeze the memory or skill library before evaluating it & 2 & companion \\
\raisebox{0pt}[0pt][0pt]{\hypertarget{erca-113}{}}\texttt{ERCA-113} & Pin which stored form receives retrieval credit & 2 & companion \\
\raisebox{0pt}[0pt][0pt]{\hypertarget{erca-114}{}}\texttt{ERCA-114} & Run ablation and no-tool controls before crediting the component & 2 & developed \\
\raisebox{0pt}[0pt][0pt]{\hypertarget{erca-115}{}}\texttt{ERCA-115} & Synthesize eval instances from structured seeds; gate through human review & 2 & companion \\
\raisebox{0pt}[0pt][0pt]{\hypertarget{erca-116}{}}\texttt{ERCA-116} & Retain raw traces as ground truth; distill into a rebuildable layer & 15 & developed \\
\raisebox{0pt}[0pt][0pt]{\hypertarget{erca-117}{}}\texttt{ERCA-117} & Model time explicitly: supersede vs version & 15 & companion \\
\raisebox{0pt}[0pt][0pt]{\hypertarget{erca-118}{}}\texttt{ERCA-118} & Don't reach for the heavy memory store by default & 15 & developed \\
\raisebox{0pt}[0pt][0pt]{\hypertarget{erca-119}{}}\texttt{ERCA-119} & Plan for memory portability and audit from day one\,$\dagger$ & 15 & companion \\
\raisebox{0pt}[0pt][0pt]{\hypertarget{erca-120}{}}\texttt{ERCA-120} & Budget for the half of the truth not in the source & 15 & companion \\
\raisebox{0pt}[0pt][0pt]{\hypertarget{erca-121}{}}\texttt{ERCA-121} & Record nondeterministic step results; replay only deterministic code & 10 & companion \\
\raisebox{0pt}[0pt][0pt]{\hypertarget{erca-122}{}}\texttt{ERCA-122} & Bound the history a recovery must replay & 10 & companion \\
\raisebox{0pt}[0pt][0pt]{\hypertarget{erca-123}{}}\texttt{ERCA-123} & State the retry contract as three invariants and verify the runtime provides them & 9 & companion \\
\raisebox{0pt}[0pt][0pt]{\hypertarget{erca-124}{}}\texttt{ERCA-124} & Make every retried step idempotent; deduplicate by invocation identity & 9 & developed \\
\raisebox{0pt}[0pt][0pt]{\hypertarget{erca-125}{}}\texttt{ERCA-125} & Checkpoint asynchronously at natural consistency boundaries, never stop-the-world & 9 & companion \\
\raisebox{0pt}[0pt][0pt]{\hypertarget{erca-126}{}}\texttt{ERCA-126} & Set checkpoint interval and granularity from measured cost and failure rate, not defaults & 9 & companion \\
\raisebox{0pt}[0pt][0pt]{\hypertarget{erca-127}{}}\texttt{ERCA-127} & Benchmark recovery with fault injection; do not read fault tolerance off architecture diagrams & 10 & developed \\
\raisebox{0pt}[0pt][0pt]{\hypertarget{erca-128}{}}\texttt{ERCA-128} & Run multi-step coordination through a durable workflow engine, not hand-rolled logic & 9 & developed \\
\raisebox{0pt}[0pt][0pt]{\hypertarget{erca-129}{}}\texttt{ERCA-129} & Treat ordering, delivery, replay, and dead-letter handling as first-class design items in event-driven coordination\,$\dagger$ & 9 & companion \\
\raisebox{0pt}[0pt][0pt]{\hypertarget{erca-130}{}}\texttt{ERCA-130} & Externalize agent and workflow state as a first-class persistent artifact & 9 & developed \\
\raisebox{0pt}[0pt][0pt]{\hypertarget{erca-131}{}}\texttt{ERCA-131} & Log every step execution transactionally, co-located with the data it touches & 9 & companion \\
\raisebox{0pt}[0pt][0pt]{\hypertarget{erca-132}{}}\texttt{ERCA-132} & Use provenance-driven re-execution when the workflow spans heterogeneous environments & 10 & companion \\
\raisebox{0pt}[0pt][0pt]{\hypertarget{erca-133}{}}\texttt{ERCA-133} & Certify restore points against committed downstream work before rolling back & 10 & companion \\
\raisebox{0pt}[0pt][0pt]{\hypertarget{erca-134}{}}\texttt{ERCA-134} & Diagnose failures and gate retries behind matched, evidence-grounded guidance & 10 & companion \\
\raisebox{0pt}[0pt][0pt]{\hypertarget{erca-135}{}}\texttt{ERCA-135} & Pair every stochastic generation with a deterministic post-condition guard and tiered recovery\,$\dagger$ & 9 & companion \\
\raisebox{0pt}[0pt][0pt]{\hypertarget{erca-136}{}}\texttt{ERCA-136} & Hide failure handling behind checkpoint-recovery only if observations are monotonic & 10 & companion \\
\raisebox{0pt}[0pt][0pt]{\hypertarget{erca-137}{}}\texttt{ERCA-137} & Start event stores with versioned events and weak schema; graduate as the system ages & 9 & companion \\
\raisebox{0pt}[0pt][0pt]{\hypertarget{erca-138}{}}\texttt{ERCA-138} & Design domain operations for tolerance to partial execution order & 9 & companion \\
\raisebox{0pt}[0pt][0pt]{\hypertarget{erca-139}{}}\texttt{ERCA-139} & Let agents propose remediations, but execute only through a typed, validated policy path & 8 & companion \\
\raisebox{0pt}[0pt][0pt]{\hypertarget{erca-140}{}}\texttt{ERCA-140} & Treat shared agent state as a transactional resource under isolation control & 9 & companion \\
\raisebox{0pt}[0pt][0pt]{\hypertarget{erca-141}{}}\texttt{ERCA-141} & Crystallize repeated agent plans into declarative blueprints executed without the agent\,$\dagger$ & 9 & companion \\
\raisebox{0pt}[0pt][0pt]{\hypertarget{erca-142}{}}\texttt{ERCA-142} & Record agent workflow steps in an append-only, hash-chained ledger & 17 & companion \\
\raisebox{0pt}[0pt][0pt]{\hypertarget{erca-143}{}}\texttt{ERCA-143} & Persist large transient outputs as files; keep only pointers in context & 14 & developed \\
\raisebox{0pt}[0pt][0pt]{\hypertarget{erca-144}{}}\texttt{ERCA-144} & Keep always-loaded context minimal; budget it, lazy-load the rest, and reset poisoned sessions & 14 & companion \\
\raisebox{0pt}[0pt][0pt]{\hypertarget{erca-145}{}}\texttt{ERCA-145} & Widen autonomy only on a measured per-action track record & 17 & developed \\
\raisebox{0pt}[0pt][0pt]{\hypertarget{erca-146}{}}\texttt{ERCA-146} & Design overseer roles for skill retention, not residual monitoring & 16 & companion \\
\raisebox{0pt}[0pt][0pt]{\hypertarget{erca-147}{}}\texttt{ERCA-147} & Calibrate any confidence signal before displaying it; do not expect calibration alone to add oversight & 16 & companion \\
\raisebox{0pt}[0pt][0pt]{\hypertarget{erca-148}{}}\texttt{ERCA-148} & Put cognitive forcing functions at the accept decision, targeted at acceleration mode & 16 & developed \\
\raisebox{0pt}[0pt][0pt]{\hypertarget{erca-149}{}}\texttt{ERCA-149} & Assign accountability only where control actually exists, and audit it on real users & 17 & developed \\
\raisebox{0pt}[0pt][0pt]{\hypertarget{erca-150}{}}\texttt{ERCA-150} & Budget verification as the dominant cost of AI-assisted coding\,$\dagger$ & 16 & companion \\
\raisebox{0pt}[0pt][0pt]{\hypertarget{erca-151}{}}\texttt{ERCA-151} & Measure your own pipeline's security delta before importing controls\,$\dagger$ & 8 & companion \\
\raisebox{0pt}[0pt][0pt]{\hypertarget{erca-152}{}}\texttt{ERCA-152} & Never accept self-attestation of AI-assisted code quality & 8 & companion \\
\raisebox{0pt}[0pt][0pt]{\hypertarget{erca-153}{}}\texttt{ERCA-153} & Build verification surfaces and make verifying cheaper than blind acceptance & 16 & developed \\
\raisebox{0pt}[0pt][0pt]{\hypertarget{erca-154}{}}\texttt{ERCA-154} & Audit every human gate for effectiveness; replace person-gates with validated institutional oversight & 17 & developed \\
\raisebox{0pt}[0pt][0pt]{\hypertarget{erca-155}{}}\texttt{ERCA-155} & Re-create collective accountability when agents replace peer review & 17 & companion \\
\raisebox{0pt}[0pt][0pt]{\hypertarget{erca-156}{}}\texttt{ERCA-156} & Separate control (ex-ante) from oversight (ex-post) and staff both\,$\dagger$ & 17 & companion \\
\raisebox{0pt}[0pt][0pt]{\hypertarget{erca-157}{}}\texttt{ERCA-157} & Escalate only monitor-flagged cases to humans who own the verdict; scaffold the monitors & 16 & developed \\
\raisebox{0pt}[0pt][0pt]{\hypertarget{erca-158}{}}\texttt{ERCA-158} & Block epistemically-absent oversight: require understanding before shipping\,$\dagger$ & 17 & companion \\
\raisebox{0pt}[0pt][0pt]{\hypertarget{erca-159}{}}\texttt{ERCA-159} & Write an oversight card: who oversees what, with what access and authority\,$\dagger$ & 17 & companion \\
\raisebox{0pt}[0pt][0pt]{\hypertarget{erca-160}{}}\texttt{ERCA-160} & Use plain reliance disclaimers before fancier interventions\,$\dagger$ & 17 & companion \\
\raisebox{0pt}[0pt][0pt]{\hypertarget{erca-161}{}}\texttt{ERCA-161} & Curate what reviewers see: withhold low-value suggestions and highlight edit-likely tokens & 16 & companion \\
\raisebox{0pt}[0pt][0pt]{\hypertarget{erca-162}{}}\texttt{ERCA-162} & Keep override authority real --- experts do catch bad machine scores\,$\dagger$ & 17 & companion \\
\raisebox{0pt}[0pt][0pt]{\hypertarget{erca-163}{}}\texttt{ERCA-163} & Measure oversight with decomposed metrics, never a single aggregate & 17 & companion \\
\raisebox{0pt}[0pt][0pt]{\hypertarget{erca-164}{}}\texttt{ERCA-164} & Shield high-stakes review from time pressure & 16 & companion \\
\raisebox{0pt}[0pt][0pt]{\hypertarget{erca-165}{}}\texttt{ERCA-165} & Label AI-generated code explicitly --- reviewers cannot tell and behave better when told & 17 & developed \\
\raisebox{0pt}[0pt][0pt]{\hypertarget{erca-166}{}}\texttt{ERCA-166} & Scale review capacity to contributor experience, not just contribution volume & 17 & companion \\
\raisebox{0pt}[0pt][0pt]{\hypertarget{erca-167}{}}\texttt{ERCA-167} & Write an agent-contribution policy covering six dimensions & 17 & companion \\
\raisebox{0pt}[0pt][0pt]{\hypertarget{erca-168}{}}\texttt{ERCA-168} & Set autonomy defaults per task type --- accountability and identity predict where humans want the wheel & 17 & companion \\
\raisebox{0pt}[0pt][0pt]{\hypertarget{erca-169}{}}\texttt{ERCA-169} & Budget the verifier tax; pair runtime blocks with recovery paths & 4 & companion \\
\raisebox{0pt}[0pt][0pt]{\hypertarget{erca-170}{}}\texttt{ERCA-170} & Render agent actions as live, step-level visual state for mid-flight steering\,$\dagger$ & 16 & companion \\
\raisebox{0pt}[0pt][0pt]{\hypertarget{erca-171}{}}\texttt{ERCA-171} & Re-decide cheaply on a control loop; ship the best schedule in hand at the time cap & 19 & developed \\
\raisebox{0pt}[0pt][0pt]{\hypertarget{erca-172}{}}\texttt{ERCA-172} & Keep one exact method that can certify provably-missed work & 5 & companion \\
\raisebox{0pt}[0pt][0pt]{\hypertarget{erca-173}{}}\texttt{ERCA-173} & Accept on a vector, not a scalar; keep the acceptance judge independent of the generator and the optimization target\,$\dagger$ & 6 & companion \\
\raisebox{0pt}[0pt][0pt]{\hypertarget{erca-174}{}}\texttt{ERCA-174} & Index the repository as a typed knowledge graph and localize by traversal & 13 & developed \\
\raisebox{0pt}[0pt][0pt]{\hypertarget{erca-175}{}}\texttt{ERCA-175} & Extend the graph below function level when the task needs line-level precision & 13 & companion \\
\raisebox{0pt}[0pt][0pt]{\hypertarget{erca-176}{}}\texttt{ERCA-176} & Route entity queries to cheap graph lookups; reserve LLM exploration for natural-language queries & 13 & companion \\
\raisebox{0pt}[0pt][0pt]{\hypertarget{erca-177}{}}\texttt{ERCA-177} & Treat the indexed representation as a first-class design choice & 13 & companion \\
\raisebox{0pt}[0pt][0pt]{\hypertarget{erca-178}{}}\texttt{ERCA-178} & A/B-test context files before treating them as best practice & 14 & developed \\
\raisebox{0pt}[0pt][0pt]{\hypertarget{erca-179}{}}\texttt{ERCA-179} & Formalize work requests in a common constraint formalism so a solver can allocate globally & 19 & companion \\
\raisebox{0pt}[0pt][0pt]{\hypertarget{erca-180}{}}\texttt{ERCA-180} & Treat objective terms and weights as policy artifacts: make them explicit and review them like code & 17 & companion \\
\raisebox{0pt}[0pt][0pt]{\hypertarget{erca-181}{}}\texttt{ERCA-181} & Design the online rescheduling loop: horizon, time-step, triggers, and closed-loop-aware constraints & 19 & companion \\
\raisebox{0pt}[0pt][0pt]{\hypertarget{erca-182}{}}\texttt{ERCA-182} & Pay for a probe only above a threshold of uncertainty\,$\dagger$ & 19 & companion \\
\raisebox{0pt}[0pt][0pt]{\hypertarget{erca-183}{}}\texttt{ERCA-183} & Document every scheduling decision so selection effects can be modeled later\,$\dagger$ & 19 & companion \\
\raisebox{0pt}[0pt][0pt]{\hypertarget{erca-184}{}}\texttt{ERCA-184} & Split fleet scheduling into a global allocation level and a local per-worker level & 19 & companion \\
\raisebox{0pt}[0pt][0pt]{\hypertarget{erca-185}{}}\texttt{ERCA-185} & Gate optimizer output behind human review; do not ship on metric improvement & 17 & companion \\
\raisebox{0pt}[0pt][0pt]{\hypertarget{erca-186}{}}\texttt{ERCA-186} & Pick the algorithm with short pilots, then commit the long budget & 19 & companion \\
\raisebox{0pt}[0pt][0pt]{\hypertarget{erca-187}{}}\texttt{ERCA-187} & Route each task to the cheapest model predicted to suffice, learned under budget & 19 & developed \\
\raisebox{0pt}[0pt][0pt]{\hypertarget{erca-188}{}}\texttt{ERCA-188} & Treat escalation-to-human as a budgeted online decision, not a fixed threshold & 16 & companion \\
\raisebox{0pt}[0pt][0pt]{\hypertarget{erca-189}{}}\texttt{ERCA-189} & Score each agent action by reversibility and blast radius before choosing an autonomy level & 8 & companion \\
\raisebox{0pt}[0pt][0pt]{\hypertarget{erca-190}{}}\texttt{ERCA-190} & Enumerate what competes for the window, budget it, and audit the assembled context & 14 & companion \\
\raisebox{0pt}[0pt][0pt]{\hypertarget{erca-191}{}}\texttt{ERCA-191} & Replay recorded traces under competing policies before changing scheduling behavior & 19 & developed \\
\raisebox{0pt}[0pt][0pt]{\hypertarget{erca-192}{}}\texttt{ERCA-192} & Treat the allocation layer as a first-class, audited, shared component\,$\dagger$ & 19 & companion \\
\raisebox{0pt}[0pt][0pt]{\hypertarget{erca-193}{}}\texttt{ERCA-193} & Fence stale worker authority at the mutation boundary & 18 & lead \\
\raisebox{0pt}[0pt][0pt]{\hypertarget{erca-194}{}}\texttt{ERCA-194} & Separate logical work identity from attempt identity & 9 & lead \\
\raisebox{0pt}[0pt][0pt]{\hypertarget{erca-195}{}}\texttt{ERCA-195} & Reconcile unknown external effects instead of assuming & 9 & lead \\
\raisebox{0pt}[0pt][0pt]{\hypertarget{erca-196}{}}\texttt{ERCA-196} & Test the controller under stale and missed state & 10 & lead \\
\raisebox{0pt}[0pt][0pt]{\hypertarget{erca-197}{}}\texttt{ERCA-197} & Place fault injections at named commit barriers & 10 & lead \\
\raisebox{0pt}[0pt][0pt]{\hypertarget{erca-198}{}}\texttt{ERCA-198} & Test the recovery action itself & 10 & lead \\
\raisebox{0pt}[0pt][0pt]{\hypertarget{erca-199}{}}\texttt{ERCA-199} & Budget retries as load & 19 & lead \\
\raisebox{0pt}[0pt][0pt]{\hypertarget{erca-200}{}}\texttt{ERCA-200} & Jitter recovery redelivery & 19 & lead \\
\raisebox{0pt}[0pt][0pt]{\hypertarget{erca-201}{}}\texttt{ERCA-201} & Reserve or prioritize capacity for recovery work & 19 & lead \\
\raisebox{0pt}[0pt][0pt]{\hypertarget{erca-202}{}}\texttt{ERCA-202} & Separate global capacity from local dispatch & 19 & lead \\
\raisebox{0pt}[0pt][0pt]{\hypertarget{erca-203}{}}\texttt{ERCA-203} & Version verifier results & 9 & lead \\
\raisebox{0pt}[0pt][0pt]{\hypertarget{erca-204}{}}\texttt{ERCA-204} & Measure recovery under live load & 10 & lead \\
\raisebox{0pt}[0pt][0pt]{\hypertarget{erca-205}{}}\texttt{ERCA-205} & Detect retry storms and recovery starvation & 19 & lead \\
\raisebox{0pt}[0pt][0pt]{\hypertarget{erca-206}{}}\texttt{ERCA-206} & Bound admitted demand, isolate demand domains, and release recovery traffic at a controlled rate & 19 & developed \\
\end{longtable}
\endgroup

\backmatter
\markboth{Reference material}{}
\gdef\currentparttitle{}
\chapter*{Glossary}
\addcontentsline{toc}{chapter}{Glossary}
\textbf{Ablation.} A comparison that removes, disables, or weakens one component while holding the rest of the evaluated system as constant as the design permits.

\textbf{Arrival trace.} An ordered record of work entering a system, including time, eligibility, resources, decisions, outcomes, and later review signals. It supports counterfactual scheduling replay.

\textbf{Blast radius.} The resources and effects that one identity, process, or failure can reach before a new authorization decision is required.

\textbf{Calibration.} Measurement of how an instrument's outputs relate to a reference outcome under stated conditions. Agreement is one calibration result; it is not correctness by itself.

\textbf{Compaction.} Replacement of a larger execution history or context with a smaller derived representation. Compaction is lossy unless the source remains available.

\textbf{Compliance theater.} A review or approval step that records assent but lacks the information or authority needed to change the consequential execution path.

\textbf{Corroborating evidence.} A case report, practitioner account, or convergent observation that establishes plausibility without estimating prevalence.

\textbf{Directional evidence.} Evidence that supports a mechanism, threat model, comparison design, or direction of effect without establishing the complete recommendation, magnitude, or broad transfer.

\textbf{Durable execution.} Execution whose control state survives worker or process loss so incomplete work can be resumed from recorded progress.

\textbf{Effective context length.} The longest tested context at which the weakest required task still meets its declared performance criterion. It can be shorter than the advertised token capacity.

\textbf{ERCA-NNN.} The stable identifier of a practice record in the companion catalog, from the initials of this book's title. \texttt{\erca{076}} names the same record in the manuscript, the evidence ledger, the chapter crosswalk, the protocols, and the skills. Look one up in \href{https://sjarmak.ai/books/engineering-reliable-coding-agents/companion}{the companion catalog} or in \href{https://github.com/sjarmak/engineering-reliable-coding-agents/blob/main/companion/catalog.json}{\texttt{companion/catalog.json}}. Identifiers are not reused: a later edition may retire, split, or merge a record, but its number never names a different practice.

\textbf{Evaluation harness.} The code, configuration, tools, permissions, prompts, task versions, and graders that turn a system under test into recorded outcomes.

\textbf{Fail-open gate.} A gate whose error, timeout, or missing evidence permits work to continue. A fail-closed gate blocks or escalates instead.

\textbf{First upstream failure.} The earliest event in the observed causal chain that the available trace can support as necessary to the downstream failure.

\textbf{Grounding.} Connection of a claim or decision to evidence outside the producing model's self-report, such as repository state, tool output, execution results, or independent review.

\textbf{Idempotency key.} A stable identity attached to an external effect so a repeated delivery can return or reconcile the original result instead of repeating the effect.

\textbf{Independence key.} An identifier that groups sources or records sharing one originating incident or claim, preventing repeated retellings from being counted as independent corroboration.

\textbf{Indirect prompt injection.} Instructions embedded in retrieved or tool-supplied content that attempt to redirect an agent beyond the authority or purpose of the retrieval.

\textbf{Inter-rater reliability.} The reproducibility of labels assigned by different reviewers. Cohen's kappa applies to two raters; Fleiss's kappa extends the calculation to more than two.

\textbf{Local artifact.} A record from an author-operated system used as an illustration or local measurement. It is not independent external evidence.

\textbf{Minimum detectable effect.} The smallest effect an evaluation design has a stated probability of detecting at its chosen error rates and sample structure.

\textbf{Monitorability tax.} The additional state, instrumentation, and review work required to make a more complex architecture diagnosable.

\textbf{Null or conflicting evidence.} A result that did not support the expected effect or that materially limits another claim.

\textbf{Operating point.} The threshold, workload, prevalence, and cost conditions under which a grader, monitor, or gate produces its reported error tradeoff.

\textbf{Paired design.} A comparison in which both systems receive the same evaluation items and analysis uses the within-item difference.

\textbf{\(\mathrm{pass}@k\).} The probability that at least one of (k) sampled attempts succeeds under an oracle capable of recognizing success.

\textbf{pass\^{}k.} The probability that all (k) attempts succeed. It measures consistency rather than sampled coverage.

\textbf{Power analysis.} Calculation of the sample size or detectable effect implied by the variance, decision threshold, and chosen false-positive and false-negative rates.

\textbf{Proxy.} A measurable signal used in place of the outcome an engineering decision ultimately values.

\textbf{Quality assessment.} The review step that assigns an evidence group to a scoped source claim. The group applies to the claim, not the venue or paper as a whole.

\textbf{\(\mathrm{Recall}@k\).} The fraction of relevant items recovered among the first (k) results, relative to the relevant set defined by the evaluation.

\textbf{Reciprocal-rank fusion.} A rank-combination method that adds a decreasing score based on each result's position in multiple retrieval lists.

\textbf{Replay harness.} A system that re-executes recorded tasks, arrivals, or events against a candidate policy while preserving the comparison boundary.

\textbf{Saturation point.} The tested point after which additional context, samples, components, or compute no longer produce a decision-relevant gain.

\textbf{Strong evidence.} An on-claim controlled comparison, validated benchmark result, or comparably specific measurement within stated conditions.

\textbf{Temporal holdout.} Evaluation material created after the training or development cutoff relevant to the evaluated system and kept outside adaptation.

\textbf{Typed knowledge graph.} A graph whose nodes and edges carry explicit semantic types, provenance, and revision identity rather than only untyped similarity.

\renewcommand{\bibname}{References}

\chapter*{Data and materials availability}
\addcontentsline{toc}{chapter}{Data and materials availability}
The version-controlled manuscript source and companion research artifact are available at \url{https://github.com/sjarmak/engineering-reliable-coding-agents}. The companion contains the machine-readable catalog of 206 reliability records, evidence ledger, chapter crosswalk, benchmark catalog, schemas, source snapshots, thread protocols and source identities, update-screening decisions, software-engineering coverage probe, DBLP title-census replacement evidence, blinded external-grading packet, provenance record, and checksums. The repository also packages six named runnable protocols and five reusable agent skills derived from selected practices, with practice-level evidence maps; these workflows are implementation artifacts rather than independent evidence. The exact companion-file digests are listed in \href{https://github.com/sjarmak/engineering-reliable-coding-agents/blob/main/companion/SHA256SUMS}{\texttt{companion/SHA256SUMS}}, and the pinned compilation record binds the arXiv source and preview bytes in \href{https://github.com/sjarmak/engineering-reliable-coding-agents/blob/main/companion/methodology/release-verification/arxiv-compile-report.json}{\texttt{arxiv-compile-report.json}}. A browser-based catalog is available at \url{https://sjarmak.ai/books/engineering-reliable-coding-agents/companion}, and the web edition is available at \url{https://sjarmak.ai/books/engineering-reliable-coding-agents}. The evidence ledger is scheduled for annual review, with out-of-cycle releases for material factual or citation corrections. The retrieval evaluation reported in Chapter 12 is CodeScaleBench, and it is public. The frozen suite \texttt{csb-v1-mixed371}, from which the 370 paired analysis tasks are drawn, is available at \url{https://github.com/sourcegraph/CodeScaleBench} under tag \texttt{v1-mixed371} with an Apache-2.0 license, and the problem layer is mirrored at \url{https://huggingface.co/datasets/sgjarmak/CodeScaleBench}. Repository contents are version-pinned public mirrors in the \texttt{sg-evals} organization, so each task resolves to a fixed commit. Frozen result snapshots under \texttt{runs/snapshots/} carry per-task traces, reward scores, timing, and cost, and full agent trajectories are attached to the corresponding release. Chapter 12 still reports its results as an author-system illustration, because the author built and operates the benchmark at the vendor whose retrieval product supplies one of its two arms.

The 199-trace diagnostic corpus and the 1,286-item fleet ledger are not redistributed. The trace corpus records complete agent runs against private repositories, including prompts, tool outputs, and file contents captured for operational diagnosis rather than publication. The fleet ledger is the author's own work-item history from an open-source software-factory project, and carries the service, pool, and operational identifiers of that deployment. De-identification would remove state and grouping information required to audit the reported causal and dependence structure. Their aggregate results are therefore identified in the text as author-system illustrations rather than independently reproducible external evidence.

\end{document}